\documentclass[11pt]{article}
\pdfoutput=1

\usepackage{setspace}
\usepackage{cite}

\usepackage[utf8]{inputenc}
\usepackage{geometry}
\usepackage[USenglish]{babel}
\usepackage{verbatim}
\usepackage{prettyref}
\usepackage{amsmath}
\usepackage[normalem]{ulem}
\usepackage{amssymb}
\usepackage{graphicx}
\usepackage{setspace}
\usepackage{esint}
\usepackage{empheq}
\usepackage[labelfont=bf]{caption}
\usepackage[T1]{fontenc}

\usepackage{scalerel,stackengine}
\stackMath
\newcommand\reallywidehat[1]{%
\savestack{\tmpbox}{\stretchto{%
  \scaleto{%
    \scalerel*[\widthof{\ensuremath{#1}}]{\kern.1pt\mathchar"0362\kern.1pt}%
    {\rule{0ex}{\textheight}}
  }{\textheight}%
}{2.4ex}}%
\stackon[-6.9pt]{#1}{\tmpbox}%
}

\usepackage{tikz}
\newcommand*{\boxcolor}{black}
\makeatletter
\renewcommand{\boxed}[1]{\textcolor{\boxcolor}{%
\tikz[baseline={([yshift=-1ex]current bounding box.center)}] \node [rectangle, minimum width=1ex,rounded corners,draw] {\normalcolor\m@th$\displaystyle#1$};}}
 \makeatother

\usepackage{color}
\definecolor{darkgreen}{rgb}{0,0.5,0}
\definecolor{darkblue}{rgb}{0,0,0.6}
\definecolor{purple}{rgb}{0.4,.2,0.7}
\definecolor{orange}{rgb}{0.95, 0.5, 0.3}
\definecolor{darkorange}{rgb}{0.958,0.186,0.131}

\usepackage[colorlinks=true,citecolor=darkorange,linkcolor=darkorange,urlcolor=darkorange]{hyperref}

\numberwithin{equation}{section}
\numberwithin{table}{section}

\renewcommand{\title}[1]{\vbox{\center\LARGE{#1}}\vspace{5mm}}
\renewcommand{\author}[1]{\vbox{\center#1}\vspace{5mm}}
\newcommand{\address}[1]{\vbox{\center\em#1}}

\newcommand{\St}{\bar{s}}
\newcommand{\Et}{\bar{e}}
\newcommand{\Ct}{\bar{c}}
\newcommand{\st}{v}

\begin{document}
\begin{spacing}{1.2}

\thispagestyle{empty}

\begin{center}

{\LARGE \bf {The quantum $p$-spin glass model:\\\vspace{.2cm} A  user manual for holographers}}
\end{center}

\bigskip \noindent

\bigskip

\begin{center}

\author{Tarek Anous$^1$ and Felix M.\ Haehl$^2$}

\address{1) Institute for Theoretical Physics and $\Delta$-Institute for Theoretical Physics,\\
University of Amsterdam, Science Park 904, 1098 XH Amsterdam, The Netherlands}

\address{2) School of Natural Sciences, Institute for Advanced Study\\
1 Einstein Drive, Princeton, NJ 08540, USA}
\vspace{0.5in}
    
{\tt t.m.anous@uva.nl, haehl@ias.edu}

\bigskip

\vspace{1cm}

\end{center}

\begin{abstract}
\noindent We study a large-$N$ bosonic quantum mechanical sigma-model with a spherical target space subject to disordered interactions, more colloquially known as the $p$-spin spherical model. Replica symmetry is broken at low temperatures and for sufficiently weak quantum fluctuations, which drives the system into a spin glass phase. The first half of this paper is dedicated to a discussion of this model's thermodynamics, with particular emphasis on the \emph{marginally stable spin glass}. This phase exhibits an emergent conformal symmetry in the strong coupling regime, which dictates its thermodynamic properties. It is associated with an extensive
number of nearby states in the free energy landscape. 
We discuss in detail an elegant approximate solution to the spin glass equations, which interpolates between the conformal regime and an ultraviolet-complete short distance solution. In the second half of this paper we explore the real-time dynamics of the model and uncover quantum chaos as measured by out-of-time-order four-point functions, both numerically and analytically. We find exponential Lyapunov growth, which intricately depends on the model's couplings and becomes strongest in the quantum critical regime. We emphasize that the spin glass phase also exhibits quantum chaos, albeit with parametrically smaller Lyapunov exponent than in the replica symmetric phase. An analytical calculation in the marginal spin glass phase suggests that this Lyapunov exponent vanishes in a particular infinite coupling limit. We comment on the potential meaning of these observations from the perspective of holography.
\end{abstract}

\newpage

\setcounter{tocdepth}{3}
{}
\vfill
\tableofcontents

\newpage

\section{Introduction}
\subsection{Motivation from holography}
The question of low dimensional holography stands at a crossroads. A plethora of microscopic constructions exist for AdS$_{d\geq 3}$, whereas until Kitaev's seminal work \cite{KitaevTalks}, the case for $d=2$ remained crucially out of reach. The difficulty in dealing with AdS$_2$ stems from the fact that finite energy excitations tend invariably to destroy the AdS$_2$ asymptotics. And besides wilting in the presence of these finite-energy excitations, AdS$_2$ also exhibits a \emph{fragmentation} instability \cite{Maldacena:1998uz} meaning that a single AdS$_2$ throat can break up into multiple distinct throats via quantum tunneling, each with its own (near-)extremal horizon. One particularly simple example of this phenomenon involves a single electrically charged black hole with mass equal to its charge in Planck units, which has the potential to break up in any of the multi-horizoned geometries discovered independently by Majumdar and Papapetrou \cite{Majumdar:1947eu,papapetrou1945static} with the same total charge.

Despite these difficulties, many papers have been dedicated to uncovering the microscopic underpinnings responsible for the observed dynamics found in the near-horizon region of extremal black holes \cite{Strominger:1998yg,Azeyanagi:2007bj,Castro:2008ms,Sen:2011cn,Anninos:2013nra} with a boon of renewed interest in the topic following Kitaev's discovery \cite{Polchinski:2016xgd,Maldacena:2016hyu,Gross:2016kjj,Sachdev:2015efa,Jensen:2016pah,Anninos:2016szt,Jevicki:2016bwu,Anninos:2017cnw,Castro:2019vog,Castro:2018ffi,Kitaev:2017awl,Kitaev:2018wpr}. Most of the current set of microscopic models are of the following type: they involve a fermionic quantum mechanical system with a large-$N$ flavor index and a disordered multi-body interaction. All models exhibits an emergent time-reparametrization symmetry in the infrared and at strong coupling. They have successfully elucidated several unique features of AdS$_2$ quantum gravity, from explicitly matching the boundary dynamics describing the soft breaking of the AdS$_2$ asymptotics\cite{Maldacena:2016upp} to reproducing the multi-boundary Euclidean gravity path integral \cite{Cotler:2016fpe,Saad:2018bqo,Saad:2019lba,Altland:2020ccq}. But it bears mentioning that, since the concern in all of these works is in reproducing the dynamics in the near-horizon region of a \emph{single} AdS$_2$ throat, none have yet addressed the question of fragmentation. What additional features must we add to the microscopic models in order to describe this kind of physics? Our paper aims to take a step in this direction.

\subsection*{Aside on string constructions}
In quantum gravity, it is usually a good idea to ask questions within the framework of string theory. String theoretic black holes also fragment. The clearest example involves the landscape of rigid multi-centered solutions to $\mathcal{N}=2$ supergravity in four dimensions \cite{Denef:2000nb,LopesCardoso:2000qm,Bates:2003vx}.  Single-horizon configurations are, at best, local free energy minima of the supergravity ensemble at low temperatures, and any system prepared in such a single-horizon state should be expected to fragment dynamically into a multi-centered configuration at low temperatures. This is due to the fragmented configurations being the globally preferred minima of the ensemble. Each individual black hole in these galaxies admits an AdS$_2$ near horizon region, and moreover the entire collection of black holes also fits inside one larger AdS$_2$ throat.  In certain limits, these bound systems of extremal black holes have quantum mechanical duals originating in brane constructions from string theory. We will refer to these brane models as quiver quantum mechanics (QQM) \cite{Douglas:1996sw,Denef:2002ru,Denef:2007vg}. 

Let us briefly review the essential features of QQM systems. First, they exhibit $\mathcal{N}=4$ supersymmetry, matching the number of supersymmetries preserved by the black hole geometries. The degrees of freedom can be split up into two types: 
\begin{itemize}
\item  Chiral multiplet degrees of freedom representing open strings stretched between stacks of D-branes $(\phi_i^\alpha,\psi_i^\alpha, F_i^\alpha)$~. 
\item Vector multiplet degrees of freedom representing the worldvolume degrees if freedom of D-branes in a stack: $(\{A, \vec{\mathbf{X}}\},\lambda, D)$~. 
\end{itemize}
The bottom components $\phi,\{A, \vec{\mathbf{X}}\}$ of these multiplets  are bosons, followed by their respective fermionic superpartners $(\psi,\lambda)$. The top components $F$ and $D$ are auxiliary bosons required for closure of the supersymmetry algebra. They are non-dynamical and do not participate in the interaction potential. The chiral matter is bifundamentally charged under \emph{pairs} of vector multiplet degrees of freedom and for this reason, the index $a$ labels pairs of D-brane (stacks) the open strings can end on. The index $i=1,\dots, N$ is a flavor index, allowing for an interesting large-$N$ limit, different from the large-$N$ of D-branes in a single stack usually taken in AdS/CFT. 

The chiral matter interactions are governed by a superpotential $W$, which is a polynomial of the chiral bosons. Because of the bifundamental nature of the chiral matter,  the lowest order gauge-invariant polynomial involving the $\phi$'s is cubic: 
\begin{equation}
	W(\phi)=\Omega_{ijk}\phi_i^1\phi_j^2\phi_k^3+\dots
\end{equation}
where $\Omega_{ijk}$ captures geometric data of some compactification manifold, and higher order terms are certainly permitted. For complicated enough compactifications, we can treat $\Omega_{ijk}$ like a source of disorder \cite{Anninos:2016szt}, and can focus only on disorder-averaged quantities. 
The chiral interactions are governed by the following potential: 
\begin{equation*}
V=\sum_{i,a}\left\lvert\frac{\partial W(\phi)}{\partial \phi_i^a}\right\rvert^2+\frac{1}{g_{\rm YM}^2}\sum_a\left(\theta_a-\sum_i\left\lvert\phi_i^a\right\rvert^2\right)^2+\left(\frac{\partial^2 W(\phi)}{\partial\phi_i^a\partial\phi^b_j}\psi_i^a\epsilon\psi_j^b+h.c.\right)
\end{equation*}
where the $\theta_a$, known as Fayet-Illiopoulos parameters, are allowed by the supersymmetry algebra. We highlight two features of this interaction potential: first, the fermions only participate in many-body interactions with the bosons, and not amongst themselves. So in order to get interesting SYK-type physics in the IR, it must be the result of interactions induced by integrating out the bosons.\footnote{Recently, a novel class of $\mathcal{N}=4$ models has been constructed which do have SYK-like multi-body fermionic interactions \cite{Gates:2021jdm}, (see also \cite{Lozano:2020txg,Lozano:2020sae}).} Secondly, the ground states (with $V=0$) of this system satisfy 
\begin{equation}
	\sum_i\left\lvert\phi_i^a\right\rvert^2=\theta_a\qquad \forall a\label{eq:cpnconstraint}
\end{equation}
meaning they parametrize a complex manifold made up of products of $\mathbb{CP}^{N-1}$ \cite{Denef:2002ru}. The disorder averaged version of this model was studied in \cite{Anninos:2016szt}, in the $g_{\rm YM}\rightarrow\infty$ limit, meaning the $\left(\theta-\sum_i\left\lvert\phi_i\right\rvert^2\right)^2$ part of the potential was dropped. In this approximation, the IR physics exhibits an emergent diff$(S^1)$ symmetry in the fermionic sector softly broken to $SL(2,\mathbb{R})$ (as exhibited by a linear in $T$ specific heat) prototypical of models in the SYK universality class. However, this approximation fails to capture any hint of the fragmentation instability. 

\subsection*{The $p$-spin glass model}
It is important to see if reintroducing the constraint \eqref{eq:cpnconstraint} gives rise to new features that could potentially capture the fragmentation instability. To simplify our analysis, we will instead consider a well-studied model from the spin glass literature, namely the $p$-spin spherical model \cite{crisanti1992sphericalp} (see \cite{castellani2005spin} for a review of its statics), which shares many key features with QQM. The degrees of freedom are bosons $\sigma_i$ with $i=1,\dots,N$ interacting via the potential
\begin{equation}
	V= J_{i_1\dots i_p}\sigma_{i_1}\dots\sigma_{i_p}~,
\end{equation}
where the couplings are disordered and sampled from a probability distribution, and the `spins' $\sigma_i$ are subject to the following \emph{spherical} constraint 
\begin{equation}
 \sum_i \sigma_i^2=N~,
\end{equation}
 meaning that this a nonlinear sigma model on the group manifold of an $N$-dimensional sphere. This bosonic model clearly distills the essential features of the bosonic sector of the QQM described above.\footnote{One flaw in the analogy of course is that the $p$-spin model allows for negative directions of the potential energy, which is not the case in QQM.}  Our perspective is the following: the SYK-like physics responsible for the excitations of a single AdS$_2$ geometry is captured by the fermions of the QQM. We would like to explore whether the non-linear sigma model bosons have the potential to mediate the fragmentation instability. To this end we will study the simplified $p$-spin model in order to understand if this is the case, and if so, what is the order parameter of the fragmentation instability?

A few caveats are in order before we continue. Let us recall that the fragmentation instability mediates between single- and multi-horizoned geometries that all fit within an asymptotic AdS$_2$ throat. This means that the instability should preserve the $SL(2,\mathbb{R})$ symmetries so fruitful for SYK. However, the $p$-spin model (and by extension the bosonic sector of QQM) has an intrinsic scale: the volume of the sigma model manifold. It should then come as a surprise to everyone that the $p$-spin model exhibits correlation functions consistent with an emergent conformal symmetry in the IR and at strong coupling, as first reported in a beautiful paper by Cugliandolo, Grempel, and da Silva Santos \cite{cugliandolo2001imaginary}. Moreover, this interesting behavior is found below the model's \emph{spin glass} transition, meaning that there are gapless excitations in a system that is meant to be frustrated! 

As we will review, these two features go hand in hand. The spin glass transitions is what allows the $p$-spin model to forget about its IR cutoff and comes along with an explosion in the number of nearby states of the free energy landscape, captured by an entropic quantity, called the \emph{complexity} in the spin glass literature, which scales extensively with $N$. Our aim in this paper is to suggest, albeit we have not proven, that the spin glass transition is microscopically dual to the  fragmentation instability. The order parameter for this spin glass transition is known as the Edwards-Anderson parameter \cite{edwards1975theory}, and is diagnosed by a failure of Euclidean correlators to cluster at large separations. In order to affirm this guess, one would need to find a Lorentzian bulk observable sensitive to this order parameter, and exhibit a failure of clustering for a particular observable. We do not claim to have done this. Moreover, this idea is not new, as there are previous attempts to link the phenomenology of glasses to multi-horizoned black holes \cite{Anninos:2011vn,Anninos:2012gk,Anninos:2013nra,Anninos:2013mfa,Kent-Dobias:2020egr} (see also \cite{Kurchan:2016nju,Facoetti:2019rab,Kurchan:2021qvf}).

Beyond an in-depth review of the thermodynamics and real-time dynamics of this model aimed at high energy physicists, we also compute a real-time out-of-time-order correlation functions for the $p$-spin model and find a nonzero Lyapunov exponent in the conformal spin glass phase, meaning that certain modes scramble efficiently even beyond the spin glass phase transition. However, unlike the SYK model \cite{KitaevTalks,Maldacena:2016hyu}, the Lyapunov exponent vanishes in a particular infinite coupling limit, suggesting that the bosonic modes would not participate in scrambling deep in the spin glass phase in the putative holographic dual. Other important recent explorations of the relation between the SYK model, spin glasses, and holography include \cite{Gur-Ari:2018okm,Arefeva:2018vfp,Baldwin:2019dki,Engelhardt:2020qpv,Johnson:2020mwi}. Finally, we note that other bosonic spin models have enjoyed significant interest in recent years, and we will have the opportunity to make use of some of the techniques that were developed \cite{Fu:2016yrv,Witten:2016iux,Klebanov:2016xxf,Azeyanagi:2017drg,Giombi:2017dtl,Chang:2018sve,Giombi:2018qgp,Klebanov:2018fzb,Ferrari:2019ogc,Tikhanovskaya:2020elb,Tikhanovskaya:2020zcw}.

With the introduction now winding down, it would be remiss of us not to mention that holography only makes fleeting appearances in the remainder of this paper (see, e.g., section \ref{sec:holography}), which instead is devoted to an in-depth study of $p$-spin model in its own right, to allow for a potential holographic interpretation in the future. 
We have gone to great effort to make the paper self-contained, often at the expense of length. So let us now summarize the main results.

\begin{figure}
\begin{center}
\includegraphics[width=0.55\textwidth]{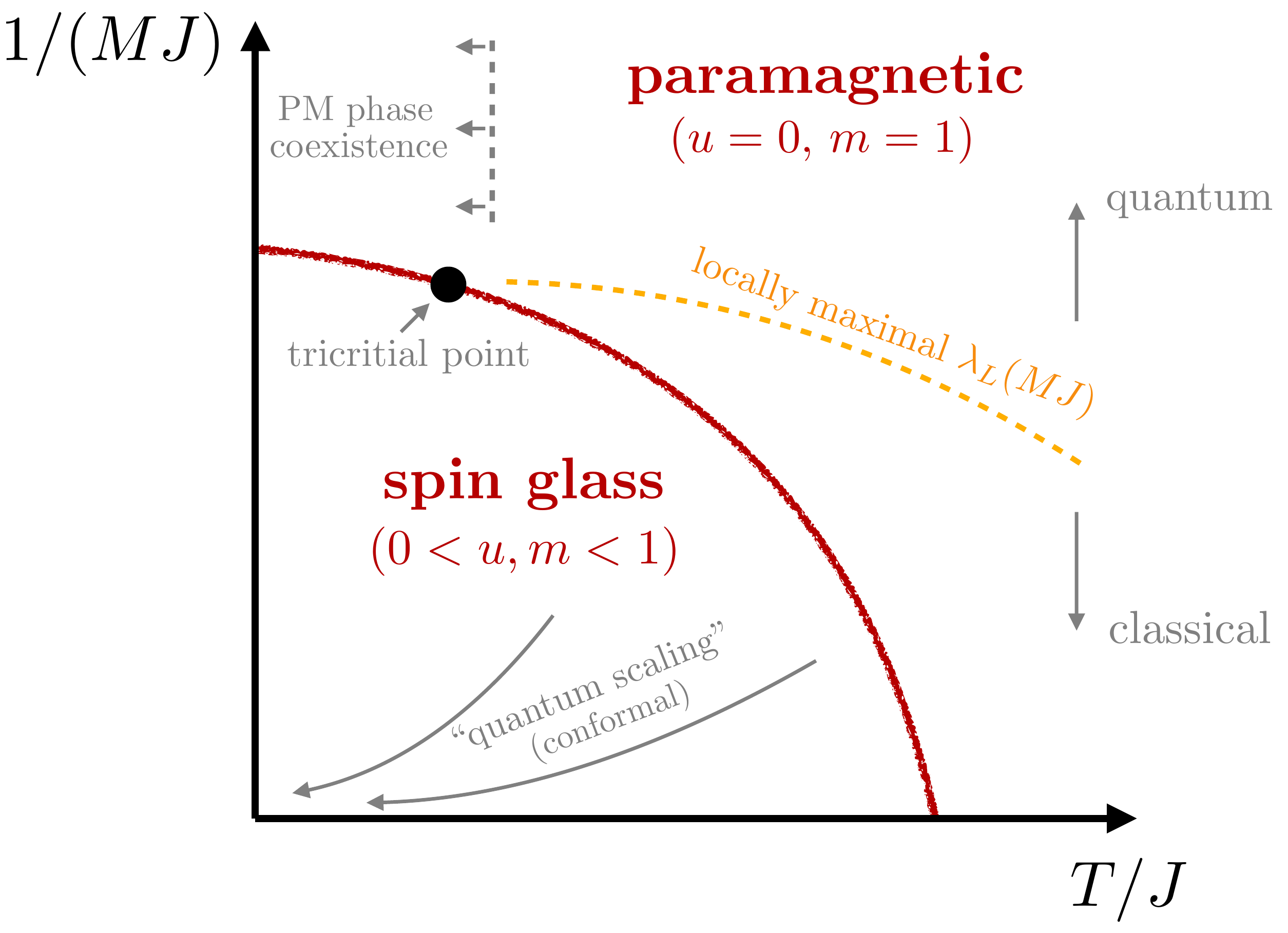}
\end{center}
\vspace{-.5cm}
\caption{{\bf Cartoon of the phase diagram of the $p$-spin spherical model.} The basic parameters are temperature $T/J \equiv 1/(\beta J)$ and $1/(MJ)$. We show the rough shape of the spin glass transition and the associated values of the order parameter $u$ as well as the break point parameter $m$. The `quantum scaling' limit we discuss is a particularly useful way to approach the strong coupling regime without eliminating the quantum features of the model. The yellow curve shows where the quantum Lyapunov exponent is found to be the largest as a function of $MJ$ for any fixed temperature.} 
\label{fig:phasesCartoon}
\end{figure}

\subsection{Summary and main results}{}
To guide the reader, we will now summarize the main points and results of our analysis (we illustrate some of these features in the cartoon figure \ref{fig:phasesCartoon}). Some of these results were already known and can be found in the references (in particular \cite{cugliandolo2001imaginary}):
\begin{itemize}
 \item The model in question has two dimensionless parameters: one measuring the coupling strength $\beta J$ and another controlling the strength of quantum fluctuations $MJ$, where the classical limit is the limit of large $MJ$. These dimensionless parameters can be combined into a third dimensionless quantity, namely $M/\beta$. 
 \item In disordered models, the spin glass transition is diagnosed by replica symmetry breaking (RSB). We will explain in detail what these replicas are, and what symmetry is being broken below, however it is important to state here that RSB is accompanied by a failure of clustering in Euclidean correlation functions at zero-temperature: 
 \begin{equation}
 	\lim_{\tau\rightarrow\infty}\langle\sigma^i(\tau)\sigma^i(0)\rangle = u
 \end{equation}
 where the strength of the clustering violation is known as the Edwards-Anderson parameter $u$ \cite{edwards1975theory}. As we will explain below, $u$ also functions to measure the strength of replica symmetry breaking (see, e.g., \eqref{eq:1rsb}) and plays a crucial role in our discussions of quantum chaos. 
\item The spin glass phase is also characterized by a non-zero cluster parameter $m$, which is sometimes called the breakpoint parameter in the spin glass literature. The quantity $m$ roughly corresponds to the sizes of clusters of states within the free-energy landscape, and indicates a nontrivial hierarchical structure on the space of thermodynamic states \cite{mezard1987spin,Denef:2011ee,castellani2005spin,de2006random}. 
 \item The  spin glass transition in this models occurs at low temperatures (large $\beta J$) and in the classical regime (large $MJ$). Both thermal and quantum fluctuations destroy the spin glass phase, so by either increasing the temperature, or by decreasing the coupling  strength $MJ$, we drive the model towards a {\it paramagnetic phase}.
 The phase diagram of the model is sketched in figure \ref{fig:phasesCartoon} (see also figure \ref{fig:PhasesMarg}).
  \item At low temperatures there exist two branches of solutions, both in the paramagnetic and in the spin glass phase. In each phase, one branch enters a conformal regime at large $\beta J$. In the case of the paramagnet, the conformal branch is unphysical (thermodynamically disfavored and unstable). In the case of the spin glass, the scaling solution is physical and {\it marginally stable}. Thermodynamically, the marginally stable spin glass behaves like a conformal system in many respects (zero temperature entropy, linear-in-temperature specific heat, gapless spectrum etc.). However, it is important to state that it exists in a different thermodynamic ensemble than the one typically considered (see, e.g., \eqref{eq:thermodef}), where $m$ becomes an externally tunable thermodynamic parameter. In practice, this implies tuning to the conformal spin glass phase requires the replica symmetry to be \emph{explicitly}, rather than spontaneously broken, exhibited by a particular choice of the parameter $m$, see \eqref{eq:replicatedstep}.  The spontaneously broken spin glass phase \textit{does not} exhibit gapless excitations.  
  \item The conformal spin glass phase exhibits an emergent diff$(S^1)$ symmetry at low energies, much like the SYK model (see \ref{sec:diffs1}). The linear in $T$ specific heat in this phase would suggest that this symmetry is broken to $SL(2,\mathbb{R})$, exhibited by a Schwarzian effective action. Some care is required in this interpretation, which we touch upon in section \ref{sec:holography}.  
  \item The conformal solution \eqref{eq:qrcSG}, describing long-wavelength fluctuations in the marginally stable spin glass phase, admits an analytic UV completion at short distances. The resulting {\it approximate spin glass solution} \eqref{eq:exactsolfourier} is analytically tractable and provides a remarkably good approximation to the exact spin glass dynamics. We identify a {\it quantum scaling} where $MJ \sim \beta J \gg 1$, which is particularly well suited for analytical investigations, see \eqref{eq:QuantScale}.
  \item The Euclidean four-point kernel (given in \eqref{eq:kernelEucl}), used to derive the Euclidean four-point function, exhibits an interesting structure in the spin glass phase. In particular it contains a four-point generalization of the Edwards-Anderson parameter $u$, which we have denoted by $v$. It would be interesting to explore the consequences of this parameter in future work. 
  \item The model is strongly quantum chaotic in the sense of exponential growth of out-of-time-order correlation functions (OTOCs). We compute the quantum Lyapunov exponent for a range of values of the couplings and find that it displays an intricate dependence on the couplings $\beta J$ and $MJ$. We also find exponential growth of OTOCs in the conformal spin glass phase, albeit with smaller Lyapunov exponents than in the paramagnetic phase, which vanishes in the infinite $\beta J$ limit with $M/\beta$ held fixed. In particular, an analytic calculation  in the conformal spin glass phase reveals:
\begin{equation*}
	\frac{\beta}{2\pi}\, \lambda_L\approx \frac{5m}{24}~,
\end{equation*}
which relates the Lyapunov exponent to the cluster parameter $m$. 
  See figures \ref{fig:lyapunov1} and \ref{fig:lyapunov2} for the values of Lyapunov exponents.
\end{itemize}

\subsection{Outline of the paper}

{\it Part \ref{part:euc}:} Much of the first part is a review and an expansion on previous works (especially \cite{cugliandolo2001imaginary}) from a holography-inspired point of view. In section \ref{sec:model} we introduce the model and review its dynamical equations. Extremizing the free energy with respect to all parameters defines the equilibrium equations of motion, whose thermodynamics we elaborate on in section \ref{sec:EquilThermo}. In section \ref{sec:MargThermo}, we reinterpret the criterion of marginal stability, which allows for a conformal solution to the equations of motion, as arising from a different choice of ensemble. An analysis of this phase leads us to identify an emergent diff$(S^1)$ symmetry at low energies, allowing for an analytical approximate solution to the equations of motion, and accompanied by rich low temperature thermodynamics.
\bigskip

\noindent{\it Part  \ref{part:RTD}:} We discuss real-time two-point correlation functions in section \ref{sec:TwoPtReal} as a precursor to computing the out-of-time-order four-point function in real time. This necessitates a Euclidean analysis of the four-point kernel. Using this, we numerically extract the Lyapunov exponents in section \ref{sec:FourPt} as a function of the various couplings. We compare the numerical calculation in the marginal spin glass with an analytical treatment, finding a nice match. We conclude with a summary and a discussion of possible future directions in section \ref{sec:conclusion}.
\bigskip 

\noindent{\it Part \ref{part:app} (appendices):} We collect some computational details and background material in the appendices. We wish to highlight appendix \ref{ap:conventions}, which contains a collection of conventions and formulae, and appendix \ref{app:confPM}, which contains original material regarding the conformal limit of an existent but somewhat unphysical branch of paramagnetic solutions.\\

\noindent {\bf Note:} While this work was nearing completion, reference \cite{Bera:2021vhx} appeared on the arXiv. Their analysis overlaps with some of our discussion, especially with section \ref{sec:FourPt}.

\newpage

\part{Euclidean approach and phase structure}\label{part:euc}

\section{The quantum spherical \texorpdfstring{$p$}{p}-spin model}
\label{sec:model}

In this section we will study the $p$-spin model, describing $N$ bosonic spins with $p$-body interactions, largely following \cite{cugliandolo2001imaginary,Cugliandolo_2000} (see also \cite{crisanti1992sphericalp,castellani2005spin,biroli2001quantum} for related studies of the model). We will work in Euclidean time, with partition function at inverse temperature $\beta$:
\begin{equation}
\label{eq:ModelDef}
	Z[J_{i_1 \ldots i_p}]=\int D\sigma_i \;\exp\left\lbrace-\int_0^\beta d\tau\, \left[\frac{M}{2}\dot{\sigma_i}(\tau)\dot{\sigma_i}(\tau)+\sum_{i_1 <  \ldots < i_p } J_{i_1\ldots i_p}\sigma_{i_1}(\tau)\ldots\sigma_{i_p}(\tau)\right]\right\rbrace~,
\end{equation}
where $M$ is a dimensionful `inertial mass' designed such that $\sigma_i$ are dimensionless in the UV, and $i = 1 , \ldots , N$. The randomized couplings are sampled from the following distribution:
\begin{equation}\label{pj}
	P(J_{i_1\ldots i_p})\propto \exp\left[- \frac{N^{p-1}}{p!}\frac{J_{i_1\ldots i_p}^2}{J^2}\right]~,
\end{equation}
where $J$ sets the width of the distribution over couplings. We have chosen the scaling with $N$ in \eqref{pj} such that we have a controlled large-$N$ limit \cite{castellani2005spin,crisanti1992sphericalp}. 
This is very similar to the SYK model \cite{KitaevTalks,Maldacena:2016hyu} with fermions replaced by bosons. One crucial difference is that the potential term will generically have negative directions, which is quite problematic for the stability of this bosonic theory. To counteract this, we will suplement this model with a \emph{spherical constraint}:
\begin{equation}
\label{eq:SphConstr}
	\sum_{i=1}^{N}\sigma_i(\tau)\sigma_i(\tau)=N \,.
\end{equation}
Thus the spin degrees of freedom lie on an $N$-dimensional sphere of radius $\sqrt{N}$, describing a nonlinear sigma model.

\paragraph{Dimensionless parameters and conventions:} The various phases of the model \eqref{eq:ModelDef}  depend on the strengths of the parameters introduced above. The dimensionless parameters in our analysis will be $\beta J$ and $MJ$~, where the parameter $MJ$ characterizes the strength of quantum effects since limit $MJ \rightarrow \infty$ corresponds to the classical limit (since fluctuations are suppressed), while $MJ \rightarrow 0$ describes a system with strong quantum fluctuations.
The reader may consult appendix \ref{ap:conventions}, where we have gathered our conventions for ease of reading. 

\subsection{Effective action and Schwinger-Dyson equations}

In this subsection we derive the large $N$ effective action using various path integral manipulations, following \cite{cugliandolo2001imaginary}. Readers wanting to skip these details may jump directly to the Schwinger-Dyson equations \eqref{eq:SDeqs} for the collective field $Q_{ab}(\tau,\tau')$ defined in equation \eqref{eq:Qabdef}.

In order impose the spherical constraint \eqref{eq:SphConstr} in the path integral, we insert:
\begin{equation}
	\delta\left(\sum_{i=1}^{N}\sigma_i(\tau)\sigma_i(\tau)-N\right)=\int Dz \exp\left[i\int_0^\beta d\tau\, z(\tau)\left(\sum_{i=1}^{N}\sigma_i(\tau)\sigma_i(\tau)-N\right)\right]~
\end{equation}
into \eqref{eq:ModelDef}~. We are now ready to perform the disorder average. 

However, in order to properly perform the disorder average with respect to the distribution \eqref{pj}, we are instructed to average the free energy, rather than the partition function. That is we want to compute (overlines denote a disorder average)
\begin{equation}
	\beta \overline{F}= -\int dJ_{i_1\ldots i_p} P(J_{i_1\ldots i_p}) \log Z[J_{i_1\ldots i_p}]~,
\end{equation}
but of course this computation requires us to have enough analytic control to compute $Z[J_{i_1\dots i_p}]$ for arbitrary couplings. Instead, we will use the following representation of the logarithm
\begin{equation}
	\log Z= \lim_{n\rightarrow0}\partial_n Z^n
\end{equation}
and take the average of $Z^n$. The typical way of doing this is by introducing an integer number of replicas of the original system, and then hope that we can trust the analytic continuation to $n=0$ at the end. In this spirit, we introduce a replica index $a=1,\dots,n$ and the averaged replicated partition function becomes: 
{\small
\begin{multline}
\label{eq:replicatedstep}
	\overline{Z^n}=\int dJ_{i_1\dots i_p} P(J_{i_1\dots i_p})\int D\sigma_i^a Dz^a \,\exp\Bigg\{-\int_0^\beta d\tau \Bigg[\frac{M}{2}\dot{\sigma}^a_i(\tau)\dot{\sigma}^a_i(\tau)+\sum_{i_1 <  \ldots < i_p } J_{i_1\dots i_p}\sigma^a_{i_1}(\tau)\dots\sigma^a_{i_p}(\tau)\Bigg]\\+i\int_0^\beta d\tau\, z^a(\tau)\left(\sigma^a_i(\tau)\sigma^a_i(\tau)-N\right)\Bigg\}~.
\end{multline}
}\normalsize
Integrating out the disorder $J_{i_1\dots i_p}$, we introduce couplings between the replicas: 
\begin{multline}
	\overline{Z^n}=\int D\sigma_i^a Dz^a \,\exp\Bigg\{-\int_0^\beta d\tau\, \left[\frac{M}{2}\dot{\sigma}^a_i(\tau)\dot{\sigma}^a_i(\tau)-i\, z^a(\tau)\left(\sigma^a_i(\tau)\sigma^a_i(\tau)-N\right)\right]\\
	+\frac{J^2 }{4N^{p-1}}\int_0^\beta \int_0^\beta d\tau\,d\tau'\,\sum_{a,b=1}^n\left(\sum_{i=1}^N\sigma_{i}^a(\tau)\sigma_{i}^b(\tau')\right)^p\Bigg\}~.
\end{multline}
We now introduce the collective variables for this problem, which we will call the ($Q$, $\lambda$) variables, where:\footnote{ In the SYK model, the analogous quantities are called ($G$, $\Sigma$).} 
\begin{equation}
  \boxed{\;\;Q_{a b}(\tau,\tau')\equiv\frac{1}{N}\sum_{i=1}^N\overline{\langle \sigma_i^a(\tau)\sigma_i^b(\tau')\rangle}~.\;}\label{eq:Qabdef} 
\end{equation}
We do this by inserting a factor of 1 into the path integral:
\begin{align}
1&=\int DQ_{ab}\;\delta\left(N\,Q_{ab}(\tau,\tau')-\sum_{i=1}^N 	\sigma_{i}^a(\tau)\sigma_{i}^b(\tau')\right)\nonumber\\
&=\int DQ_{ab}D\lambda_{ab}\,\exp \left\lbrace i\int_0^\beta\int_0^\beta d\tau d\tau'\, \lambda_{ab}(\tau,\tau')\left(N\,Q_{ab}(\tau,\tau')-\sum_{i=1}^N 	\sigma_{i}^a(\tau)\sigma_{i}^b(\tau')\right)\right\rbrace \,.\label{eq:deltafunc}
\end{align}
After inserting this identity into the path integral, we are free to replace many bilinears of $\sigma_i$ with $Q_{ab}$. Finally, integrating out the spins $\sigma_i$ altogether we get:\footnote{ The determinant is normalized according to: \begin{equation}\int_{-\infty}^\infty d^n x	\exp\left(-\frac{1}{2}\mathbf{x}\cdot \mathbf{A}\cdot\mathbf{x}\right)=\left(\text{det}\,\frac{\mathbf{A}}{2\pi}\right)^{-1/2}~. \end{equation}}
\begin{equation}
\begin{split}
	&\overline{Z^n}=\int DQ_{ab}D\lambda_{ab} Dz^a \;\text{det}^{-N/2}\left[-\frac{1}{\pi}\delta_{ab}\delta(\tau-\tau')\left(\frac{M}{2}\partial_{\tau'}^2+i\, z^a(\tau)\right)+\frac{i}{\pi}\lambda_{ab}(\tau,\tau')\right]\\
	&\quad \times \exp\left\lbrace-N\,i\sum_{a=1}^n\int_0^\beta d\tau\,  z^a(\tau)+N\int_0^\beta \int_0^\beta d\tau\,d\tau'\,\left[i\lambda_{ab}(\tau,\tau')Q_{ab}(\tau,\tau')+\frac{J^2 }{4}\sum_{a,b=1}^n Q_{ab}(\tau,\tau')^p\right]\right\rbrace
\end{split}
\end{equation}
From this we can extract the replicated effective action: 
\begin{multline}
	\frac{S_{\rm eff}}{N}=\frac{1}{2}\text{Tr}\log \left[-\frac{1}{\pi}\delta_{ab}\delta(\tau-\tau')\left(\frac{M}{2}\partial_{\tau'}^2+i\, z^a(\tau)\right)+\frac{i}{\pi}\lambda_{ab}(\tau,\tau')\right]\\+i\sum_{a=1}^n\int_0^\beta d\tau\,  z^a(\tau)-\int_0^\beta \int_0^\beta d\tau\,d\tau'\,\left[i\lambda_{ab}(\tau,\tau')Q_{ab}(\tau,\tau')+\frac{J^2 }{4}\sum_{a,b=1}^n Q_{ab}(\tau,\tau')^p\right]~.
\end{multline}
Since the effective action is proportional to $N$, the saddle point approximation is exact in the $N\rightarrow\infty$ limit. As a result, we can now completely eliminate the $\lambda$ matrix by solving its equations of motion:
\begin{equation}
	\left(\mathbf{M}+i\boldsymbol{\lambda}\right)^{-1}=\frac{1}{\pi}\mathbf{Q} \,,
\end{equation}
where we have defined the matrix $\mathbf{M}$ in replica space and imaginary time: 
\begin{equation}
	{M}_{ab}(\tau,\tau')=-\delta_{ab}\delta(\tau-\tau')\left(\frac{M}{2}\partial_\tau^2+i\, z^a(\tau)\right)~.
\end{equation}
This tells us that
\begin{equation}
	i\boldsymbol{\lambda}=-\mathbf{M}+\pi\mathbf{Q}^{-1}~,
\end{equation}
and the effective action becomes (up to an additive constant):
\begin{equation}\label{eq:effaction}
\begin{split}
	\frac{S_{\rm eff}}{N} &=-\frac{1}{2}\text{Tr}\log \left[Q_{ab}(\tau,\tau')\right]-i\sum_{a=1}^n\int_0^\beta d\tau\,  z^a(\tau)\left(Q_{aa}(\tau,\tau)-1\right)\\
	&\quad\, -\sum_{a,b=1}^n\int_0^\beta \int_0^\beta d\tau\,d\tau'\,\left[\delta_{ab}\,\delta(\tau-\tau')\frac{M}{2}\partial_\tau^2Q_{ab}(\tau,\tau')+\frac{J^2 }{4} Q_{ab}(\tau,\tau')^p\right]~.
\end{split}
\end{equation}
Notice that the effect of the $z^a$ is to enforce the condition that the diagonal elements of the matrix $\mathbf{Q}$ (in time and in replica space) are all 1. To obtain the free energy, we simply notice that
\begin{equation}
	\beta\overline{F}=\lim_{n\rightarrow 0}\partial_n S_{\rm eff}(Q_\star)
\end{equation}
where $Q_\star$ satisfies the Schwinger-Dyson equation. As we will see, there can be many such solutions, and picking the right one is the challenge. 

\paragraph{Schwinger-Dyson equations:}
From \eqref{eq:effaction} we can immediately derive the Schwinger-Dyson equations in replica space by variation with respect to $Q_{ab}$:
\begin{equation}
	-(2\mathbf{Q})^{-1}-\delta_{ab}\,\delta(\tau-\tau')\left[\frac{M}{2}\partial_\tau^2+iz^a(\tau)\right]-\frac{pJ^2 }{4} Q_{ab}(\tau,\tau')^{p-1}=0
\end{equation}
which we can rewrite in more familiar form:
\begin{equation}\label{eq:SDeqs}
	\boxed{\;-\delta_{ab}\,\left[\frac{M}{2}\partial_\tau^2+iz^a(\tau)\right]Q_{ab}(\tau,\tau')-\frac{p J^2}{4}\int_0^\beta d\tau''\, Q_{ac}^{p-1}(\tau,\tau'')Q_{cb}(\tau'',\tau')=\frac{1}{2}\delta_{ab}\,\delta(\tau-\tau')~}
\end{equation}
with boundary condition $Q_{aa}(\tau,\tau)=1$. In the following, we discuss how to solve this equation in two steps: first, we need to simplify the matrix structure of \eqref{eq:SDeqs} in replica space by making an educated ansatz for the matrix $Q_{ab}$. Only then will it be useful to write more explicit equations of motion and solve them in various limits or numerically.

\subsection{Replica symmetry breaking: 1-RSB ansatz}

To proceed, let us note the following fact, proven in \cite{bray1980replica}. The matrix elements $Q_{a\neq b}$ are \emph{time independent}. The reason is that, following the definition \eqref{eq:Qabdef}: 
\begin{equation}
	Q_{a\neq b}(\tau,\tau')=\frac{1}{N}\sum_{i=1}^N\overline{\langle \sigma_i^a(\tau)\sigma_i^b(\tau')\rangle}=\frac{1}{N}\sum_{i=1}^N\overline{\langle \sigma_i^a(\tau)\rangle\langle\sigma_i^b(\tau')\rangle}
\end{equation}
where $\langle\cdot\rangle$ denotes a thermodynamic average, whereas the overline denotes a disorder average. Since the replica index is not the same, the thermodynamic averages split up into single replica expectation values --- but the one point functions are time independent (although crucially nonzero). This means only the diagonal elements have any time dependence. The late-Euclidean time limit of this diagonal element is sometimes called the Edwards-Anderson order parameter \cite{edwards1975theory} and is the exact analog of the equilibrium Ising magnetization, which obtains a nonzero expectation value below the critical temperature. As we will see, the diagonal elements will fail to decay to zero at low temperatures, indicating that the system is in a nontrivial thermodynamic state. 

The second important fact about this model is that it has a spin glass phase with a solution that has replica symmetry breaking at 1-step (see \cite{crisanti1992sphericalp} for the static case and \cite{cugliandolo2001imaginary} for the case at hand). The basic idea behind replica symmetry breaking is that the matrix $Q_{ab}$ can have additional structure beyond the dichotomy of diagonal/off-diagonal. The overlap between replica $a$ and replica $b$ may differ from the overlap between $a$ and $c$ (and $b$ and $c$ for that matter). This suggests an interesting set of thermodynamic states at low temperatures, as was nicely reviewed in \cite{mezard1987spin,castellani2005spin,de2006random,Denef:2011ee}. To fully parametrize the solutions to this model, it is sufficient to consider an ansatz of the following form for the overlap matrix: 
\begin{equation}
\label{eq:1rsb}
	\mathbf{Q}=\begin{pmatrix} 
	q(\tau,\tau') & u & u &   &  & &\\
	u & q(\tau,\tau') &u &  &s & &\cdots\\
	u & u & q(\tau,\tau') & & & &\\
	  &  &  & q(\tau,\tau') &  u& u &\\
	  &  s&  & u & q(\tau,\tau') &u &\\
	  &  &  & u & u & q(\tau,\tau') &\\
	  & \vdots & & & & & \ddots &
	\end{pmatrix}
\end{equation}
where the $n\times n$ matrix is made up of a set of $m\times m$ blocks along the diagonal, with $m$ to be determined. This ansatz is called 1-step \emph{replica symmetry breaking}. The reason it is called replica symmetry \emph{breaking} is because we allow for configurations with $s\neq u$. The name 1-step comes from the fact that the diagonals are composed of replica \emph{symmetric} matrices. It follows that, in order to get a 2-step RSB matrix, we build a matrix like \eqref{eq:1rsb} but instead we populate the diagonal blocks with 1-RSB. The higher-step RSB matrices are defined iteratively this way, but will not be needed in the $p$-spin model.

To write the 1-RSB matrix, we will split the index $a=a_0 a_1$ as in \cite{Denef:2011ee} with $a_0=1,\dots,m$ parametrizing the place in the block and $a_1=1,\dots,\frac{n}{m}$ parametrizing which block we are in. With this, the matrix in \eqref{eq:1rsb} can be written as\footnote{We are of course working in a situation where $m$ divides $n$, but we will treat the final formulas as analytic both in $n$ and $m$ and eventually take the limit $n \rightarrow 0$.}
\begin{equation}
	\mathbf{Q}=\delta_{a_1b_1}\left[\delta_{a_0b_0}q(\tau,\tau')+u(1-\delta_{a_0b_0})\right]+s(1-\delta_{a_1b_1})~.\label{eq:rsbnotation}
\end{equation}
Our strategy will now be to plug this ansatz into the effective action and derive equations for the components $q(\tau,\tau')$, $s$, and $u$. One way to do this is to note that the matrix \eqref{eq:1rsb} has the following eigenvalues in replica space: 
\begin{align}
 \lambda_0&=q(\tau,\tau')-u&	&\text{degeneracy}: \; n\left(1-\frac{1}{m}\right)\nonumber\\
 \lambda_1&= q(\tau,\tau')-u +m(u-s)& &\text{degeneracy}:\;  \frac{n}{m}-1\nonumber\\
\lambda_2&=q(\tau,\tau')-u+m(u-s)+ns& &\text{degeneracy}: \; 1\nonumber
\end{align}
One last thing we need is that the number of times each of $q(\tau,\tau')$, $s$ and $u$ appear in the replica matrix. This is easy to compute at 1-RSB and is given by
\begin{align}
	\#\, q(\tau,\tau'): &~~n\nonumber\\
	\#\, u:&~~ n(m-1)\nonumber\\
	\#\,  s: &~~n(n-m)~.\nonumber
\end{align}
If we also assume that $z^a(\tau)$ is independent of the replica index, we find that we can finally combine all of this to obtain the {\it 1-RSB effective action}: 
\begin{align}\label{eq:1rsbeffaction}
	\frac{S_{\rm eff}}{N}=&-\frac{n}{2}\left(1-\frac{1}{m}\right)\text{Tr}\log \left[q(\tau,\tau')-u\right]-\frac{1}{2}\left(\frac{n}{m}-1\right)\text{Tr}\log \left[q(\tau,\tau')-u +m(u-s)\right]\nonumber\\&-\frac{1}{2}\text{Tr}\log\left[ q(\tau,\tau')-u+m(u-s)+ns\right]-in\int_0^\beta d\tau\,  z(\tau)\left(q(\tau,\tau)-1\right)\nonumber\\
	&-n\int_0^\beta \int_0^\beta d\tau\,d\tau'\,\left\lbrace\delta(\tau-\tau')\frac{M}{2}\partial_\tau^2q(\tau,\tau')+\frac{J^2 }{4} \left[q(\tau,\tau')^p+(m-1)u^p+(n-m)s^p\right]\right\rbrace~.
\end{align}
As we will show, the equations of motion set $s=0$, and thus the effective action is linear in $n$. This means taking a derivative with respect to $n$ and setting it to zero is trivial. Nevertheless, the block size $m$ is now a real parameter. We notice that $m\rightarrow1$ is the paramagnetic limit where $u$ drops out. On the other hand, $u$ becomes important for $m< 1$ (whence replica symmetry is spontaneously broken). It is indeed a common idea in the spin glass literature that the continuum value of $m$ lies in $0<m\leq 1$ and serves as an order parameter for replica symmetry breaking. We now turn to the 1-RSB equations of motion.\footnote{An important caveat in the theory of spin glasses is that the $n\rightarrow0$ limit forces us to consider \emph{maximizing} rather than minimizing the effective action in the replica directions \cite{parisi1979infinite,parisi1980order,crisanti1992replica,crisanti1992sphericalp}. This stems from the fact that, while the fluctuation Hessian can be shown to have positive eigenvalues, their degeneracies become negative as we take $n\rightarrow 0$. See appendix \ref{app:kernelDerivation} for details.}

\paragraph{Off-diagonal parameters:}
We begin with the equations of motion for the variables $\{u,s\}$ parameterizing the 1-RSB ansatz. We will focus on the remaining variables once we pass to Fourier-space, as the zero-modes of the problem require some care. 
Let us begin with the equation of motion for varying $s$:
\begin{align}
\text{EOM}_s: \quad\;\; 0&=\frac{pJ^2}{4}\int_0^\beta d\tau''\left[q(\tau,\tau'')^{p-1}s+s^{p-1}q(\tau'',\tau')+(m-1)\left(s\, u^{p-1}+s^{p-1}u\right)+(n-2m)s^p\right]
\end{align}
We see that this is proportional to $s$, so it is safe to set $s$ to zero. The equation for motion for varying $u$ takes a similar form (for $s=0$):
\begin{align}
\label{eq:EOMu}
\text{EOM}_u: \qquad 0&=\frac{pJ^2}{4}\int_0^\beta d\tau''\left[q(\tau,\tau'')^{p-1}u+u^{p-1}q(\tau'',\tau')+(m-2)u^p\right]
\end{align}
From this, one might also conclude that $u=0$ by similar logic as above, and indeed, this solution is valid at high temperatures. However, unlike the case with $s$, at low temperatures a new solution with $u\neq0$ appears and dominates the thermodynamics. So we will allow for generic values of $u$ for now.

\subsection{Effective action and equations of motion in frequency space}

Consider now the matrix $Q_{ab}(\tau,\tau')$ in frequency space, utilizing the fact that, by time-translation symmetry, the saddle point solutions will be functions of the difference $\tau-\tau'$:
\begin{equation}
	q(\tau,\tau')-c u=\frac{1}{\beta}\sum_{k=-\infty}^\infty e^{\frac{2\pi i\,k}{\beta}(\tau-\tau')}\left[\hat{q}(k)-\beta c u\,\delta_{k,0}\right]
\end{equation}
where $c$ is any constant.  We will be particularly interested in the cases $c=1$ and $c=m-1$. Note, furthermore, that due to time reversal symmetry, $\hat{q}(k) = \hat{q}(-k)$.
We thus have the following expression for the large $N$ effective action on the $s=0$ subspace:

\begin{align}
\label{eq:Sfinal}
	\frac{S_{\rm eff}}{nN}=&-\frac{1}{2}\sum_{k\neq 0}\log \frac{\hat{q}(k)}{\beta}-\frac{1}{2}\left(1-\frac{1}{m}\right)\log \left[\frac{\hat{q}(0)}{\beta}-  u\right]-\frac{1}{2m}\log \left[\frac{\hat{q}(0)}{\beta}+ (m-1) u\right]\nonumber\\
	&+i\hat{z}(0)\left(1-\sum_{k=-\infty}^\infty \frac{\hat{q}(k)}{\beta}\right)+\frac{M}{2}\sum_{k=-\infty}^\infty \left(\frac{2\pi k}{\beta}\right)^2\hat{q}(k)\nonumber\\
	&-\frac{\beta^2J^2 }{4}\left\{(m-1)u^p+\sum_{k_1,\dots,k_{p}}\left[\prod_{i=1}^{p}\frac{\hat{q}(l_i)}{\beta}\right]\,\delta_{k_1+\dots+k_p,0}\right\}~,
\end{align}
where we notice that $u$ and $m$ only couple to the zero mode of $q(\tau,\tau')$. Note moreover that we need only retain the zero mode of $z(\tau)$ in order to impose the spherical constraint:
\begin{equation}
 	\hat{z}(0) \equiv \int_0^\beta d\tau\, z(\tau)~.
 \end{equation} 

From \eqref{eq:Sfinal} we readily derive the equations of motion in frequency space for the field $\hat{q}(k)$:
\begin{multline}
\label{eq:qhatEOM}
\text{EOM}_q:\qquad	M\left(\frac{2\pi k}{\beta}\right)^2-2 i\frac{\hat{z}(0)}{\beta}=\frac{\hat{q}(k)+\beta\,u(m-2)\delta_{k,0}}{\hat{q}(k)^2-\beta\,u\,\delta_{k,0}\left[(m-1)\beta\,u-(m-2)\hat{q}(k)\right]}\\+p\frac{\beta^2J^2}{2}\sum_{k_2,\dots,k_p}\frac{\hat{q}(k_2)\dots\hat{q}(k_p)}{\beta^p}\delta(k+k_2+\dots+k_p)~.
\end{multline}
This equation looks quite complicated, but we will see below that some clever field redefinitions simplify this expression significantly. 
 Varying with respect to $\hat{z}(0)$ gives the spherical constraint:  
\begin{equation}
\label{eq:EOMz}
\text{EOM}_z:\qquad \frac{1}{\beta}\sum_{k=-\infty}^\infty \hat{q}(k) = 1 \,,
\end{equation}
and varying with respect to $u$ and $m$ gives:
\begin{align}
\text{EOM}_u:\qquad{} &\frac{m-1}{2}\left(\frac{\beta^2J^2}{2}p\,u^{p-1}-\frac{u}{\left(\frac{\hat{q}(0)}{\beta}-u\right)\left(\frac{\hat{q}(0)}{\beta}+(m-1)u\right)}\right)=0 \label{eq:EOMu0}\\
\text{EOM}_m^\text{(equil.)}:\qquad &\frac{\beta^2J^2}{2}u^p+\frac{1}{m}\frac{u}{\frac{\hat{q}(0)}{\beta}+(m-1)u}+\frac{1}{m^2}\log\left[\frac{\frac{\hat{q}(0)}{\beta}-u}{\frac{\hat{q}(0)}{\beta}+(m-1)u}\right]=0 \label{eq:EOMm0}
\end{align}
We have written a superscript (equil.) on \eqref{eq:EOMm0}, whose meaning will become apparent in a later section.\footnote{The truly impatient reader should know that the conformal solution in the low temperature spin glass phase does not satisfy this equation. We will comment on this at length.}
 
Equations \eqref{eq:EOMu0} and \eqref{eq:EOMm0} are the standard (static) spin glass equations, that one can compare to equation (83) of \cite{castellani2005spin}. They have nontrivial solutions for $p\geq3$. Since the parameter $u$ defines an overlap, it is important that it takes values in $0<u<1$ (it can be thought of as the dot product between two normalized vectors). Not so trivial is the idea that when $n\rightarrow 0$, then the parameter $m$ also takes values in $0<m<1$. This can be motivated from a static stability analysis, but will also be evident when we solve the equations explicitly in later sections. Finally the normalized zero mode 
\begin{equation}
	\frac{\hat{q}(0)}{\beta}=\frac{1}{\beta}\int_0^\beta d\tau\, q(\tau)~
\end{equation}
appears in the above equations, but its $\beta$ dependence is misleading: since $q(\tau,\tau')$ is dimensionless, then this time average can only depend on the dimensionless quantities $\beta J$ and $MJ$.

\paragraph{Removal of explicit Lagrange multiplier:}
As anticipated, we can further simplify the equations of motion using some judiciously chosen field redefinitions.
We first rewrite \eqref{eq:qhatEOM} as follows:
\begin{equation}
\label{eq:EOMsigma}
	\frac{\hat{q}(k)+\beta\,u(m-2)\delta_{k,0}}{\hat{q}(k)^2-\beta\,u\,\delta_{k,0}\left[(m-1)\beta\,u-(m-2)\hat{q}(k)\right]}= M \left(\frac{2\pi k}{\beta}\right)^2-2 i   \frac{\hat{z}(0)}{\beta} - J^2 \, \hat{\Lambda}(k) \,,
\end{equation}
where we define the self energy: 
\begin{equation}
\label{eq:LambdaDef0}
   \boxed{\;\;\Lambda(\tau,\tau') \equiv \frac{p}{2}\, q(\tau,\tau')^{p-1}  \,, \;}
\end{equation}
whose Fourier transform is: 
\begin{equation} 
\hat{\Lambda}(k) = \frac{p}{2\beta^{p-2}} \sum_{k_1,\dots,k_{p-2}} \hat{q}(k_1)\dots\hat{q}(k_{p-2})\hat{q}(k-k_1-\dots-k_{p-2})~.
\end{equation}
A remarkable simplification happens \cite{cugliandolo2001imaginary} if we subtract off the constant spin glass parameters by defining
\begin{equation}
\label{eq:qrSigmar}
\begin{split}
   q_r(\tau,\tau') &= q(\tau,\tau') - u \,,\qquad \Lambda_r(\tau,\tau') = \Lambda(\tau,\tau') - \frac{p}{2}\, u^{p-1} \,,\\
   \hat{q}_r(k) &= \hat{q}(k) - \beta u \delta_{k,0} \,,\qquad\;
   \hat{\Lambda}_r(k) = \hat{\Lambda}(k) - \frac{p}{2} \beta u^{p-1} \delta_{k,0}
   \end{split}
\end{equation}
with boundary conditions $q_r(\tau,\tau) = 1-u$ and $\Lambda_r(\tau,\tau) = \frac{p}{2} (1-u^{p-1})$.
The equation of motion \eqref{eq:EOMsigma} then becomes very simple for all values of $k$:
\begin{equation}
\label{eq:EOMsigma2}
\text{EOM}_q:\qquad	\frac{1}{\hat{q}_r(k)}= M \left(\frac{2\pi k}{\beta}\right)^2-2   \frac{i\hat{z}(0)}{\beta} - J^2 \, \hat{\Lambda}_r(k) \,,
\end{equation}
where we used \eqref{eq:EOMu0} to cancel the terms proportional to $\delta_{k,0}$. Evaluating this for $k=0$, we can solve for the Lagrange multiplier $\hat{z}(0)$:
\begin{equation}
\label{eq:zSol}
 	\frac{-i\hat{z}(0)}{\beta}\equiv \frac{1}{2} \left[ \frac{1}{\hat{q}_r(0)} + J^2 \hat{\Lambda}_r(0)\right]=\frac{1}{2}\left[\frac{1}{\int_0^\beta\, d\tau\, q_r(\tau)}+J^2\int_0^\beta d\tau\, \Lambda_r(\tau)\right] ~ .
 \end{equation} 
 Combined with \eqref{eq:EOMsigma2} this gives a regularized and Lagrange multiplier-independent equation of motion:
\begin{equation}
\label{eq:EOMsigma3}
\boxed{ \;\;\text{EOM}_q:\qquad 	\frac{1}{\hat{q}_r(k)}-\frac{1}{\hat{q}_r(0)}= M \left(\frac{2\pi k}{\beta}\right)^2- J^2 \, (\hat{\Lambda}_r(k)-\hat{\Lambda}_r(0)  ) \, ,\;
 }\end{equation}
which is valid both in the paramagnetic phase ($u=0$) and in the spin glass phase ($u\neq 0$). Of course, for $k=0$ this equation is empty. Moreover, one is still required to impose the spherical constraint \eqref{eq:EOMz}.
As we will see, this equation is best suited for analytic continuation: it has no more terms proportional to $\delta_{k,0}$, and the dependence on the Lagrange multiplier is replaced by dependence on zero modes. 

We note here that it will often be useful to also write the self-energy term as a function of $q_r(\tau)$:
\begin{equation}
\label{eq:LambdaDefF}
  \hat{\Lambda}_r(k)-\hat{\Lambda}_r(0) = \frac{p}{2} \int_0^\beta d\tau \, \left[ \cos \left( \frac{2\pi k \tau}{\beta}  \right) -1 \right] \left[ \left( q_r(\tau)+u \right)^{p-1} - u^{p-1} \right] \,.
\end{equation}

Finally, in this parametrization, equations \eqref{eq:EOMu0} and \eqref{eq:EOMm0} become:
\begin{equation}
\label{eq:mvar1}
\boxed{
\begin{aligned}
 \text{EOM}_u: \qquad 0&=\frac{m-1}{2}\left[\frac{(\beta J)^2}{2}p\,u^{p-1}-\frac{u}{\frac{\hat{q}_r(0)}{\beta}\left(\frac{\hat{q}_r(0)}{\beta}+mu\right)}\right] \\
\;\;\;\text{EOM}_m^\text{(equil.)}: \qquad 0&=\frac{(\beta J)^2}{2}u^p+\frac{1}{m}\frac{u}{\frac{\hat{q}_r(0)}{\beta}+mu}+\frac{1}{m^2}\log\left[\frac{\frac{\hat{q}_r(0)}{\beta}}{\frac{\hat{q}_r(0)}{\beta}+mu}\right]\;
\end{aligned}
}
\end{equation}

\section{Analysis of equilibrium states}
\label{sec:EquilThermo}

In this section we analyze the equilibrium equations, \eqref{eq:EOMsigma3} and \eqref{eq:mvar1}. The conformal spin glass solution will be the main focus of this paper, but we relegate its analyis to section \ref{sec:MargThermo}. 
In order to obtain the solutions relevant to this section we rely heavily on a numerical analysis of the saddle point equation \eqref{eq:EOMsigma3}. This gives us a window into the physics of this model for a wide range of couplings. Details regarding the numerical implementation are given in appendix \ref{app:numerics}.

\subsection{The static spin glass transition: simplified analysis}\label{sec:statictrans}
A caveat is in order before we continue. While the \emph{conformal} spin glass solution studied in later sections is not a solution to the equilibrium equations of motion, this does not mean that there is no \emph{equilibrium} spin glass solution at low temperatures. In fact, the analysis in e.g. \cite{crisanti1992sphericalp,castellani2005spin} largely focuses on studying the equilibrium equations obtained from the 1-RSB effective action in the static limit. We turn to these equations now.   

The phenomenology of the equilibrium spin glass phase transition stems from a simple analysis of the equations for the 1-RSB parameters $u$, $m$. Indeed, consider the equations \eqref{eq:EOMu0} and \eqref{eq:EOMm0}. At high temperatures ($\beta J \ll 1$) the unique solution is given by $u=0$ and $m$ arbitrary. In order to infer the existence of a replica symmetry breaking spin glass solution at lower temperatures, we look for a solution with $u \neq 0$. The following relabeling is useful to elucidate the phases of the equilibrium equations:
\begin{equation}
	u\equiv \frac{\hat{q}(0)}{\beta}\tilde{u}~, \qquad {\beta^2J^2}\left(\frac{\hat{q}(0)}{\beta}\right)^p\equiv \tilde{\jmath}^2~,
	\label{eq:tildeVarsDef}
\end{equation}
which gives: 
\begin{align}
\text{EOM}_u: \quad 0&=\frac{m-1}{2}\left(p\,\frac{\tilde{\jmath}^2}{2}\tilde{u}^{p-1}-\frac{\tilde{u}}{\left(1-\tilde{u}\right)\left[1+(m-1)\tilde{u}\right]}\right)\,,\label{eq:utvar}\\
\text{EOM}_m^{\text{(equil.)}}:\quad 0&=\frac{\tilde{\jmath}^2}{2}\tilde{u}^p+\frac{1}{m}\,\frac{\tilde{u}}{1+(m-1)\tilde{u}}+\frac{1}{m^2}\log\left[\frac{1-\tilde{u}}{1+(m-1)\tilde{u}}\right]~.\label{eq:mvar2}
\end{align}

These equations are easily solved numerically for $\tilde{u}$ and $m$ as functions of $\tilde{\jmath}$ (for any given $p$). Translating back to our original variables: for fixed $\beta J$ and $p$, we obtain $u$ and $m$ as functions of the zero mode $\hat{q}(0)$. Since the zero mode can be expressed in terms of $\hat{q}(k \neq 0)$ (see \eqref{eq:EOMz}) we are then left with a single equation \eqref{eq:EOMsigma3} to solve using other techniques.

We can learn more about the qualitative structure of the model's phase diagram by analyzing equations \eqref{eq:mvar2} and \eqref{eq:utvar} for the RSB parameters $(m,\tilde{u})$ as a function of $\tilde{\jmath}$. They take the same form as in static analyses (the non-staticity is, of course, hidden in the field redefinition), so we can study them by similar means \cite{crisanti1992sphericalp,castellani2005spin,de2006random}. For small $\tilde{\jmath}<\tilde{\jmath}_{\rm crit}$ (corresponding to high temperatures), the solutions to the equations are $\tilde{u}=0$ and $m$ arbitrary. However, for $\tilde{\jmath}>\tilde{\jmath}_{\rm crit}$ (low temperatures), there is a spin glass solution with nonzero $\tilde{u}$. We will now discuss how to find the value of $\tilde{\jmath}_{\rm crit}$. 

\begin{figure}
\begin{center}
\includegraphics[width=0.44\textwidth]{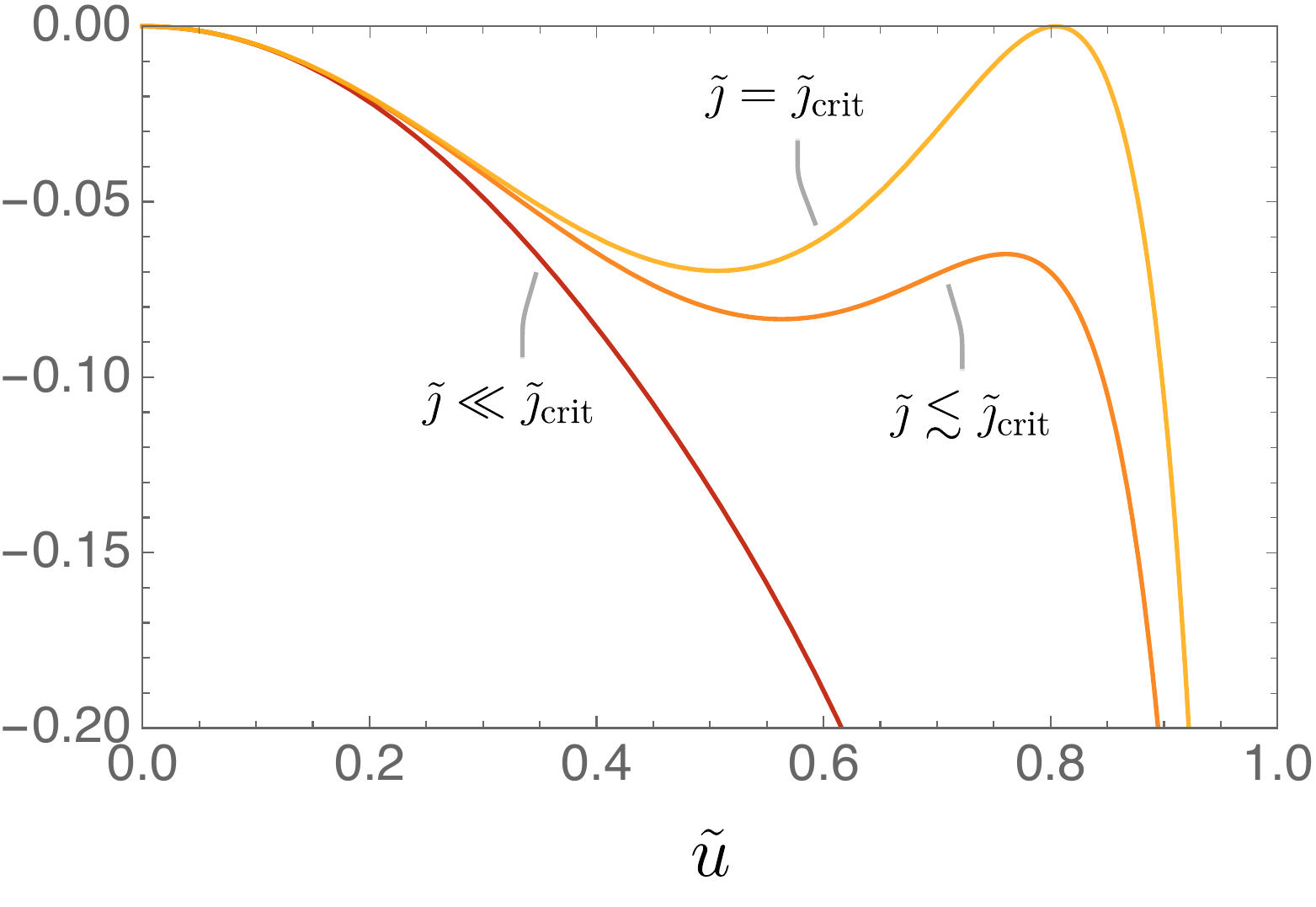}$\qquad$
\includegraphics[width=0.45\textwidth]{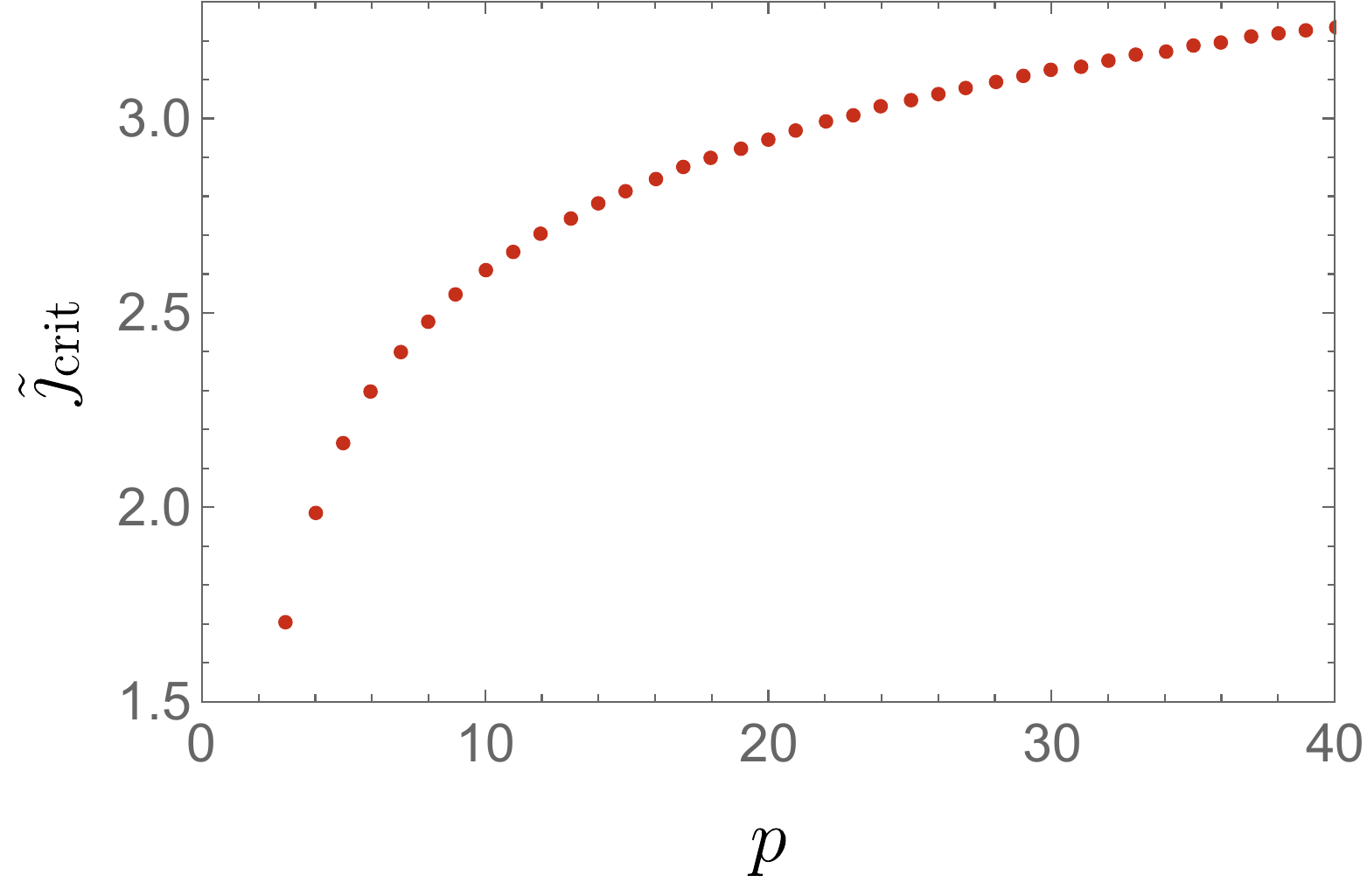}
\end{center}
\vspace{-.5cm}
\caption{\emph{Left:} Plot of the right hand side of \eqref{eq:remut}. Below $\tilde{\jmath}_{\rm crit}$ we see that the only solution is $\tilde{u}=0$. Still below the critical coupling, the curve develops a maximum. The critical value $\tilde{\jmath}_{\rm crit}$ is defined as the point where the maximum hits zero. We have displayed the case for $p=4$. \emph{Right:} Plot of $\tilde{\jmath}_{\rm crit}$ as a function of $p$.} \label{fig:critj}
\end{figure}

Studying the zero-locus of \eqref{eq:utvar} and \eqref{eq:mvar2} suggests that $m=1$ as we approach the phase transition from the high temperature regime, meaning $m=1$ at the critical point. The remaining equation is: 
\begin{equation}\label{eq:remut}
	0=\frac{\tilde{\jmath}^2}{2}\tilde{u}^p+\tilde{u}+\log\left[1-\tilde{u}\right]~,
\end{equation}
In figure \ref{fig:critj} we plot the right hand side of \eqref{eq:remut} as $\tilde{\jmath}$ is varied. 
Near the phase transition, the curve satisfying equation \eqref{eq:remut} develops a local maximum below zero. The phase transition can occur once $\tilde{\jmath}$ is tuned such that the maximum hits zero.

To find where this maximum is, we simply take a derivative of \eqref{eq:remut} with respect to $\tilde{u}$ and set it to zero. This gives us a parametric expression for $\tilde{\jmath}_{\rm crit}$ as a function of the interaction order $p$: 
\begin{equation}
 	\tilde{\jmath}_{\rm crit}^2=\frac{2\tilde{u}^{2-p}_\star}{p(1-\tilde{u}_\star)}~, \quad\quad (1-\tilde{u}_\star)[\tilde{u}_\star+\log(1-\tilde{u}_\star)]+\frac{u_\star^2}{p}=0 \,.
\end{equation} 
The nature of the equations \eqref{eq:mvar2} and \eqref{eq:utvar}  is that increasing $\tilde{\jmath}$ above $\tilde{\jmath}_{\rm crit}$ will push the value of $m$ to zero and hence the value of $\tilde{u}$ to 1. This guarantees that $\tilde{u}\geq\tilde{u}_\star$ once we've passed the phase transition.

This analysis of the critical line is oversimplified since we have hidden a dependence on $\hat{q}(0)$ in the coupling $\tilde{\jmath}$  (which in turn depends on $MJ$ and $\beta J$ in a complicated way). To fully understand the phases of the equilibrium model, we need to solve the equations of motion numerically.  We discuss these solutions and their detailed features next.

\subsection{Thermodynamics and phase diagram}
\label{sec:phases}

To study the thermodynamics, we start by evaluating the on-shell effective action. Using \eqref{eq:Sfinal}, the effective action can be written as:
\begin{align}
\label{eq:Sdivergent}
	\frac{S_{\rm eff}}{nN}=&-\frac{1}{2}\sum_{k \neq 0}\log \frac{\hat{q}(k)}{\beta} -\frac{1}{2}\left(1-\frac{1}{m}\right)\log \left[\frac{\hat{q}(0)}{\beta}-  u\right]-\frac{1}{2m}\log \left[\frac{\hat{q}(0)}{\beta}+ (m-1) u\right]\nonumber\\
	&-\frac{M}{2} \sum_{k=-\infty}^\infty  \left( \frac{2\pi i}{\beta} k \right)^2 \hat{q}(k)
	- \frac{(\beta J)^2 }{4} \left[ (m-1)u^p + \frac{1}{\beta} \int_0^\beta  d\tau\, q(\tau)^p \right]
\end{align}
This expression is naively UV divergent as it contains two separate infinite sums with high frequency divergences. 
To regularize it, we observe that the solution to the equation of motion \eqref{eq:EOMsigma3} for large $|k|$ is just 
\begin{equation}
\label{eq:asympSol}
   \frac{\hat{q}_r(k)}{\beta} \longrightarrow  \frac{\hat{q}_r^\text{UV}(k)}{\beta} \equiv \left[ \frac{M}{\beta} \left( 2\pi k \right)^2 + \frac{\beta}{\hat{q}_r(0)} + \beta J^2 \hat{\Lambda}_r(0) \right]^{-1} \equiv  \left[ \frac{M}{\beta} \left(2\pi k \right)^2 - 2i\hat{z}(0) \right]^{-1} \,,
\end{equation}
where $i\hat{z}(0)$ is real and was given in \eqref{eq:zSol}.
This is similar to the free solution ($J=0$) that we discuss in appendix \ref{sec:freeSol}. By adding and subtracting $S_{\rm eff}$ evaluated on the UV solution to and from \eqref{eq:Sdivergent}, we get the following {\it regularized effective action}:
\begin{equation}
\label{eq:freediv}
\begin{split}
\beta \phi \equiv \frac{S_{\rm eff}^{\rm reg}}{nN}&=-\frac{1}{2}\sum_{k}\log \left\{   \left[ \frac{M}{\beta} \left( 2\pi k \right)^2-2i\hat{z}(0) \right] \frac{\hat{q}_r(k)}{\beta}\right\}  
 +\frac{1}{2m}\log \left[\frac{\frac{\hat{q}_r(0)}{\beta}}{\frac{\hat{q}_r(0)}{\beta}+ m u}  \right]\\
&\quad -\frac{1}{2} \sum_k \left\{1-\left[ \frac{M}{\beta} \left( 2\pi k \right)^2-2i\hat{z}(0) \right]  \frac{\hat{q}_r(k)}{\beta} \right\}
	- \frac{(\beta J)^2 }{4} \left[ (m-1)u^p + \frac{1}{\beta} \int_0^\beta  d\tau\, q(\tau)^p \right] \\
&\quad + \frac{1}{2} \sum_k  \left\{ 1 + \log\left[ \frac{M}{\beta} \left( 2\pi k \right)^2-2i\hat{z}(0) \right] \right\} + i \hat{z}(0) \sum_k \frac{\hat{q}_r(k)}{\beta} 
\end{split}
\end{equation}
The sums in the first two lines are now convergent at large $|k|$. The evaluation of the divergent sum in the last line can be accomplished using standard $\zeta$-function regularization (see appendix \ref{app:regularization} for details). We find:
\begin{equation}
\label{eq:freeregularized}
\begin{split}
\beta \phi &=-\frac{1}{2}\sum_{k}\log \left\{  \left[ \frac{M}{\beta} \left( 2\pi k \right)^2-2i\hat{z}(0) \right]\frac{\hat{q}_r(k)}{\beta}  \right\}  
 +\frac{1}{2m}\log \left[\frac{\frac{\hat{q}_r(0)}{\beta}}{\frac{\hat{q}_r(0)}{\beta}+ m u}  \right]\\
 &\quad -\frac{1}{2} \sum_k \left\{1-\left[ \frac{M}{\beta} \left( 2\pi k \right)^2-2i\hat{z}(0) \right]  \frac{\hat{q}_r(k)}{\beta} \right\}
	- \frac{(\beta J)^2 }{4} \left[ m u^p + \frac{1}{\beta} \int_0^\beta  d\tau\, \left( q(\tau)^p - u^p \right)\right] \\
&\quad + \log \left[ 2 \sinh \left( \sqrt{\frac{-i\hat{z}(0) \beta}{2M}} \right) \right] +  i \hat{z}(0) \,(1-u)
\end{split}
\end{equation}

Later, when we discuss the conformal spin glass in section \ref{sec:MargThermo}, it will be important to distinguish $\phi$ from the usual thermodynamic free energy per unit site $\bar{f}$. But for now, this distinction is not important in the paramagnetic and the equilibrium spin glass phases, and the free energy is then given by the on-shell effective action as usual:
\begin{equation}
\label{eq:fEquilDef}
  \beta \bar{f} = \frac{S_{\rm eff}^{\rm reg}(Q_\star)}{nN} =\beta\phi \,.
\end{equation} 
where $Q_\star$ denotes evaluation on a solution to the equations of motion.
Again, this definition assumes that all parameters (including $m$) are determined by an extremization procedure.

From \eqref{eq:fEquilDef} we can derive other thermodynamic quantities:
\begin{equation}
\label{eq:ThermoDefs}
	\St =-\partial_T \bar{f}~, \qquad \Et=\partial_\beta (\beta \bar{f})~,\qquad \Ct=\partial_T \Et~,
\end{equation}
where $\St$ is the entropy, $\Et$ is the energy and $\Ct$ is the specific heat of the thermodynamic state per unit site. Using these relations, it is easy to see that we can obtain the entropy from: 
\begin{equation}
\label{eq:entropyFormula}
	\St= \beta \Et-\beta \bar{f}~.
\end{equation}
In computing $\Et$, the only derivatives that survive are with respect to the \emph{explicit} dependence on $\beta$. That is because terms such as: 
\begin{equation}
	\frac{\partial S_{\rm eff}}{\partial u} \frac{\partial u}{\partial T} 
\end{equation}
vanish because we have extremized $S_{\rm eff}$ with respect to $u$. The combination $\beta \bar{f}$ then depends on $\beta$ only through the dimensionless quantities $\beta J$ and $M/\beta$ and we can write $\beta \Et= J \partial_J(\beta \bar{f}) - M \partial_M (\beta \bar{f})$.

Repeated use of the equation of motion \eqref{eq:EOMsigma3}, the identity \eqref{eq:LambdaDefF}, and the first line of \eqref{eq:mvar1}, allows us to simplify the Matsubara sum and find a simple exact expression for the internal energy (see appendix \ref{sec:AppendixE} for details):
\begin{equation}
\label{eq:Eposition}
\begin{split}
   \beta \Et 
       &= 
     \frac{1}{2} \frac{\beta}{\hat{q}_r(0)}  + (\beta J)^2 \, \left\{ \frac{p}{4\beta } \int_0^\beta d\tau \, \left[ q(\tau)^{p-1} - u^{p-1}  \right] - \frac{p+2}{4} \, mu^p - \frac{p+2}{4\beta} \int_0^\beta  d\tau\, \left[q(\tau)^p - u^p \right] 
       \right\} 
\end{split} 
\end{equation}
where $\hat{q}_r(0) =  \int d\tau \, q(\tau) - \beta u$. This expression simplifies further when replica symmetry is unbroken ($u=0$). 

\begin{figure}
\begin{center}
\includegraphics[width=.503\textwidth]{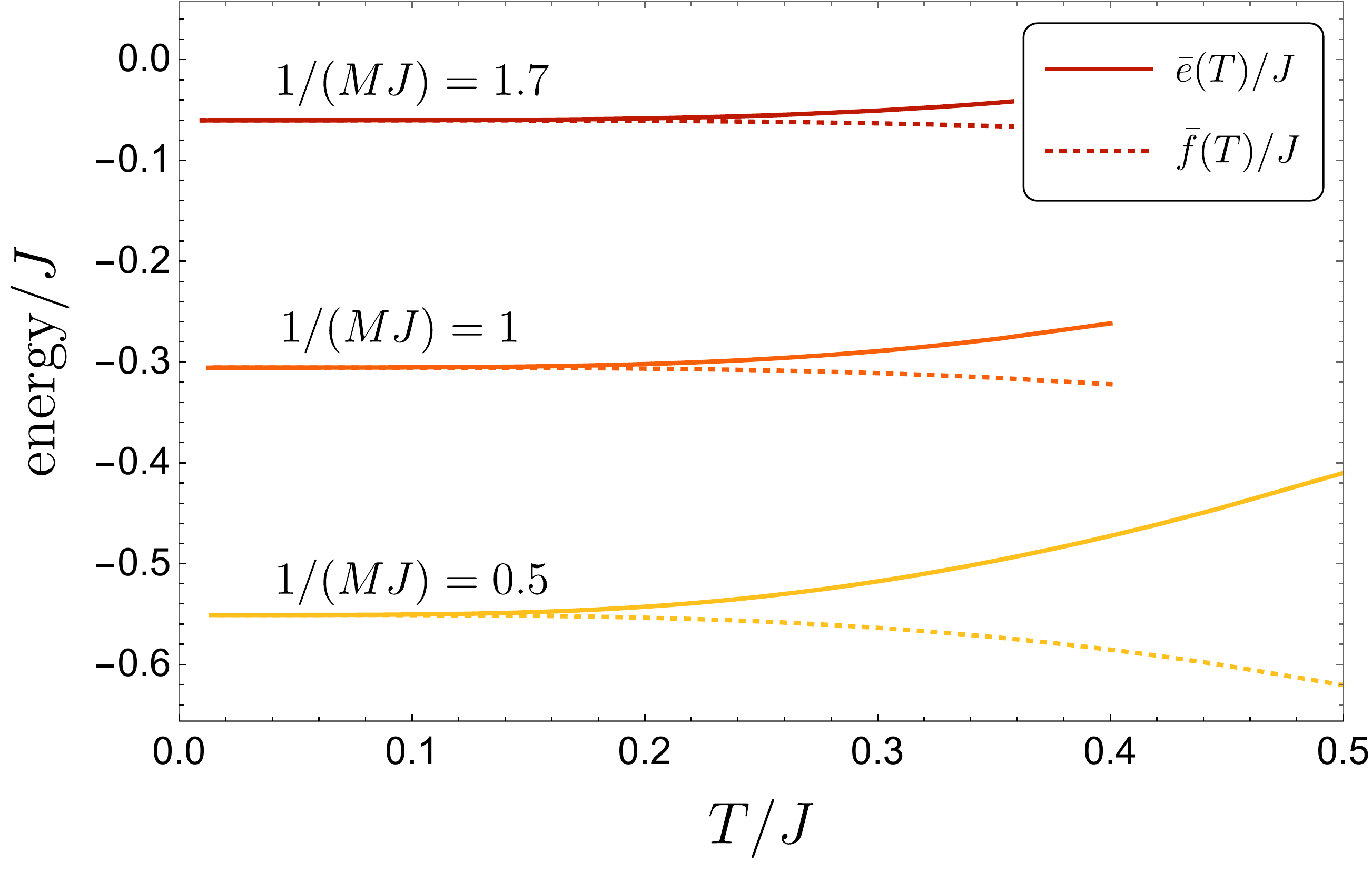}
\includegraphics[width=.49\textwidth]{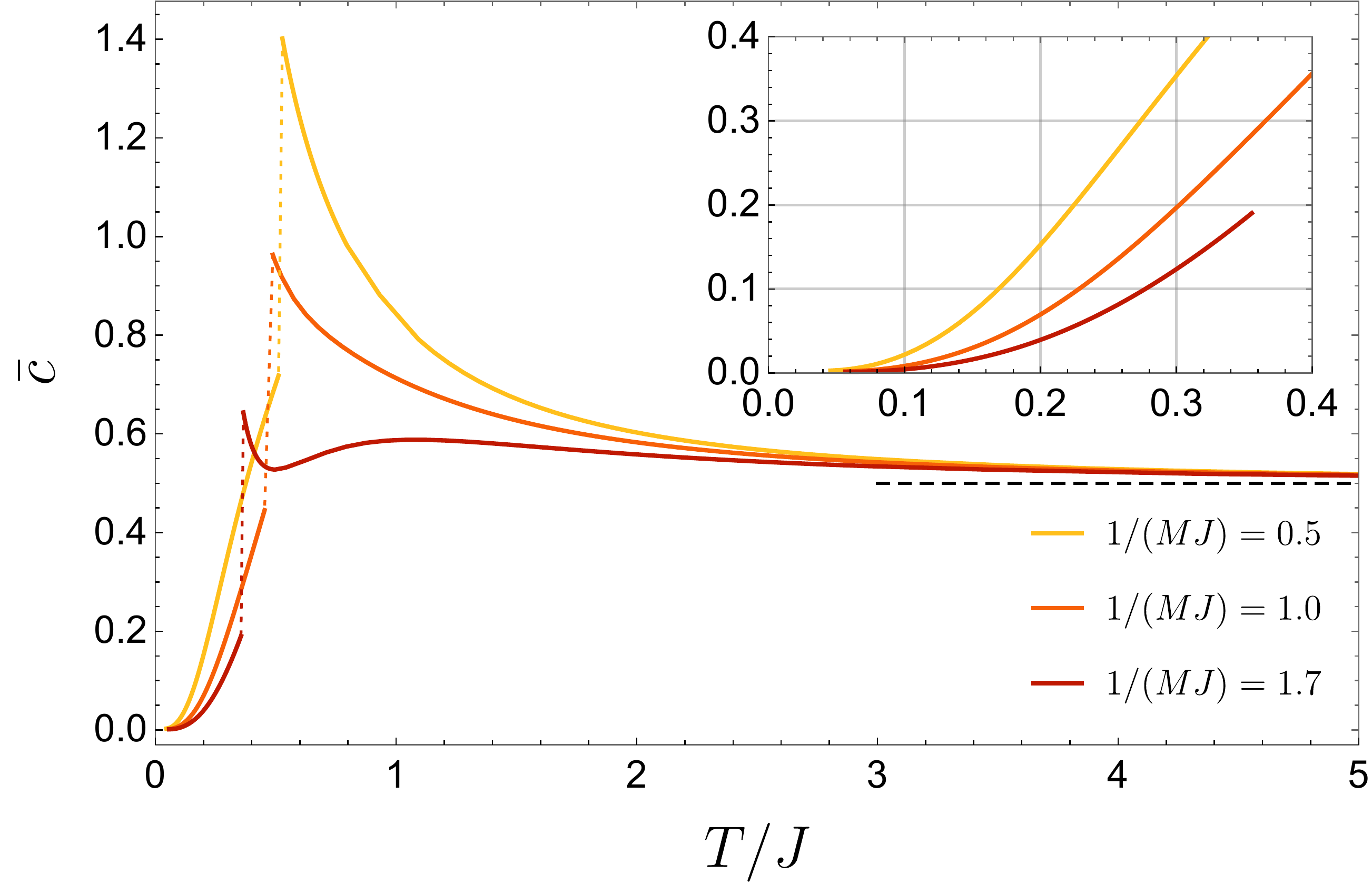}
\end{center}
\vspace{-.3cm}
\caption{{\bf Thermodynamics of the equilibrium spin glass and paramagnet (fixed $MJ$).} {\it Left:} Low-temperature thermodynamic energies of the equilibrium spin glass. The internal energies (upper curves) and free energies (lower curves) per spin approach the same constant values at zero temperature. Correspondingly, the entropy $\St=\beta \Et-\beta \bar{f}$ vanishes at zero temperature. {\it Right:} Specific heat for the equilibrium spin glass at low temperatures and for the paramagnetic solution at high temperatures. As $T\rightarrow \infty$ the specific heat approaches the universal value $1/2$. Compare also with the corresponding plots in case of the marginal spin glass, figure \ref{fig:lowTthermoMarg}.} \label{fig:lowTthermoEquil}
\end{figure}

\paragraph{High temperature limit:}
For small $\beta J$, the system is always in the paramagnetic phase and we can approximate the thermodynamic quantities using the `free' approximation, which is discussed in appendix \ref{sec:freeSol}. We find for the internal energy and entropy:
\begin{equation}
\begin{split}
  \beta \Et_0 &= \frac{\beta}{2\hat{q}_0(0)} + \ldots\,,\qquad\quad
 \St_0  =  \left\{\frac{\beta}{\hat{q}_0(0)} -\log \left[ 2 \sinh \left(  \frac{\beta}{2\sqrt{M \hat{q}_0(0)}} \right) \right] \right\}+ \ldots
\end{split}
\end{equation}
with $\hat{q}_0(0)/\beta$ determined by \eqref{eq:freeSols} such that this leading term in the entropy is purely a function of $M/\beta$. Specifically, for $\beta J \ll 1$ and fixed $MJ$, solving \eqref{eq:freeSols} perturbatively gives $\frac{\hat{q}_0(0)}{\beta} \sim 1 - \frac{\beta}{12M} + \ldots$. We can then immediately infer the high temperature asymptotics of thermodynamic quantities:
\begin{equation}
  \beta \bar{e}_0 \;\longrightarrow\; \frac{1}{2} \,,\quad \qquad \bar{c}_0 \;\longrightarrow \; \frac{1}{2}\qquad\quad (\beta J\rightarrow 0, \; MJ = \text{fix}) \,. 
\end{equation}
We confirm this behavior for the specific heat in figure \ref{fig:lowTthermoEquil}.

\paragraph{Phase diagram:}
We are now in a position to explain the phase diagram of the model based on free energy considerations of equilibrium states.

\begin{figure}
\begin{center}
\includegraphics[height=7cm]{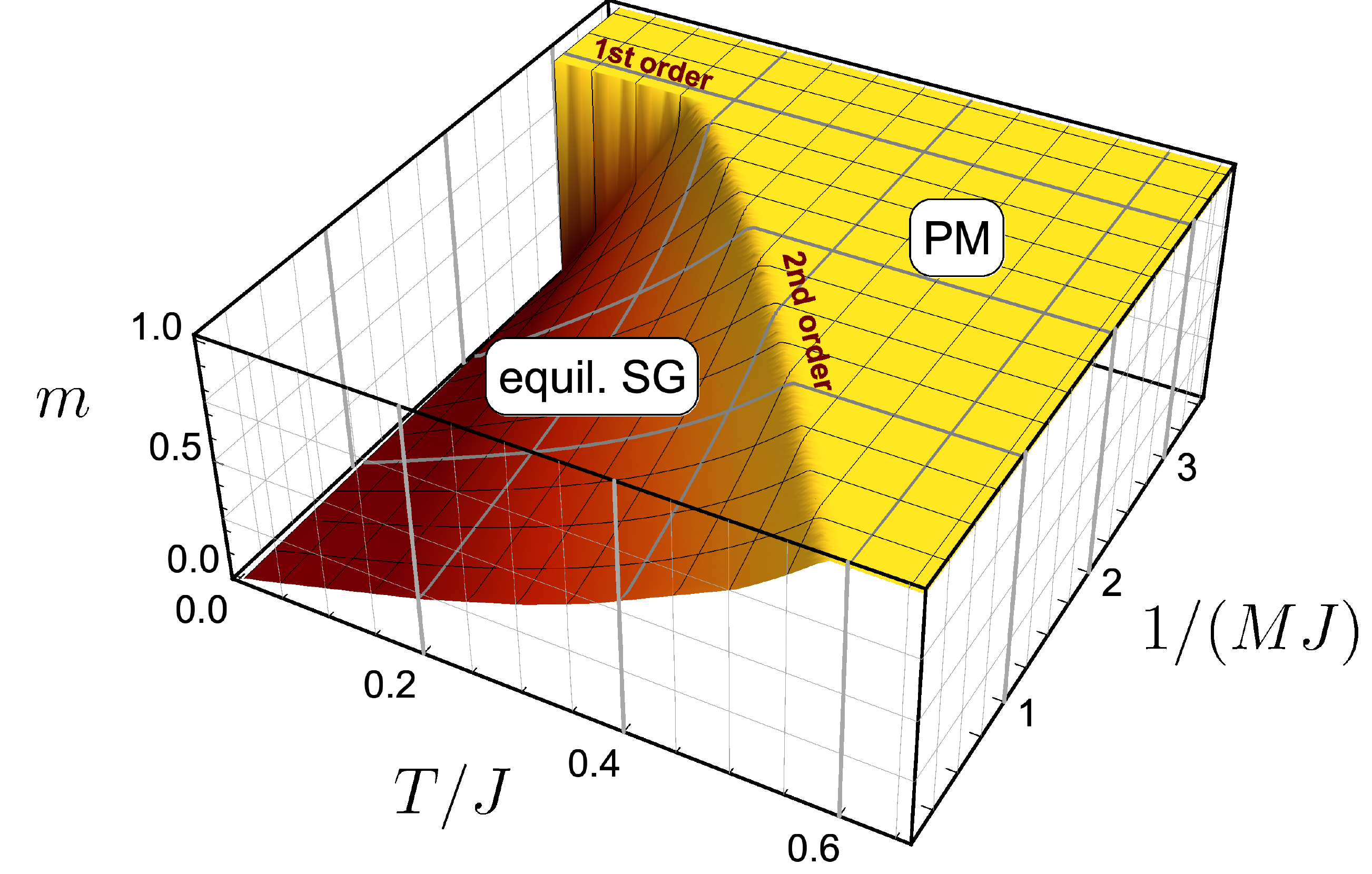}
\end{center}
\vspace{-.3cm}
\caption{{\bf Phase diagram for the equilibrium spin glass transition.} We show the value of $m$ as a function of temperature and $1/M$ (both measured in units of $J$) for $p=3$. The plateau where $m=1$ corresponds to the replica symmetric `paramagnetic' phase (PM). The spin glass phase (SG) is characterized by $m<1$. The parameter $m$ is continuous across the second order transition, but discontinuous across the first order line. The plot for the order parameter $1-u$ looks very similar, except for the fact that it is discontinuous across both the first and second order lines. A similar phase diagram for the marginally stable spin glass is shown in figure \ref{fig:PhasesMarg}.} \label{fig:Phases}
\end{figure}

Figure \ref{fig:Phases} shows the equilibrium phase diagram, where we discard unphysical solutions when necessary (meaning dropping subdominant thermodynamic solutions). We contrast this with the \emph{dynamic} phase transition which will be discussed in the next section (see figure \ref{fig:PhasesMarg}).
The structure of the solutions to the equations of motion giving rise to the static phase diagram is as follows:
\begin{itemize}
\item For  temperatures $T \gtrapprox 0.19 J$ (for $p=3$), there exist two possible solutions: a paramagnetic solution (where $m=1$ and $u=0$) and a spin glass solution depending on the value of $MJ$. They are separated by a second order phase transition as $MJ$ is tuned. The crossover between the phases is characterized by a continuous (discontinuous) change in $m$ ($u$) across the phase transition line. Our numerical analyses in the following sections will mostly focus on this region. 
  \item For temperatures $T \lessapprox 0.19 J$, there can exist up to three relevant solutions depending on the value of $MJ$: two paramagnetic ones and a replica symmetry breaking solution. For most values of $MJ$ in this regime, one of the paramagnetic solutions can be discarded as it is continuously connected to an unphysical ground state.\footnote{ However, for a very narrow range of temperatures both paramagnetic solutions are relevant and exchange dominance before the system enters the spin glass phase, see \cite{cugliandolo2001imaginary} for a detailed discussion.} In order to obtain the phase diagram in figure \ref{fig:Phases} we compared the free energies of the spin glass solution vs.\ the relevant paramagnetic one. Both solutions coexist in a small region surrounding a first order transition line, but said line demarcates the location where the two solutions exchange free energy dominance. The parameters $m$ and $u$ are both discontinuous across this line.
\end{itemize}

\begin{figure}
\begin{center}
\includegraphics[width=.48\textwidth]{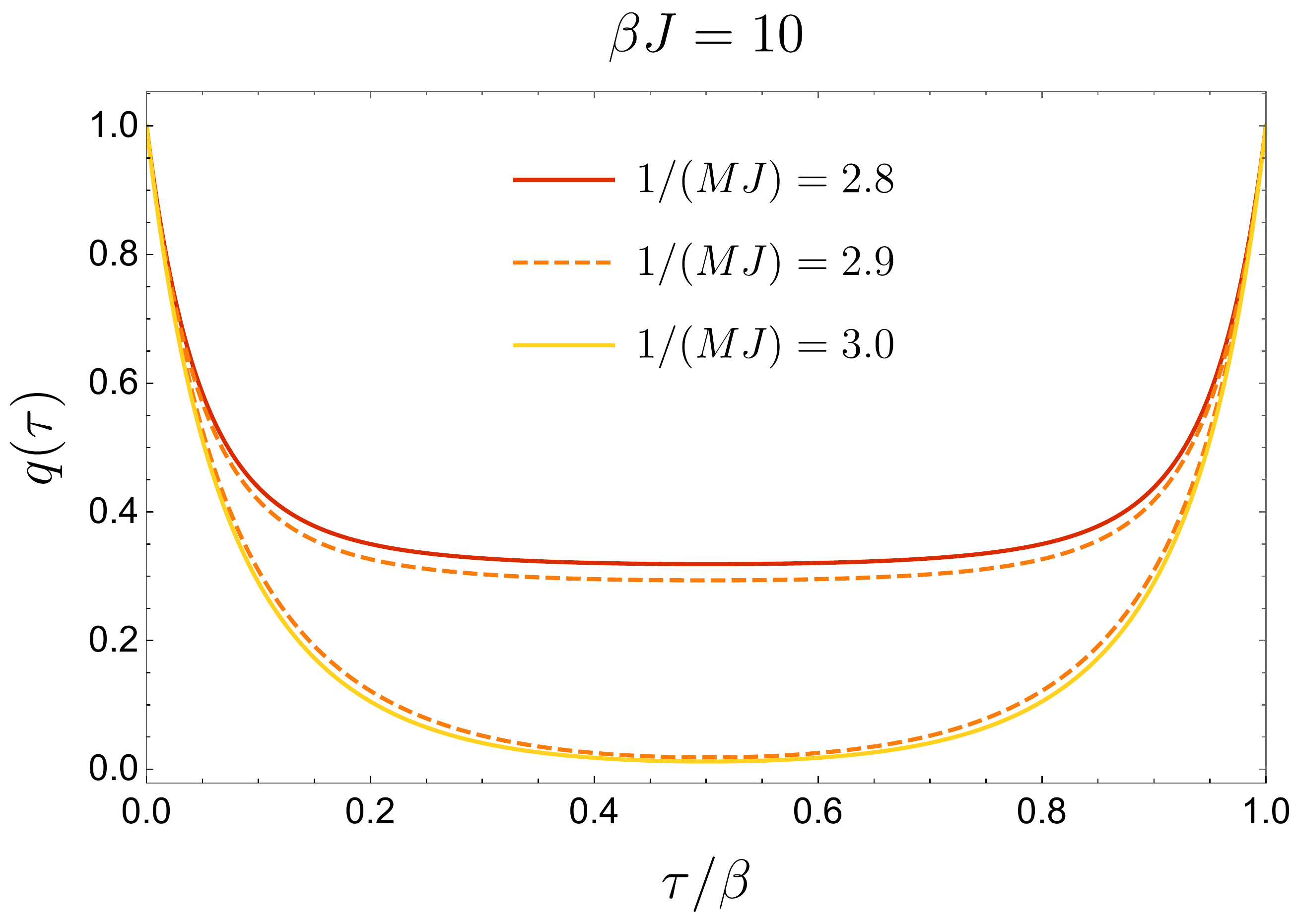}
$\;\;$
\includegraphics[width=.48\textwidth]{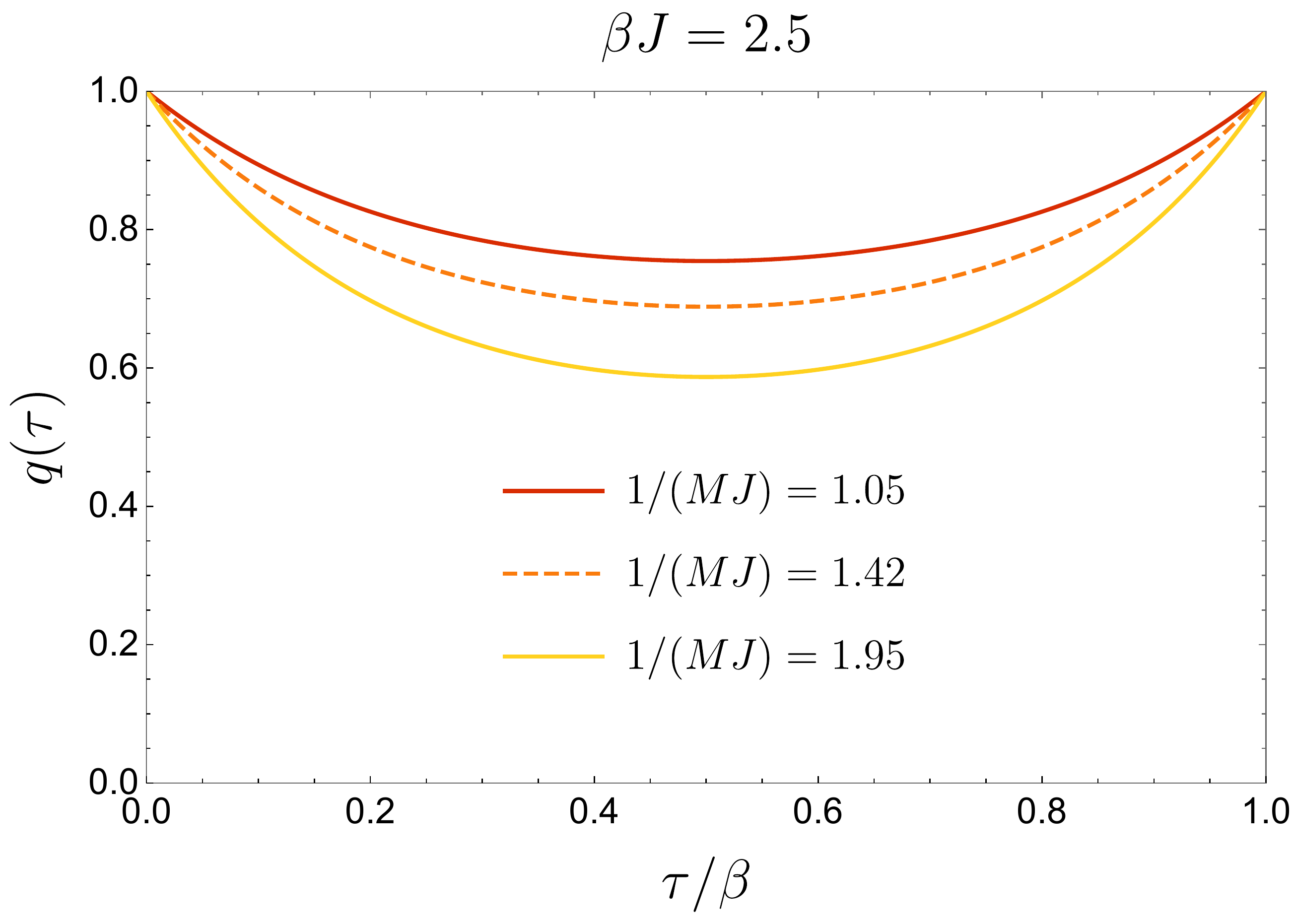}
\end{center}
\vspace{-.3cm}
\caption{{\bf Two-point functions at the phase transition.} {\it Left:} At $T/J = 0.1$ a first order transition occurs at $1/(MJ) \approx 2.9$. For $1/(MJ) < 2.9$ the thermodynamically preferred phase is the spin glass with $u>0$, while for $1/MJ< 2.9$ the paramagnetic phase is preferred. The dashed orange lines show the existence of solutions from both the spin glass and the paramagnetic branch at the critical value; here, the extremization of free energy suggests a first oder transition between these two types of solutions (both continue to exist into the other phase). {\it Right: } At $T/J = 0.4$ a second order phase transition occurs at $1/(MJ) \approx 1.42$. The spin glass and the paramagnetic solutions continuously morph into one another. The transition is second order and the parameter $m$ is continuous.} \label{fig:Transition}
\end{figure}

\begin{figure}
\begin{center}
\includegraphics[height=5.5cm]{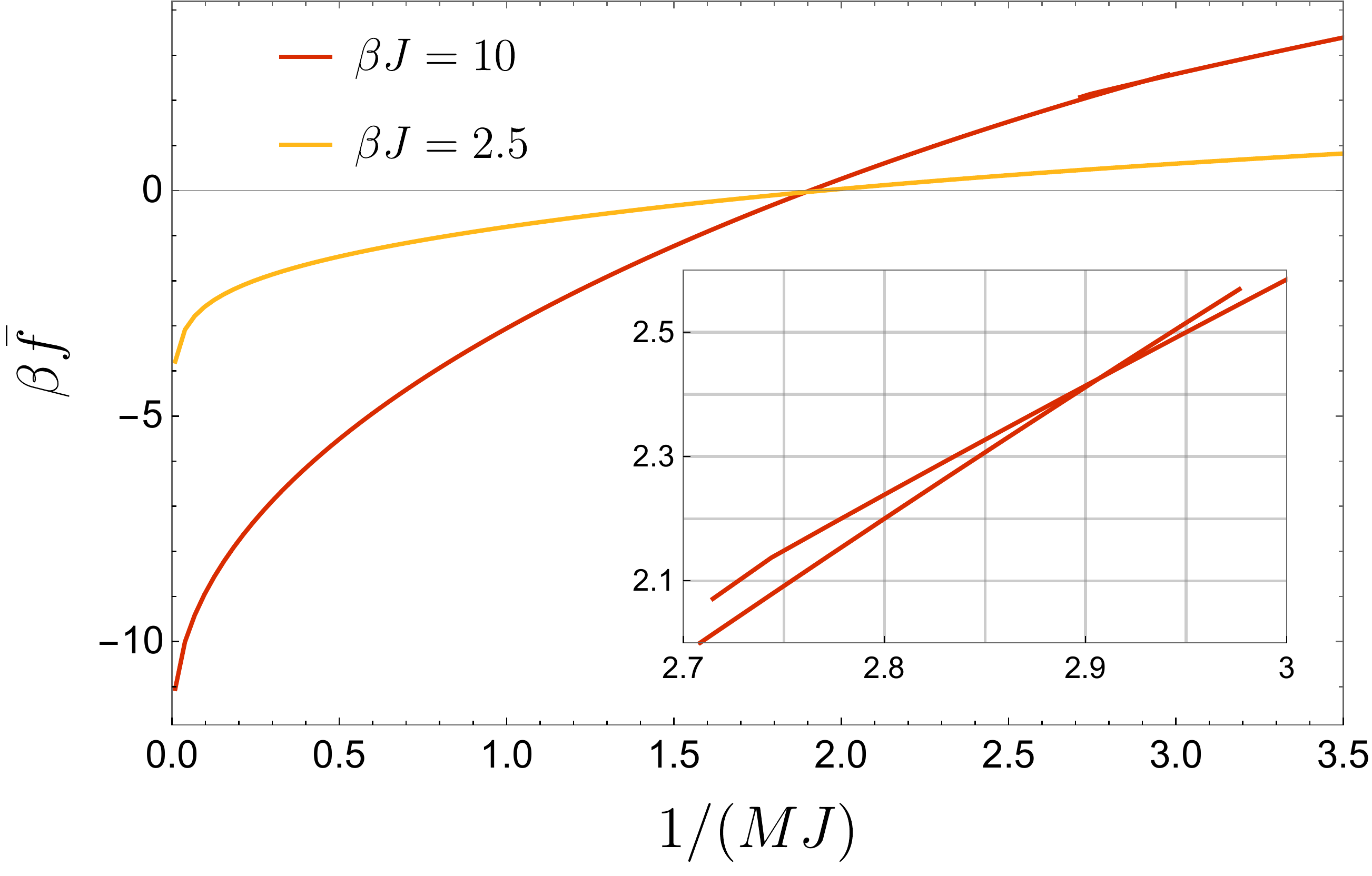}
\end{center}
\vspace{-.3cm}
\caption{{\bf Free energy across first and second order equilibrium spin glass transitions.} Free energies as a function of $1/(MJ)$ at the same temperatures as in figure \ref{fig:Transition}. For the low temperature case the inset shows the crossing of free energies of the coexisting paramagnetic and spin glass solutions near the location of the first order phase transition, $1/(MJ) \approx 2.9$. For the high temperature case the transition is second order.} \label{fig:FreeEnergy}
\end{figure}

Figure \ref{fig:Transition} shows properties of the Euclidean equilibrium solutions near the first and second order transitions (for $p=3$). In the left panel we illustrate the coexistence of two types of solutions near the first order transition at low temperatures: the branches of paramagnetic and spin glass solutions extend into the respective other phase, but they exchange thermodynamic dominance at a value of $1/(MJ) \approx 2.9$.
This is also illustrated by the corresponding free energies at these temperatures, shown in figure \ref{fig:FreeEnergy}.

\paragraph{Low temperature thermodynamics:}
To study the low temperature thermodynamics, we fix the value of $MJ$ and compute various thermodynamic quantities as a function of $\beta J$.
Figure \ref{fig:lowTthermoEquil} shows the free energy $\bar{f}$ and the internal energy $\Et$, both in units of $J$ and as functions of temperature $1/(\beta J)$. We can see that the two energies approach the same finite value at zero temperature. The entropy $\St=\beta \Et-\beta \bar{f}$ therefore vanishes at zero temperature. Furthermore, note that both $\Et/J$ and $\bar{f}/J$ are monotonic and have vanishing slope at zero temperature. This is an indication that the 1-RSB ansatz is thermodynamically consistent. For comparison, in other spin glass models, such as the Sherrington-Kirkpatrick model \cite{PhysRevLett.35.1792}, the free energy of the 1-RSB ansatz becomes non-monotonic at low temperatures. This pathological behavior is then fixed by introducing $k$-step RSB. The solution valid all the way to $T=0$  requires $k\rightarrow\infty$. In the model we study here, this is not necessary.

\section{The conformal spin glass phase}
\label{sec:MargThermo}

One of the most interesting discoveries in \cite{cugliandolo2001imaginary} was that of an exact analytic solution to \eqref{eq:EOMsigma3} that exhibits a power-law decay at large-Euclidean time separation and low temperature. The decay power was found to be 2, which we would like to associate to an operator of conformal dimension $\Delta=1$. We will review the derivation of this solution below, but before doing so, we must explain a key feature of this solution. That being that this solution does not satisfy the equilibrium equations of motion for the variable $m$.  

The historical argument, found in \cite{cugliandolo2001imaginary}, is that the equilibrium criterion used to set $m$, \eqref{eq:mvar1}, can be replaced in cases where the dynamics are important. This observation is based on the analysis of \cite{kirkpatrick1987p,horner1992dynamics,horner1992dynamics2,crisanti1993sphericalp,PhysRevLett.71.173} where it was observed that in dynamical simulations, the parameter $m$ settles instead on configurations where there exists a vanishing eigenvalue among the fluctuations in the replica directions.  Similar observations were made in \cite{Georges_2000} for the Heisenberg spin glass model. We review these replica fluctuations in appendix \ref{ap:statfluct} and we direct readers interested in the details to that section, particularly the analysis leading to \eqref{eq:mvarMarg}, which gives the criterion leading to a vanishing eigenvalue in the replica directions. Since parameters are chosen such that the transverse eigenvalue in replica space $\lambda_T^{\rm diag}=0$, this state is sometimes called the \emph{marginal} spin glass state. We will use the terms marginal and conformal interchangeably when referring to this particular thermodynamic state. 

Our perspective is different. Since the conformal solution is present when $m$ fails to satisfy the equilibrium condition \eqref{eq:EOMm0}, perhaps it is best to think of it as an external parameter which we can tune. One simple way to achieve this in the $p$-spin model, following \cite{monasson1995structural,mezard1999compute} (see \cite{zamponi2010mean} for an elucidating explanation), is to \emph{explicitly} clone the system $m$ times at the outset, before disorder averaging, meaning we are interested in the following grand-canonical free energy: 
\begin{equation}
	\beta m\overline{\Phi}= -\int dJ_{i_1\ldots i_p} P(J_{i_1\ldots i_p}) \log Z[J_{i_1\ldots i_p}]^m~, \label{eq:cloneddisorder}
\end{equation}
and $m$ is a parameter that we may tune as we please. To deal with this power of $m$, we will first assume it is an integer and replicate the system $m$ times by introducing a replica index $a=1,\dots,m$:
\begin{multline}
	Z[J_{i_1\ldots i_p}]^m=\int D\sigma_i^a Dz^a \,\exp\Bigg\{-\int_0^\beta d\tau \Bigg[\frac{M}{2}\dot{\sigma}^a_i(\tau)\dot{\sigma}^a_i(\tau)+\sum_{i_1 <  \ldots < i_p } J_{i_1\dots i_p}\sigma^a_{i_1}(\tau)\dots\sigma^a_{i_p}(\tau)\Bigg]\\+i\int_0^\beta d\tau\, z^a(\tau)\left(\sigma^a_i(\tau)\sigma^a_i(\tau)-N\right)-\epsilon\sum_{a,b=1}^m\sum_i\sigma^a_i\sigma^b_i\Bigg\}~.
\end{multline}
where we have introduced a small coupling $\epsilon$ in order to \emph{explicitly} break the replica symmetry, such that the $m$ replicas are pinned to the same thermodynamic state. In order to compute the disorder-averaged grand-canonical free energy $\overline{\Phi}$, we use the replica trick once more to again rid ourselves of the $\log$ by noting that: 
\begin{equation}
	\log Z^m= \lim_{n\rightarrow0}\partial_n Z^{n\times m}~. 
\end{equation}
In practice, this can be achieved by extending the range of the replica index $a=1,\dots,m\times n$. In doing so, we must make sure that the small coupling $\epsilon$ only pins replicas in a single $m\times m$ block to the same state. It is worth noting that such an ensemble can be physically realized without clones by coupling our system to a pair of baths such as in \cite{contucci2020stationarization}.

The remainder of the steps in calculating $\overline{\Phi}$ follow those in section \ref{sec:model} verbatim. Now, however, the 1-RSB solution is the result of the explicit, rather than spontaneous, breaking of the replica symmetry induced by the small coupling $\epsilon$, which we take to zero at the end of the computation. In this ensemble, $m$ acts as a bias, similarly to the external magnetic field in the Ising ferromagnet, which picks out a particular thermodynamic state. The consequence of this, is that we no longer have to consider the equation of motion for $m$, and may set it to whatever pleases us. 

 The final result is:
 \begin{equation}
 	\beta m\overline{\Phi}= \beta m \phi
 \end{equation}
 where $\phi$ is the regularized free energy \eqref{eq:freediv}. In this ensemble, $m$ is a free external parameter, much like the inverse temperature $\beta$ and we will describe its thermodynamic dual observable in the discussion below. Since the effective action in this ensemble is the same up to an overall factor of $m$, all the equations of motion remain the same. Thus we are now free to consider the same equations as before, but can happily ignore the equilibrium condition for $m$, by dropping \eqref{eq:EOMm0} from our consideration. 

 We recall that $m$ drops out of all equations when the Edwards-Anderson parameter $u$ is zero, so we are only sensitive to the value of $m$ in the spin glass phase and it only truly makes sense to consider tuning it once we are beyond the spin glass phase transition. 

 \subsection{Conformal solution to the equations of motion}
 \label{sec:confSG}
 
 We will now derive the aforementioned conformal solution explicitly. The original derivation in \cite{cugliandolo2001imaginary} is presented in a way that makes the emergent time-reparametrization symmetry of the equations of motion manifest. In fact, in this section we will derive two important analytical solutions: one which we denote as the `approximate' solution $q_r^\approx$, which solves the equations for the marginally stable spin glass approximately for both short and long times (but nevertheless in some perturbative scheme), and a `conformal' scaling solution $q_r^c$, which captures the long Euclidean-time limit of $q_r^\approx$.
 
 To do this, let us analyze \eqref{eq:EOMsigma3} in position space: 
\begin{equation}\label{eq:EOMfull}
	-\left[ \delta(\tau,\tau')-\frac{q_r(\tau,\tau')}{\hat{q}_r(0)}\right]=M \partial_\tau^2 q_r(\tau,\tau')+{J^2} \int_0^\beta d\tau''\Lambda_r(\tau,\tau'')\left[q_r(\tau'',\tau')-q_r(\tau,\tau')\right]~,
\end{equation}
with
\begin{equation}
	\Lambda_r(\tau,\tau') \equiv \frac{p}{2}\left[\left(q_r(\tau,\tau')+u\right)^{p-1} - u^{p-1}\right]~,
\end{equation}
and we recall 
\begin{equation}
	\hat{q}_r(0)=\int_0^\beta d\tau \, q_r(\tau)~.
\end{equation}
Let us now expand \eqref{eq:EOMfull} for small $q_r(\tau)\ll u$. Since $q_r(\tau)$ interpolates between $1-u$ and zero, this approximation is valid if $u\approx 1$, deep in the spin glass phase. This approximation leads to: 
\begin{multline}
\label{eq:qExpandRR}
	-\left[ \delta(\tau,\tau')-\frac{q_r(\tau,\tau')}{\hat{q}_r(0)}\right]=M \partial_\tau^2 q_r(\tau,\tau')\\
	+{\mathcal{J}^2}u^{p-2} \int_0^\beta d\tau''\left[q_r(\tau,\tau'')+\frac{p-2}{2u}q_r(\tau,\tau'')^2+\dots\right]\left[q_r(\tau'',\tau')-q_r(\tau,\tau')\right]~,
\end{multline}
where we have defined a new coupling:  
\begin{equation}
	\mathcal{J}\equiv J\sqrt{\frac{p(p-1)}{2}}~.\label{eq:calJdef}
\end{equation}

It was noted in \cite{cugliandolo2001imaginary} that a conformal solution exists when there is a vanishing transverse eigenvalue in the replica directions. In practice, this means looking at the quadratic fluctuations in the off-diagonal components of replica matrix, and characterizing the spectrum of these fluctuations (see appendix \ref{ap:statfluct}). There is a particular eigenvalue $\lambda_T^{\rm diag}$, corresponding to a particular set of \emph{transverse} fluctuations, which can be made to vanish if $u$ and $\hat{q}_r(0)$ take the following values (see equation \eqref{eq:mvarMarg}):
\begin{equation}
\boxed{\;\;
	u=\left[\frac{p-2}{\beta \mathcal{J} m}\right]^{2/p}~,\qquad \frac{\hat{q}_r(0)}{\beta}=\frac{m u}{p-2}=\left(\frac{m }{p-2}\right)\left[\frac{p-2}{\beta \mathcal{J} m}\right]^{2/p} \;}~.\label{eq:mvarMargMain}
\end{equation}
We can plug these values into the $u$ equation of motion \eqref{eq:EOMu0} and see that it is satisfied. However, these relations fail to satisfy the equilibrium equation for $m$ \eqref{eq:EOMm0}. As explained above, we take this as an indication that this solution exists in a different ensemble than the one previously considered. Equation \eqref{eq:mvarMargMain} implies a very interesting identity, relating the order parameter to the zero mode:
\begin{equation}\label{eq:conformalmagicrel}
	\mathcal{J}^2u^{p-2}=\left(\hat{q}_r(0)\right)^{-2}
\end{equation}
which will allow us to further simplify the equations of motion in this phase. Moreover, this relation allows us to trade the zero mode with the effective coupling, letting the model forget its intrinsic IR scale, and allowing for the conformal solution to exist. 

If we drop the subleading terms in the small $q_r(\tau)$ expansion and use the magic relation \eqref{eq:conformalmagicrel}, the equation of motion becomes: 
\begin{multline}
	-\left[ \delta(\tau,\tau')-\frac{q_r(\tau,\tau')}{\hat{q}_r(0)}\right]=M \partial_\tau^2 q_r(\tau,\tau')+\left(\hat{q}_r(0)\right)^{-2}\int_0^\beta d\tau''q_r(\tau,\tau'')\left[q_r(\tau'',\tau')-q_r(\tau,\tau')\right]~.
\end{multline}
We can massage this expression, and use translation invariance to write it as
\begin{equation}
-M\partial_\tau^2 q_r(\tau,\tau')_=\int_0^\beta d\tau''\left[\delta(\tau,\tau'')-\frac{q_r(\tau,\tau'')}{\hat{q}_r(0)}\right]\left[\delta(\tau'',\tau')-\frac{q_r(\tau'',\tau')}{\hat{q}_r(0)}\right]\,.~\label{eq:eommassaged}
\end{equation}
In momentum space, this equation of motion takes the following very simple form:
\begin{equation}
\label{eq:approxEOM0}
	M\omega^2\hat{q}_r(\omega)=\left(\frac{\hat{q}_r(\omega)-\hat{q}_r(0)}{\hat{q}_r(0)}\right)^2~.
\end{equation}
The solution to this equation is: 
\begin{equation}
	\frac{\hat{q}^{\approx}_r(\omega)}{\hat{q}_r(0)}=1+2\gamma^2\omega^2-2\sqrt{\gamma^2\omega^2+\gamma^4\omega^4}~, \qquad \gamma\equiv\sqrt{\frac{M\hat{q}_r(0)}{4}}~,\label{eq:exactsolfourier}
\end{equation}
which shares many features with the two point function in the random rotor model studied in \cite{PhysRevB.52.384,Cheng:2019nxy,Mao:2019xvt}. 
To get intuition about what we will do shortly, let us expand the solution \eqref{eq:exactsolfourier} for small frequencies: 
\begin{equation}
	\frac{\hat{q}^{\approx}_r(\omega)}{\hat{q}_r(0)}=1-2\gamma|\omega|+\dots
\end{equation}
Fourier transforming we see: 
\begin{equation}
	q^\approx_r(\tau)=\hat{q}_r(0)\delta(\tau)+\frac{A}{\tau^2}+\dots\qquad\qquad A\equiv \frac{8\gamma^3}{M\pi}\label{eq:Asol}
\end{equation}
which leads us to interpret the term subleading to the $\delta$-function contact term as a `conformal'  solution when viewed from the perspective of the low-frequency expansion. At finite temperature, this conformal approximation takes the form 
\begin{equation}
\label{eq:qrcSG}
  q_r^c(\tau) = A \left( \frac{\pi}{\beta \, \sin \left( \frac{\pi \tau}{\beta} \right) } \right)^2 \,.
\end{equation}

\subsubsection{Symmetries of the deep spin glass equations} \label{sec:diffs1}
As we have come to appreciate, the appearance of a conformal solution at low temperatures is typically accompanied by the presence of soft modes associated to a certain symmetry breaking pattern \cite{KitaevTalks,Kitaev:2017awl,Maldacena:2016hyu}. Thus we are instructed to identify the symmetries preserved by the low energy equations of motion.  This intuition will guide the analysis of our EOM, which we will rewrite as:
\begin{equation}\label{eq:eommassaged1}
-4\gamma^2\partial_\tau^2 \left(\frac{q_r(\tau,\tau')}{\hat{q}_r(0)}\right)=\int_0^\beta d\tau''\left[\delta(\tau,\tau'')-\frac{q_r(\tau,\tau'')}{\hat{q}_r(0)}\right]\left[\delta(\tau'',\tau')-\frac{q_r(\tau'',\tau')}{\hat{q}_r(0)}\right]~.
\end{equation}
If we take \eqref{eq:Asol} as guidance and remove the zero mode,
\begin{equation}
  q_r(\tau,\tau') \equiv \hat{q}_r(0)\delta(\tau,\tau') + x_r(\tau,\tau') \,,
\end{equation}
this turns \eqref{eq:eommassaged1} into
\begin{equation}
4\gamma^2\partial_\tau\partial_{\tau'}  \left[\delta(\tau,\tau')+\frac{x_r(\tau,\tau')}{\hat{q}_r(0)}\right]=\mathcal{J}^2u^{p-2}\int_0^\beta d\tau''\,x_r(\tau,\tau'')x_r(\tau'',\tau')~,\label{eq:eommassaged2}
\end{equation}
where we have used \eqref{eq:conformalmagicrel}.
Due to the derivatives of the delta-function on the left hand side, we notice that this equation does not transform nicely under time reparametrizations of the bilocal field $x_r(\tau,\tau')$. Let us nevertheless define a field $y_r(\tau,\tau')$ such that:\footnote{ We thank J.\ Maldacena for this suggestion.}
\begin{equation}
	x_r(\tau,\tau')\equiv\left(\partial_\tau-\partial_{\tau'}\right)y_r(\tau,\tau')~,
\end{equation}
then \eqref{eq:eommassaged2}, to lowest order in derivatives, becomes: 
\begin{equation}
-  \delta(\tau,\tau')=\frac{\mathcal{J}^2u^{p-2}}{\gamma^2}\int_0^\beta d\tau''\,y_r(\tau,\tau'')y_r(\tau'',\tau')~.\label{eq:eommassaged3}
\end{equation}
The same equation of motion governs the paramagnetic phase of the $q=2$ SYK model. Under the diff$(S^1)$ transformation
\begin{equation}
	\tau\rightarrow f(\tau)~, \qquad y_r(\tau_1,\tau_2)\rightarrow \left[f'(\tau_1)f'(\tau_2)\right]^\Delta y_r(f_1,f_2)~, 
\end{equation}
the equation of motion transforms to: 
\begin{align}
&-\left[f'(\tau_1)f'(\tau_2)\right]^{1/2}\delta(f_1,f_2)=\frac{\mathcal{J}^2u^{p-2}}{\gamma^2}\left[f'(\tau_1)f'(\tau_2)\right]^\Delta\int_{f(0)}^{f(\beta)} df_3(f'(\tau_3))^{2\Delta-1}\,y_r(f_1,f_3)y_r(f_3,f_2)~, 
\end{align}
and we notice that the equation of motion is invariant for $\Delta=1/2$. In short, the model exhibits an invariance under diff$(S^1)$ at low energies, as exhibited by the equations of motion in the marginal spin glass phase. However, the bilocal field of interest $x_r$, which measures the response of the microscopic spins $\sigma_i$ does not transform as the two-point function of primary operators. Instead, the conformal solution we obtain is:
\begin{equation}
	y_r(\tau,\tau')=-\frac{4\gamma^3}{M\pi}\frac{1}{|\tau-\tau'|}~, 
\end{equation}
and under a general diff$(S^1)$ transformation, this becomes
\begin{equation}
\label{eq:saddleReparam}
	y_r\longrightarrow-\frac{4\gamma^3}{M\pi}\frac{f'(\tau)^{1/2}f'(\tau')^{1/2}}{|f(\tau)-f(\tau')|}~. 
\end{equation}
The transformation of the physical field $x_r$  is induced from that of $y_r$. In this sense, we should think of $x_r$ as computing a correlation function involving descendant fields.

\subsubsection{Properties of the approximate solution}\label{sec:qApprox}

We now discuss the `approximate solution', which we call $q_r^\approx$, in more detail. This solution interpolates smoothly between a scaling solution at large imaginary-time separations, and a `free' UV solution for short times. This approximate solution will be the most complete analytical solution to the equations of motion and it is a powerful tool for gaining precise insights into the spin glass dynamics.

\paragraph{Zero temperature approximate solution.}
The approximate solution $\hat{q}_r^\approx(\omega)$ is defined at zero temperature by \eqref{eq:exactsolfourier}, which we repeat here:
\begin{equation}
	\hat{q}^\approx_{r,\beta=\infty}(\omega)=\hat{q}_r(0)\left(1+2\gamma^2\omega^2-2\sqrt{\gamma^2\omega^2+\gamma^4\omega^4}\right)\,,\qquad\quad
  \gamma \equiv \sqrt{\frac{M \hat{q}_r(0)}{4}} \,.
\end{equation}
This has the nice property of interpolating between the constant $\hat{q}_r(0)$ (given in \eqref{eq:mvarMargMain}) for small $\omega$, and $\hat{q}^{\text{UV}}_r(\omega) \equiv (M\omega^2)^{-1}$ for large $\omega$.

At zero temperature, we can readily Fourier transform to position space, similarly to \cite{Anninos:2017cnw,Tikhanovskaya:2020elb}, which yields:
\begin{equation}
\label{eq:qApproxSol}
\boxed{\;\;
  q_{r,\beta=\infty}^\approx(\tau) 
  = \frac{4\gamma}{M} \left\{ \frac{2}{3\pi} + \frac{\gamma}{|\tau|} \left[ \mathbf{L}_2\left( \frac{|\tau|}{\gamma}  \right) - {I}_2\left(\frac{|\tau|}{\gamma}   \right) \right]\right\}  
    \;\;
   }
\end{equation}
where $\mathbf{L}_\nu(x)$ and $I_\nu(x)$ are the modified Struve and Bessel functions, respectively, and we have dropped contact terms.
For large $\tau$, the approximate solution unsurprisingly approaches the scaling solution, $q_{r,\beta=\infty}^\approx \rightarrow \frac{A}{\tau^{2}}$ with $A$ as in \eqref{eq:Asol}. Moreover, as we will see, this approximate solution is a remarkably good approximation to the exact numerical solutions even for small $\tau$.

\paragraph{Finite temperature corrections.}
While the above solution was obtained for zero temperature, we can define a corresponding finite temperature solution by the Poisson-resummation formula, known more colloquially as the sum over images: 
\begin{equation}
	q^\approx_r(\tau)\equiv\sum_{n=-\infty}^\infty q_{r,\beta=\infty}^\approx(\tau+n\beta)~.
\end{equation}
The spherical constraint can be imposed on this approximate solution consistently. It requires:
\begin{equation}
	1-u=q^\approx_r(0)=\sum_{n=-\infty}^\infty q_{r,\beta=\infty}^\approx(n\beta) \,,
\end{equation}
which can be solved perturbatively in $1/(\beta J)$.
We can get the leading finite temperature correction by noting that for large argument \cite{abramowitz1988handbook}: 
\begin{equation}
	\mathbf{L}_\nu\left( z  \right) - {I}_{-\nu}\left( z\right)\approx\frac{1}{\pi}\sum_{k=0}^\infty\frac{(-1)^{k+1}\Gamma\left(k+\frac{1}{2}\right)}{\left(\frac{z}{2}\right)^{2k-\nu+1}\Gamma\left(\nu+\frac{1}{2}-k\right)}~, 
\end{equation}
leading to 
\begin{equation*}
	1-u=\frac{8\gamma}{3 M\pi}\left(1+\sum_{n=1}^\infty\frac{6\gamma^2}{n^2\beta^2}+\dots\right)=\frac{8\gamma}{3 M\pi}\left(1+\frac{\pi^2\gamma^2}{\beta^2}+\dots\right)~.
\end{equation*}

The Edwards-Anderson parameter $u$ is determined as a function of the thermodynamic variables $(m, \beta J, MJ)$ according to \eqref{eq:mvarMargMain}, so this equation can be interpreted as a global condition on $m$ such that we have a consistent solution that obeys the UV spherical constraint. Plugging in the expression for $u$ from \eqref{eq:mvarMargMain} tells us that, to approach the marginal spin glass phase, $m$ must be tuned to the following value: 
\begin{equation}\label{eq:mlowtSG}
 	m=\frac{p-2}{\beta \mathcal{J}}u_*^{-p/2}\left(1+\frac{1}{(\beta \mathcal{J})^2} \,\frac{8p\, u_*^{2-p}}{9(1-u_*)\left[(p+2)u_*-(p-2)\right]}+\dots\right)
 \end{equation} 
 where we suppress terms of order $(\beta {\cal J})^{-4}$ and $u_*$ is determined by 
 \begin{equation}\label{eq:ustardef}
 	0=1-u_*-\frac{4}{3\pi}\frac{u_*^{\frac{2-p}{4}}}{\sqrt{M \mathcal{J}}}~. 
 \end{equation}
 This simply says that $m$ must approach zero linearly in $T$ as the temperature is lowered. We conclude that the marginal spin glass is consistent with a scaling limit for large $\beta {\cal J}$.
 Via \eqref{eq:mvarMargMain} we also obtain the low temperature expansion of $u$,
\begin{equation}
\label{eq:ulowtSG}
	u= u_*-\frac{16}{9(\beta \mathcal{J})^2u_*^{p-3}(1-u_*)\left[(p+2)u_*-(p-2)\right]}+\dots \,,
\end{equation}
which approaches $u_*$ quadratically in $T$ as $T\rightarrow 0$. We also use \eqref{eq:mvarMargMain} to calculate the zero mode: 
 \begin{equation}\label{eq:qr0lowtempexp}
 	\frac{\hat{q}_r(0)}{\beta}=\frac{m u}{p-2}=\frac{u_*^{1-p/2}}{\beta \mathcal{J}}\left(1+\frac{1}{(\beta \mathcal{J})^2} \,\frac{8(p-2) u_*^{2-p}}{9(1-u_*)\left[(p+2)u_*-(p-2)\right]}+\dots\right)
 \end{equation}
 and we see that $\frac{1}{\beta}\int_0^\beta d\tau\, q_r(\tau) $ is tending to zero linearly in $T$.

\begin{figure}
\begin{center}
\includegraphics[width=0.6\textwidth]{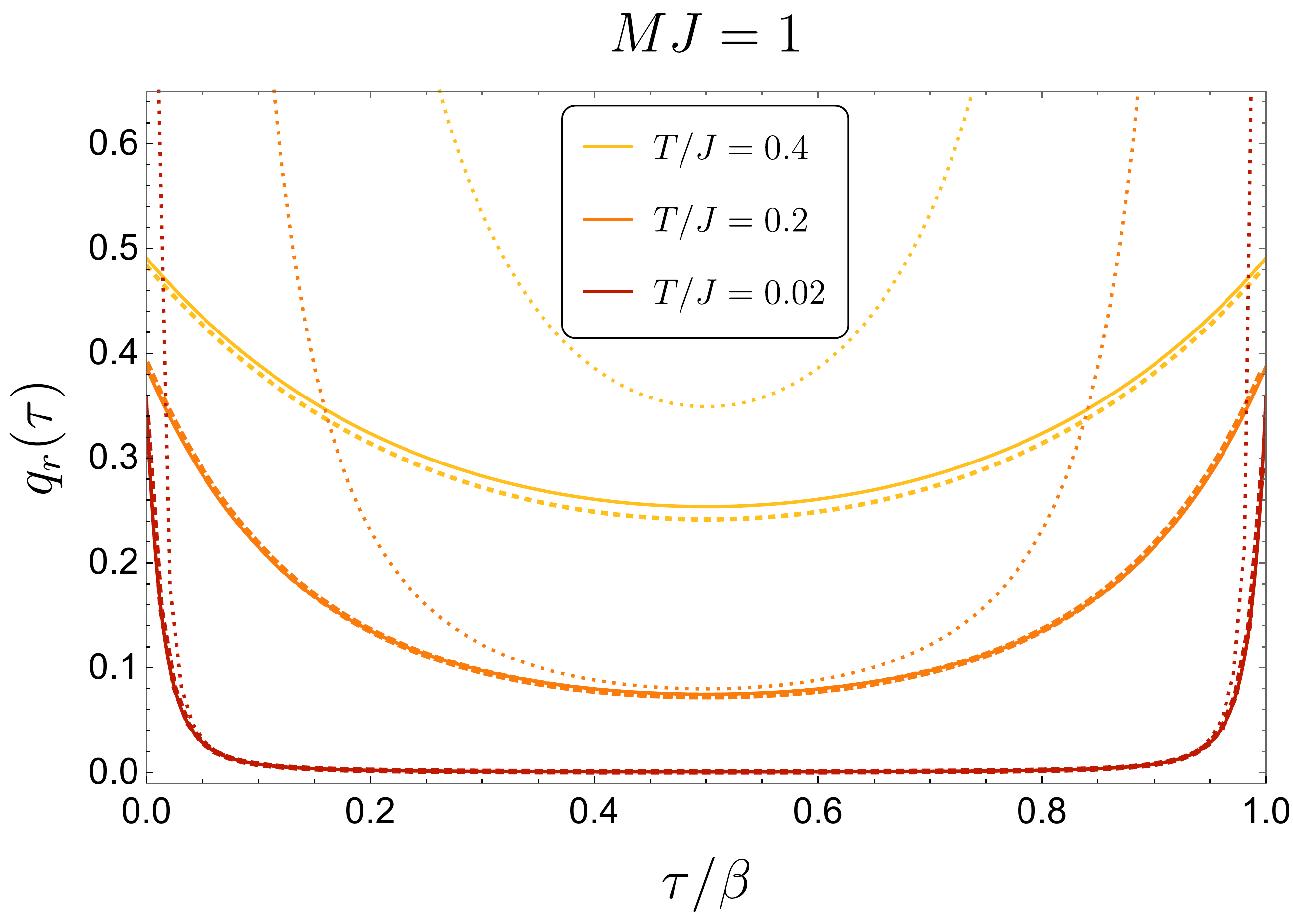}
\end{center}
\vspace{-.5cm}
\caption{{\bf Marginally stable spin glass solutions and analytical approximations.} Full lines are the exact numerical solutions at various temperatures for fixed $MJ = 1$. Thick dashed lines are the approximate solution $q^\approx_r$ with thermodynamic parameters determined to first subleading order in the $1/(\beta J)$ expansion (for small temperatures these become essentially indistinguishable from the exact solutions). Thin dotted lines are the conformal solution $q_{r}^c$ (which diverges at short distances).} \label{fig:MarginalComparison}
\end{figure}

\paragraph{Spectral representation:}
It is useful for later to define a zero-temperature spectral representation $\rho_r(\omega)$ of the approximate solution. At zero temperature:
\begin{equation}
	q_{r,\beta=\infty}^\approx(\tau)=  \int_{0}^{1/\gamma} \frac{d\omega}{2\pi}   \, \rho^\approx_{r}(\omega)  \, e^{-\omega |\tau|}  
	\,,\qquad \quad 
      \rho^\approx_{r}(\omega) = \frac{16\gamma^2}{M} \, \gamma\omega\sqrt{1-\gamma^2\omega^2} \,.
\end{equation}

As before, the finite temperature solution can be written as a sum over thermal images:
\begin{align*}
	q^\approx_r(\tau)&=\sum_{n=-\infty}^{\infty}\int_{0}^{1/\gamma} \frac{d\omega}{2\pi}   \,  \rho^\approx_{r}(\omega)  \, e^{-\omega |\tau+n\beta|} 
	=\int_{0}^{1/\gamma} \frac{d\omega}{2\pi}   \, \rho_{r}^\approx(\omega)\, \frac{\cosh\left[\left(|\tau|-\frac{\beta}{2}\right)\omega\right]}{\sinh\left(\frac{\beta\omega}{2}\right)}~, 
\end{align*}
which is valid for both positive and negative $\tau$. Defining $\omega=\frac{1}{\gamma}\sin\theta$, we can write this as
\begin{equation}
\label{eq:qApproxFinal}
\boxed{\;\;
	q^\approx_r(\tau)=\frac{8\gamma}{M\pi}\int_0^{\frac{\pi}{2}}d\theta\,\cos^2\theta\, \sin\theta\, \frac{\cosh\left[\left(\frac{|\tau|}{\gamma}-\frac{\beta}{2\gamma}\right)\sin\theta\right]}{\sinh\left(\frac{\beta}{2\gamma}\sin\theta\right)} \;\;}
\end{equation}
which gives us the following late time expansion of the thermal correlation function: 
\begin{multline}
	q_r^\approx(\tau)\approx \frac{8\pi\gamma^3}{M\beta^2\sin^2\left(\frac{\pi\tau}{\beta}\right)}\Bigg[\left(1+\frac{2\pi^2\gamma^2}{\beta^2}-\frac{2\pi^4\gamma^4}{\beta^4}+\dots\right)\\
	-\frac{3\pi^2\gamma^2}{\beta^2\sin^2\left(\frac{\pi\tau}{\beta}\right)}\left(1-\frac{5\pi^2\gamma^2}{\beta^2}+\dots\right)-\frac{15\pi^4\gamma^4}{\beta^4\sin^4\left(\frac{\pi\tau}{\beta}\right)}\left(1-\dots\right)+\dots\Bigg]~.
\end{multline}
The representation \eqref{eq:qApproxFinal} is very efficient for explicit evaluations. It also leads to immediate consistency checks. For example, we recognize the first term in the expansion as the conformal result. Further, one can readily show from this expression that the zero mode is correctly reproduced:
\begin{equation}
	\int_0^\beta d\tau \,q^\approx_r(\tau)=\frac{16\gamma^2}{M\pi}\int_0^{\frac{\pi}{2}}d\theta\, \cos^2\theta=\frac{4\gamma^2}{M}=\hat{q}_r(0)~.
\end{equation}
Let us emphasize that our finite temperature expression is not exact, and gets a number of corrections coming from adding terms neglected in the derivation of the conformal solution. At any finite temperature, the true scaling exponent differs from the precise value of $\Delta=1$ found earlier \cite{gotze2008complex}. 

In figure \ref{fig:MarginalComparison} we display a few marginally stable spin glass solutions $q_r(\tau)$, obtained numerically, and compare them with the approximate solution \eqref{eq:qApproxFinal} and with its conformal long-time limit \eqref{eq:qrcSG}. This figure illustrates that $q^\approx_r$ is an extraordinarily good approximation both at short and at long distances. We can improve the match with the exact numerical results even further if, instead of determining $\gamma$ from the low-temperature values \eqref{eq:qr0lowtempexp} etc., we tune $m$ to the ideal value to obtain the best possible match -- doing so makes the approximate solution essentially indistinguishable from the exact solution for a very wide range of temperatures and couplings $MJ$. We also display the conformal solutions $q_r^c$, which provide a reasonable approximation for $0 \ll \tau \ll \beta$ and sufficiently small temperatures.

 \subsubsection{Perturbative double expansion and `quantum scaling'}
 \label{sec:QSdef} 
 
 We can use the approximate solution $q^\approx_r$ to solve for the prameters $u, m, \hat{q}_r(0)$ perturbatively. To this end, recall that initially we made the assumption that $q_r \ll u$, which holds for large values of $MJ$. Let us therefore consider a double expansion for large $\beta {\cal J}$ and large $M{\cal J}$.
When $M{\cal J} \gg 1$, the value of $u_*$ can be determined iteratively from \eqref{eq:ustardef}:
 \begin{equation}\label{eq:uexpansion}
 u_*=1-\frac{4}{3\pi\sqrt{M\mathcal{J}}}-\frac{p-2}{4}\left(\frac{4}{3\pi\sqrt{M\mathcal{J}}}\right)^2\\-\frac{(p-2)(3p-2)}{32}\left(\frac{4}{3\pi\sqrt{M\mathcal{J}}}\right)^3+\dots
 \end{equation}
 In this double expansion, the time-independent parameters \eqref{eq:mlowtSG}, \eqref{eq:ulowtSG}, and \eqref{eq:qr0lowtempexp} take the following form:
 \begin{equation}
 \begin{split}
 	m&=\frac{p-2}{\beta\mathcal{J}}\left(1+\frac{2p}{3\pi \sqrt{M\mathcal{J}}}+\dots\right)+\frac{p(p-2)}{6(\beta \mathcal{J})^3}\left(\pi\sqrt{M\mathcal{J}}+\frac{2}{3}(3p-2)+\dots\right)+\dots \\
	u&=\left(1-\frac{4}{3\pi\sqrt{M\mathcal{J}}}+\dots\right)-\frac{1}{3(\beta\mathcal{J})^2}\left(\pi\sqrt{M\mathcal{J}}+\frac{4}{3}(p-2)+\dots\right)+\dots\\
	\frac{\hat{q}_r(0)}{\beta}&=\frac{1}{\beta\mathcal{J}}\left(1+\frac{2(p-2)}{3\pi \sqrt{M\mathcal{J}}}+\dots\right)+\frac{(p-2)}{6(\beta \mathcal{J})^3}\left(\pi\sqrt{M\mathcal{J}}+\frac{2}{3}(3p-4)+\dots\right)+\dots
\end{split}
 \end{equation}
 In order for this double expansion to converge we need to make sure that the terms  $\frac{(M\mathcal{J})^{n-1/2}}{ (\beta\mathcal{J})^{2n+1}}$ remain small, and diminish in size as $n$ is increased (in particular, $\beta {\cal J} \gg \sqrt{M{\cal J}} \gg 1$). To ensure this, we will take what we call the \emph{quantum scaling}: 
 \begin{equation}
 \label{eq:QuantScale}
\boxed{\;\; \beta \mathcal{J} = \Lambda^2 \; \widehat{\beta \mathcal{J}}~,\qquad M\mathcal{J}= \Lambda^2\; \widehat{M\mathcal{J}} \qquad \text{with} \qquad \Lambda \gg 1~, \quad\widehat{\beta \mathcal{J}}\sim\widehat{M\mathcal{J}}\sim {\cal O}(1) \,. \;\; }
 \end{equation}
Working perturbatively in $1/\Lambda$, we find:
 \begin{equation}
 \label{eq:muqhatQuantum}
 \begin{split}
 m&=\frac{1}{\Lambda^2} \, \frac{p-2}{\widehat{\beta \mathcal{J}}}+\frac{1}{\Lambda^3} \, \frac{2p(p-2)}{3\pi \, \widehat{\beta \mathcal{J}}\sqrt{\widehat{M \mathcal{J}}}}
 +\frac{1}{\Lambda^4} \, \frac{4p^2(p-2)}{9\pi^2\, \widehat{\beta \mathcal{J}}\,\widehat{M \mathcal{J}}}
 + \dots\\
 u&=1-\frac{1}{\Lambda} \, \frac{4}{3\pi\sqrt{\widehat{M\mathcal{J}}}}-\frac{1}{\Lambda^2} \, \frac{4(p-2)}{9\pi^2\widehat{M\mathcal{J}}}+\dots\\
 \frac{\hat{q}_r(0)}{\beta}&=\frac{1}{\Lambda^2} \, \frac{1}{\widehat{\beta\mathcal{J}}}+\frac{1}{\Lambda^3} \, \frac{2(p-2)}{3\pi \widehat{\beta \mathcal{J}}\sqrt{\widehat{M \mathcal{J}}}}+\dots
 \end{split}
 \end{equation}
In this scaling $u$ is parametrically close to 1, meaning that $q_r(\tau)$ is very small since it interpolates between $1-u$ and 0. This explains why $\frac{\hat{q}_r(0)}{\beta}$ is getting parametrically small in this limit. 

While we can be certain that the derivation of the approximate analytical solution holds in the quantum scaled regime, we often find that evaluating the analytical expressions using the exact in $MJ$ expressions \eqref{eq:mlowtSG}-\eqref{eq:qr0lowtempexp} match the numerics for a wide range of temperatures within the spin glass phase. As such all numerical versus analytical comparisons are made in this way.

\subsection{The transition to a marginally stable spin glass}
In order for the analysis of the marginal-spin glass state to be consistent, we must be at sufficiently low temperatures such that $u\neq0$, meaning that above a certain temperature, the marginal spin glass state ceases to exist and our definition of the cloned ensemble is no longer consistent. Thus we must identify the region in coupling space where the marginal spin glass state exists and identify the nature of the transition between the paramgnetic and the marginal spin glass. In order to do so, we repeat the analysis of section \ref{sec:statictrans}, but replace the equation for $m$ with the condition of vanishing replicon eigenvalue \eqref{eq:vanishinglambdaT} obtained in appendix \ref{ap:statfluct}. 
We will again use the rescaled dimensionless variables \eqref{eq:tildeVarsDef}. 

Let us reiterate the perspective here, since it is perhaps not exactly correct to call this a phase transition, per se. In the marginal spin glass phase, we are \emph{tuning} $m$ such that we have a vanishing transverse eigenvalue $\lambda_T^{\rm diag}$ among fluctuations in the replica directions (appendix \ref{ap:statfluct}). To do so, we have to identifying the region in parameter space ($\beta J$, $M J$) where this is consistent. Outside of this region we will take $m$ arbitrary, but since $u=0$, this does not matter, and the solution will be the equilibrium paramagnetic configuration. The nature of the ``transition'' depicted in figure \ref{fig:PhasesMarg} shows how $m$ varies as we cross from the paramagnetic phase, where no marginal solution exists, to the region of coupling space where the marginal spin glass does indeed.

Thus, to find the location of the phase boundary, the relevant constraint equations for $(\tilde u, m)$ become:
\begin{align}
\text{EOM}_u: \quad 0&=p\,\frac{\tilde{\jmath}^2}{2}\tilde{u}^{p-1}-\frac{\tilde{u}}{\left(1-\tilde{u}\right)\left[1+(m-1)\tilde{u}\right]}\,,\label{eq:utvarMarg}\\
\text{EOM}_m^{\text{(marg.)}}:\quad m&=\frac{1}{\tilde{\jmath}} \, \frac{(p-2) \sqrt{2}}{\sqrt{p(p-1)\tilde{u}^p}}~.\label{eq:mvar2Marg}
\end{align}
where we have removed an overall factor of $(m-1)$ in the first equation since we are now only looking for a spin glass solution with $m \neq 1$. Plugging the second equation into the first, we find an equation that determines $\tilde{u}$ as a function of $\tilde{\jmath}$ and $p$:
\begin{equation}
  \frac{p}{2} \, \tilde{\jmath}^2 \tilde{u}^{p-1} (1-\tilde u)^2 + \tilde{\jmath} \,  \frac{\sqrt{p}\, (p-2)}{\sqrt{2(p-1)}}\, \tilde{u}^{p/2}(1-\tilde u) - \tilde u = 0 \,.
\end{equation}  
 The behavior of this relation is qualitatively very similar to the equilibrium analysis, shown in the left panel of figure \ref{fig:critj}. However, due to the absence of the logarithm, the present case is easier to treat analytically. The physical solution satisfying $\tilde{\jmath} \geq 0$ for $0<\tilde{u} <1$ (when it exists) can thus be written explicitly and reads as follows:
\begin{equation}
\tilde{\jmath}^2 = \frac{2}{p(p-1)} \, \frac{\tilde{u}^{2-p}}{(1-\tilde u)^2} \,,\qquad m = (p-2) \, \frac{1-\tilde u}{\tilde u} \,.
\label{eq:tildejMarg}
\end{equation}  
To find the critical value where the marginal spin glass solution ceases to exist, one has to investigate what values of $\tilde{\jmath}^2$ the above solution can produce. To this end, one observes that $\tilde{\jmath}^2$ as a function of $\tilde{u}$ has a minimum at the critical value $\tilde{u}_\star$, characterized by:
\begin{equation}
\label{eq:CritMarg}
   \tilde{u}_\star = \frac{p-2}{p} \,,\qquad  \tilde{\jmath}_\text{crit}^2 \equiv \tilde{\jmath}^2(\tilde{u}_\star)= \frac{p^{p-1}}{2 (p-1)(p-2)^{p-2}} \,.
\end{equation}
The value of $m$ along the critical line \eqref{eq:CritMarg} is $m=2$.
For $\tilde{\jmath}^2 < \tilde{\jmath}^2_\text{crit}$ the marginally stable spin glass solution does not exist. Note, however, that it may cease to exist earlier, depending on the ability to satisfy the remaining equations of motion. Furthermore, it certainly will not be thermodynamically preferred over the disordered solution all the way up to $m=2$, as we will see in the next subsection. In fact we will argue that a more stringent physical criterion is that $m$ should never exceed $1$. Demanding $m=1$ as the boundary of the region where the marginally stable spin glass solution makes sense physically, we find that the maximal allowed values are in fact:
\begin{equation}
\label{eq:meq1}
 m=1 \qquad \Rightarrow \qquad  \tilde{u} = \frac{p-2}{p-1} \,,\qquad  \tilde{\jmath}^2 =  \frac{2}{p} \frac{(p-1)^{p-1}}{(p-2)^{p-2}} \,.
\end{equation}

The value of $\tilde{\jmath}$ as a function of $p$ is again qualitatively similar to the equilibrium case shown in the right panel of figure \ref{fig:critj}, as long as $p$ is not too large. However, it is interesting to note that in the large $p$ limit, the critical value $\tilde{\jmath}^2_\text{crit}$ in the equilibrium case diverges with a complicated $p$ dependence, whereas in the present case it is given by $\tilde{\jmath}^2_\text{crit} \rightarrow \frac{1}{2}e^2$ (as $p \rightarrow \infty$). Similarly, the physical boundary \eqref{eq:meq1} approaches $\tilde{\jmath}^2 \rightarrow 2e$.
Note that solving \eqref{eq:tildejMarg} for $\tilde{u}$ yields two branches of solutions: one for $\tilde{u} \geq \tilde{u}_\star$ and one for $\tilde{u} \leq \tilde{u}_\star$. It is clear that the first case corresponds to physical solutions, as it is the only one consistent with $\tilde{u} \rightarrow 1$ in the ``deep spin glass limit'' $\tilde{\jmath} \rightarrow \infty$.

Again, as in the equilibrium case, the above analysis is somewhat misleading because the rescaled tilde-variables hide all the dependence on the value of $MJ$. In order to gain a full understanding of the phase diagram corresponding to the marginal stability criterion, we now consider numerical solutions again.

\subsection{Thermodynamics and phase diagram}
\label{sec:ThermoMarg}

Let us now discuss the detailed thermodynamics of the marginally stable spin glass states. Our first goal will be to find an appropriate definition of the thermodynamic quantities. We begin with the regularized expression for the effective action, \eqref{eq:freeregularized}.
For equilibrium configurations, it is natural to identify $\phi$ with the free energy per unit site in the $n\rightarrow0$  limit.

However, as we have already discussed, below the spin glass phase transition it is natural to pass to another ensemble where the replica symmetry is \emph{explicitly}, rather than spontaneously broken \cite{monasson1995structural,mezard1999compute}, which in practice is done by explicitly coupling $m$ replicas of the $p$-spin model before disorder averaging. This allows $m$ to be thought of as an external parameter, therefore allowing us to consider $m$ as its own thermodynamic variable (see section II.B of \cite{zamponi2010mean} for an elucidating review and \cite{nieuwenhuizen1995maximize} for related comments).  As discussed in \cite{biroli2001quantum}, this is equivalent to computing the Thouless-Anderson-Palmer (TAP) free energy \cite{thouless1977solution} of the system. There is a subtlety in the $n\rightarrow 0$ limit of the replica trick which makes it so that this procedure is only valid for values of $m$ in the spin glass phase. It is important to recall, as alluded to earlier in this section, that the reason we may want to tune $m$ in this way is to connect to the dynamics of the model \cite{kirkpatrick1987p,horner1992dynamics,horner1992dynamics2,crisanti1993sphericalp,PhysRevLett.71.173}, which shows that $m$ settles on the marginally stable value under classical evolution.

Following the procedure of allowing for $m$ replicas to be pinned to the same state by an explicit breaking, as in \eqref{eq:cloneddisorder}, one finds for the effective action per unit site, again in the $n\rightarrow 0$ limit: 
\begin{equation}
	\beta m \overline{\Phi}=\frac{\tilde{S}_{\rm eff}^{\rm reg}}{N n}=\beta m \phi~, 
\end{equation}
where $\phi$ is the same as in \eqref{eq:freeregularized}. 
In this ensemble, the various thermodynamic quantities are given as follows \cite{zamponi2010mean}: 
\begin{equation}\label{eq:thermodef}
	\boxed{\;\;\bar{f}=\partial_m (m\phi)~, \qquad \St=-\partial_T \bar{f}~, \qquad\Sigma =-\partial_{1/m} (\beta \phi)~,\qquad \Et=\partial_\beta (\beta \bar{f})~,\qquad \Ct=\partial_T \Et~,\;}
\end{equation}
where $\bar{f}$ is the free energy per unit site, $\St$ is the entropy, $\Et$ is the energy and $\Ct$ is the specific heat of the thermodynamic state per unit site. Note that $m$ plays the role of thermodynamic potential dual to free energy much like $\beta$ is the thermodynamic potential dual to energy. Also notice that there is a new `entropy'-like quantity $\Sigma$ which counts the number of non-equilibrium states that are within a particular free energy band whose width is given by $1/m$, much like how the entropy counts the number of equilibrium states within a coarse grained energy band whose width is given by the temperature. 

From the definition of the free energy above we notice: 
\begin{equation}
\label{eq:fMargDef}
	 \beta \bar{f} \equiv \beta\phi+ m \, \partial_m \left( \beta\phi \right)=\beta\phi + \frac{1}{m}\,\Sigma~.
\end{equation}
This leads to the following interpretation of the on-shell action:
\begin{equation}
	\beta m\phi=\beta m \Et- (m \St+\Sigma)\,,
	\label{eq:newensemble}  
\end{equation}
meaning that $m\Et$ serves as the effective energy, while $m\St+\Sigma$ contribute as the microscopic entropy normally would. Interestingly, even if the microscopic entropy $\St$ is zero, as we will show, there is still a thermodynamically large number of metastable states coming from the complexity $\Sigma$. We can now evaluate the thermodynamic quantities explicitly.

\paragraph{Free energy:}
We begin with the simple evaluation of the free energy. According to \eqref{eq:fMargDef}, in this ensemble: 
\begin{align}
 \beta \bar{f} 
 &=\beta\phi-\frac{m}{2}\left(\frac{(\beta J)^2}{2}u^p+\frac{1}{m}\frac{u}{\frac{\hat{q}_r(0)}{\beta}+mu}+\frac{1}{m^2}\log\left[\frac{\frac{\hat{q}_r(0)}{\beta}}{\frac{\hat{q}_r(0)}{\beta}+mu}\right]\right)~.
\end{align}
As expected, we find the free energy at equilibrium, plus a contribution from the complexity $\Sigma$. The complexity would vanish on equilibrium values of $m$ satisfying the would-be equations of motion for $m$, \eqref{eq:mvar1}. However, since we are now free to tune $m$ as we please, this extra term will contribute to the free energy in this ensemble. Combining the above result with \eqref{eq:freeregularized} gives
\begin{align}
\beta\bar{f}=&-\frac{1}{2}\sum_{k=-\infty}^\infty\log \left[\left(\frac{M}{\beta} \left({2\pi k}\right)^2-2   {i\hat{z}(0)}\right)  \frac{\hat{q}_r(k)}{\beta}  \right]  +\frac{1}{2}\sum_{k=-\infty}^\infty\left\{-1+ \left[M\left(\frac{2\pi k}{\beta}\right)^2-2i\frac{\hat{z}(0)}{\beta}\right]\hat{q}_r(k)\right\}
 \nonumber\\
& +i \hat{z}(0)\left(1-u\right) -\frac{\beta^2J^2 }{2}\left\{mu^p+\frac{1}{2\beta}\int_0^\beta d\tau\,(q(\tau)^p-u^p)\right\}\nonumber\\
&+\log\left[2\sinh\sqrt{\frac{-i\hat{z}(0)\beta}{2M}}\right]-\frac{1}{2}\left(\frac{u}{\frac{\hat{q}_r(0)}{\beta}+mu}\right)~.
\end{align}
We can massage this expression using the on-shell identity \eqref{eq:replace}:
\begin{align}\label{eq:fSG}
\beta\bar{f}=& -\frac{1}{2}\frac{\beta}{\hat{q}_r(0)}-\frac{1}{2}\sum_{k=-\infty}^\infty\log \left[\left(\frac{M}{\beta} \left({2\pi k}\right)^2-2   {i\hat{z}(0)}\right)  \frac{\hat{q}_r(k)}{\beta}  \right] 
 \nonumber\\
&+\left[\frac{(\beta J)^2}{4}\left\lbrace (p-2)m  u^{p}+\frac{p-1}{\beta}\int_0^\beta d\tau\, \left[q(\tau)^{p}-u^{p}\right]-\frac{p}{\beta}\int_0^\beta d\tau\, \left[q(\tau)^{p-1}-u^{p-1}\right]\right\rbrace\right] \nonumber\\
&+\log\left[2\sinh\sqrt{\frac{-i\hat{z}(0)\beta}{2M}}\right]-\frac{1}{2}\left(\frac{u}{\frac{\hat{q}_r(0)}{\beta}+mu}\right)~.
\end{align}

\paragraph{Complexity:}
We can explicitly compute the on-shell complexity in the marginal spin glass state (evaluated using \eqref{eq:mvarMargMain}) and we find:
\begin{equation}
	\Sigma=\frac{1}{2}\log(p-1)-\frac{p-2}{p}~.\label{eq:complexityvalue}
\end{equation}
Interestingly the complexity is zero for $p=2$ (which is known to evade the spin glass transition), and positive for $p>2$. It is also independent of any couplings or the temperature. This suggests that it will not contribute to the energy of the spin glass phase.

\paragraph{Energy:} 
To compute the energy $\Et$, we use the definition
\begin{equation}
\label{eq:EmargDef0}
	\Et=\partial_\beta (\beta \bar{f})=\partial_\beta[1+m\partial_m]\beta\phi=[1+m\partial_m]\partial_\beta(\beta\phi)~. 
\end{equation}
To calculate the first term, which only consists of a single derivative acting on $\beta\phi$, we can use the fact that the equations of motion are obtained by varying $\beta\phi$ with respect to most variables:
\begin{equation}
	\frac{\partial }{\partial \hat{q}_r(k)}(\beta\phi)=\frac{\partial }{\partial u}(\beta\phi)=\frac{\partial }{\partial \hat{z}(0)}(\beta\phi)=0~.
\end{equation}
In this ensemble, $m$ is an external parameter, so there is no meaning to considering terms such as $\partial m/\partial\beta$ while varying the free energy. This is because we are selecting $m$ externally such that we land on the conformal solution. In computing the second term of \eqref{eq:EmargDef0}, we must evaluate the mixed derivatives \emph{before} plugging in the on-shell conditions. 
Carefully keeping track of these contributions, we find: 
 \begin{align}
\beta \Et=&\,\frac{1}{2}\sum_{k=-\infty}^\infty\left[ 1 - \frac{M}{\beta} (2\pi k)^2 \, \frac{\hat{q}_r(k)}{\beta} \right]-\frac{\beta^2J^2 }{2}\left\{m u^p+\frac{1}{\beta}\int_0^\beta d\tau\left[q(\tau)^p-u^p\right]\right\}\nonumber\\
&+\frac{m}{2}\left[-(\beta J)^2u^p\left(1+\frac{p}{2}\frac{\beta\,\partial_\beta u}{u}\right)+\left(\frac{u}{\frac{\hat{q}_r(0)}{\beta}+m u}\right)^2\left(1-\frac{\beta\,\partial_\beta \hat{q}_r(0)}{\hat{q}_r(0)}+\frac{\beta\,\partial_\beta u}{u}\right)\right].
\end{align}
Evaluated on the marginal-spin glass parameters \eqref{eq:mvarMargMain}, the second line of the above expression vanishes. This was not guaranteed, but it does confirm the results of \cite{cugliandolo2001imaginary}.
We could have anticipated the vanishing of the second line of \eqref{eq:mvarMargMain} for the marginal spin glass, since the on-shell complexity is temperature independent, so its explicit $\beta $ derivative is certain to vanish. 
Finally, some massaging using the identity \eqref{eq:replace} allows us to simplify the first line:  
\begin{equation}
\label{eq:EpositionMarg}
   \beta \Et = 
     \frac{1}{2} \frac{\beta}{\hat{q}_r(0)}  + \frac{(\beta J)^2}{4} \, \left\{ \frac{p}{\beta } \int_0^\beta d\tau \, \left[ q(\tau)^{p-1} - u^{p-1}  \right] - (p+2) \, mu^p - \frac{p+2}{\beta} \int_0^\beta  d\tau\, \left[q(\tau)^p - u^p \right] 
       \right\} \,.
\end{equation}
We thus conclude that the expression for the energy in the marginal spin glass phase is the same expression as in equilibrium \eqref{eq:Eposition}. This is consistent with \cite{cugliandolo2001imaginary} although our reasoning is different. We show that this is a consequence of the ensemble, and the particular SG solution. 

\paragraph{Entropy:} 
To compute the entropy note that by definition, it is given by 
\begin{equation}
	\St= \beta \Et- \beta\bar{f}\,.
\end{equation}
It therefore follows immediately from \eqref{eq:Eposition} and \eqref{eq:fSG} (see \eqref{eq:Sappendix} for an explicit expression).

\subsection{Phase diagram}

\begin{figure}
\begin{center}
\includegraphics[height=7cm]{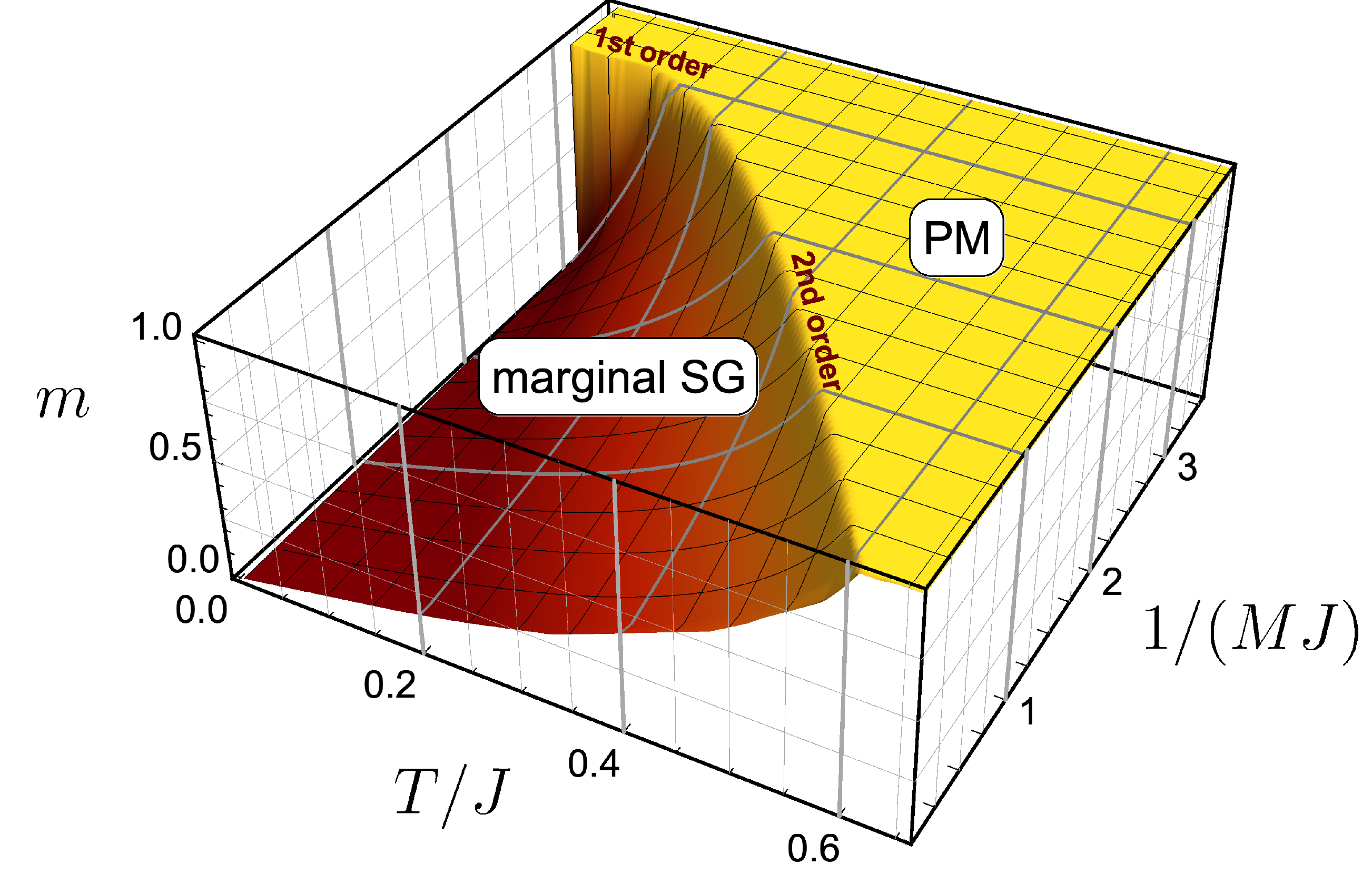}
\end{center}
\vspace{-.3cm}
\caption{{\bf Phase diagram for the transition to the marginally stable spin glass.} We show the value of $m$ as a function of temperature and $1/M$ (both measured in units of $J$) for $p=3$. The qualitative features are the same as in the equilibrium case (figure \ref{fig:Phases}): the parameter $m$ is continuous (discontinuous) along the second (first) order line. Note that the phase transition happens at slightly larger values of $1/(MJ)$ compared to the equilibrium case.} \label{fig:PhasesMarg}
\end{figure}

The phase diagram for the transition between the marginally stable spin glass state and the paramagnetic ``disordered'' state is shown in figure \ref{fig:PhasesMarg}. The boundary of the spin glass region is defined as the line where either the marginally stable spin glass solution ceases to exist (first order transition), or it continues to exist but the value of $m$ begins to take unphysical values $m>1$ (second order transition). The qualitative features of the phase diagram are the same as in the equilibrium analysis (c.f., figure \ref{fig:Phases}): the parameter $m$ is continuous (discontinuous) across the second order (first order) transition. The order parameter $u$ is discontinuous across both. The main difference compared to figure \ref{fig:Phases} is that the transition happens at slightly larger values of $1/(MJ)$. 

\begin{figure}
\begin{center}
\includegraphics[width=.496\textwidth]{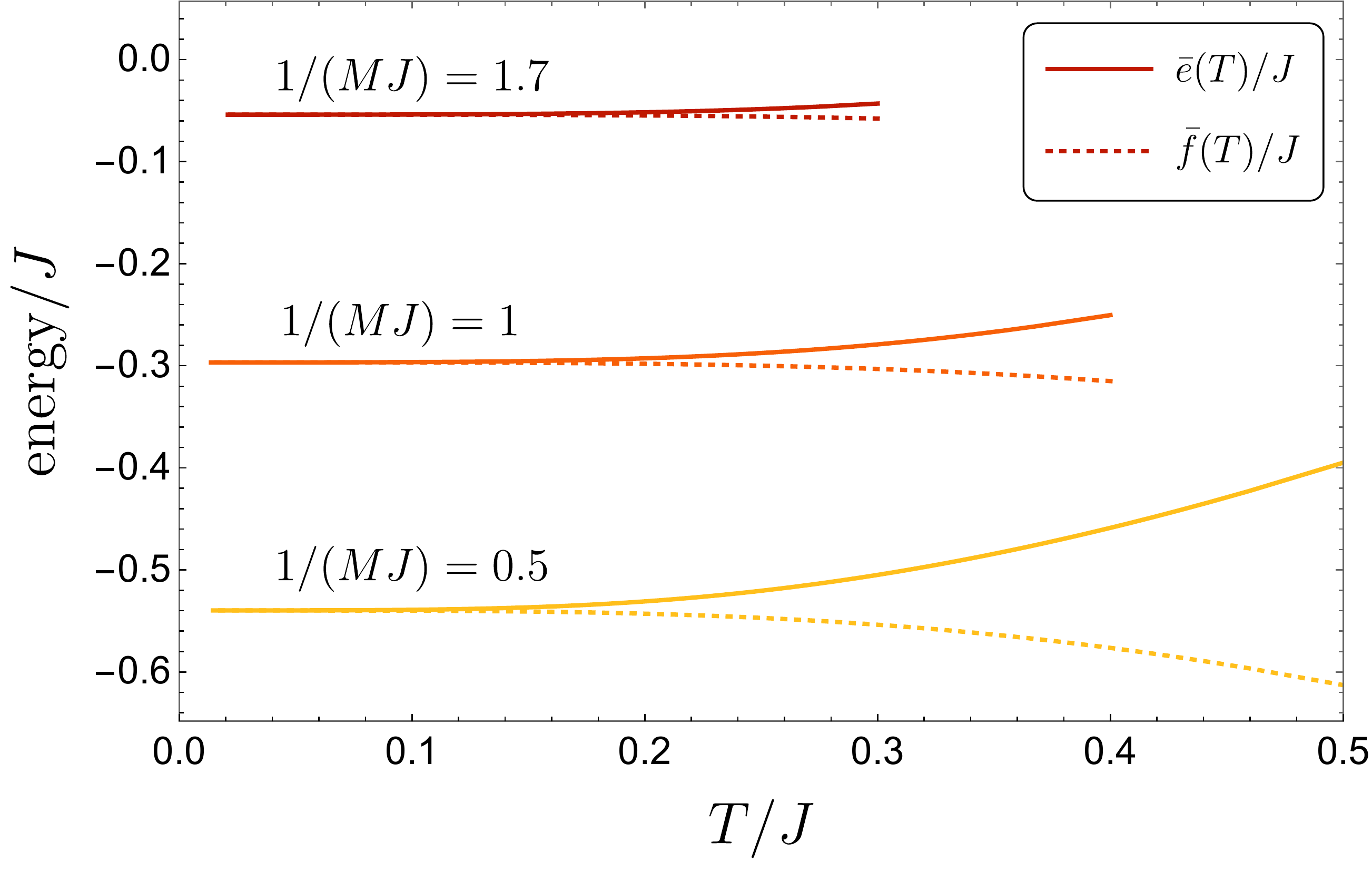}
\includegraphics[width=.49\textwidth]{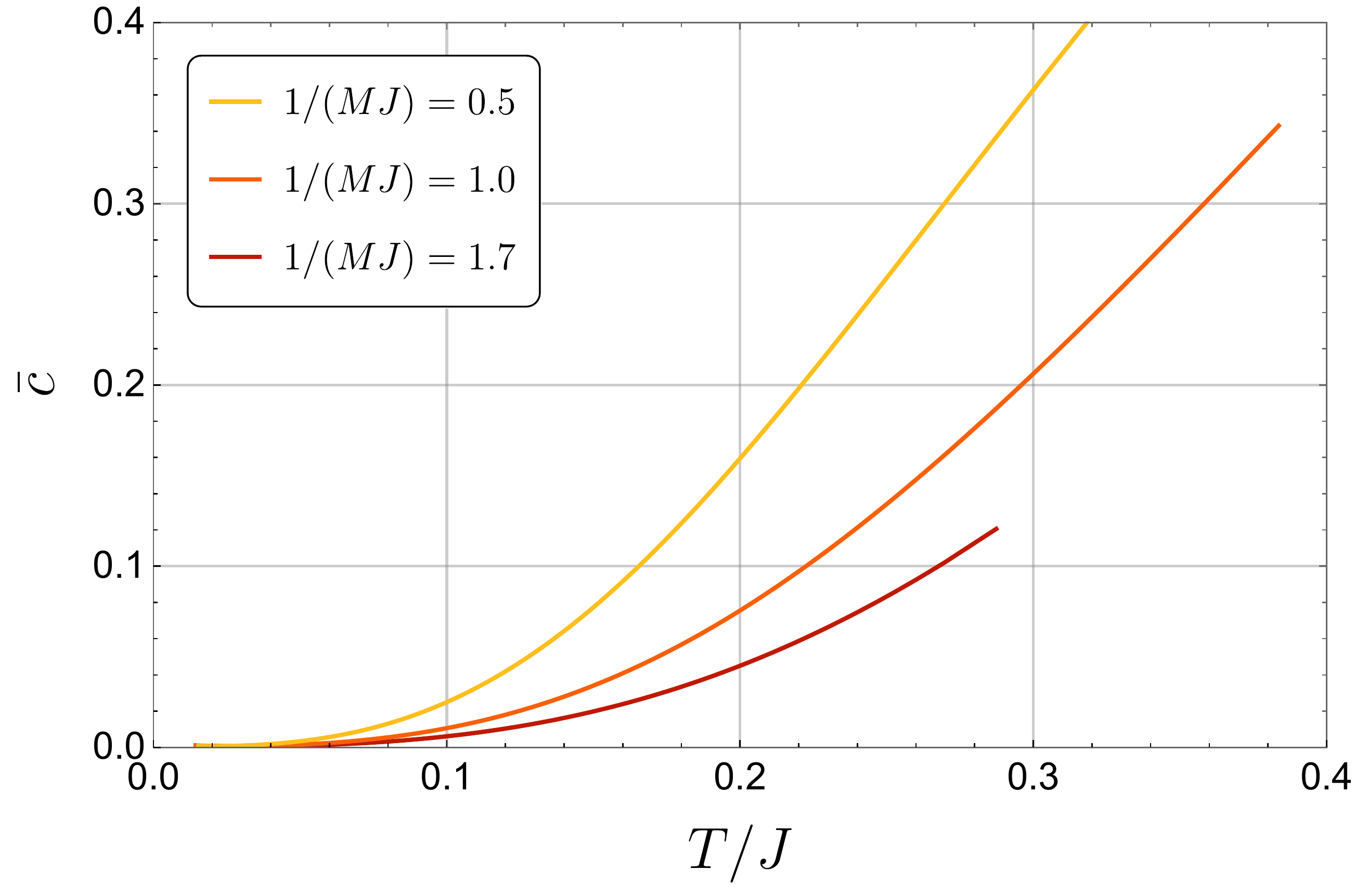}
\end{center}
\vspace{-.3cm}
\caption{{\bf Low temperature thermodynamics of the marginal spin glass ($MJ$ fixed).} {\it Left:} The internal energies (upper curves) and free energies (lower curves) per spin approach the same constant value at zero temperature. {\it Right:} The specific heat $\bar{c}$ approaches zero linearly. Compare also with the corresponding plot in case of the equilibrium spin glass, figure \ref{fig:lowTthermoEquil}.} \label{fig:lowTthermoMarg}
\end{figure}

In figure \ref{fig:lowTthermoMarg} we further present numerical results for the thermodynamic functions in the marginally stable spin glass phase. Particularly interesting is the low temperature limit, where we have analytic control. Let us discuss this limit in some detail.

\subsection{Low temperature expansion (quantum scaling)}
\label{sec:quantumSc}

We can understand much of the thermodynamics in the marginally stable spin glass analytically by evaluating thermodynamic quantities on the approximate solution $q^\approx_r$ discussed in section \ref{sec:qApprox}. We give a more detailed discussion of the large $\beta {\cal J}$ expansion in appendix \ref{app:thermo}. Here, we simply summarize the results in the `quantum scaling' \eqref{eq:QuantScale}, where both $\beta {\cal J}$ and $M{\cal J}$ are parametrically large.

\begin{figure}
\begin{center}
\includegraphics[width=.49\textwidth]{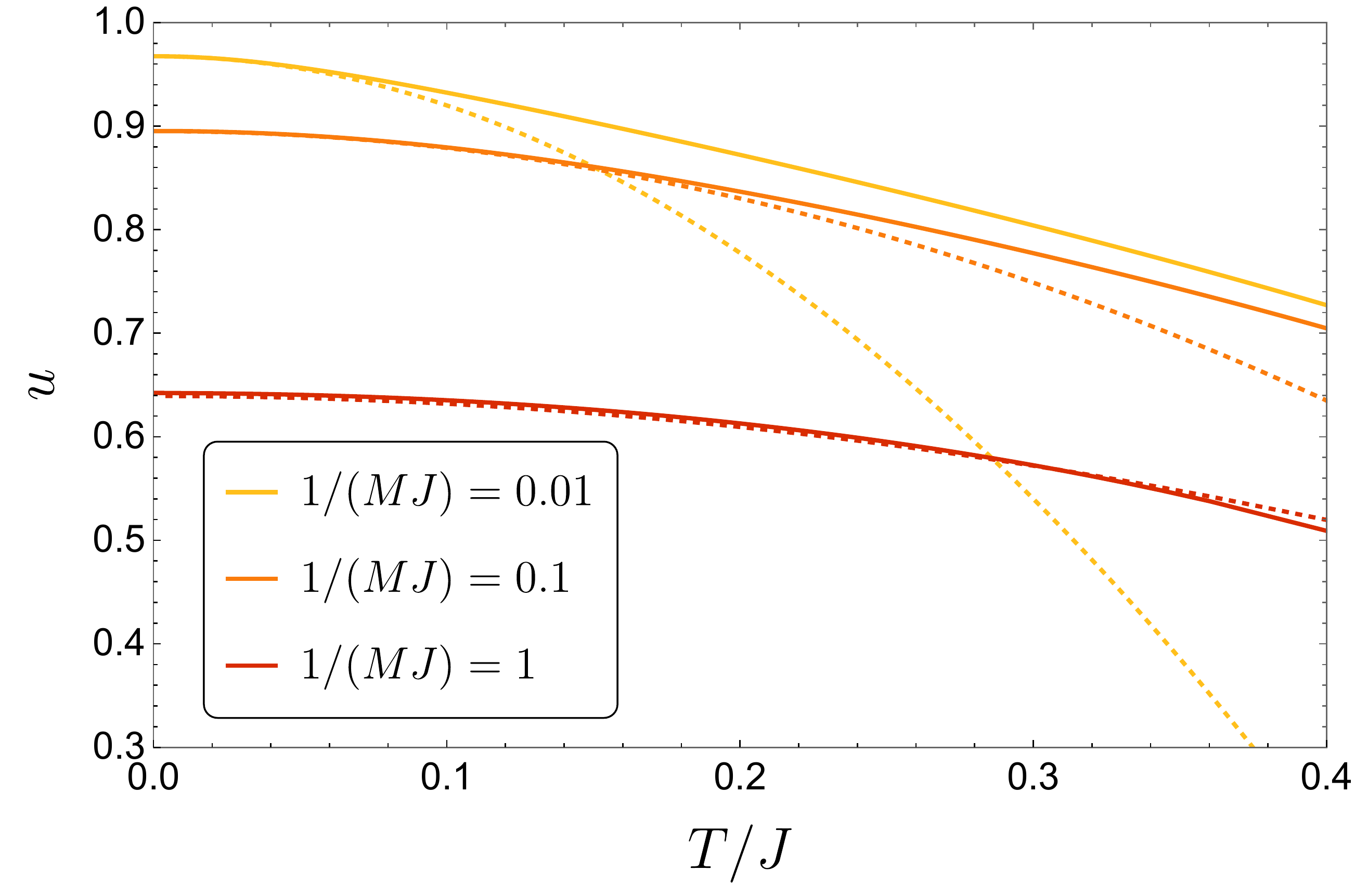}$\;\;$
\includegraphics[width=.49\textwidth]{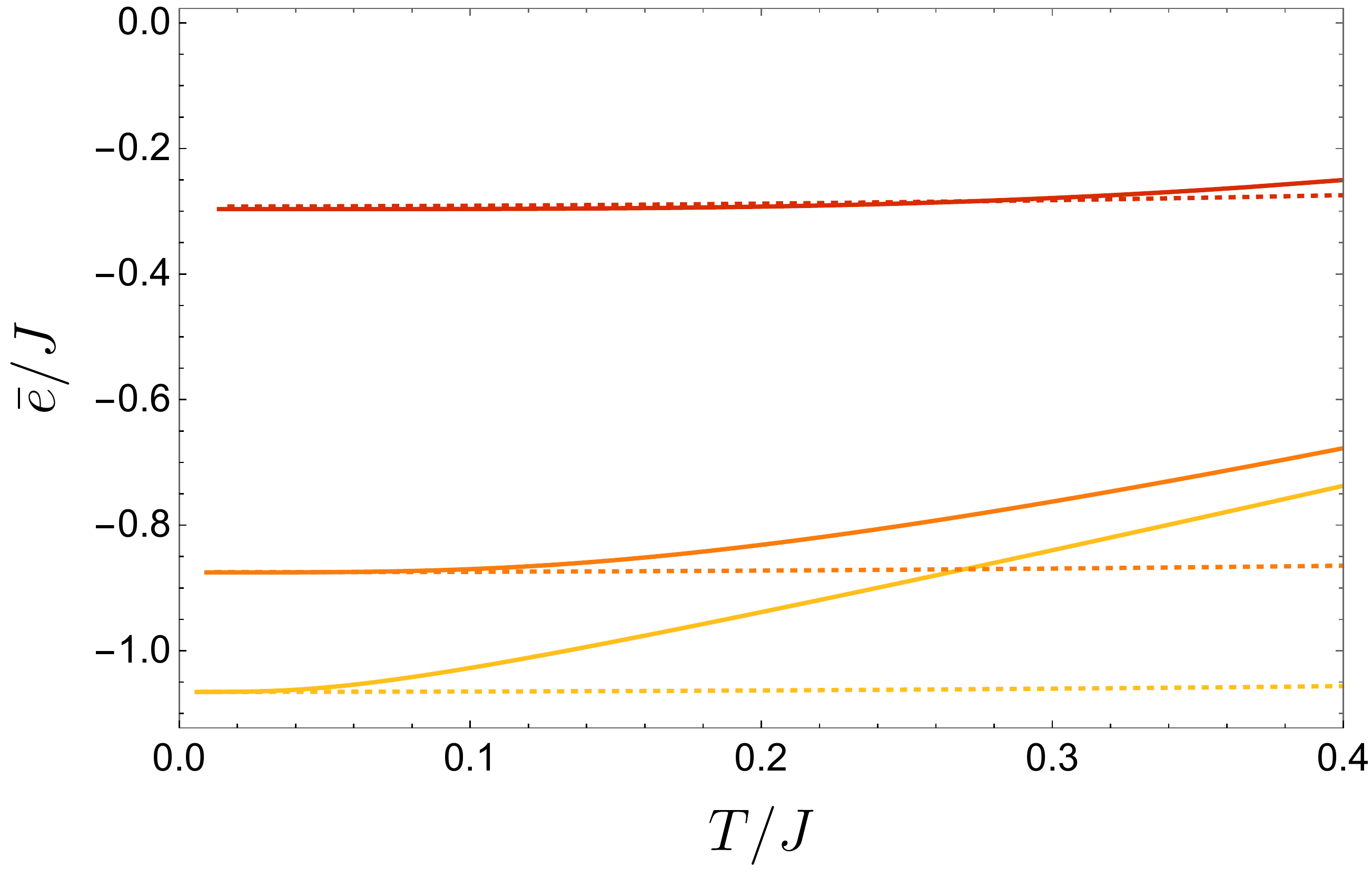}
\end{center}
\vspace{-.5cm}
\caption{{\bf Marginal spin glass thermodynamics ($MJ$ fixed).} We show the Edwards-Anderson parameter $u$ and the energy $\Et$ as functions of temperature. Dotted lines correspond to the analytical approximations \eqref{eq:ulowtSG} and \eqref{eq:Equantum}. The approximation is best for sufficiently large $MJ$ and sufficiently low temperatures. See also figure \ref{fig:thermo_marg_quantum} for comparison with the `quantum scaled' regime.} \label{fig:thermo_marg_fixedMJ}
\vspace{.5cm}
\begin{center}
\includegraphics[width=.49\textwidth]{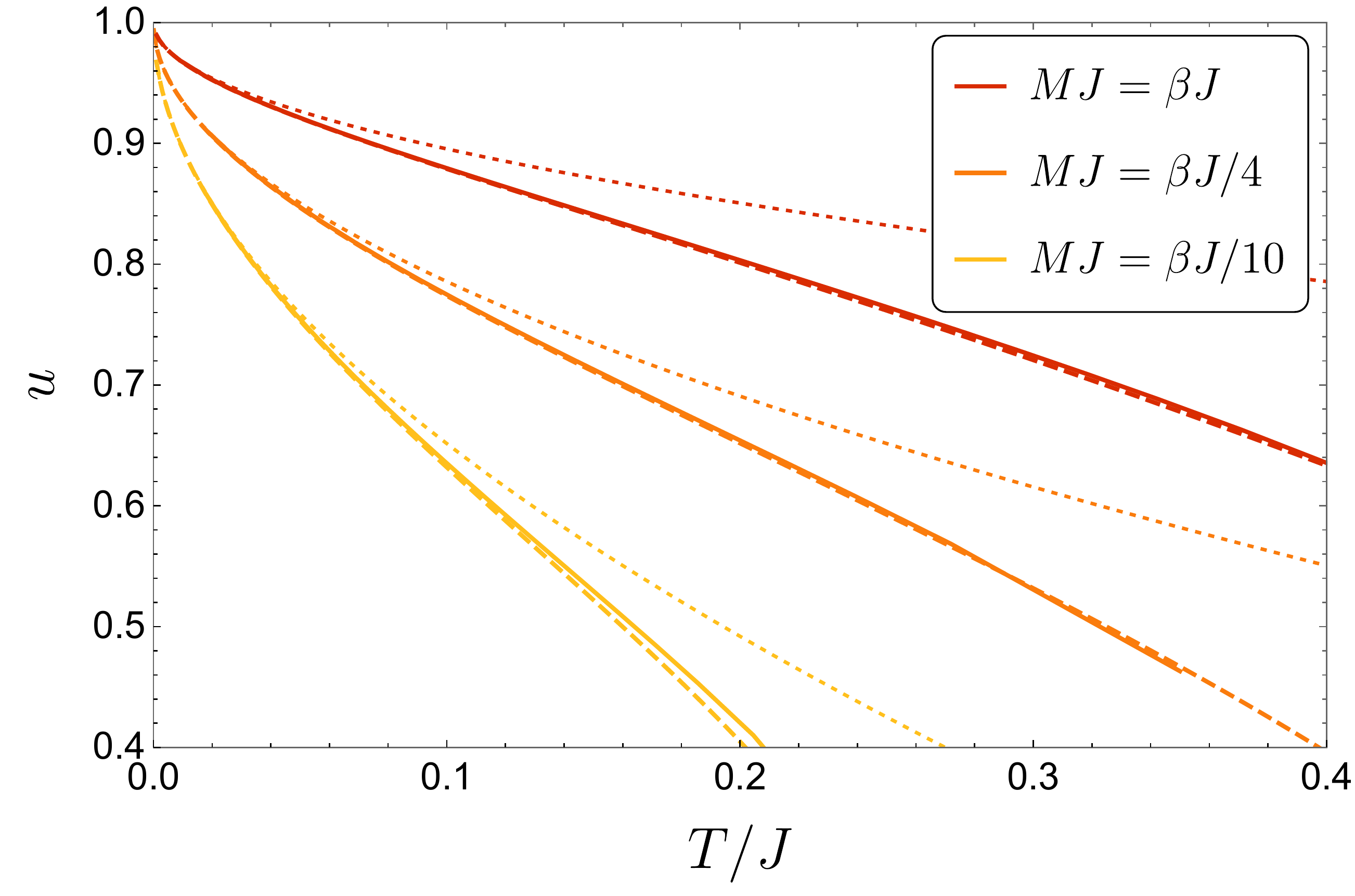}$\;\;$
\includegraphics[width=.49\textwidth]{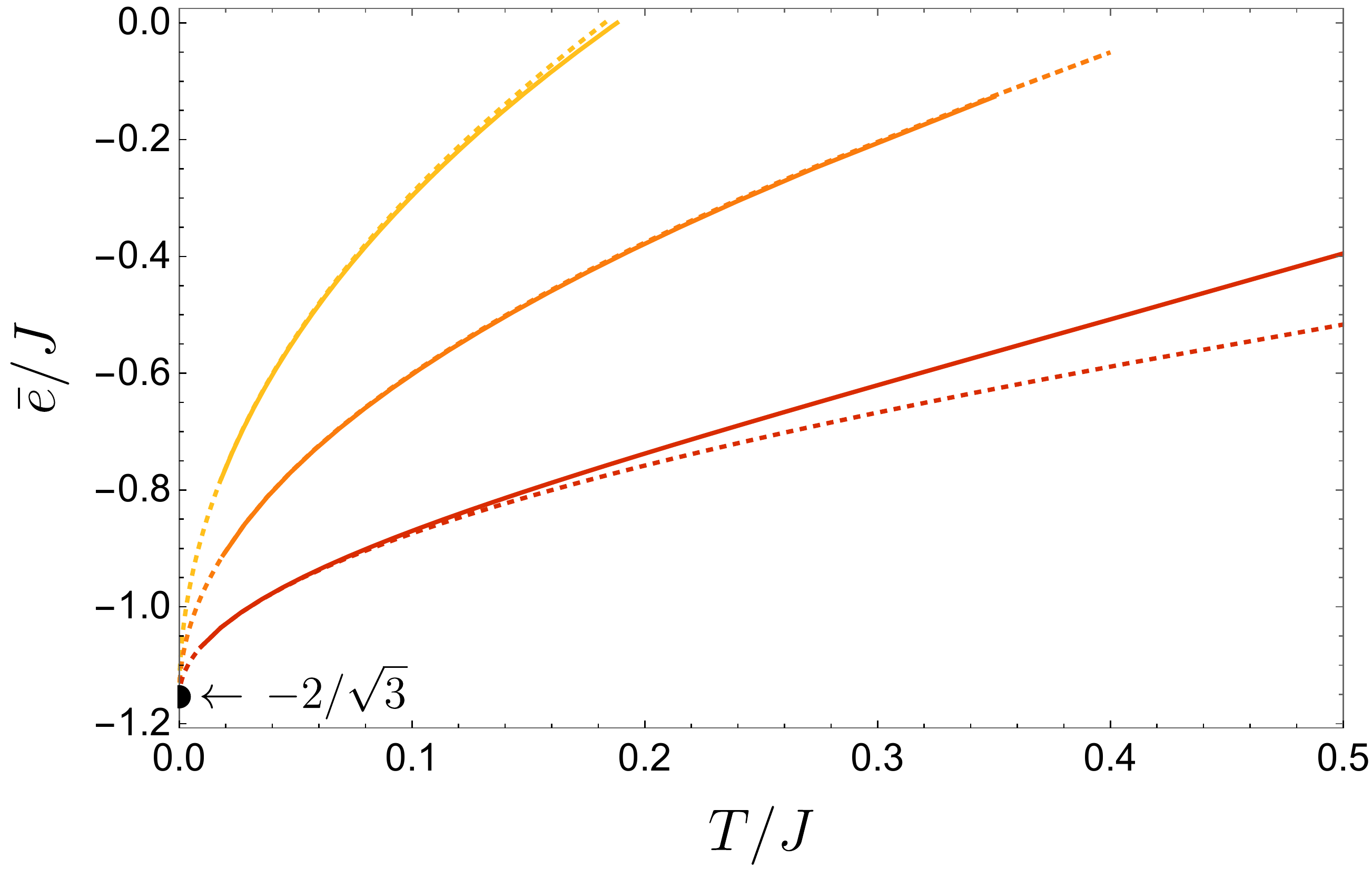}
\end{center}
\vspace{-.5cm}
\caption{{\bf Marginal spin glass thermodynamics (quantum scaling: $MJ \propto \beta J \gg 1$).} We show the Edwards-Anderson order parameter $u$ and the energy $\Et$. The analytical approximations based on \eqref{eq:ulowtSG} and \eqref{eq:Equantum} (thick dashed lines) match the numerics (full lines) very well. For $u$ we show the perturbative expression  \eqref{eq:muqhatQuantum} separately (thin dotted lines). For $\Et$, the difference between \eqref{eq:Equantum} and \eqref{eq:betaElowTres} is very small and not shown. At $T=0$ all the `quantum scaled' internal energies approach the value $-2/\sqrt{3}$, which follows from \eqref{eq:Equantum} for $p=3$ (note that we plot $\bar{e}/J$, whereas \eqref{eq:Equantum} gives $\bar{e}/ {\cal J}$).} \label{fig:thermo_marg_quantum}
\end{figure}

For the energy, we find in the quantum scaling with large $\Lambda$:
\begin{multline}
\label{eq:Equantum}
	\frac{\Et}{\mathcal{J}}=-\frac{2 }{p}+\frac{32}{15\pi\Lambda\sqrt{\widehat{M\mathcal{J}}}}-\frac{4(p-2)(1+20n_3)}{45\pi^2\Lambda^2\widehat{M\mathcal{J}}}+\frac{8(p-2)(p-10-24n_3(p-8))}{405\pi^3\Lambda^3\left(\widehat{M\mathcal{J}}\right)^{3/2}}\\+\frac{1}{\Lambda^4}\left(\frac{2(p-2)(7-10n_3')}{45\left(\widehat{\beta \mathcal{J}}\right)^2}-\frac{4(p-2)\left[(p+2)(p-24)-48n_3(p(p-2)-28)\right]}{1215\pi^4\left(\widehat{M\mathcal{J}}\right)^2}\right)+\dots
\end{multline}
where $n_3$, $n_3'$ are ${\cal O}(1)$ numbers (see \eqref{eq:zetacube}).
In this scaling, we see immediately that $\Et$ is finite at zero temperature, approaching the value $\bar{e}/{\cal J} = -2/p$, and the specific heat is linear in $T$. 
In appendix \ref{app:thermo} we also derive the entropy in the quantum scaling, and show that it vanishes to leading order:
\begin{equation}
\St = 0 + {\cal O}(\Lambda^0) \,.
\end{equation}
This is consistent with what we find numerically: the entropy vanishes at zero temperature (c.f., figure \ref{fig:lowTthermoMarg}). However, recall that in the grand-canonical ensemble we find ourselves in, where $m$ is its own a thermodynamic potential, the generalized free energy is fixed to \eqref{eq:newensemble}, which has two entropic contributions. This means that the entropy-like quantity is 
\begin{equation}
\label{eq:mSSigma}
\boxed{\;\;
	m \St + \Sigma= \left[ \frac{1}{2}\log(p-1)-\frac{p-2}{p} \right] + {\cal O}(\Lambda^{-2})~.
	\;}
\end{equation}
since this is the entropy per unit site, the actual entropy grows with $N$.

It is interesting to use the thermodynamic quantities as a benchmark for the accuracy of the approximate solution $q_r^\approx$. To this end, we compare the exact numerical values for $u$ and $\Et$ with the analytical values to first subleading order in the low-temperature expansion (expressions \eqref{eq:muqhatQuantum} and \eqref{eq:Equantum}). Figure \ref{fig:thermo_marg_quantum} demonstrates good agreement when we scale $M{\cal J}$ linearly with $\beta {\cal J}$ (as is natural to focus on the quantum scaling). On the other hand, working with fixed values of $M{\cal J}$, the analytical expressions are only good at very small temperatures (c.f., figure \ref{fig:thermo_marg_fixedMJ}). We also compared with the expressions obtained by expanding in $1/(\beta J)$, but solving for the $MJ$ dependence exactly according to \eqref{eq:ulowtSG} (and a similar expression \eqref{eq:betaElowTres} for $\Et$). At least for $u$, this improves the approximation further (see figure \ref{fig:thermo_marg_fixedMJ}, left panel).
When comparing figures \ref{fig:thermo_marg_fixedMJ} and \ref{fig:thermo_marg_quantum} it is useful to mentally overlay them: the curves plotted in figure \ref{fig:thermo_marg_quantum} only go through regions of figure \ref{fig:thermo_marg_fixedMJ} where the analytical approximation is relatively good. This is the purpose of the quantum scaling.

\subsection{Thermodynamic functions, soft-mode actions, and holography}\label{sec:holography}

At this stage we are ready to assemble the various thermodynamic quantities computed in the previous sections and give them a holographic interpretation. To do so, we use the established links between thermodynamic functions and static solutions to dilaton gravity as described in \cite{Kitaev:2017awl} (see also \cite{Anninos:2017hhn,Jiang:2019pam,Witten:2020ert,Witten:2020wvy,Maxfield:2020ale,Anninos:2020cwo}). 

There are two important on-shell thermodynamic functions that should be highlighted in the marginal spin glass phase. First is the free energy, which admits the following expansion in the quantum scaling regime:\footnote{Recall that $\bar{f}$ and $\Et$ are the free energy and energy per site, respectively, whereas capitalized quantities are extensive in $N$.} 
\begin{equation}
	\beta(\bar{F}-\bar{E}_0)=N\left(-\frac{2(p-2)(7-10n_3')}{45\beta \mathcal{J}}+\dots\right)\label{eq:freeEexp}
\end{equation}
where $\bar{E}_0$ is the zero-temperature energy, which admits an expansion in $1/\sqrt{M\mathcal{J}}$ as can be deduced from \eqref{eq:Equantum}.  The second is the generalized free energy of the cloned ensemble, also known as the TAP free energy \cite{biroli2001quantum} (see section \ref{sec:ThermoMarg}): 
\begin{equation}
	\beta m (\bar{\Phi}-\bar{E}_0)=N\left(-\Sigma -\frac{2(p-2)^2(7-10n_3')}{45\left(\beta \mathcal{J}\right)^2}+\dots\right)\label{eq:FTAPexp}
\end{equation}
where, on the right hand side, we have used that $m$ itself admits a low temperature expansion in the quantum scaling regime according to \eqref{eq:muqhatQuantum}, in order to tune to the conformal spin glass phase. The complexity $\Sigma$ per unit site is given in \eqref{eq:complexityvalue}. 

How and to which of these thermodynamic functions should we ascribe a holographic meaning? While it is difficult to be certain, let us propose the following interpretation: On the one-hand, \eqref{eq:freeEexp} suggests that the free energy of a \emph{single} clone of the disordered $p$-spin model has a soft-mode action which is a Schwarzian with a very particular coefficient 
\begin{equation}
	\alpha_S= \frac{(p-2)(7-10n_3')}{45\pi^2}\approx0.0048(p-2)~.
\end{equation}
This can be diagnosed by fact that the power $(\beta\mathcal{J})^{-1}$ in the free energy has its origins in the particular Schwarzian-type breaking of diff$(S^1)$ down to $SL(2,\mathbb{R})$. The microscopic description of a single clone, however, does not have an extensive entropy in $N$, suggesting a limitation to this interpretation as the soft breaking of the asymptotic geometry within a single black hole throat.   

The quantity \eqref{eq:FTAPexp}, governing the grand canonical free energy of a collection of clones of the $p$-spin model also admits a power series in $(\beta\mathcal{J})^{-1}$ and has an extensive `entropy' in $N$, namely the complexity $N\Sigma$. This entropy counts the number of nearby free-energy states in the landscape of this cloned model. However, the absence of a term proportional to $(\beta\mathcal{J})^{-1}$ suggests that the holographic dual to this ensemble of clones will be inherently non-local, since there is no contribution accounting for the breaking of the symmetries of AdS$_2$. Indeed the contribution giving rise to the piece proportional to $(\beta\mathcal{J})^{-2}$ was computed in \cite{Kitaev:2017awl} and is explicitly bilocal in time. Perhaps then, we should view this ensemble as a collection of black holes and the TAP free energy in \eqref{eq:FTAPexp} shows that the low energy dynamics governing the breaking of diff$(S^1)$ in this ensemble is inherently non-local. It would be nice to explore a connection between the complexity $\Sigma$ and the number of fragmented horizons.  

It would be interesting if the more complicated, string-inspired, models alluded to in the introduction, containing both bosons with a `spherical' constraint as well as dynamical fermions, has a more local description in the replica symmetry broken phase, if one exists. We leave this exploration for future work.


\newpage
\part{Real time dynamics and quantum chaos}\label{part:RTD}

\section{Two-point functions in real time}
\label{sec:TwoPtReal}

Having understood the thermodynamic properties of the model based on the Euclidean two-point functions both in equilibrium and in the marginally stable spin glass, it is now time to turn to dynamical questions. We begin by formulating the Schwinger-Dyson equations in real time.

\subsection{Schwinger-Keldysh approach}
\label{eq:realTime2pt}

We begin by defining the spectral function $\rho_r(\omega)$ in terms of the Euclidean two-point function $q(\tau)$:
\begin{equation}
\label{eq:rhoDef}
   q_r(\tau) \equiv  \int_{-\infty}^\infty \frac{d\omega}{2\pi}  \, \frac{e^{-\omega \tau}}{1-e^{-\beta \omega}} \, \rho_r(\omega) =  \int_0^\infty \frac{d\omega}{2\pi}  \, \frac{\cosh \left[ \left( \frac{\beta}{2} - \tau \right) \omega \right]}{\sinh \left( \frac{\beta}{2} \, \omega \right)} \, \rho_r(\omega)
   \,,
\end{equation}
where we used $\rho_r(-\omega) = - \rho_r(\omega)$ in the second step (for bosonic theories). 
 In addition to being odd, the spectral function furthermore has the property $\omega \rho_r(\omega) \geq 0$.

In real time, we obtain the Wightman functions as analytic continuations of the Euclidean correlator:
\begin{equation}
\begin{split}
   q^>_r(t) &\equiv   q_r(\tau = it + \epsilon) \,,\qquad\qquad \hat{q}^>_r(\omega) = \int_{-\infty}^{\infty} dt \, e^{-i\omega t} \, q^>_r(t) = n_B(\omega) \rho_r(\omega) \\
  q^<_r(t) &\equiv q_r(\tau = -it + \epsilon) \,,\qquad\quad\; \hat{q}^<_r(\omega) = \int_{-\infty}^{\infty} dt \, e^{-i\omega t} \, q^<_r(t) = \left(1+n_B(\omega) \right)\rho_r(\omega) 
\end{split}
\end{equation}
where we provided explicit expressions in terms of the spectral function $\rho_r(\omega)$ using the Bose distribution $n_B(\omega) = (e^{\beta\omega} - 1)^{-1}$. The KMS condition (a.k.a.\ fluctuation-dissipation theorem) is the statement that $\hat{q}^<_r(\omega) = e^{\beta\omega} \hat{q}^>_r(\omega)$.

We will further require the retarded correlator, which can be given in terms of the Wightman function:
\begin{equation}
\label{eq:qrRdef}
\begin{split}
 i q^R_r(t) & =  \Theta(t) \left[ q^>_r(t) - q^<_r(t) \right] \,,\qquad\quad\;\; i\hat{q}^R_r(\omega) = -i\int_{-\infty}^{\infty} \frac{d\omega'}{2\pi} \frac{\rho_r(\omega')}{\omega+\omega'-i0^+} =  -i \hat{q}_r(-i\omega - 0^+) \,,
\end{split}
\end{equation}
where the last step can be proven easily for Matsubara frequencies $i\omega \in \frac{2\pi}{\beta} \, \mathbb{Z}$ and then {\it defines} the analytic continuation for arbitrary $\omega \in \mathbb{C}$.
From this integral expression, one can also easily verify how the spectral function can be extracted from the retarded correlator:
\begin{equation}
   \rho_r(\omega) = 2\, \text{Im} \, \hat{q}^R_r(\omega) 
   \,.
\end{equation}
Finally, we will need the analytically continued `left-right' autocorrelation function defined as
\begin{equation}
 q^>_r(t-i\beta/2) = \int_{-\infty}^{\infty} \frac{d\omega}{2\pi} \, e^{i\omega(t-i\beta/2)} \, n_B(\omega) \rho_r(\omega) \,.
\end{equation}
This satisfies the normalization condition $q^>_r(0-i\beta/2) = q_r(\tau = \beta/2)$.

Before moving on to the analytically continued equations of motion, we note here the role of the offset played by the Edwards-Anderson parameter $u$ in the analytic continuation to real time. One would presume that the offset survives in the analytic continuation of $q(\tau)\equiv q_r(\tau)+u$ to real times. This is certainly true of the Wightman functions:  
\begin{equation}\label{eq:wightmanoffset}
	q^>(t)\equiv q^>_r(t)+u~,\qquad\qquad q^<(t)\equiv q^<_r(t)+u~.
\end{equation}
However, since the retarded correlator is defined as the difference of these Wightman functions, the $u$ dependence drops out: 
\begin{equation}
	q^R(t)\equiv q_r^R(t)~.
\end{equation}

Let us now return to the equation of motion in the form \eqref{eq:EOMsigma3}, which is already written in a form amenable to analytic continuation.
Since the analytic continuations of $\hat{q}_r(k)$ and $\hat{\Lambda}_r(k)$ under $k\rightarrow -i \omega-\epsilon$ are directly related to the retarded propagator, we obtain the following relation in frequencies conjugate to real time:
\begin{equation}
\label{eq:EOMret}
	\frac{1}{\hat{q}_r^R(\omega)}-\frac{1}{\hat{q}_r^R(0)}= M \omega^2 - J^2 \, \left( \hat{\Lambda}_r^R(\omega)  - \hat{\Lambda}_r^R(0)  \right) \,,
\end{equation}
where 
\begin{equation}
\label{eq:selfEnergyTime}
\begin{split}
\hat{\Lambda}_r^R(\omega)  - \hat{\Lambda}_r^R(0) 
  &\equiv - \left\{ \hat{\Lambda}_r(-i\omega)  - \hat{\Lambda}_r(0)  \right\} \\
      &= - \frac{ip}{2}\int_0^\infty dt \, \left( e^{-i\omega t} - 1 \right) \,  \left[ \left(  q_r^>(t) +  u \right)^{p-1} - \left(  q_r^>(-t) +  u \right)^{p-1}  \right] 
\end{split}
\end{equation}
Finally, the normalization condition $q_r(\tau = 0) = 1-u$  translates via \eqref{eq:rhoDef} into an additional normalization condition for the spectral function:
\begin{equation}
\label{eq:normRho}
 \int_{-\infty}^\infty \frac{d\omega}{2\pi}  \, \frac{\rho_r(\omega)}{1-e^{-\beta \omega}}   =1-u \,.
\end{equation}
The spectral function for $q(\tau)$, which we would denote as $\rho(\omega) \equiv \rho_r(\omega) + 2\pi u\, \delta(\omega)$, satisfies a similar normalization condition with $1$ on the right hand side.

\begin{figure}
\begin{center}
\includegraphics[width=0.485\textwidth]{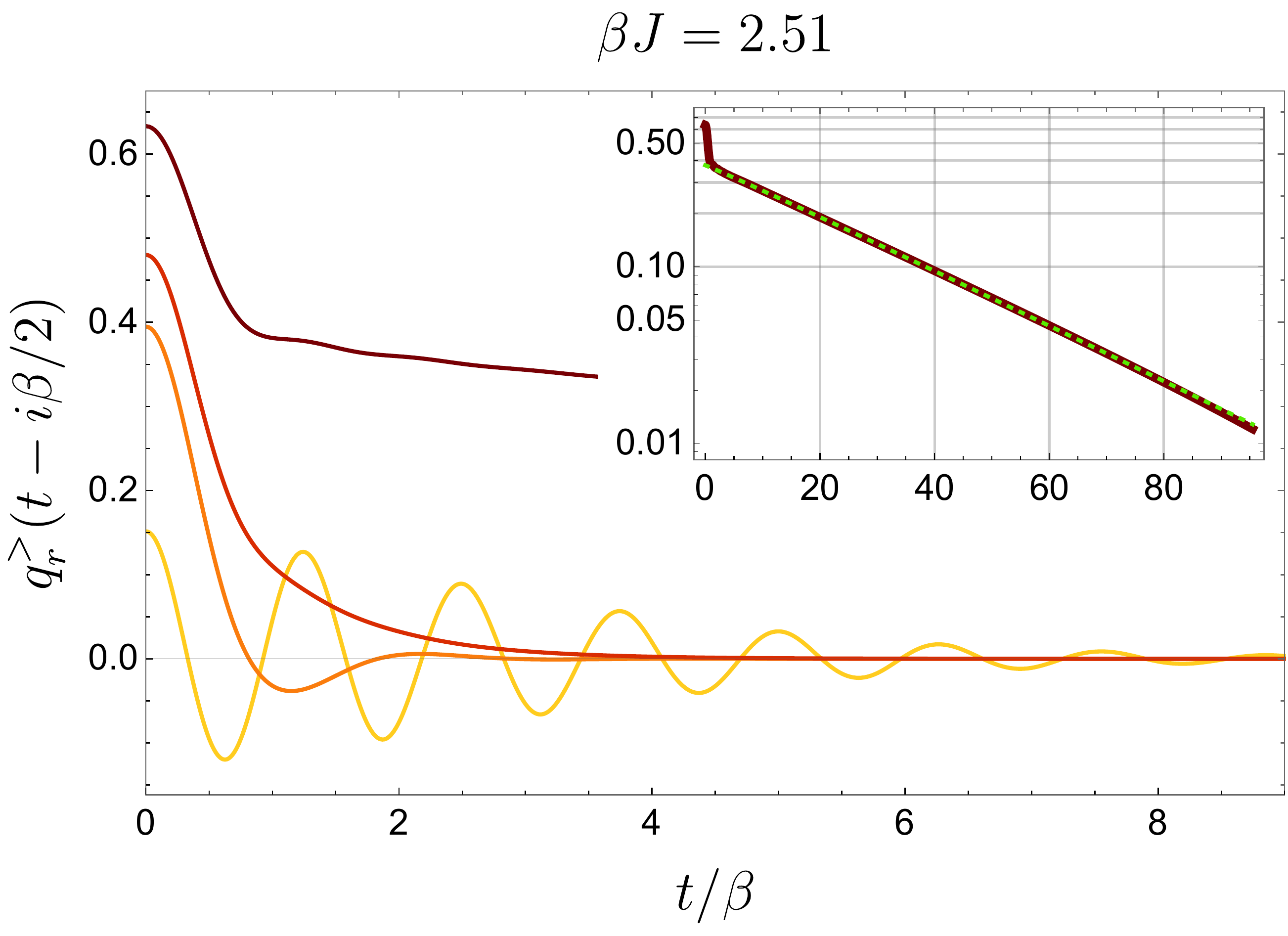}$\;\;$
\includegraphics[width=0.494\textwidth]{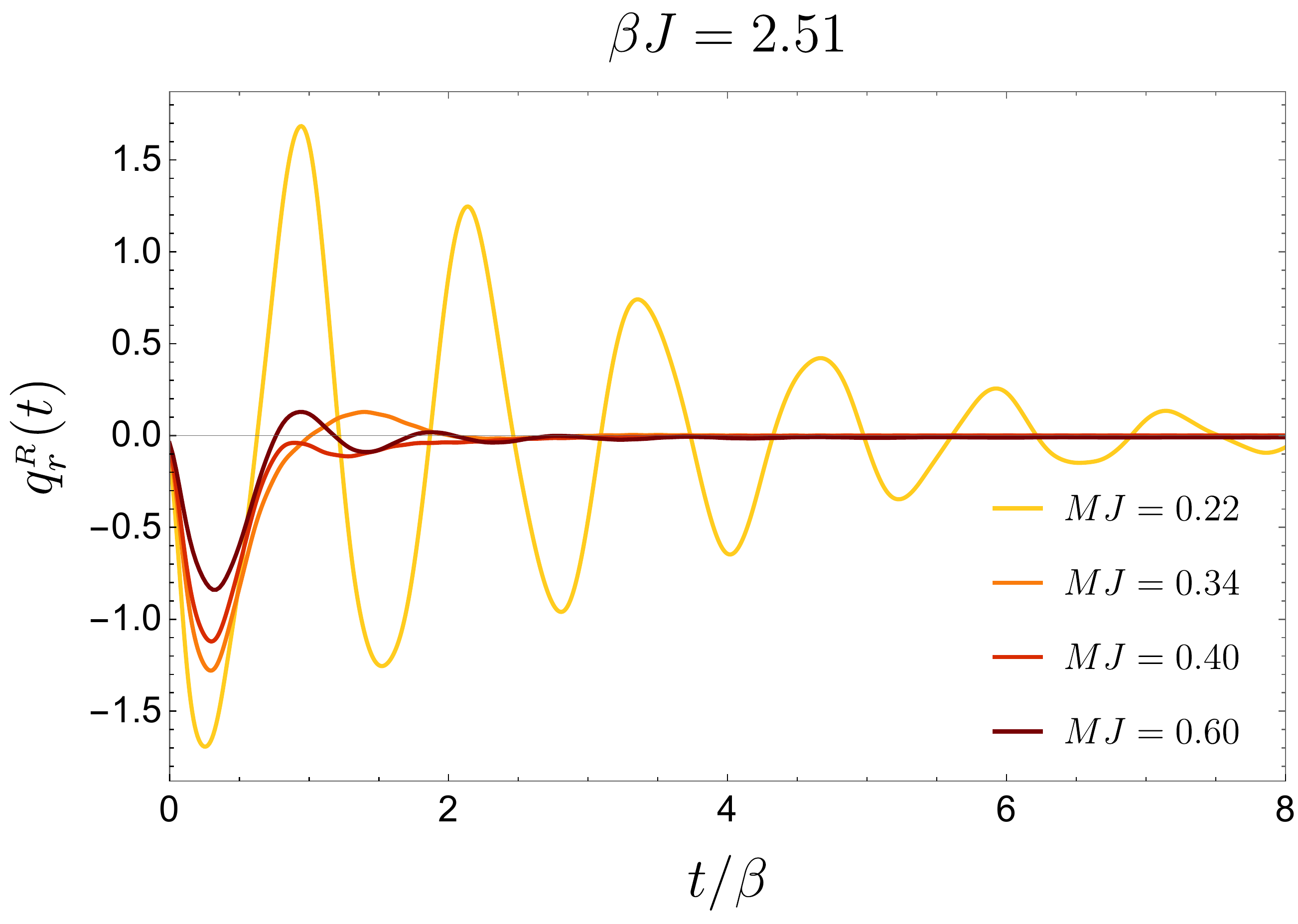}
\end{center}
\vspace{-.5cm}
\caption{{\bf Lorentzian two-point functions (paramagnetic phase).} Wightman (autocorrelation) and retarded (response) functions at fixed coupling $\beta J = 2.51$ and for various values of $MJ$ in the paramagnetic phase. Darker lines are closer to the phase transition. For small $MJ$ we observe oscillatory (`quantum') behavior. For large $MJ$ the Wightman function shows monotonic (`classical') decay over larger time scales.
As we approach the spin glass transition ($MJ \approx 0.63$) the decay becomes increasingly slow. 
The decay fits an exponential, as we show in the inset by fitting the case of $MJ=0.6$ to the function $\sim \exp[{-0.031\, (t/\beta)^{1.03}}]$ (green dotted line) on a log-plot.
The corresponding spectral functions are shown in figure \ref{fig:RhoEucl}.}
\vspace{.5cm}
\label{fig:WightRet}
\begin{center}
\includegraphics[width=0.48\textwidth]{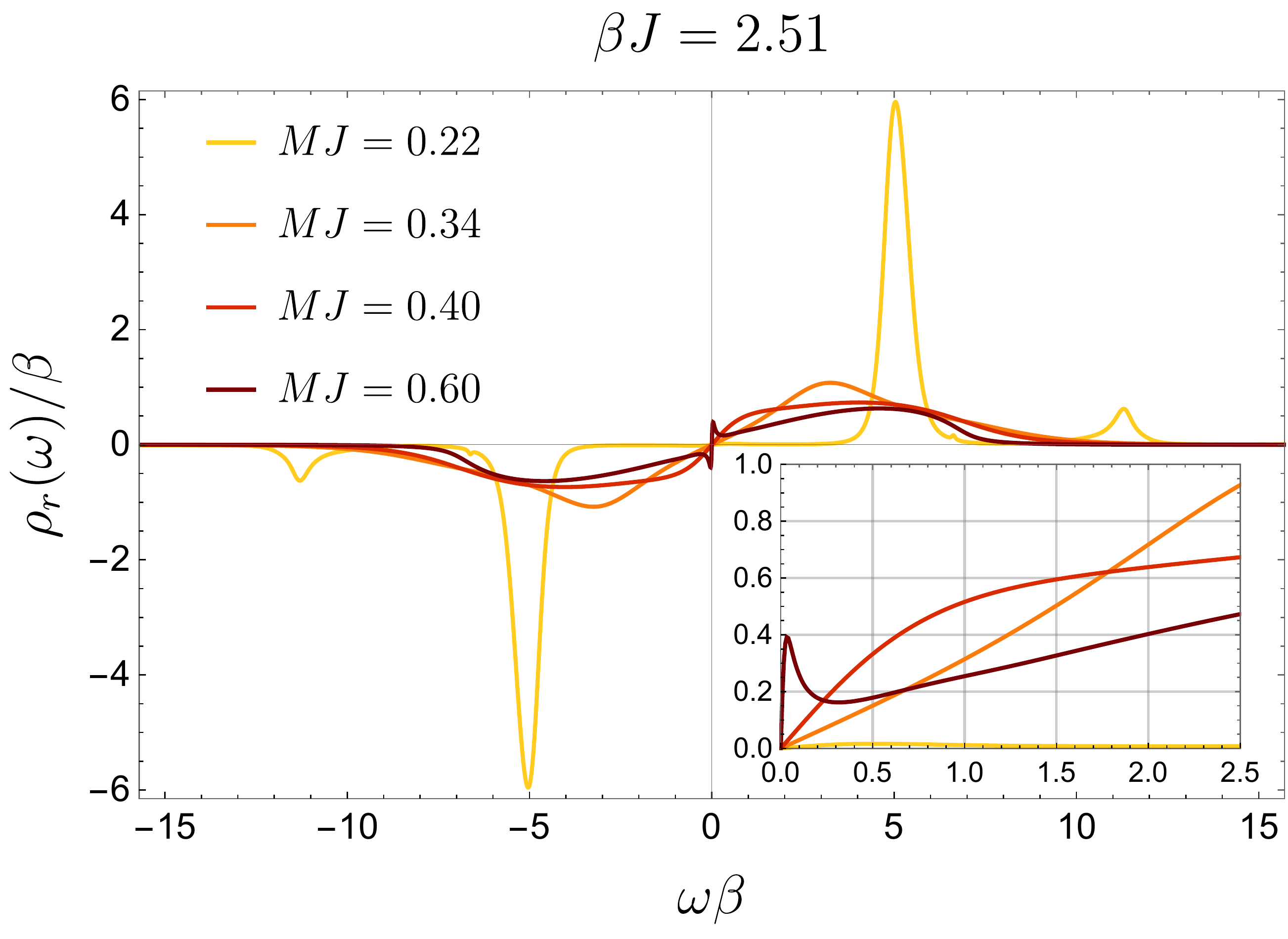}$\;\;$
\includegraphics[width=0.49\textwidth]{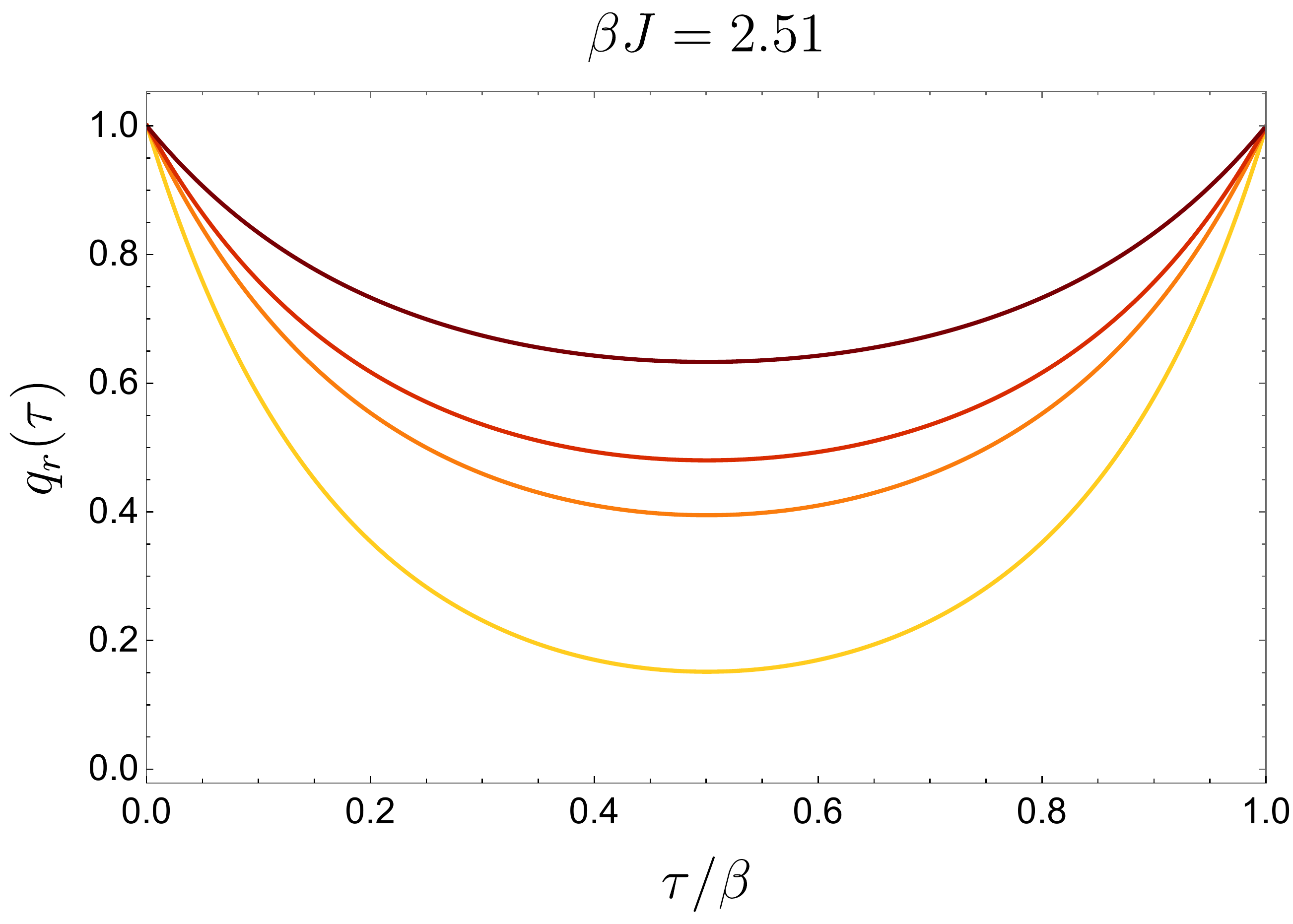}
\end{center}
\vspace{-.5cm}
\caption{{\bf Spectral properties of paramagnetic solutions.} Spectral functions $\rho_r(\omega)$ and Euclidean two-point functions $q_r(\tau)$ for the same values of $\beta J$ and $MJ$ as in figure \ref{fig:WightRet}. The spectral function develops a gap for small values of $MJ$. For large values of $MJ$ (as we get into a more classical regime and approach the spin glass transition), low frequency excitations are apparent. The Euclidean correlator develops exponential decay for small $MJ$ (corresponding to the oscillations visible in the real-time correlation functions of figure \ref{fig:WightRet}).}
\label{fig:RhoEucl}
\end{figure}

We can numerically solve \eqref{eq:EOMret} by a similar iterative method as described for the Euclidean case. This is described in appendix \ref{app:numerics}. In the remainder of this section we discuss some salient features of the two-point functions thus obtained.

\subsection{Features of the paramagnetic correlation functions}

For illustration, we show some real-time two-point functions in figure \ref{fig:WightRet}. The corresponding spectral function and the Euclidean correlators are shown in figure \ref{fig:RhoEucl}.\footnote{ These were obtained with the algorithm of appendix \ref{app:numerics}, using discretization of the relevant time windows into ${\cal O}(5000)$ elements, and discretization of the relevant frequency windows into ${\cal O}(10^6)$ elements.}
Interesting features can readily be seen from the plots at fixed value of $\beta J$ as we vary $MJ$:\footnote{ For similar observations, see also \cite{Cugliandolo_1998,Cugliandolo_1999} and more recently \cite{Tulipman:2020abw}.} at small values of $MJ$, the spectral function shows a gapped spectrum. Correspondingly, the correlation functions behave very `quantum': the Euclidean two-point function develops exponential decay, which in turn implies oscillatory behavior in the real-time correlators due to inertial effects.\footnote{It is worth mentioning that the classical dynamics of the $p$-spin model also exhibits oscillations in the two-point function  \cite{cugliandolo2017non}, so one should not conclude that classical oscillations are prohibited from this intuitive picture.} As we increase $MJ$, we observe the competition of the kinetic term and the coupling strength: at larger values of $MJ$, the real-time oscillations disappear and the gap in the spectrum closes, indicating weakly coupled physics. As we approach the spin glass transition point ($MJ \approx 0.70$ for the shown value of $\beta J$), the real-time correlations display very slow relaxation over long time scales. This is illustrated for $MJ=0.60$ in the logarithmic plot (inset of figure \ref{fig:RhoEucl}). Before even reaching the spin glass regime, the system therefore undergoes a transition from a `quantum' to a `classical' paramagnet. As we will see in section \ref{sec:OTOCs}, this transition manifests itself in four-point functions, as well. Some of these features are also illustrated in the schematic phase diagram, figure \ref{fig:phasesCartoon}. It is important that we point out, at this stage, that we are only interested in the initial relaxation process for the spin glass phase (also known as  $\beta$-relaxtion \cite{gotze1992relaxation,gotze2008complex}), whereas we probe a two-step relaxation process within the paramagnetic phase (see the inset of the left panel of figure \ref{fig:WightRet}).

\subsection{Features of the spin glass correlation functions}

In principle we can use the same definitions as in the previous section for the spin glass phase. For example, \eqref{eq:selfEnergyTime} is already written appropriately for the spin glass case ($u > 0$). The new ingredient, which we would like to discuss here, is the availability of a simple analytic approach to the real-time correlation functions in the marginally stable spin glass. 

\begin{figure}
\begin{center}
\includegraphics[width=0.485\textwidth]{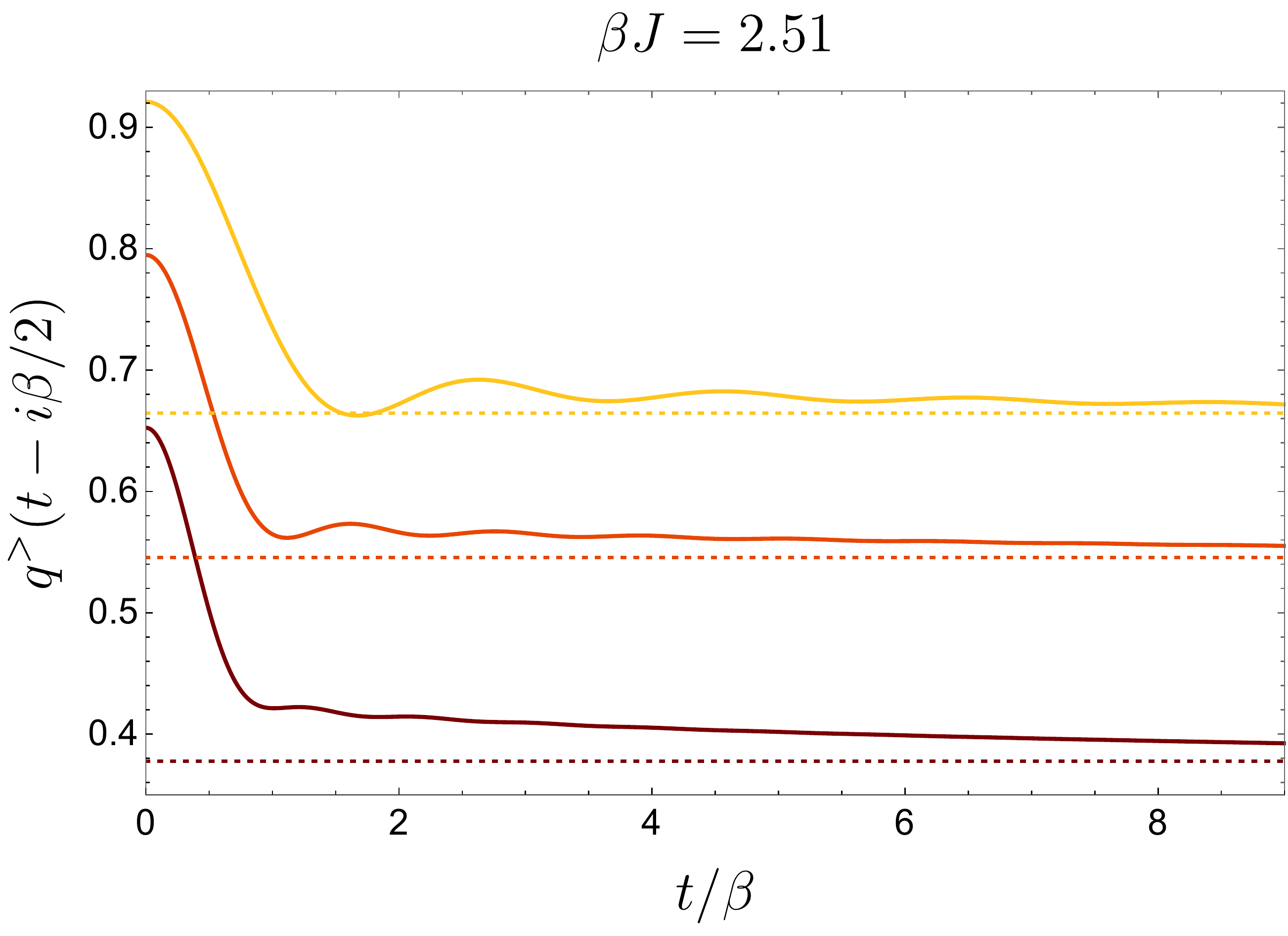}$\;\;$
\includegraphics[width=0.494\textwidth]{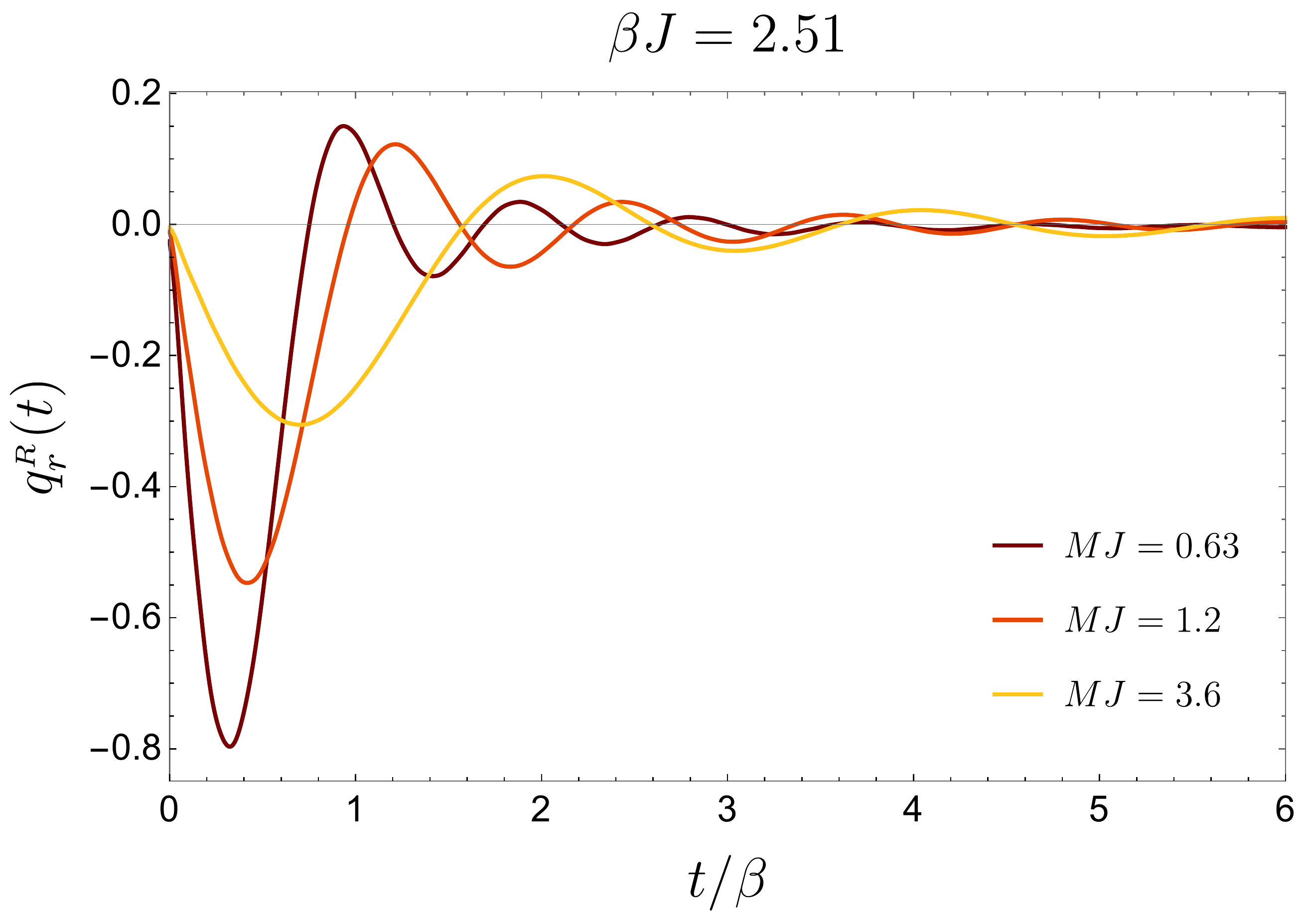}
\end{center}
\vspace{-.5cm}
\caption{{\bf Lorentzian two-point functions (spin glass phase).} Wightman (autocorrelation) and retarded (response) functions at fixed coupling $\beta J = 2.51$ and for various values of $MJ$ in the marginally stable spin glass phase. Darker lines are closer to the phase transition. The behavior is qualitatively similar to the correlators in the paramagnetic phase (figure \ref{fig:WightRet}) near the phase transition. Note however that the Wightman function in the spin glass is offset by a finite amount $u$ and never decays to zero (dotted lines).}
\vspace{.5cm}
\label{fig:WightRetSG}
\begin{center}
\includegraphics[width=0.48\textwidth]{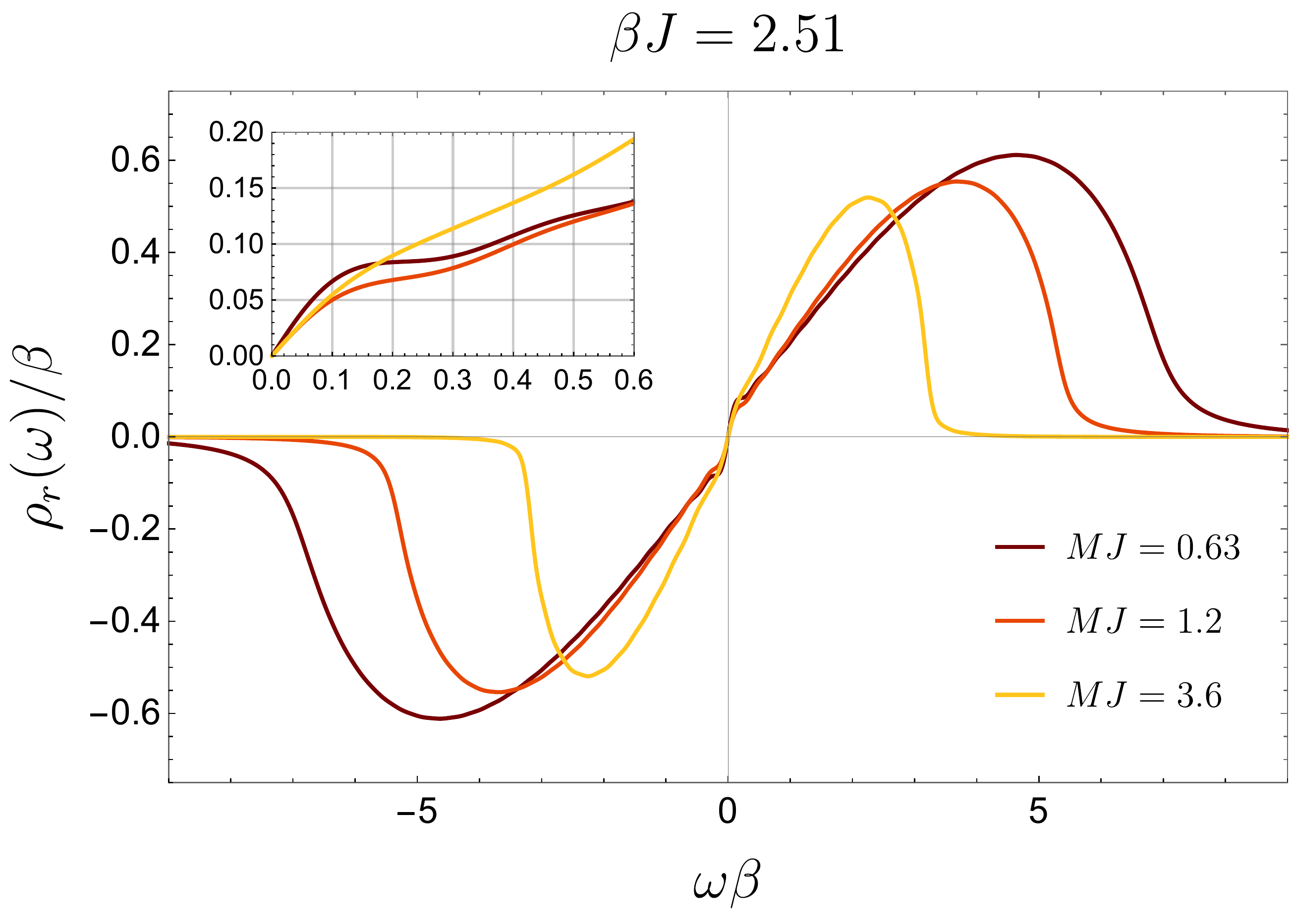}$\;\;$
\includegraphics[width=0.49\textwidth]{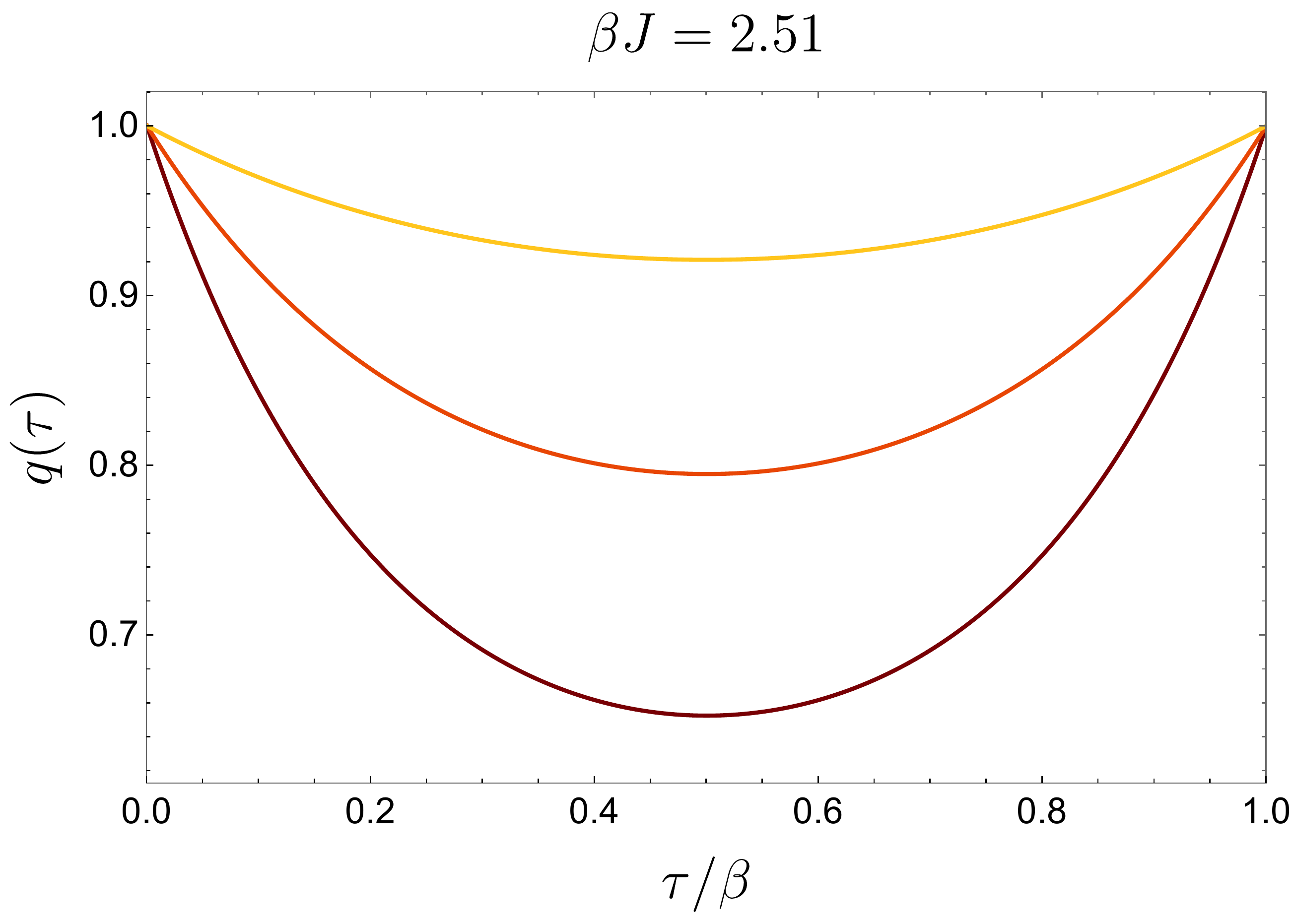}
\end{center}
\vspace{-.5cm}
\caption{{\bf Spectral properties of spin glass solutions.} Spectral functions $\rho_r(\omega)$ and Euclidean two-point functions $q(\tau)$ in the marginally stable spin glass state for the same values of $\beta J$ and $MJ$ as in figure \ref{fig:WightRetSG}. The spectral function is always gapless in this phase. The Euclidean correlators approach the finite order parameter value $u$ at $\tau=\beta/2$.}
\label{fig:RhoEuclSG}
\end{figure}

Consider the approximate solution for the marginal spin glass, discussed in section \ref{sec:qApprox}. We recall here the spectral representation of the Euclidean solution at finite temperature:
\begin{equation}
 q_r^\approx(\tau) = \int_{0}^{\infty} \frac{d\omega}{2\pi}  \, \rho^\approx_r(\omega) \, \frac{\cosh \left[ \left( |\tau| - \frac{\beta}{2} \right) \omega\right]}{\sinh \left( \frac{\beta\omega}{2} \right)} \,,\qquad
 \rho^\approx_r (\omega) = \frac{16\gamma^2}{M} \, \Theta\left( \frac{1}{\gamma}  - |\omega| \right) \, \gamma \omega\, \sqrt{1- \gamma^2\omega^2}
\end{equation}
Using the definitions from the previous subsections, we immediately obtain integral expressions for the approximate real-time two-point functions. For instance, the Wightman functions are:
\begin{equation}
\begin{split}
  \big( q_r^\approx \big)^>(t) &= \int_{-\infty}^{\infty} \frac{d\omega}{2\pi} \, e^{i\omega t} \, n_B(\omega) \rho_r^\approx(\omega) = \frac{8 \gamma}{\pi M} \int_{-\pi/2}^{\pi/2} d\theta \, \cos^2\theta \sin \theta \, \frac{e^{\frac{it}{\gamma} \sin \theta}}{e^{\frac{\beta}{\gamma}\sin\theta}-1}  \\
    \big( q_r^\approx \big)^<(t) &= \int_{-\infty}^{\infty} \frac{d\omega}{2\pi} \, e^{i\omega t} \, (1+n_B(\omega)) \rho_r^\approx(\omega) = \frac{8 \gamma}{\pi M} \int_{-\pi/2}^{\pi/2} d\theta \, \cos^2\theta \sin \theta \, \frac{e^{\frac{it}{\gamma} \sin \theta}}{1-e^{-\frac{\beta}{\gamma}\sin\theta}}  \,.
\end{split}
\end{equation}
Later on (for the purpose of computing quantum chaos), we need the retarded and the analytically continued Wightman functions, which are particularly simple and manifestly real:
\begin{equation}
\begin{split}
    \big( q_r^\approx \big)^R(t) & = -i\Theta(t) \left[ \big( q_r^\approx \big)^>(t) - \big( q_r^\approx \big)^<(t) \right]
    = - \frac{8\gamma^2}{M} \, \frac{\Theta(t)}{t} \, J_2 \left( \frac{t}{\gamma} \right)\,,\\
  \big( q_r^\approx \big)^>(t-i\beta/2) &=  \frac{8 \gamma}{\pi M} \int_{0}^{\pi/2} d\theta \, \cos^2\theta \sin \theta \; \frac{\cos \left(\frac{t}{\gamma} \,\sin \theta \right)}{\sinh \left( \frac{\beta}{2\gamma} \, \sin \theta \right)}  \,.
\end{split}
\end{equation}
where $J_2$ is a Bessel function. For large Lorentzian times $t \gg \beta$, these behave as
\begin{equation}
\begin{split}
    \big( q_r^\approx \big)^R(t) & \sim \frac{16\gamma}{M\sqrt{2\pi}} \left( \frac{\gamma}{t} \right)^{3/2} \cos \left( \frac{t}{\gamma}  - \frac{\pi}{4} \right)\,,\\
  \big( q_r^\approx \big)^>(t-i\beta/2) &\sim  - \frac{8\gamma}{M\sinh \left( \frac{\beta}{2\gamma}  \right)\sqrt{e\pi}} \left( \frac{\gamma}{t} \right)^{3/2} \cos \left( \frac{t}{\gamma}  +\frac{\pi}{4} \right)\,.
\end{split}
\end{equation}

\begin{figure}
\begin{center}
\includegraphics[width=0.485\textwidth]{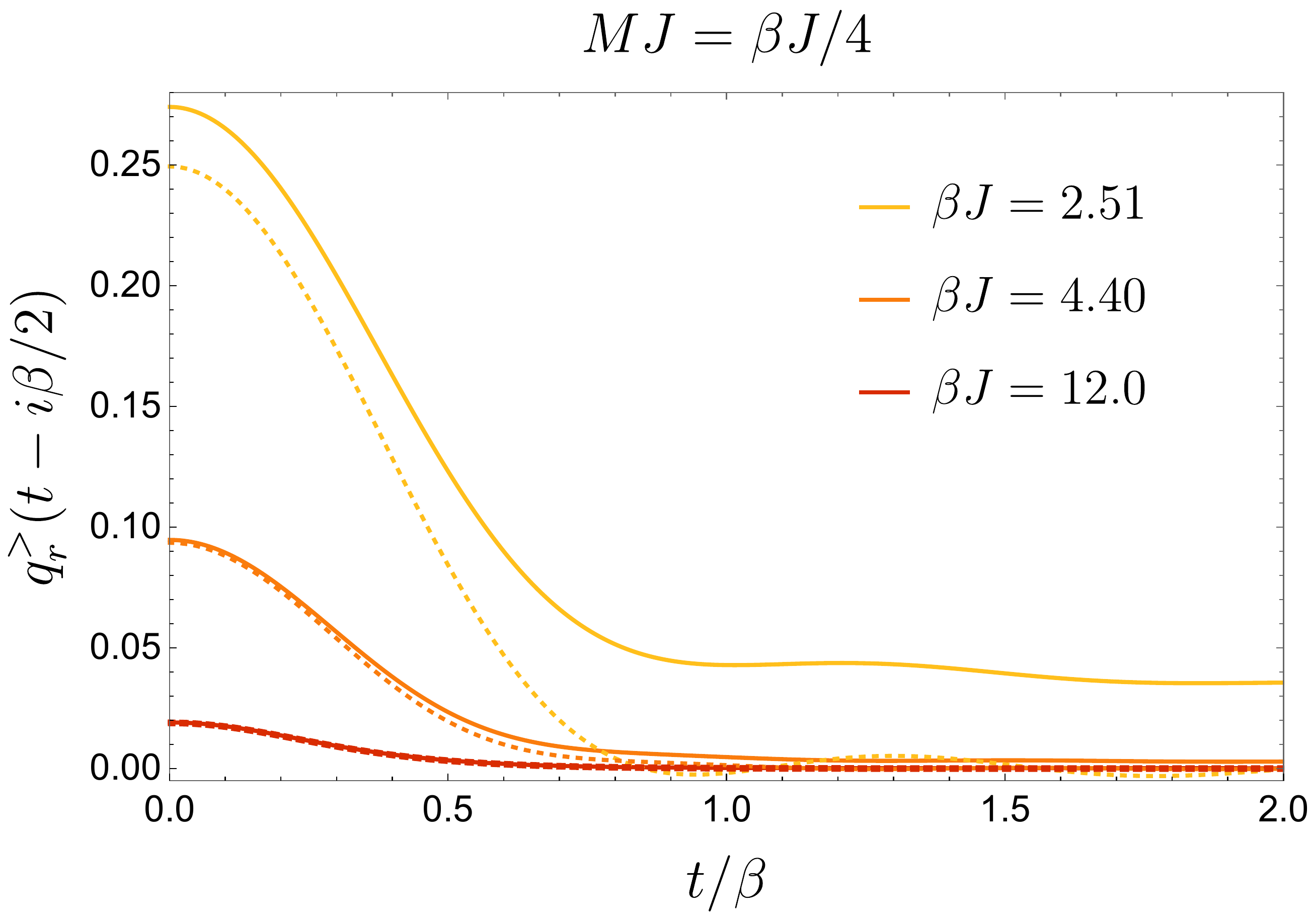}$\;\;$
\includegraphics[width=0.485\textwidth]{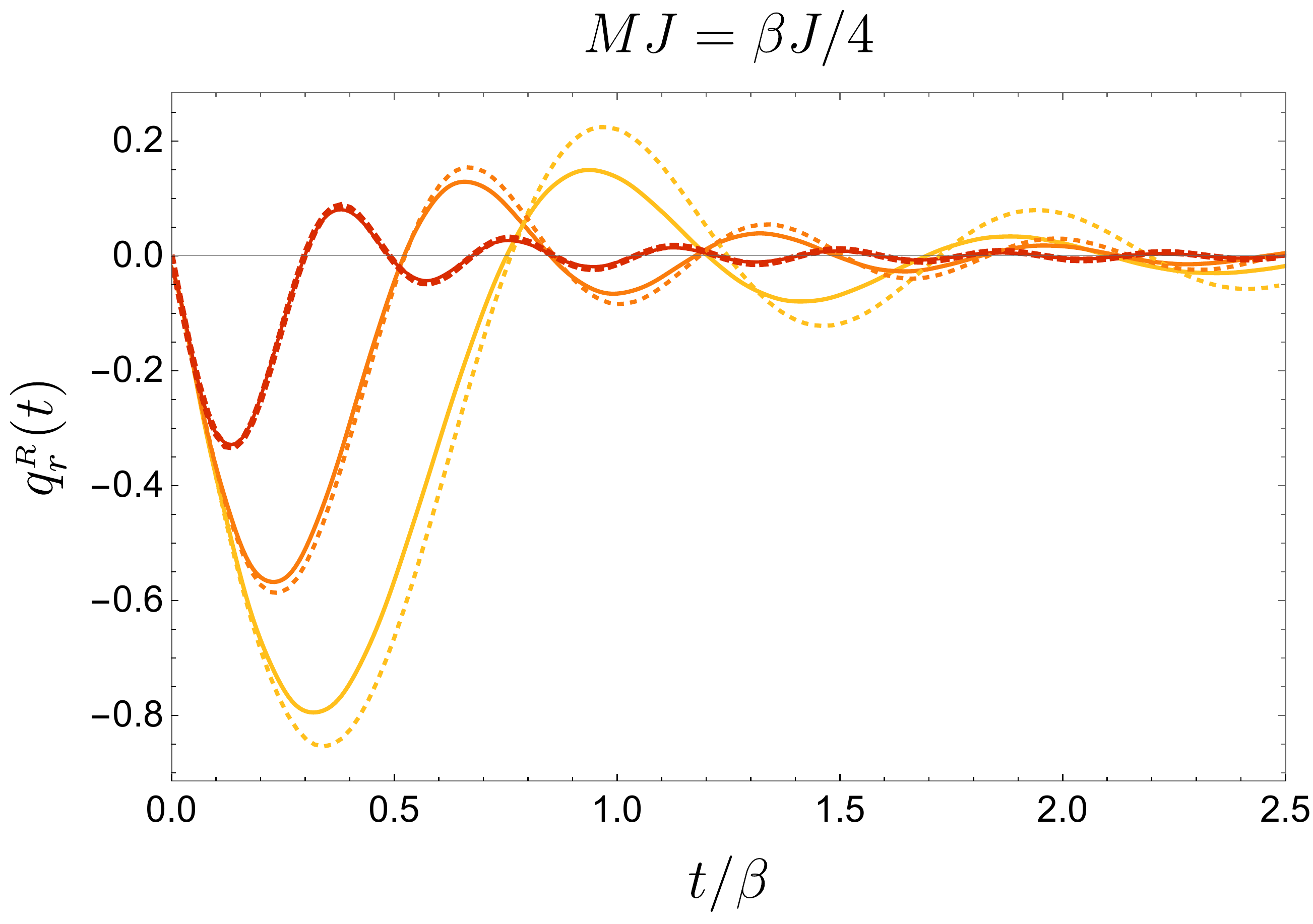}
\end{center}
\vspace{-.5cm}
\begin{center}
\includegraphics[width=0.48\textwidth]{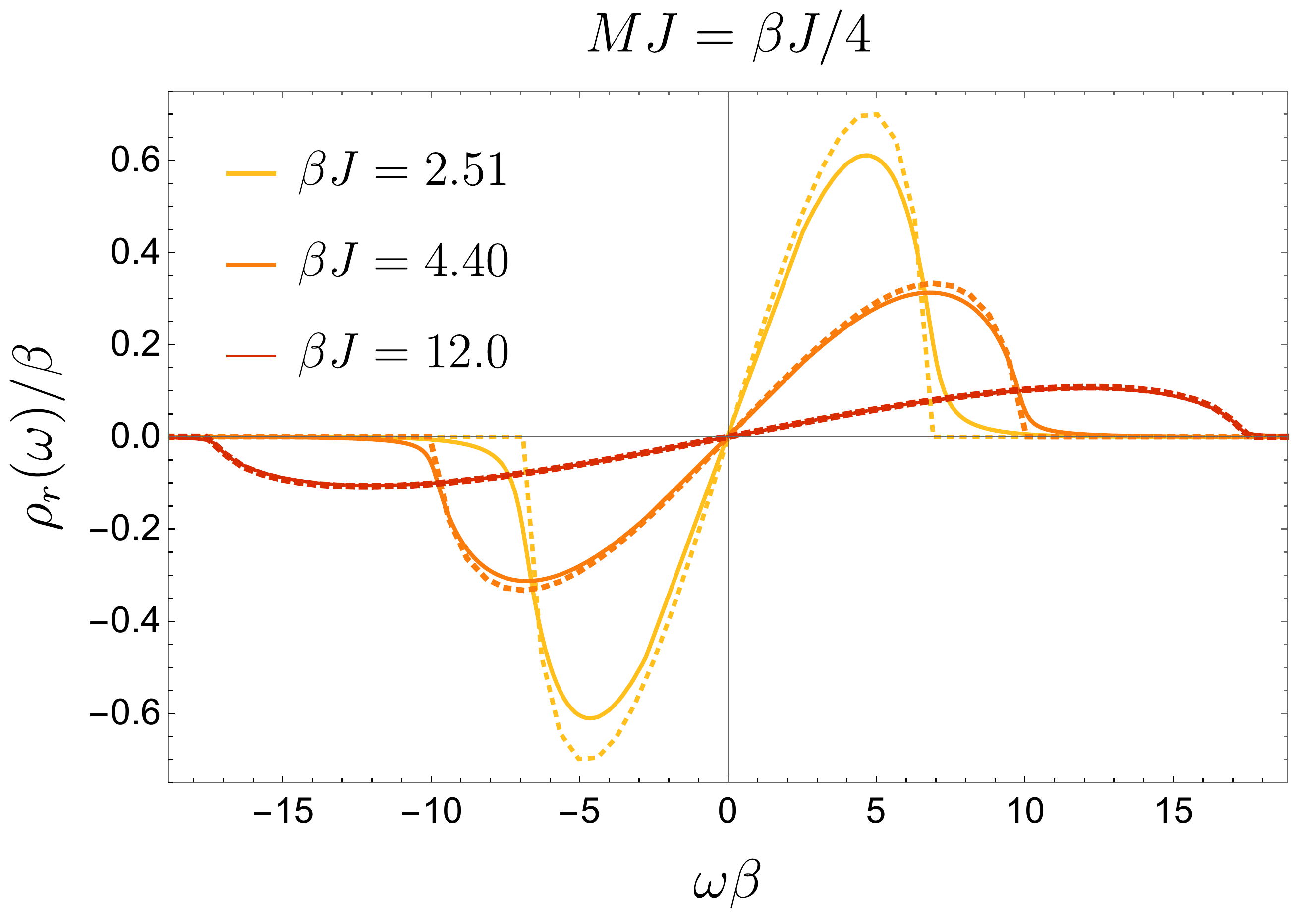}$\;\;$
\includegraphics[width=0.49\textwidth]{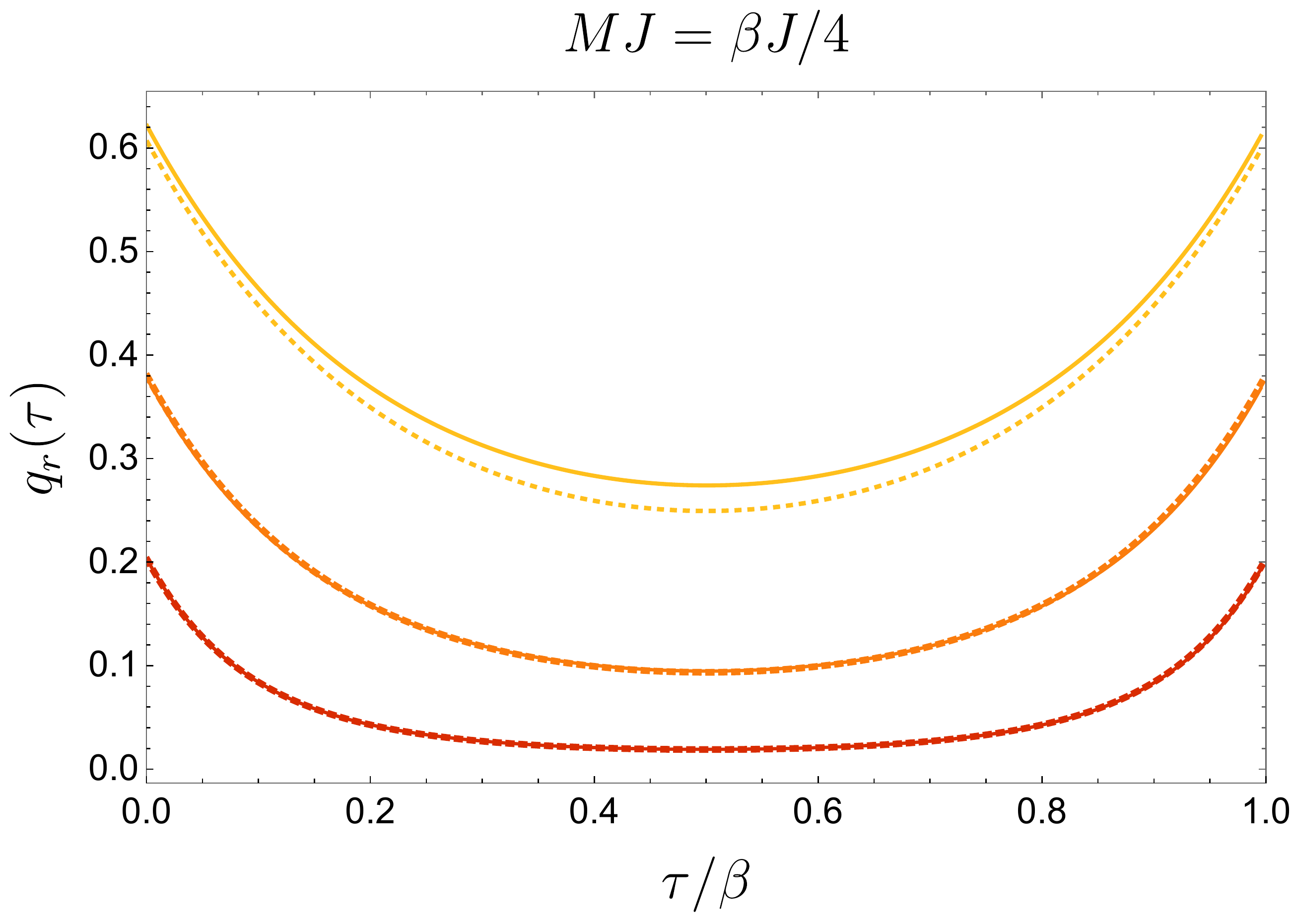}
\end{center}
\vspace{-.5cm}
\caption{{\bf Comparison of marginal spin glass solutions and analytical approximation.} Wightman (autocorrelation), retarded (response), spectral, and Euclidean two-point functions at fixed ratio $\beta / M = 4$ and for various values of $\beta J$. In case of the Wightman and Euclidean functions, we only show the non-constant parts $q_r^>(t-i\beta/2)$ and $q_r(\tau)$. Their full form $q = q_r+u$, where $u = \{0.378,\, 0.622,\, 0.797\}$ for the cases shown. The spectral function is always gapless. Dotted lines denote the approximate analytical solutions, evaluated perturbatively to first subleading order in $1/(\beta J)$ using \eqref{eq:mlowtSG}-\eqref{eq:qr0lowtempexp}. The approximate solutions are remarkably good at low temperatures and capture all the nontrivial features.}
\label{fig:RhoEuclMargComp}
\end{figure}

We show the various marginal spin glass two-point functions and the spectral functions in figures \ref{fig:WightRetSG} and \ref{fig:RhoEuclSG} for fixed temperatures. Several points are evident from these figures and in stark contrast with the paramagnetic case of the previous subsection: the Wightman and retarded correlation functions are very long lived. Indeed, their decay is polynomial and, in addition, the full Wightman function $q^>(t-i\beta/2) = q^>_r(t-i\beta/2) +u$ does not decay to zero at all, but rather to the order parameter $u$. The same is true for the Euclidean two-point functions. Finally, the spectral functions $\rho_r(\omega)$ are gapless for all temperatures shown. This is consistent with the scaling behavior of these functions at low temperatures. Note also that the full spectral function $\rho(\omega) = \rho_r(\omega) + 2\pi u \, \delta(\omega)$ has an additional peak at $\omega=0$.

In figure \ref{fig:RhoEuclMargComp} we show similar data, but for varying temperatures and fixed values of $M/\beta$. As this slice of parameter space is expected to be well approximated by the approximate analytical solution, we plot both. When evaluating the approximate solution, we determine parameters such as $m$ and $u$ using the second-order expansion of section \ref{sec:qApprox} (in particular, we use \eqref{eq:mlowtSG} and \eqref{eq:ulowtSG}). Also note that we consider the case where $\beta J/ (MJ) = 4$ is constant, which is appropriate for the `quantum scaling' introduced in section \ref{sec:quantumSc}. The same case was considered in the context of thermodynamics (c.f., figure \ref{fig:thermo_marg_quantum}). The fact that the real-time correlators at low temperatures are very well approximated by the analytical approximation, thus confirms that not only thermodynamics, but also dynamical aspects are well captured by $q_r^\approx$.

\section{The four-point function and quantum chaos}
\label{sec:FourPt}

In this section we discuss four-point functions. We begin in Euclidean signature and then analytically continue, using the Schwinger-Keldysh approach, in order to evaluate the out-of-time-order four-point correlation function, which serves as a diagnostic of quantum chaos.

\subsection{The ladder kernel}

We are interested in the four point function
\begin{equation}
\frac{1}{N^2}\langle \sigma_i(\tau_1)\sigma_i(\tau_2)\sigma_j(\tau_3)\sigma_j(\tau_4)\rangle=\langle q(\tau_1,\tau_2)q(\tau_3,\tau_4)\rangle = q_\star(\tau_{12}) q_\star(\tau_{34}) \, \left[ 1+ \frac{1}{N} \, {\cal F}(\tau_1,\tau_2,\tau_3,\tau_4) \right] \,.
\end{equation} 
where $q_\star$ is a solution to the SD equation \eqref{eq:SDeqs}. The connected piece ${\cal F}$ can be extracted from the effective action \eqref{eq:1rsbeffaction}. Following \cite{Maldacena:2016hyu,Sarosi:2017ykf}, we expand (see appendix \ref{ap:dynfluct} for details on the derivation):
\begin{equation}
	q(\tau,\tau')=q_\star(\tau,\tau')+ q_\star(\tau,\tau')^{\frac{2-p}{2}}\,\tilde{r}(\tau,\tau')~.
\end{equation}
In order to find the two-point function of the fluctuation field $\tilde{r}(\tau,\tau')$ we look at the quadratic action for fluctuations, which takes the following form:
\begin{equation}
\label{eq:S2eucl}
\begin{split}
\frac{\delta S_2}{nN}&=\frac{(\beta\mathcal{J})^2}{4}\int_0^\beta\frac{d\tau_1}{\beta}\cdots\frac{d\tau_4}{\beta}\, \tilde{r}(\tau_1,\tau_2)\left[\tilde{K}^{-1}\left(\tau_1,\tau_2;\tau_3,\tau_4\right)- \beta\delta(\tau_{13})\,\beta\delta(\tau_{24})\right]\tilde{r}(\tau_3,\tau_4) \\
&\equiv \frac{(\beta\mathcal{J})^2}{4} \; \tilde{r} \cdot \left[ \tilde{K}^{-1} - 1 \right] \cdot \tilde{r}\,.
\end{split}
\end{equation}
In the second line we introduce a formal notation, where $\tilde{r}$ is interpreted as a vector in the space of functions of two arguments, $\tilde{K}$ as a matrix in the same space, and the product of $\delta$-functions as the identity matrix. Matrix multiplication means integration over the shared indices.
The kernel $\tilde{K}$ is derived in appendix \ref{app:kernelDerivation}. It is given by the following combination of propagators:
\begin{equation}
\label{eq:kernelEucl}
\boxed{\;\;
 \tilde{K}(\tau_1,\tau_2;\tau_3,\tau_4) = (\beta {\cal J})^2 \, \left[ q_{r\star}(\tau_{12}) + u \right]^{\frac{p-2}{2}} \, \left[ q_{r\star}(\tau_{13}) \, q_{r\star}(\tau_{24}) + \st \right] \, \left[ q_{r\star}(\tau_{34}) + u \right]^{\frac{p-2}{2}} \;\; }
\end{equation}
where 
\begin{equation}
\label{eq:sMain}
 \st \equiv
  \left\{ \begin{aligned} &\;\;\;\,0 \qquad\qquad\qquad\qquad\qquad\qquad\;\, \text{(paramagnet)} \\ 
 &\frac{p}{p-2} \,\frac{m^2u^2}{m(p-1)^2 - p(p-2)} \qquad \text{(marginal spin glass)}
 \end{aligned} \right.~.
\end{equation}
For a more general expression of $\st$ that is also valid in the equilibrium spin glass, see \eqref{eq:sdef}. Interestingly the parameter $\st$ appears as a four-point generalization of the Edwards-Anderson parameter $u$ for the two-point function: it serves as an order parameter of replica symmetry breaking in four-point functions and measures the amount by which the four-point kernel never decays to zero in the spin glass phase. It would be interesting to understand the connection between these obvservations and attempts to identify a growing length scale across the spin glass transtion \cite{dasgupta1991there,lavcevic2003spatially}.

From \eqref{eq:S2eucl} we can immediately write a formal expression for the Euclidean four-point function by inverting the kernel $ \tilde{K}^{-1} - 1$:
\begin{equation}
\label{eq:fourPtEucl}
  {\cal F}(\tau_1,\tau_2;\tau_3,\tau_4) 
   = \frac{1}{(\beta {\cal J})^2}\sum_{n\geq 1} {\cal F}_n(\tau_1,\tau_2;\tau_3,\tau_4)
\end{equation}
where the ladder diagrams are constructed recursively:
\begin{equation}
\label{eq:FnRecurse}
   {\cal F}_{n+1}(\tau_1,\tau_2;\tau_3,\tau_4) = \int \frac{d\tau}{\beta}\frac{d\tau'}{\beta} \, 
  \tilde{K}(\tau_1,\tau_2;\tau,\tau') \, {\cal F}_n(\tau,\tau';\tau_3,\tau_4) \,, \qquad {\cal F}_1 = \tilde{K} \,.
\end{equation} 
Pictorially: 
\begin{equation*}
\includegraphics[width=0.82\textwidth]{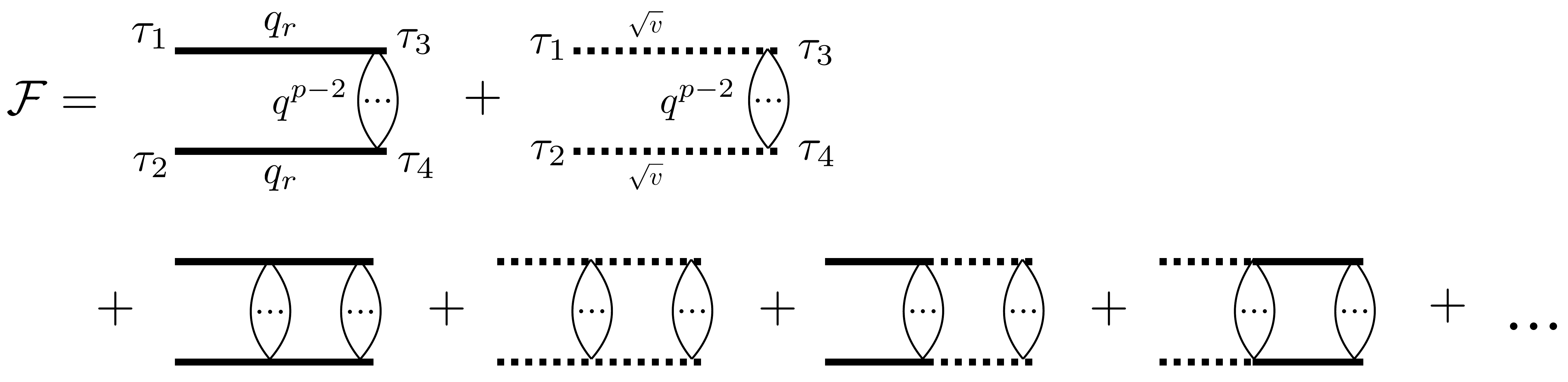}
\end{equation*}
For the paramagnetic case ($q = q_r$, $\st=0$), the ladder diagrams with constant `rails' (dashed lines) vanish and this structure is completely analogous to the SYK model. However, in the spin glass phase we see a new feature by inspecting  the kernel \eqref{eq:kernelEucl}: the `rails' of the ladders are either $q_r$ fields or constants with value $\sqrt{\st}$, while the `rungs' are given by fields $q = q_r + u$, which are also offset by a constant due to the Edwards-Anderson parameter.\footnote{ For simplicity, the diagrams we show are for the equivalent `asymmetric' kernel $\tilde{K}' \equiv q(\tau_{12})^{-\frac{p-2}{2}} \, q(\tau_{34})^{\frac{p-2}{2}} \, \tilde{K}$.\label{foot:ladders}} We discuss this feature again below in the context of the real-time four-point function.

\subsection{Out-of-time-order correlators}
\label{sec:OTOCs}

We have now all the ingredients to compute the out-of-time-order four-point function and extract the quantum Lyapunov exponent for the $p$-spin model. We begin by describing the general method, which resembles the approach described in \cite{Maldacena:2016hyu}, but comes with extra subtleties due to replica symmetry breaking and bosonic zero modes.

\subsubsection{General formalism}
\label{sec:OTOCgeneral}

We define the OTOC of interest as follows:
\begin{equation}
\begin{split}
   {\cal F}(t_1,t_2,t_3,t_4) &\equiv \frac{1}{N^2}\sum_{i,j}\left\langle \sigma_i \left(t_1\right) \, \sigma_j\left(t_3\right)\, \rho_\beta^{1/2}\, \sigma_i\left(t_2\right)\, \sigma_j(t_4) \, \rho_\beta^{1/2}\right\rangle 
   \qquad\quad (t_1 \approx t_2 \gg t_3 \approx t_4)\,.
\end{split}
\end{equation}
The time $t$ is real, whereas the imaginary shifts along the thermal circle serve to regularize the correlator. As explained in \cite{Maldacena:2016hyu}, (the connected part of) this particular analytic continuation of the Euclidean four-point function is computed by the following {\it retarded kernel}:
\begin{equation}
\label{eq:kernelLor}
 \tilde{K}_{\rm ret}'(t_1,t_2;t_3,t_4) = (\beta {\cal J})^2 \,  q_{r\star}^R(t_{13}) \, q_{r\star}^R(t_{24}) \, q^>_\star(t_{34}-i\beta/2)^{p-2} \,.
\end{equation}
This is simply the analytic continuation of \eqref{eq:kernelEucl}, where we replace the rails of the ladder diagrams with retarded two-point functions, and the rungs with analytically continued Wightman functions. Recall from \eqref{eq:wightmanoffset} that this Wightman function is offset by the Edwards-Anderson parameter $u$. We also redefined the kernel slightly with respect to \eqref{eq:kernelEucl} by putting all the rungs on one pair of external operators (see footnote \ref{foot:ladders}). Note that the constant shift $\st$ of the `rail propagators' has disappeared according to the retarded boundary conditions.

The real-time out-of-time-order correlation function is computed by inversion of this kernel, just as in the Euclidean case, c.f., \eqref{eq:FnRecurse}. Since we are interested in identifying Lyapunov behavior of the OTOC, we will consider the following exponential growth condition:
\begin{equation}
\label{eq:OTOCinteq}
	\mathcal{F}_{\rm conn.}(t_1,t_2;t_3,t_4)= \frac{1}{\beta^2}\int dt dt' \, \tilde{K}'_{\rm ret}(t_1,t_2;t,t')\, \mathcal{F}_{\rm conn.}(t,t';t_3,t_4)\,.
\end{equation}
In order to extract the Lyapunov exponent, we consider $\mathcal{F}_{\rm conn.}(t_1,t_2;0,0) \equiv \mathcal{F}_{\rm conn.}(t_1,t_2)$, and make a growth ansatz
\begin{equation}
   \mathcal{F}_{\rm conn.}(t_1,t_2) = e^{\lambda_L (t_1+t_2)/2} \, f(t_1-t_2).
\end{equation}
The condition for exponential growth then reads:
\begin{equation}
\label{eq:OTOCeq}
	f(t_1 - t_2) \, e^{\lambda_L(t_1+t_2)/2}=\frac{1}{\beta^2}\int dt dt' \, \tilde{K}'_{\rm ret}(t_1,t_2;t,t') \, f(t-t')\, e^{\lambda_L(t+t')/2}\,.
\end{equation}
Determining the Lyapunov exponent $\lambda_L$ thus corresponds to the task of finding an eigenfunction $f$ of the integral operator written above with eigenvalue 1. Given the retarded kernel $\tilde{K}'_{\rm ret}$, solving this eigenvalue problem is straightforward to do numerically: we start by computing the real-time retarded and Wightman two-point functions using the algorithm described above. These feed into the expression \eqref{eq:kernelLor} for retarded kernel. We discretize time in the integral equation and interpret it as a matrix equation for a matrix with row index $(t-t'\,,\,(t+t')/2)$ and column index $(t_1-t_2\,,\,(t_1+t_2)/2)$. Since the equation has to hold for any value of $(t_1+t_2)/2$, we can w.l.o.g.\ set this combination to some arbitrary value. 
Numerically, it is more efficient to solve the eigenvalue problem \eqref{eq:OTOCeq} in frequency space. The reason is that the position space two-point functions making up the kernel can sometimes fluctuate on vastly different timescales, especially near the phase transition. For more details on the numerical implementation, see appendix \ref{app:numerics}.

\subsubsection{OTOC at finite coupling: numerical results}

\begin{figure}
\begin{center}
\includegraphics[width=0.95\textwidth]{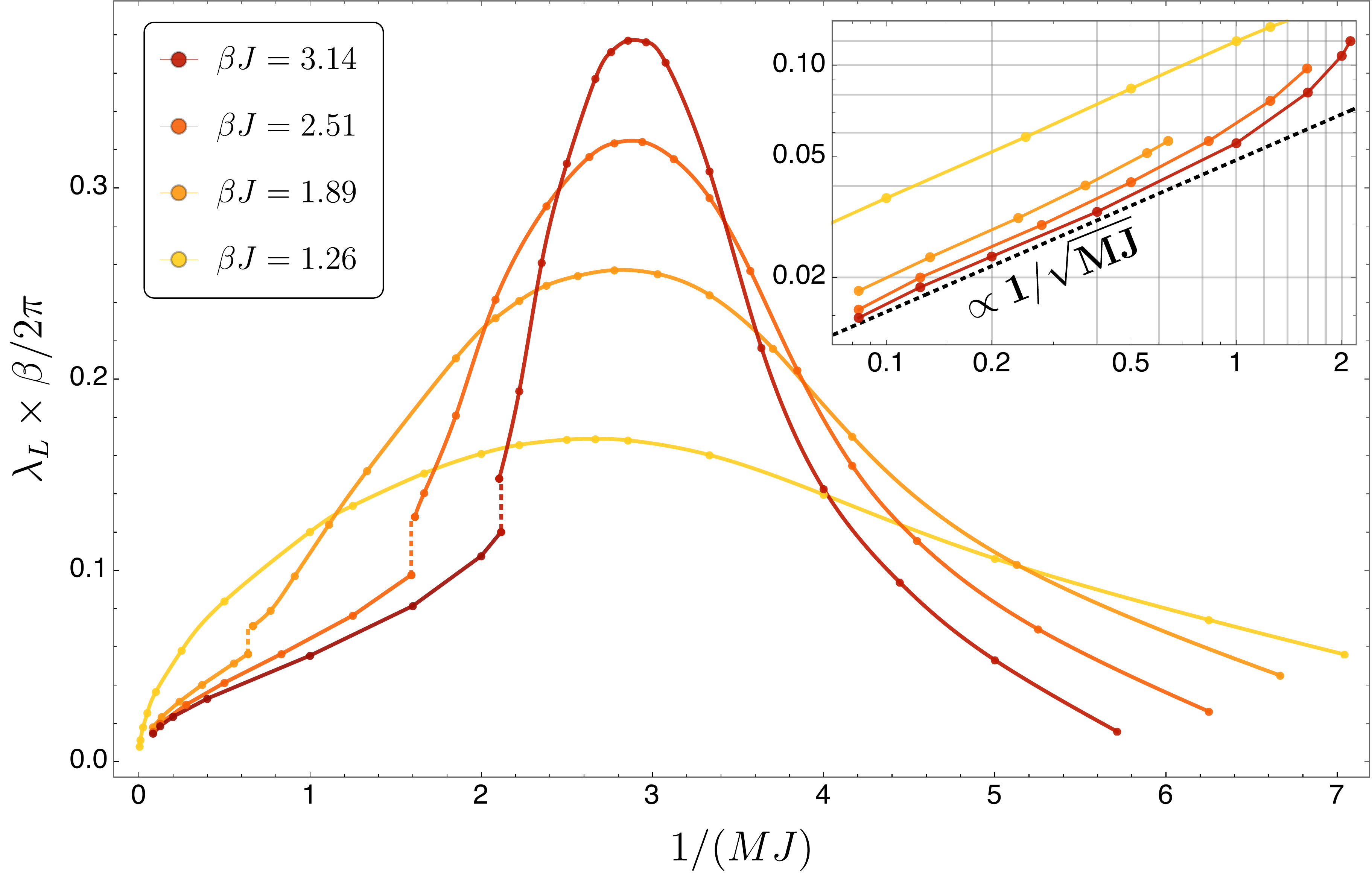}
\end{center}
\vspace{-.5cm}
\caption{{\bf Lyapunov exponents across the paramagnetic classical-to-quantum transition and across the spin glass transition (fixed $\beta J$).} We show values of the Lyapunov exponent as a function of $MJ$ for different values of $\beta J$. The phase transition (for the three darker curves) is characterized by a sharp change in slope (dotted lines). In case of the lightest curve ($\beta J$ = 1.26) no spin glass phase exists for any value of $MJ$ and we can see that the Lyapunov exponent decays both for large and small $MJ$. We observe a local maximum in the Lyapunov exponent for a critical value $M=M_c(\beta J)$. These peak locations roughly coincide with the location where the qualitative behavior of replica symmetric solutions changes: for $M\ll M_c$, the Wightman function is highly oscillatory and the spectral function displays a clear gap; for $M\gg M_c$, the real time two-point functions display increasingly monotonic decay and the spectral function develops a peak at zero frequency (c.f.\ figures \ref{fig:WightRet} and \ref{fig:RhoEucl}). {\it Inset:} we give a log-log plot of the Lyapunov exponents at large $MJ$, demonstrating an approach to zero that is to a good approximation proportional to $1/\sqrt{MJ}$ both in the spin glass and in the paramagnet.}
\label{fig:lyapunov1}
\end{figure}

If replica symmetry is unbroken, i.e., $u=0$, then the retarded kernel is very simple.
In figure \ref{fig:lyapunov1} we show the Lyapunov exponents for which the retarded kernel has an eigenfunction with eigenvalue 1. 
In this figure each dot was obtained by solving the real-time equations of motion numerically for the given values of $\beta J$ and $MJ$. Then, the method described above was applied to the solution in order to extract the exponentially growing modes. It is worth pointing out that the Lyapunov exponents we display are the {\it only} exponentially growing modes. In other words, the second largest value of $\lambda_L$ which solves the integral equation \eqref{eq:OTOCeq} is always 0 (and subsequent ones are negative, i.e., decaying); see appendix \ref{app:numerics} for more details.

\begin{figure}
\begin{center}
\includegraphics[width=0.4\textwidth]{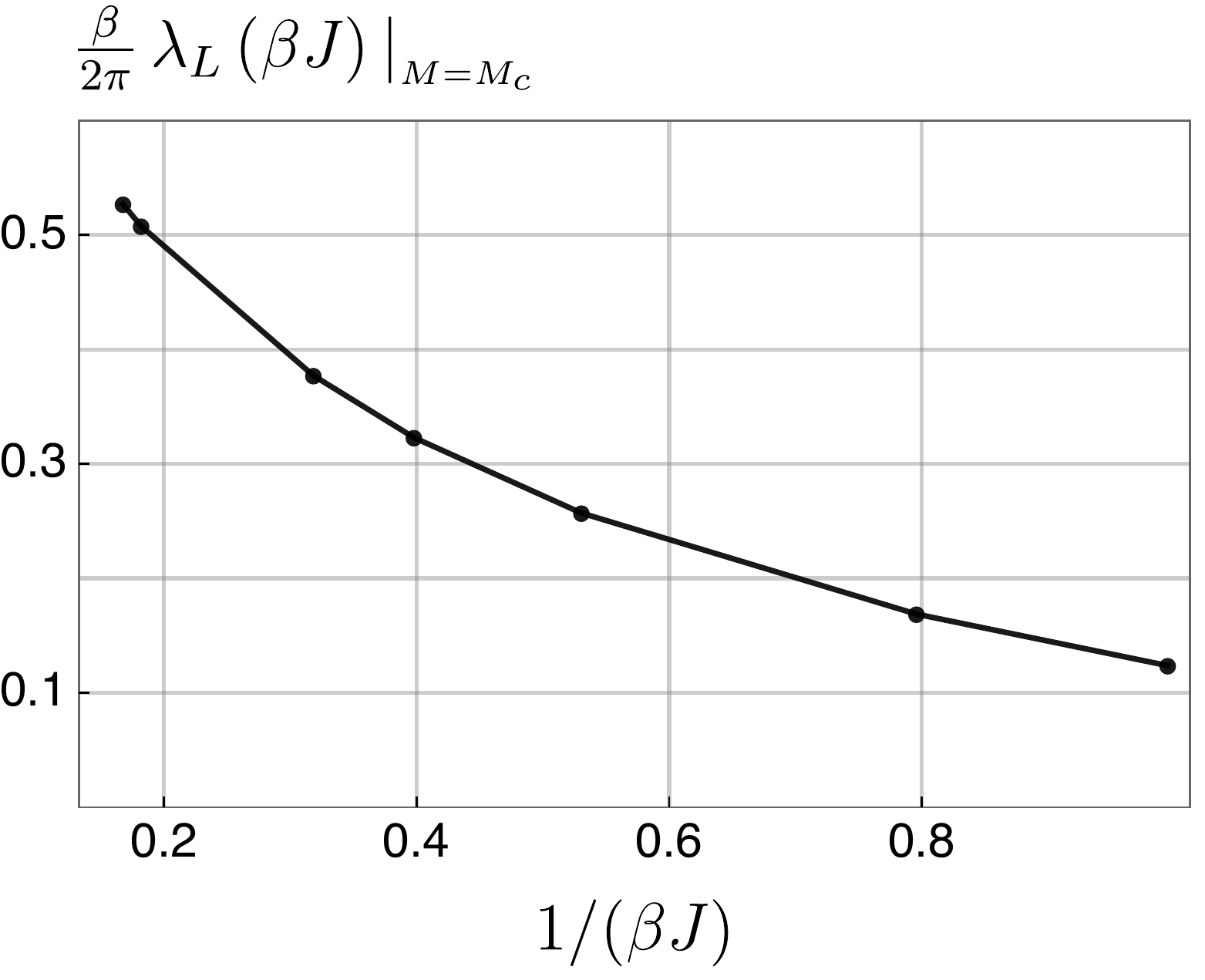}$\qquad$
\includegraphics[width=0.4\textwidth]{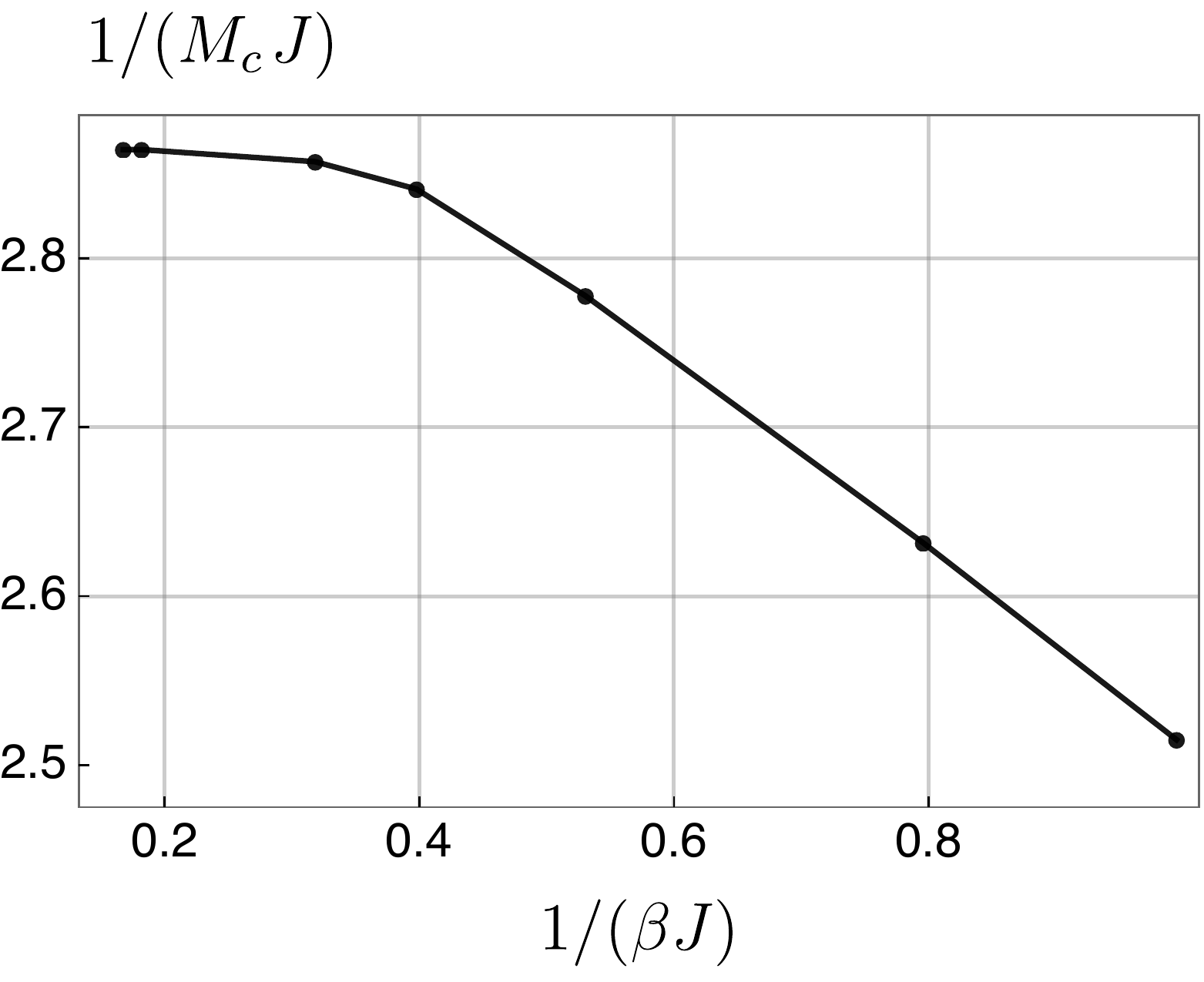}
\end{center}
\vspace{-.5cm}
\caption{{\bf Peak values of Lyapunov exponents as a function of temperature.} In the left panel we show the maximal attainable values of $\frac{\beta}{2\pi}\lambda_L(\beta J)$ at any given temperature, i.e., we evaluate $\lambda_L(\beta J,MJ)$ at its peak value $M=M_c$. On the right we show these peak values $M_c$ as a function of temperature.}
\label{fig:lyapunov1bb}
\end{figure}

An interesting aspect of figure \ref{fig:lyapunov1} is the strong dependence of the Lyapunov exponent $\lambda_L$ on the coupling $MJ$ at any fixed temperature. As a function of $MJ$, the Lyapunov exponent has two interesting features:

\paragraph{Classical-to-quantum transition:} Within the paramagnetic phase, $\lambda_L(MJ)$ displays a peak at the critical value, which is only mildly dependent on temperature in the range of temperatures shown.
At this value, the Lyapunov exponents reach their maximal value attainable for any fixed value of $\beta J$. In the phase diagram figure \ref{fig:phasesCartoon} (or figure \ref{fig:PhasesMarg}), the location of these `locally maximal' exponents corresponds to a line, which appears to approach the tri-critical point as we lower the temperature (see figure \ref{fig:lyapunov1bb}). It would be interesting to study the vicinity of this point further. For now, we only make the following remark: the location of the maximal Lyapunov exponent for fixed temperature $1/(\beta J)$ correlates with the transition from `classical' to `quantum' paramagnetic solutions. Indeed, as we saw previously (for example in figure \ref{fig:WightRet}), different values of $MJ$ at fixed temperature lead to vastly different qualitative behavior of the real-time correlation functions: small $MJ$ leads to oscillatory behavior (classical regime), while large $MJ$ leads to correlators that decay over a long time scale (quantum regime). Similarly, the spectral density is characterized by whether or not it has a gap (c.f., figure \ref{fig:RhoEucl}). The fixed-temperature maximal chaos exponent roughly coincides with the transition between these regimes, at least for the range of temperatures shown.\footnote{We thank the anonymous referee for pointing out that the maximal-Lyapunov line and the quantum-to-classical line seem very analogous to the Widom and Fisher-Widom lines in a supercritical fluid, which delineate, respectively, where the specific heat is maximal, and where the density-density correlation function changes from oscillatory to monotonic in a supercritical fluid. It would be interesting to explore this analogy further. }

\paragraph{Spin glass transition:} The Lyapunov exponent is non-zero in the marginal spin glass. However, it is parametrically smaller than in the paramagnetic phase and displays a sharp change in slope at the phase transition (dotted lines in figure \ref{fig:lyapunov1}).\footnote{ Our numerics are not quite accurate enough to determine whether or not the drop of $\lambda_L$ across the phase transition is discontinuous.} At fixed $\beta J$, we observe that the Lyapunov exponent approaches zero in the spin glass as $\lambda_L \sim 1/\sqrt{MJ}$. In figure \ref{fig:lyapunov2} we investigate the Lyapunov exponents in the spin glass phase further by fixing the ratio $M/\beta$. Such scaling is expected to be captured by the quantum scaling and thus amenable to analytic calculations, as we discuss next.

\subsubsection{OTOC in the marginal spin glass phase: perturbation theory}
\label{sec:OTOCperturbation}

In the conformal limit we have analytical control over the four-point functions. For the conformal paramagnet this is explained in appendix \ref{app:PMototc} and is analogous to the SYK model. Indeed, we find a maximal quantum Lyapunov exponent $\lambda_L = \frac{2\pi}{\beta}$ in that case. However, since the conformal paramagnetic solution is never dominant in the thermodynamic ensemble, we cannot conclude that the model is maximally chaotic in the paramagnetic phase (see also \cite{Tikhanovskaya:2020elb}).

 Let us now turn to a similar analysis in the conformal limit of the marginally stable spin glass, which {\it is} the physical solution at strong coupling. In particular we will use the approximate solution (section \ref{sec:qApprox}) and the quantum scaling ansatz (section \ref{sec:QSdef}) to obtain analytical results at low temperatures.\footnote{ The technical details of our calculation are inspired by similar computations in the paramagnetic phase of $p=2$ quantum rotors in \cite{PhysRevB.52.384,Cheng:2019nxy,Mao:2019xvt}. This is due to the fact that the exact solution to the paramagnetic equations of the $p=2$ model take a similar form as our approximate solution to the marginal spin glass equations for general $p$.}

Note that the Wightman correlator is dominated by its $u$-offset, while the retarded correlator has no such constant contribution. 
Consequently the retarded kernel in the deep spin glass regime can be analyzed systematically by expanding the factor $(q^>_{\star})^{p-2}\equiv(q^>_{r\star}+u)^{p-2}$ for $u\gg q_{r\star}^>$:
\begin{equation}
\label{eq:kernelLor2}
\begin{split}
& \tilde{K}_{\rm ret}'(t_1,t_2;t_3,t_4) = \tilde{K}_{\rm ret}'^{(0)}(t_1,t_2;t_3,t_4) +  \tilde{K}_{\rm ret}'^{(1)}(t_1,t_2;t_3,t_4) + \ldots\,,\\
& \text{where:} \quad \tilde{K}_{\rm ret}'^{(0)}(t_1,t_2;t_3,t_4)=  (\beta {\cal J})^2 \,u^{p-2} \left[ q_{r\star}^R(t_{13}) \, q_{r\star}^R(t_{24}) \right] \,,\\
  &\qquad\quad\;\;\; \tilde{K}_{\rm ret}'^{(1)}(t_1,t_2;t_3,t_4) =   \tilde{K}_{\rm ret}'^{(0)}(t_1,t_2;t_3,t_4)  \times  \frac{p-2}{u} \,q^>_{r\star}(t_{34}-i\beta/2) \,.
 \end{split}
\end{equation}
This expansion truncates, and the shown terms are complete in the case $p=3$.
Diagrammatically, we can understand this expansion in small $q_r^>$ as constructing each step of the ladder from a set of different rung diagrams:
\begin{equation*}
\includegraphics[width=.85\textwidth]{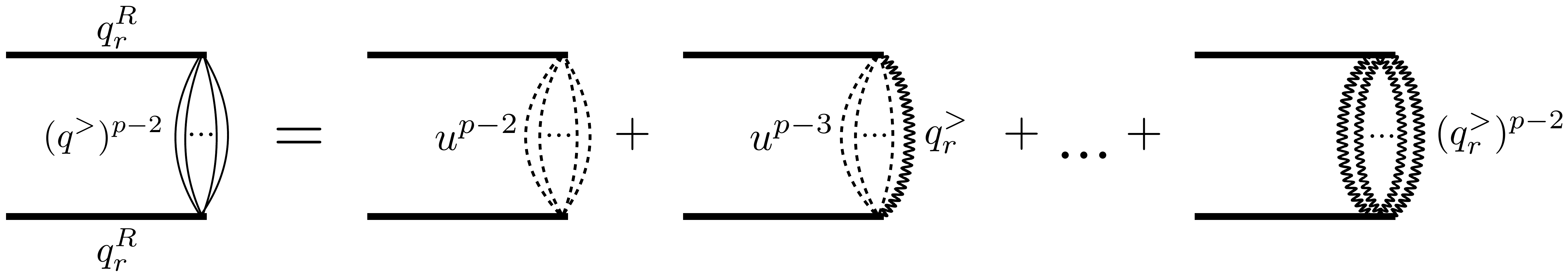}
\end{equation*}

\begin{figure}
\begin{center}
\includegraphics[width=0.6\textwidth]{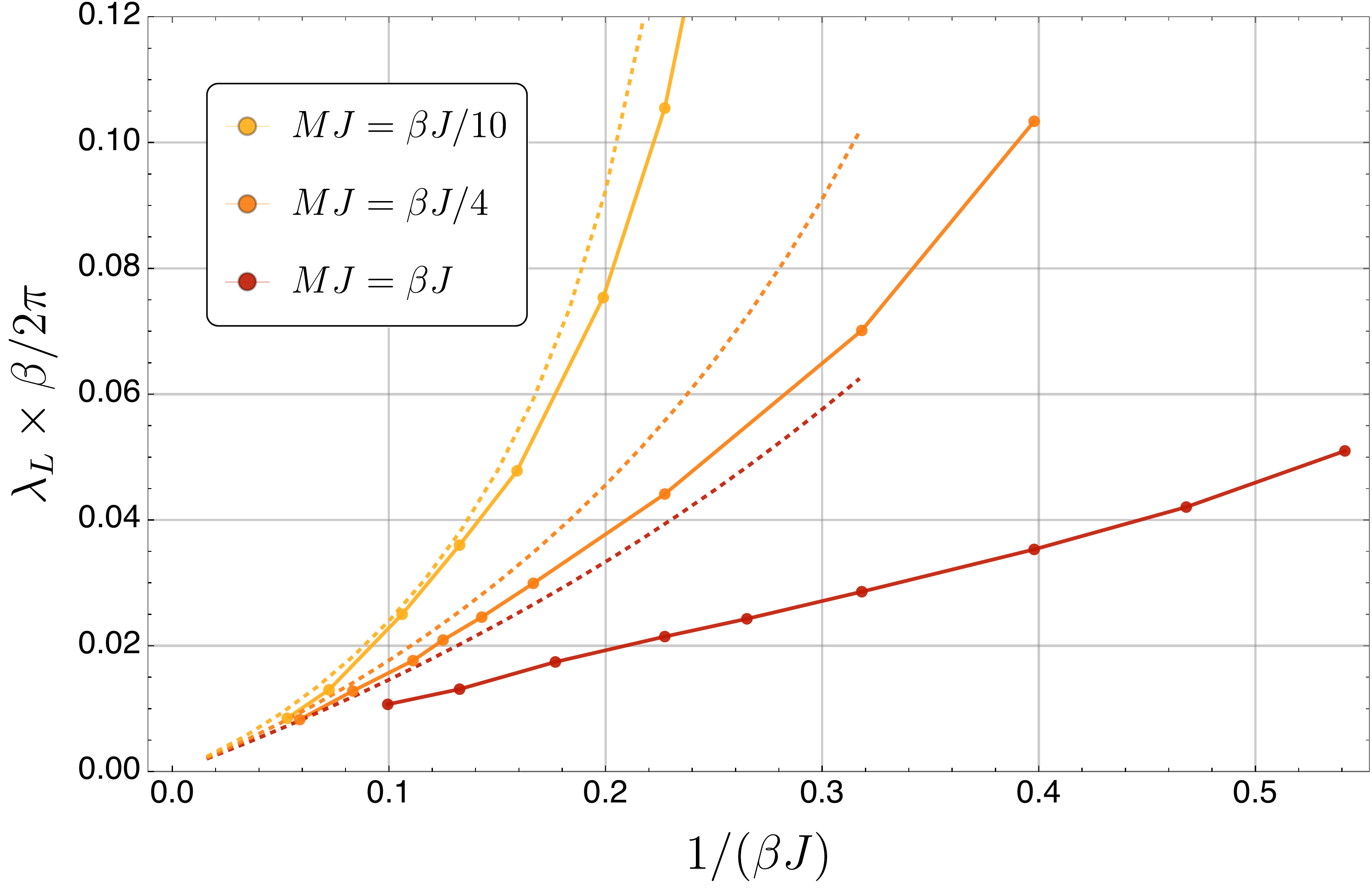}
\end{center}
\vspace{-.5cm}
\caption{{\bf Lyapunov exponents for $MJ \propto \beta J$ and in the quantum scaling limit.} As discussed in section \ref{sec:quantumSc}, holding $M/\beta$ fixed is appropriate for studying the `quantum scaling' $\beta J \sim MJ  \gg 1$ (in particular, compare with figure \ref{fig:thermo_marg_quantum}). Full lines are numerical results. Dashed lines show the analytical result \eqref{eq:SGlyapunovRes} evaluated using \eqref{eq:mlowtSG}.}
\label{fig:lyapunov2}
\end{figure}

The condition for exponential growth \eqref{eq:OTOCeq} can similarly be expanded perturbatively. This is best done in frequency space. We give a detailed account of the required nontrivial Fourier transforms in appendix \ref{app:kernelDer}. To a first approximation (using only $\tilde{K}_{\rm ret}'^{(0)}$), we find the following eigenvalue condition in frequency space:
\begin{equation}
\label{eq:expCondMS}
  \hat{f}\left(\omega\right) 
  \approx{\cal J}^2 u^{p-2} \, 
  \hat{q}^R_{r}\left(\omega - \frac{i\lambda_L}{2}\right) \, \hat{q}^R_{r}\left(-\omega - \frac{i\lambda_L}{2}\right)
  \, \hat{f}\left(\omega \right) 
\end{equation}
where we drop the star ``${}_\star$'' subscripts for ease of reading. 
Recall now the approximate solution for the retarded spin glass correlator:
\begin{equation}
\label{eq:RetApprox3}
 \left( \hat{q}_{r}^{\approx}\right)^R\!(\omega) = -  \hat{q}_r^\approx(-i\omega-\varepsilon)  = - \hat{q}_r(0) \left[ 1 - 2i \gamma \omega  - 2 \gamma^2\omega^2 + \ldots \right]
\end{equation}
The constant term in $\left( \hat{q}_{r}^{\approx}\right)^R\!(\omega)$ precisely solves the eigenvalue equation \eqref{eq:expCondMS} for any non-zero frequencies (using once again the relation \eqref{eq:conformalmagicrel}). At the next order in a small-$(\gamma\omega)$ expansion, the condition \eqref{eq:expCondMS} reduces to $\lambda_L = 0$. We conclude that to this order the Lyapunov exponent vanishes.

Let us then consider the first correction to the retarded kernel, coming from $\tilde{K}_{\rm ret}'^{(1)}$. The condition \eqref{eq:expCondMS} generalizes to the following  (see appendix \ref{app:kernelDer}):
\begin{equation}
\label{eq:expCondMS2}
\begin{split}
 & \hat{f}\left(\omega \right)
  \approx {\cal J}^2 u^{p-2}  \,
    \hat{q}^R_{r}\left(\omega - \frac{i\lambda_L}{2}\right) \, \hat{q}^R_{r}\left(-\omega - \frac{i\lambda_L}{2}\right)  \int \frac{d\omega'}{2\pi} \, \left[ 2\pi \delta(\omega') + \frac{p-2}{u} \, e^{\frac{\beta\omega'}{2}}\, \hat{q}_{r}^>(\omega') \right]  \hat{f}\left(\omega -\omega'\right) 
 \end{split}
\end{equation}
After using the spectral representation of $\hat{q}_r^>(\omega)$ and shifting the integration variable, this can be written as follows:
\begin{equation}
\label{eq:expCondMS2c}
\begin{split}
\frac{\hat{q}_r(0)^2-  \,
   \hat{q}^R_{r}\left(-\frac{i\lambda_L}{2} +{\omega}\right) \, \hat{q}^R_{r}\left(-\frac{i\lambda_L}{2} - {\omega}\right)}{
   \hat{q}^R_{r}\left(-\frac{i\lambda_L}{2} +{\omega}\right) \, \hat{q}^R_{r}\left(-\frac{i\lambda_L}{2} - {\omega}\right)} \, \times\hat{f}(\omega) \approx \frac{p-2}{u} \int \frac{d\omega'}{2\pi} \,\frac{\rho_{r}({\omega}-\omega')}{ 2\sinh \big( \frac{\beta}{2} (\omega-\omega')\big)} \, \hat{f}(\omega') \,.
 \end{split}
\end{equation}
Note that the retarded correlators appearing here are also determined by the spectral function through
\begin{equation}
\label{eq:qRrqr}
\begin{split}
 \hat{q}^R_{r}\left(-\frac{i\lambda_L}{2} \pm {\omega}\right) \equiv - \hat{q}_r\left( - \frac{\lambda_L}{2}\mp i \omega  \right)&= - \int \frac{d\omega'}{2\pi} \, \frac{\rho_r(\omega')}{\omega' \pm \omega - \frac{i\lambda_L}{2}} \,.
\end{split}
\end{equation}
Equation \eqref{eq:expCondMS2c} is well suited for a numerical approach: we merely need to discretize the integral and interpret the equation as a matrix eigenvalue problem. Due to the fact that the support of $\rho_r(\omega)$ to a very good approximation is confined to a finite interval, this procedure is efficient. Some results of such numerical analysis are shown as full lines in figure \ref{fig:lyapunov2}. We observe that the Lyapunov exponent approaches zero as we increase $\beta J$, while holding the ratio $M/\beta$ fixed. Let us now turn to an analytical perturbative solution of \eqref{eq:expCondMS2c}, which will allow us to quantify this observation in more detail.

\paragraph{Perturbative solution of \eqref{eq:expCondMS2c}:} To analyze the eigenvalue equation \eqref{eq:expCondMS2c} analytically, we would like to turn it into a differential equation for the eigenfunction $f(t)$ and study its spectrum. We will achieve this using the approximate solution $\hat{q}_r^\approx(\omega)$, as well as the approximate spectral function $\rho_r^\approx$.
As we will show, a consistent treatment requires us to include a correction to the approximate solution $q_r^\approx$. Recall the approximate equation of motion in the marginally stable spin glass, \eqref{eq:approxEOM0}. Including one higher order term in our perturbative scheme \eqref{eq:qExpandRR}, we find the following corrected equation of motion (where $\omega$ are Matsubara frequencies):
\begin{equation}
\label{eq:approxEOM1}
	M\omega^2\hat{q}_r(\omega)= \left(\frac{\hat{q}_r(\omega)-\hat{q}_r(0)}{\hat{q}_r(0)}\right)^2 +\, \frac{p-2}{2u} \, \frac{\hat{q}_r(\omega)}{\hat{q}_r(0)^2}  \left[ \hat{\Lambda}_{(2)}(\omega) - \hat{\Lambda}_{(2)}(0) \right]\,,
\end{equation}
where the effective self-energy sourcing the perturbation is given by
\begin{equation}
  \hat{\Lambda}_{(2)}(\omega) \equiv 
  \int_0^\beta d\tau \; e^{-i\omega\tau} \, q_r(\tau)^2 \,,\qquad \omega = \frac{2\pi k}{\beta} \,,\qquad k \in \mathbb{Z} \,.
\end{equation}
Using the approximate conformal solution \eqref{eq:qrcSG} in the low-temperature expansion, we find:\footnote{ We use the identity \eqref{eq:qrconfFiniteT} and took a low frequency limit.}
\begin{equation}
\hat{\Lambda}_{(2)}^\approx(\omega) - \hat{\Lambda}^\approx_{(2)}(0) 
\equiv  \int_0^\beta d\tau \;\left( e^{-i\omega\tau} -1 \right) \, q_r^c(\tau)^2 
 = \frac{32 \gamma^6}{3\pi M^2}  \left[  |\omega|^3 - \left(\frac{2\pi}{\beta}\right)^2 |\omega| + \ldots \right] \,. 
\end{equation}
Feeding this into the equation of motion \eqref{eq:approxEOM1}, we find the correction to $\hat{q}_r^\approx(\omega)$:
\begin{equation}
\label{eq:qrNew}
  \hat{q}_r(\omega) =   \hat{q}_r^\approx(\omega) + \delta \hat{q}_r^\approx(\omega) \,,\qquad  \delta \hat{q}_r^\approx(\omega)  = \hat{q}_r(0) \, \frac{4\pi(p-2)\gamma^3}{3Mu\beta^2} \left[ - 1 + 2\gamma  \, |\omega| + \frac{\beta^2}{4\pi^2} \, \omega^2 + \ldots \right]  
  \end{equation}
with $\hat{q}_r^\approx(\omega)$ as in \eqref{eq:exactsolfourier}.
In order to analytically continue this expression for use in \eqref{eq:expCondMS2c}, we will replace $|\omega|\rightarrow \sqrt{\omega^2}$. Therefore, $\lambda_L$ will set the prescription allowing us to pick the correct branch of the square root in \eqref{eq:qrNew}. This can be done, e.g., by rescaling $\lambda_L\rightarrow \epsilon^2\lambda_L$, $\omega\rightarrow \epsilon\, \omega$, $\beta\rightarrow \beta/\epsilon$, and $\gamma \rightarrow \epsilon \gamma$, and taking $\epsilon\rightarrow 0$. This allows us to approximate the LHS of \eqref{eq:expCondMS2c} consistently as:
\begin{align}
	\left[2\gamma \lambda_L+\frac{8(p-2)\pi\gamma^3}{3M u\beta^2} \left( 1 + \frac{\beta^2}{4\pi^2} \, \omega^2\right)+\dots\right]\hat{f}(\omega) &	=\frac{p-2}{u} \int_{-\infty}^{\infty} \frac{d\omega'}{2\pi} \, \,  \, \frac{ \rho_r(\omega')}{2\sinh\frac{\beta\omega'}{2}}\,{\hat{f}(\omega-\omega')}~.\label{eq:almosteigapprox}
\end{align}
To simplify the right hand side, our final step is to use the low-energy approximation
\begin{equation}
	\rho_r(\omega)\approx\rho_r^\approx(\omega)=\frac{16\gamma^2}{M} \gamma\omega\sqrt{1-\gamma^2\omega^2}\approx \frac{16\gamma^3}{M} \omega\,,
\end{equation}
such that we can simply Fourier transform \eqref{eq:almosteigapprox} back to the time domain: 
\begin{equation}
	\left(2\gamma \lambda_L+\frac{8(p-2)\pi\gamma^3}{3M u\beta^2}-\frac{2(p-2)\gamma^3}{3\pi M u}\partial_t^2\right)f(t)	= \frac{8(p-2)\pi\gamma^3}{M u\beta^2}\; \frac{1}{\cosh^2 \left(\frac{\pi t}{\beta}\right) } \; f(t)~.
\end{equation}
Defining a dimensionless coordinate $x=\frac{\pi t}{\beta}$, this differential equation takes the following form:
\begin{equation}
	-\frac{1}{2}\, f''(x)-\frac{6}{\cosh^2x}\;f(x)=-\left(2+\frac{3 M u \beta^2\lambda_L}{2(p-2)\pi\gamma^2}\right)f(x)~. 
\end{equation}
This is the well-known Schr\"{o}dinger problem with a P\"{o}schl-Teller potential:
\begin{equation}
 	-\frac{1}{2}\psi''(x)-\frac{\mu(\mu+1)}{2 \cosh^2 x}\psi(x)=E\, \psi(x)~,\qquad \mu\in \mathbb{Z}^+
 \end{equation} 
 whose eigenfunctions are $\psi(x)=P_\mu^k(\tanh(x))$ and eigenvalues are given by $E=-\frac{k^2}{2}$, with $k=1,2,\dots,\mu$. The case at hand corresponds to $\mu=3$, which translates to the following possibilities for $\lambda_L$:
\begin{equation}
 	\lambda_L= \frac{(p-2)\pi\gamma^2}{3 M u\beta^2}\left(k^2-4\right)=\frac{m\pi}{12\beta}\left(k^2-4\right)~\quad\qquad k=1,2,3\,.
 \end{equation}
In the second equality we have substituted the following relations valid in the marginal spin glass phase: 
\begin{equation}
 	\gamma=\sqrt{\frac{M\hat{q}_r(0)}{4}}~,\qquad\frac{\hat{q}_r(0)}{\beta}=\frac{m u}{p-2}~.
 \end{equation} 
Crucially, only $k=3$ gives a positive Lyapunov exponent, whereas $k=2$ gives a vanishing one.\footnote{Since the derivation of the kernel equation \eqref{eq:expCondMS2c} assumed $\lambda_L\geq0$ we must disregard the eigenvalue with $k=1$.} This is analogous to what we observed in the numerical analysis (for example, figure \ref{fig:lyapunovEVs}). This analytical calculation of the Lyapunov exponent thus captures all the qualitative aspects of the exact solution.
In the quantum scaling, we can expand our result as follows: 
 \begin{equation}
 \label{eq:SGlyapunovRes}
 	\boxed{ \;\;
	\frac{\beta}{2\pi}\, \lambda_L\approx \frac{5m}{24}=5(p-2)\left(\frac{1}{24\beta\mathcal{J}}+\frac{p}{36\pi\beta\mathcal{J}\sqrt{M\mathcal{J}}}+\dots\right)
	\;\; }
 \end{equation}
Crucially, this vanishes at infinite coupling $\beta {\cal J}$, as well as for $p=2$. We also emphasize the intriguing property that the Lyapunov exponent in units of $\beta$ is simply proportional to the replica symmetry breaking parameter $m$. Recall also that $m$ in the marginally stable spin glass considered here plays the role of a thermodynamic quantity conjugate to the free energy. We plotted $m\equiv m(J,M)$ for the marginal spin glass in figure \ref{fig:PhasesMarg}.

\paragraph{Comparison of analytical and numerical results:}
In figure \ref{fig:lyapunov1} we show numerical results for $\lambda_L$ at fixed $\beta J$. The inset shows that for low temperatures in the marginal spin glass phase, the Lyapunov exponent scales approximately as $1/\sqrt{MJ}$ at least for the temperatures considered. This is indeed consistent with the analytical result \eqref{eq:SGlyapunovRes} for the case where $\beta J$ is held fixed. This is perhaps even surprising because such a scaling is only expected to be describable by \eqref{eq:SGlyapunovRes} for very low temperatures. 

In figure \ref{fig:lyapunov2} we show both the numerical results (full lines) and the analytical approximation \eqref{eq:SGlyapunovRes} (dotted lines) for Lyapunov exponents in the case where $\beta J \propto MJ$. As we have discussed at several occasions, such scaling is expected to be well described by the quantum scaling. And indeed, we observe a good match for a large range of parameters in figure \ref{fig:lyapunov2}.

\section{Conclusion}
\label{sec:conclusion}

In this paper we studied bosonic quantum spins with random interactions and subject to a spherical constraint. In Part \ref{part:euc}, we reviewed the essential features of replica symmetry breaking in the spin glass phase, the phase diagram, and thermodynamics, while extending and clarifying some of the previous work in this direction. We particularly focused on the `classical' low temperature regime of the marginally stable spin glass phase, where a scaling solution exists. This scaling regime is characterized by the existence of an analytical approximate solution, which interpolates between a high frequency `UV solution' and a low frequency scaling limit. This approximate solution helps elucidate many physical features such as the low-temperature thermodynamics. In Part \ref{part:RTD} we studied real-time correlation functions and in particular out-of-time-order configurations, which serve as a diagnostic of quantum chaos. In the paramagnetic phase we found an interesting dependence of the Lyapunov exponent on the parameter $MJ$, in addition to the usual temperature dependence: dialing $MJ$ at fixed temperature corresponds to transitioning from a classical to a highly quantum regime, and the Lyapunov exponent offers a sharp indicator for the transition between the two: it takes a maximal peak value at the transition. Surprisingly, we also found signatures of quantum chaos in the spin glass phase, indicating slow but nevertheless scrambling dynamics. The approximate deep spin glass solution again allowed for an analytical treatment. These findings provide a novel point of view on discussions of ergodicity and its breaking in glassy systems.\\

\noindent The richness of the model calls for a number of future investigations:
\begin{itemize}
 \item \textbf{Aging:} In our discussion of the spin glass phase we assumed time translation invariant dynamics and the usual fluctuation-dissipation theorem (FDT). Our analysis thus captures the nontrivial aspects of replica symmetry breaking, slow dynamics, and inability of the system to reach equilibrium (as measured by the Edwards-Anderson order parameter $u$). However, as is well known, glassy dynamics does allow for eventual relaxation (see \cite{Cugliandolo_1998,Cugliandolo_1999} for a discussion in the context of the $p$-spin model). 
 Furthermore, the glass phase exhibits {\it aging}, i.e., its eventual relaxation depends on its history (e.g., perturbations, or the nature of a weak coupling to an environment) in a way that is not time translation invariant. For example, the eventual decay of the two-point autocorrelation function is characterized by a breaking of time-translation invariance and a violation of the assumptions underlying the FDT. The full real-time correlation functions are assumed to be captured by the following ansatz:
 \begin{equation}
  g(t,t') = g_r(t-t') + g_\text{ag.}(t,t')\,,
 \end{equation}
 where $g_r(t)$ could be the autocorrelation function $q_r^>(t-i\beta/2)$ or the retarded correlator $q_r^R(t)$. 
 Our spin glass analysis only captures the dynamics of the first part $g_r$ and sets the aging contributions to their early-time constant values ($q^>_\text{ag.} \rightarrow u$ and $q^R_\text{ag.} \rightarrow 0$.). It will be important in the future to extend our analysis of four-point functions (and quantum chaos) by taking into account that for $t-t' \gg t'$ the second term $g_\text{ag.}$ undergoes non-trivial and time-translation-non-invariant dynamics. The idea of an effective temperature \cite{PhysRevE.55.3898} might be useful.
 \item \textbf{Chaos in glasses:} Our analysis indicates that quantum chaos (as measured by the out-of-time-order correlator) survives the transition to a spin glass. In the strongly coupled spin glass we found that the Lyapunov exponent is approximately proportional to $\frac{m\beta}{2\pi}$, where $m$ characterizes the properties of the replica symmetry breaking ansatz. This interplay between quantum chaos and replica symmetry breaking deserves further analysis. Even in paramagnetic phase we found intricate dependence of the Lyapunov exponent on the quantum parameter $MJ$ in addition to $\beta J$. The model is thus suited for a detailed analysis of quantum chaos and scrambling. Our numerical analysis suggests that the Lyapunov exponent in thermal units increases in the quantum critical region. It would be interesting to analyze this further and find the maximal quantum Lyapunov exponent that this model can achieve (in both its phases). We expect the largest attainable exponent to be close to the chaos bound \cite{Maldacena:2015waa}, but not to saturate it due to the transition into a spin glass. Finally, it will be interesting to investigate the dependence of other chaos characteristics (such as spectral correlation functions \cite{Cotler:2016fpe}) on the parameters of the model and relate them to ergodicity breaking due to the complicated free energy landscape.
  \item \textbf{Conformal perturbation theory in the marginally stable spin glass:} We discussed at length the conformal solution to the approximate equations of motion in the marginally stable spin glass state, and how it allows for a consistent UV completion. It will be interesting in the future to investigate this strong coupling solution further, using methods from conformal field theory. In particular, conformal perturbation theory (see, e.g., \cite{Maldacena:2016hyu,Gross:2016kjj,Kitaev:2017awl,Gu:2019jub,Tikhanovskaya:2020elb}) provides an algorithm for systematically improving the conformal solution. In the spin glass this is more complicated than in SYK due to the need to carefully account for zero modes. However, our perturbative calculation in section \ref{sec:OTOCperturbation} establishes that this approach is feasible and yields interesting results. It would be interesting to develop this more systematically. (See also appendices \ref{sec:PMspectrum} and \ref{sec:CPT}, where we explore this approach for the conformal paramagnetic solution.)
 \item \textbf{Paramagnetic conformal solution:} While the scaling solution was most useful in the marginally stable spin glass, the paramagnetic phase also allows for a conformal limit at large $\beta J$. We discuss this solution in appendix \ref{app:confPM}. However, this branch of solutions appears to be unstable and thermodynamically disfavored compared to the physical branch of paramagnetic solutions that does not exhibit scaling behavior at low temperatures. It would be very interesting to find a mechanism to tune the system so that it actually exhibits this paramagnetic solution with its emergent reparametrization symmetry and conformal thermodynamics. Such a mechanism could involve, for example, fine tuning a set of couplings $J_{i_1 \cdots i_p}$ for various values of $p$.\footnote{ We thank S.\ Sachdev and G.\ Tarnopolskiy for related comments.} As shown in the appendix, this branch of solutions leads to a maximally chaotic model at large $\beta J$ and might therefore also admit a simple holographic interpretation in terms of dilaton gravity in AdS${}_2$ spacetime (see also below). In particular it may capture superradiance of black holes \cite{Anninos:2017cnw,Anninos:2019oka}. 
 \item \textbf{Large $p$ limit:} Most of our analytical expressions were valid for any value of $p$, while our numerics assumed $p=3$. However, we have seen hints that the system becomes even more tractable at large $p$. In particular, the marginally stable spin glass phase showed signs of allowing for a nice large $p$ limit. While this has been known (see, e.g., \cite{derrida}), it will be worthwhile revisiting this idea in light of recent developments in the SYK model, where a similar limit leads to great analytical control (see, for example, \cite{Maldacena:2016hyu,Cotler:2016fpe,Gross:2017hcz,Berkooz:2018jqr}). Also the conformal paramagnetic solution from the previous bullet point should admit a large $p$ limit, as it shares many features with SYK.
 \item \textbf{AdS Holography:} The fermionic SYK model has done a great deal to elucidate our understanding of the AdS/CFT duality in two dimensions. Our goal in studying the $p$-spin model was to explore the possibility of whether the rich physics of replica symmetry breaking provides an avenue towards understanding the AdS$_2$ fragmentation instability \cite{Maldacena:1998uz}. For this it was necessary to consider a model where the microscopic degrees of freedom are bosons, since RSB requires non-trivial one-point functions in the spin-glass phase. It would be interesting to explore if any of the spin-glass order parameters have a holographic analog in low dimensional gravity in a particular ensemble. 
 \item \textbf{dS holography:} Besides our initial motivation, the particular model we consider in this paper is that of a non-linear sigma model on the $N$-dimensional sphere with a disordered all-to-all interaction. Recall that Euclidean de Sitter space is also a sphere. A nice argument due to Polyakov \cite{Polyakov:2007mm} that precludes string theories for de Sitter space comes from the fact that the sigma-model on the sphere target space with local (two-derivative) interactions is always gapped in the IR. The analysis of the $p$-spin model suggests that gapless physics can re-emerge for the sphere target space in a non-trivial way at strong coupling and in a complicated thermodynamic state. Perhaps this serves as a motivation to explore relations between the sphere, glasses, and de Sitter in the future in a similar vein to \cite{Anninos:2011kh,Anninos:2020hfj,Law:2020cpj}.
\end{itemize}

\subsection*{Acknowledgments}

We would like to thank D. Anninos and J.\ Maldacena for stimulating discussions, and J. Kurchan for detailed comments on the draft. 
T.A. is supported by the Delta ITP consortium, a program of the Netherlands Organisation for Scientific Research (NWO) that is funded by the Dutch Ministry of Education, Culture and Science (OCW). F.H.\ gratefully acknowledges support from the DOE grant DE-SC0009988 and from the Paul Dirac and Sivian Funds.

\newpage
\appendix

\part{Appendix}
\label{part:app}

\section{Conventions, notation, and lexicon of important equations}
\label{ap:conventions}
We use this section as a convenient index to collect notation and conventions. The reader may refer back here in order to facilitate a thorough reading of this paper.\\

\noindent {\bf Units:} Our conventions set $\hbar=k_B=1$, implying the following units for our external couplings: 
\begin{equation}
	[\beta]=[\tau]=[M]=[J]^{-1}=[\text{length}]~, 
\end{equation}
where $\beta=1/T$ is the inverse temperature, $M$ is the strength of the kinetic term of the spherical fields, and $J$ is the width of the distribution over couplings. 
It follows that the fundamental fields have the following units: 
\begin{equation}
	[\sigma_i(\tau)]=[q(\tau)]=[u]=[m]=[\text{length}]^0~, \qquad [z(\tau)]=[\text{length}]^{-1}~. 
\end{equation}
We always denote by a  $~\hat{}~$ the Fourier transform: 
\begin{equation}
	f(\tau)=\frac{1}{\beta}\sum_{k=-\infty}^\infty e^{\frac{2\pi i k\tau}{\beta}}\hat{f}(k)~,\qquad\qquad \hat{f}(k)=\int_0^\beta d\tau\, e^{-\frac{2\pi i k\tau}{\beta}}f(\tau)
\end{equation}
and similarly for Lorentzian time $t$ and frequency $\omega$.
This implies the following relationship: 
\begin{equation}
	[f(\tau)]=[\text{length}]^x\longrightarrow [\hat{f}(k)]=[\text{length}]^{x+1}~.
\end{equation}

\noindent{\bf Couplings and fields:} 
Dimensionless combinations that often appear are: 
\begin{equation}
	\beta J~,\qquad M J~,\qquad \frac{M}{\beta} ~. 
\end{equation}
These are, of course, not independent, and phases of the model are parameterized by any two of these parameters.
We also often use the following coupling:
\begin{equation}
	\mathcal{J}\equiv J\sqrt{\frac{p(p-1)}{2}}\,.
\end{equation}
The standard dynamical field is $q(\tau,\tau')$, which satisfies the boundary condition $q(\tau,\tau)=1$, stemming from the spherical constraint. 
The self energy is defined as: 
\begin{equation}
  \Lambda(\tau,\tau') \equiv \frac{p}{2}\, q(\tau,\tau')^{p-1}  \,,
\end{equation}
and its Fourier transform is: 
\begin{equation} 
\hat{\Lambda}(k) = \frac{p}{2\beta^{p-2}} \sum_{k_1,\dots,k_{p-2}} \hat{q}(k_1)\dots\hat{q}(k_{p-2})\hat{q}(k-k_1-\dots-k_{p-2})~.
\end{equation}
We also collect here the definitions of the ``subtracted'' spin glass fields
\begin{equation}
\begin{split}
   q_r(\tau,\tau') &= q(\tau,\tau') - u \,,\qquad \Lambda_r(\tau,\tau') = \Lambda(\tau,\tau') - \frac{p}{2}\, u^{p-1} \,,\\
   \hat{q}_r(k) &= \hat{q}(k) - \beta u \delta_{k,0} \,,\qquad\;
   \hat{\Lambda}_r(k) = \hat{\Lambda}(k) - \frac{p}{2} \beta u^{p-1} \delta_{k,0}\,,
   \end{split}
\end{equation}
which satisfy the boundary conditions $q_r(\tau,\tau) = 1-u$ and $\Lambda_r(\tau,\tau) = \frac{p}{2} (1-u^{p-1})$ as a result of the spherical constraint. The parameter $\gamma$ appearing in the low-temperature exact solution is defined as
\begin{equation}
	\gamma\equiv\sqrt{\frac{M\hat{q}_r(0)}{4}}~. 
\end{equation}

\noindent{\bf Maginal spin glass parameters:}
The conformal spin glass solution exists when $u$ and $\hat{q}_r(0)$ are set to the values
\begin{equation}
u=\left[\frac{p-2}{\beta \mathcal{J} m}\right]^{2/p}~,\qquad \frac{\hat{q}_r(0)}{\beta}=\frac{m u}{p-2}=\left(\frac{m }{p-2}\right)\left[\frac{p-2}{\beta \mathcal{J} m}\right]^{2/p} \,,
\end{equation}
which implies the following identity:
\begin{equation}
	\mathcal{J}^2u^{p-2}=\left(\hat{q}_r(0)\right)^{-2}~
\end{equation}
in the conformal spin glass phase. This identity allows for an elegant low-energy approximate solutions to the equations of motion at zero temperature: 
\begin{equation}
	\hat{q}^\approx_{r,\beta=\infty}(\omega)=\hat{q}_r(0)\left(1+2\gamma^2\omega^2-2\sqrt{\gamma^2\omega^2+\gamma^4\omega^4}\right) \,.
\end{equation}
This approximate zero-solution can equivalently be written as:
\begin{equation}
	q_{r,\beta=\infty}^\approx(\tau)=  \int_{0}^{1/\gamma} \frac{d\omega}{2\pi}   \, \rho^\approx_{r}(\omega)  \, e^{-\omega |\tau|}  
	\,,\qquad \quad 
      \rho^\approx_{r}(\omega) = \frac{16\gamma^2}{M} \, \gamma\omega\sqrt{1-\gamma^2\omega^2} \,,
\end{equation}
which allows us to write the finite temperature equivalent:
\begin{align}
	q^\approx_r(\tau)&=\sum_{n=-\infty}^{\infty}\int_{0}^{1/\gamma} \frac{d\omega}{2\pi}   \,  \rho^\approx_{r}(\omega)  \, e^{-\omega |\tau+n\beta|} 
	=\int_{0}^{1/\gamma} \frac{d\omega}{2\pi}   \, \rho_{r,\beta=\infty}^\approx(\omega)\, \frac{\cosh\left[\left(|\tau|-\frac{\beta}{2}\right)\omega\right]}{\sinh\left(\frac{\beta\omega}{2}\right)}~. 
\end{align}
After defining $\omega=\frac{1}{\gamma}\sin\theta$, we can write this as
\begin{equation}
	q^\approx_r(\tau)=\frac{8\gamma}{M\pi}\int_0^{\frac{\pi}{2}}d\theta\,\cos^2\theta\, \sin\theta\, \frac{\cosh\left[\left(\frac{|\tau|}{\gamma}-\frac{\beta}{2\gamma}\right)\sin\theta\right]}{\sinh\left(\frac{\beta}{2\gamma}\sin\theta\right)} \;\;.
\end{equation}

\section{The free solution}
\label{sec:freeSol}

The simplest analytical solution to the equations of motion is the free solution for $J=0$, which we label with a subscript $0$: the constraints \eqref{eq:mvar1} are solved by $u_0 = 0$ for arbitrary $m$. The frequency space solution then reads:
\begin{equation}
\hat{q}_0(k) = \left[ M \left( \frac{2\pi k}{\beta}\right)^2 + \frac{1}{\hat{q}_0(0)} \right]^{-1}\,.
\end{equation}
 From \eqref{eq:EOMz} we then obtain the following constraint determining the zero mode $\frac{1}{\beta}\hat{q}_0(0)$ as a function of $M/\beta$:
\begin{equation}
\label{eq:freeSols}
   1 = \frac{1}{2} \sqrt{\frac{\hat{q}_0(0)}{M}} \, \coth \left( \frac{\beta}{2 \sqrt{ M \hat{q}_0(0)}} \right) \,.
\end{equation}
Finally, we find the position space expression $q_0(\tau)$ by Fourier transform:
\begin{equation}
\label{eq:freeSol}
\begin{split}
  q_0(\tau) &=   \frac{\hat{q}_0(0)}{2\beta} \Big[{}_2F_1(1,ia,1+ia,x) + {}_2F_1(1,ia,1+ia,\bar{x}) -1 \Big] \; +\;  \text{c.c.}  \,,
   \qquad\; a \equiv \frac{1}{ 2\pi} \frac{\beta}{\sqrt{M \hat{q}_0(0)}} 
  \end{split}
\end{equation}
where $x = e^{\frac{2\pi i}{\beta} \tau}$. This satisfies the boundary condition $q_0(\tau = 0) = 1$.

\section{The conformal paramagnetic solution}
\label{app:confPM}

In this appendix we discuss the conformal approximation in the paramagnetic phase, which is an extremum of the free energy above the spin glass. We will determine the operator spectrum that appears in the operator product expansion (OPE) and how it leads to perturbations of the conformal Green's function $q_r^c(\tau)$. Then, we show that this branch of solutions exhibits maximal chaos at zero temperature. Our discussion is not exhaustive and merely serves as a guide towards illustrating the relevant techniques.

We must note that the conformal paramagnetic solution is {\it not} what we discuss in our numerical analysis of the disordered state in \S\ref{sec:phases} at large $\beta J$. At low temperatures (and small enough values of $MJ$ such that we do not enter the spin glass phase), there exist two paramagnetic solutions. The phase diagram in \S\ref{sec:phases} refers to the thermodynamically preferred one, which also has a free energy that is continuously connected to the spin glass at large $MJ$. The conformal solution discussed in this appendix describes the low temperature behavior of the other -- thermodynamically unfavored -- paramagnetic solution, which indeed shows scaling behavior (a gapless spectrum, linear specific heat etc.) and exhibits a full reparametrization symmetry at the level of its action. In \S\ref{sec:PMspectrum} we will present a different point of view on why this solution branch is disfavored: the presence of a complex mode in its spectrum, which can be interpreted as leading to a dynamical instability.

\subsection{Derivation of the conformal paramagnetic solution (\texorpdfstring{$1 \ll \beta J \ll N$}{1<<bj<<N})}
\label{sec:conformal1}

It will be very important for us to identify regimes of parameters where the model has emergent conformal symmetry. A scaling regime can indeed be found for small temperatures and strong coupling (large $\beta J$).

We would like to identify a scaling regime valid at large $\beta J$, where the solution obeys a `conformal' scaling,
\begin{equation}
\label{eq:ConfSol}
q^c(\tau) \equiv u + q^c_r(\tau) \,,\qquad  q^c_r(\tau) = A \left| \frac{\pi}{\beta \, \sin \left(\frac{\pi\tau}{\beta} \right)} \right|^{2\Delta}\;\; \stackrel{\beta \gg \tau}{\approx} \;\;\frac{A}{|\tau|^{2\Delta}}\,.
\end{equation}
For non-vanishing parameter $u$ (i.e., in the spin glass phase) the conformal solution $q_r^c(\tau)$ only describes the dynamics on top of $u$. This is indeed the behavior of the approximate analytical solution to the spin glass equations (see section \ref{sec:confSG}). In this appendix we will focus on the paramagnetic case, i.e., $u=0$.
 
In the low temperature limit the Matsubara frequencies become effectively continuous, so we will write
\begin{equation}
\label{eq:zeroTempW}
	\omega\equiv \frac{2\pi k}{\beta}~.
\end{equation}
To compute the scaling exponent $\Delta$ and the normalization $A$, we need to find the Fourier transform, $\hat{q}^c_r(\omega) - \hat{q}^c_r(0)$. In other bosonic tensor and SYK-like models, one encounters an IR divergence in $\hat{q}^c_r(\omega)$ and a UV divergence in $\hat{\Lambda}^c_r(\omega)$ \cite{Azeyanagi:2017drg,Azeyanagi:2017mre,Chang:2018sve}. However, since we can work at finite temperature and the zero mode gets subtracted in all expressions (for example \eqref{eq:EOMsigma3}), there is in fact no ambiguity. This is abstractly clear, but it will be very helpful to make it more explicit. We therefore make the following schematic ansatz for the approximate solution at large $\beta J$:

\begin{equation}
\label{eq:qApproxDef}
   q_r(\tau) \approx \Theta\left(\frac{1}{\mu}- |\tau| \right) \, q^\text{UV}_r(\tau) + \Theta\left( |\tau|-\frac{1}{\mu}  \right) \, q^c_r(\tau)
\end{equation}
where the first term captures ``UV'' modes, which will turn out to contribute in an interesting way even at large $\beta J$, and the second term describes the conformal solution, which only has support on time separations larger than some cutoff $1/\mu$. The approximate solution should satisfy the equation of motion \eqref{eq:EOMsigma3} perturbatively at large $\beta J$.

Let us now discuss the conformal piece, $q_r^c(\tau)$.
Its IR divergence is easily regularized by working at finite temperature.  
The finite temperature Fourier transform of \eqref{eq:ConfSol} is
\begin{equation}
\label{eq:qrconfFiniteT}
  \hat{q}_r^c(k) = \int_0^\beta d\tau \, e^{-\frac{2\pi i}{\beta} k \tau}  \, q_r^c(\tau) =A\;  \frac{\beta}{2\cos(\pi \Delta)\Gamma(2\Delta)} \left( \frac{2\pi}{\beta} \right)^{2\Delta} \, \frac{\Gamma\left( \Delta + |k| \right)}{\Gamma\left(1- \Delta + |k| \right) } \,,
\end{equation}
which is valid for $0<\Delta<\frac{1}{2}$ and can then be analytically continued to other values. Consequently, the zero mode is
\begin{equation}
   \hat{q}_r^c(0) 
   = A \,\frac{ \pi^{2\Delta - \frac{1}{2}} \beta^{1-2\Delta}  \Gamma\left( \frac{1}{2} - \Delta \right)}{\Gamma(1-\Delta)} \,.
\end{equation}
Plugging \eqref{eq:zeroTempW} into \eqref{eq:qrconfFiniteT}, we can now also take the zero temperature limit:\footnote{ This also follows directly from 
\begin{equation}
  \int_{-\infty}^\infty \frac{e^{i\omega t}}{|\tau|^{2\Delta}} = 2 \sin(\pi \Delta) \Gamma(1-2\Delta) |\omega|^{2\Delta-1} \qquad\quad (0 < \Delta < 1/2)\,.
\end{equation} 
}
\begin{equation}
\label{eq:qrcSolFreq}
  \hat{q}_r^c(\omega)  \; \stackrel{\beta \rightarrow \infty}{\approx} \; 2A\,\sin(\pi\Delta)\Gamma(1-2\Delta) \, |\omega|^{2\Delta-1} \,.
\end{equation} 

We can similarly approximate the expression \eqref{eq:LambdaDefF} for the self-energy. 
Due to the similarity with the analysis of the SYK model, we expect to find $\Delta = \frac{1}{p}$. 
Working at finite temperature, we have from \eqref{eq:LambdaDefF}:
\begin{equation}
\begin{split}
   \hat{\Lambda}_r(k) &\approx \frac{p}{2} \int_0^\beta d\tau \, e^{-\frac{2\pi i}{\beta} k \tau}  \, q_r(\tau)^{p-1}
   \approx  \hat{\Lambda}_r^{\text{UV}}(k) + \frac{p}{2} \int_{\frac{1}{\mu}}^{\beta-\frac{1}{\mu}} d\tau \, e^{-\frac{2\pi i}{\beta} k \tau}  \, q_r^c(\tau)^{p-1} \,,
   \end{split}
\end{equation}
where we used that the conformal solution is only valid above some cutoff, $\tau > \mu^{-1}$, while the UV self-energy accounts for short-distance contributions to the integral. This procedure is necessary because the integral suffers from a UV divergence. For the conformal piece, the integral can be performed by analytically continuing \eqref{eq:qrconfFiniteT} in $\Delta$:
\begin{equation}
\begin{split}
  \hat{\Lambda}^c_r(k) &\equiv \frac{p}{2} \int_{\frac{1}{\mu}}^{\beta-\frac{1}{\mu}} d\tau \, e^{-\frac{2\pi i}{\beta} k \tau}  \, q_r^c(\tau)^{p-1}
   =  \frac{p\, A^{p-1}\beta  \big( \frac{2\pi}{\beta} \big)^{2(p-1)\Delta}}{4\cos(\pi(p-1) \Delta)\Gamma(2(p-1)\Delta)} \, \frac{\Gamma\left( (p-1)\Delta + k \right)}{\Gamma\left(1- (p-1)\Delta + k \right) } + {\cal O}\big(\frac{1}{\mu} \big)
   \end{split}
\end{equation}
Using similar calculations as above, we obtain at zero temperature:
\begin{equation}
\label{eq:LambdacSolPM}
\begin{split}
 \hat{\Lambda}_r^c(\omega) & \; \stackrel{\beta \rightarrow \infty}{\approx} \;p\,A^{p-1}\,\sin(\pi(p-1)\Delta)\Gamma(1-2(p-1)\Delta) \, |\omega|^{2(p-1)\Delta-1} 
 \,.
\end{split}
\end{equation}

Let us now return to the main equation of motion \eqref{eq:EOMsigma3} and discuss it at zero temperature.
The conformal zero mode $ \hat{\Lambda}_r^c(0)$ vanishes in the zero temperature limit.
Any UV dependence of the self-energy (including a potential divergence) is manifestly regulated in \eqref{eq:EOMsigma3} because the self-energy only appears in the ``renormalized'' combination $\hat{\Lambda}_r(k)-\hat{\Lambda}_r(0) \approx \hat{\Lambda}^c_r(k)-\hat{\Lambda}^c_r(0)$.
Regarding $\hat{q}^c_r$, let us now assume $\Delta<\frac{1}{2}$ such that $\hat{q}^c_r(0)$ is a divergent constant in the low temperature limit; its inverse can thus be dropped in \eqref{eq:EOMsigma3}. With these considerations, the equation of motion at zero temperature reads as:
\begin{equation}
\label{eq:confPMmatch}
\frac{1}{\hat{q}_r^c(\omega)+\hat{q}_r^\text{UV}(\omega)} = M\omega^2 -J^2 \, \hat{\Lambda}_r^c(\omega) \,.
\end{equation}
For small frequencies we can ignore both $\hat{q}_r^\text{UV}(\omega)$ and the term $M\omega^2$. Then, plugging in the explicit expressions \eqref{eq:qrcSolFreq} and \eqref{eq:LambdacSolPM}, we see that this equation is solved by the following set of parameters:
\begin{equation}
\label{eq:Aparamagnetic}
\boxed{
\;\; \text{conformal paramagnet:} \qquad\quad
\Delta = \frac{1}{p} \,, \qquad A = \left[ \frac{p-2}{\pi J^2 p^2} \, \cot\left( \frac{\pi}{p} \right)\right]^{\frac{1}{p}} \;\;
}
\end{equation}
This case is obviously very similar to the SYK model. The value $\Delta = \frac{1}{p}$ is consistent with the assumptions we made in the derivation.

\paragraph{Reparametrization invariance:}
To summarize the above discussion succinctly, we note that the equation of motion \eqref{eq:confPMmatch} in the small frequency limit can be written as:
\begin{equation}
	\delta(\tau,\tau') \approx - J^2 \int_0^\beta d\tau''\Lambda_{r}^c(\tau,\tau'')\,q_r^c(\tau'',\tau') \,,\qquad\quad \Lambda_{r}^c(\tau,\tau')  = \frac{p}{2} \, q_r^c(\tau,\tau')^{p-1} \,,
\end{equation}
where we returned to finite temperature for clarity.
This obviously exhibits a reparametrization invariance:
\begin{equation}
\begin{split}
   q^c_r(\tau,\tau') &\rightarrow \left[ f'(\tau) f'(\tau') \right]^{\frac{1}{p}} q^c_r(f(\tau),f(\tau')) \,,\\
    \Lambda_{r}^c(\tau,\tau') &\rightarrow \left[ f'(\tau) f'(\tau') \right]^{1-\frac{1}{p}}  \Lambda_{r}^c(f(\tau),f(\tau')) \,.
 \end{split}
\end{equation}
This is again similar to the SYK model. We can in fact show that the soft mode action governing the leading reparametrization symmetry breaking effects is the Schwarzian. To this end, consider the reparametrized saddle point solution,
\begin{equation}
  q_{r,\star}^c(\tau,\tau')  \rightarrow q_{r,f}^c(\tau,\tau') \equiv A \left( \frac{f'(\tau) f'(\tau')}{(f(\tau) - f(\tau'))^2} \right)^{\frac{1}{p}} \,.
\end{equation}
Now recall the full effective action \eqref{eq:1rsbeffaction}. Plugging the above expression into the action produces a non-zero result due to the conformal symmetry breaking term $\sim \partial_\tau\partial_{\tau'} \delta(\tau,\tau')$:
\begin{equation}
\begin{split}
  \frac{S^\text{breaking}_\text{eff}}{Nn} &= \frac{M}{2}  \int_0^\beta d\tau \int_0^\beta d\tau d\tau' \, \delta(\tau-\tau'-\varepsilon)\, \partial_\tau \partial_{\tau'} q_{r,f}^c(\tau,\tau') \\
  &= \frac{MA}{2}  \int_0^\beta d\tau \left[ -\frac{1}{\varepsilon^{2+2/p}} \frac{2(p+2)}{p^2}- \frac{1}{\varepsilon^{2/p}} \, \frac{(p-1)(p-2)}{3p^3} \, \{ f(\tau) , \tau \} + \ldots \right]
\end{split}
\end{equation}
where we introduced a regulator $\varepsilon$. The most divergent non-constant term in the symmetry breaking part of the action is indeed the Schwarzian derivative of the soft mode $f(\tau)$. Determining its coefficient requires UV information.

\paragraph{UV solution:}
We can also extract the UV piece of the solution following a procedure as worked out in \cite{Tulipman:2020abw}. To this end, note that for large frequencies ($|\omega| > \mu$) the equation of motion \eqref{eq:confPMmatch} simply gives:
\begin{equation}
\label{eq:qrUV}
  \hat{q}_r^\text{UV}(\omega) \approx \Theta( |\omega| - \mu ) \, \frac{1}{M\omega^2} 
  \qquad \Rightarrow \qquad 
  q_r^\text{UV}(\tau) = \int_{|\omega| > \mu} \frac{d\omega}{2\pi} \frac{e^{i\omega \tau}}{M\omega^2} \approx \frac{1}{M\mu \pi}\,.
\end{equation}
The cutoff $\mu$ could now be determined by a matching condition such as $\hat{q}_r^\text{UV}(\mu) = \hat{q}_r^c(\mu)$. In practice it is more accurate to determine $\mu$ numerically such that the low temperature approximation matches with numerical data. In this sense, the cutoff $\mu$ is a parameter of the model, which has to be determined numerically.

\subsection{Spectrum of the conformal paramagnetic phase}
\label{sec:PMspectrum}

We now consider the conformal field theory corresponding to the approximation to the paramagnetic physics at large $\beta J$ of the previous section, and discuss its spectrum (see, e.g., \cite{Tikhanovskaya:2020elb,Gross:2016kjj,Giombi:2017dtl}). We consider a three-point function with an operator ${\cal O}_h$ of conformal dimension $h$:
\begin{equation}
  v(\tau_0,\tau_1,\tau_2) = \frac{1}{N} \sum_i \langle {\cal O}_h(\tau_0) \sigma_i(\tau_1) \sigma_i(\tau_2) \rangle 
    = \frac{C_{h\Delta\Delta}}{|\tau_{01}|^{h} |\tau_{02}|^{h} |\tau_{12}|^{\frac{2}{p}-h}} \,,
\end{equation} 
where we gave the explicit expression for a conformal three-point function, using $\Delta_\sigma = 1/p$ in the paramagnetic phase.

\begin{figure}
\begin{center}
\includegraphics[width=.4\textwidth]{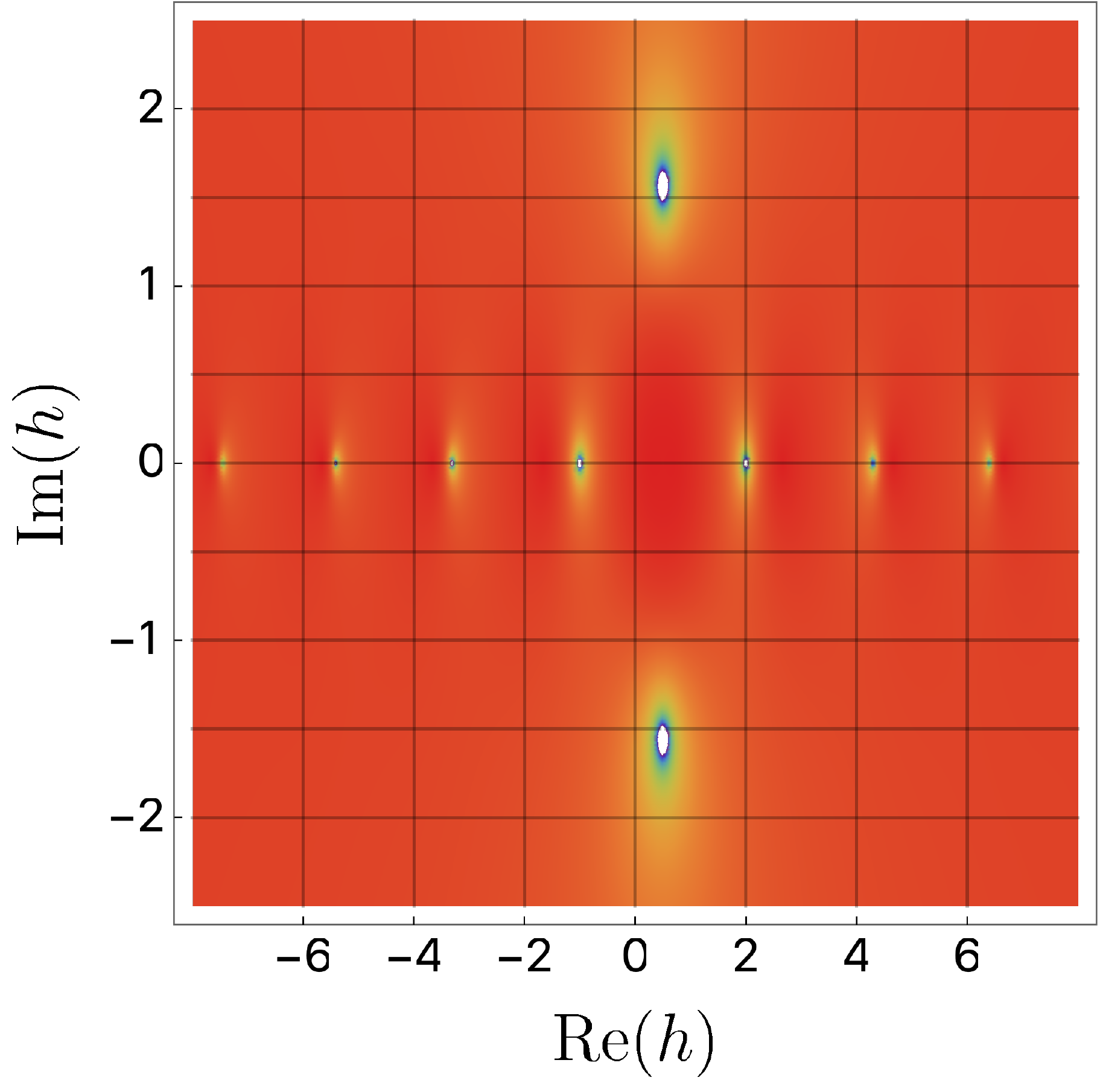}
\end{center}
\vspace{-.5cm}
\caption{{\bf Spectrum of the conformal paramagnet.} Plot of $|k(h)-1|^{-1}$ where $k(h)$ is the eigenvalue \eqref{eq:khPM} of the conformal ladder kernel for $p=3$. The spikes at the values written in \eqref{eq:hsolsPM} and at $h= \frac{1}{2} \pm 1.560 \, i$ correspond to locations where $k(h)=1$, i.e., the dimensions of operators appearing in the OPE. The complex dimension signals an instability.}
\label{fig:spectrum1}
\end{figure}

The three point function should be an eigenfunction of the ladder kernel. Schematically, we write this as ${\bf K} \cdot {\bf v} =  k(h)\,{\bf v}$ with eigenvalue $k(h)=1$. More explicitly:
\begin{equation}
\frac{J^2p(p-1)}{2} \int d\tau_3 d\tau_4 \, \left[  q^c(\tau_{12})^{\frac{p-2}{2}} \, q^c(\tau_{13}) \, q^c(\tau_{24}) \, q^c(\tau_{34})^{\frac{p-2}{2}}  \right]\, v(\tau_0,\tau_3,\tau_4) =  k(h) \, v(\tau_0,\tau_1,\tau_2)\,.
\end{equation}
In other words, the allowed operator dimensions $h$ are such that $v$ is an eigenfunction of the ladder kernel with eigenvalue $1$.
Inserting the conformal solution \eqref{eq:ConfSol}, we obtain an explicit nested integral, which can be evaluated using the following identity twice:
\begin{equation}
   \int d\tau_a \, \frac{1}{|\tau_{a1}|^\alpha |\tau_{a2}|^\beta |\tau_{a3}|^\gamma} = \sqrt{\pi} \, \frac{\Gamma\big(\frac{1-\alpha}{2}\big)\Gamma\big(\frac{1-\beta}{2}\big)\Gamma\big(\frac{1-\gamma}{2}\big)}{\Gamma\big(\frac{\alpha}{2}\big)\Gamma\big(\frac{\beta}{2}\big)\Gamma\big(\frac{\gamma}{2}\big)} \, \frac{1}{|\tau_{12}|^{1-\gamma} |\tau_{13}|^{1-\beta} |\tau_{23}|^{1-\alpha}} \,,
\end{equation}
valid for $\alpha+\beta+\gamma=1$. This procedure gives for the eigenvalue
\begin{equation}
\label{eq:khPM}
k(h) = - (p-1) \, \frac{\Gamma\big(1-\frac{1}{p}\big) \Gamma\big(\frac{1}{2}-\frac{1}{p}\big) \Gamma\big(\frac{1}{p}-\frac{h}{2}\big) \Gamma\big(\frac{1}{p}+\frac{h-1}{2}\big)}{\Gamma\big(\frac{1}{p}\big) \Gamma\big(\frac{1}{p}-\frac{1}{2}\big)\Gamma\big(1-\frac{1}{p}-\frac{h}{2}\big) \Gamma\big(\frac{1}{2} - \frac{1}{p}+\frac{h}{2}\big)} \,.
\end{equation}
The condition $k(h) = 1$ needs to be solved numerically. In figure \ref{fig:spectrum1} we plot $|k(h)-1|^{-1}$ in the complex $h$ plane to demonstrate the two important features:
\begin{itemize}
\item The spectrum consists of operators with a tower of real dimensions starting precisely at $h_0=2$ and increasing in steps of approximately $2$. For $p=3$ these dimensions are:
  \begin{equation}
  \label{eq:hsolsPM}
      h_0 = 2, \qquad h_{n=1,2,3,\ldots} = 4.303, \; 6.404, \; 8.456, \; 10.489 , \; 12.511, \, \ldots
  \end{equation}
  In the limit $p\rightarrow \infty$ these dimensions approach their integer values ($h = 2k + {\cal O}(\frac{1}{p})$ for $k=1,2,3,\ldots$).
 \item In addition there exists one operator with complex dimension $h = \frac{1}{2} \pm 1.560\, i$ (for the case $p=3$). In the limit $p \rightarrow \infty$ its imaginary part approaches $\frac{1\pm i \sqrt{7}}{2}$ asymptotically. This operator indicates that the conformal branch of the paramagnetic solution is unstable. As we discussed before, there are indeed {\it two} paramagnetic solutions at low temperatures. The thermodynamically preferred one is the one which does {\it not} have a conformal limit.
\end{itemize}

\subsection{Conformal perturbation theory}
\label{sec:CPT}

Having identified the spectrum of operators appearing in the OPE of the conformal solutions, we can now ask how these operators modify the Euclidean Green's functions. We consider the leading corrections to the conformal paramagnetic solution  and work with the following ansatz (at zero temperature):
\begin{equation}
\label{eq:qrPertAnsatz}
   q_r(\tau) = q_r^c(\tau) \left[ 1 - \left( \frac{\alpha_{0,1}}{|J\tau|}+\frac{\alpha_{0,1}^2\alpha_{0,2}}{|J\tau|^2} +\frac{\alpha_{0,1}^3\alpha_{0,3}}{|J\tau|^3} + \ldots \right)  - \sum_{n\geq 1} \left( \frac{\alpha_{n,1}}{|J\tau|^{h_n-1}} +  \frac{2\alpha_{0,1}\alpha_{n,1}\alpha_{n,2}}{|J\tau|^{h_n}} + \ldots \right)  \right]  
\end{equation}
where $q_r^c(\tau)= \frac{A}{|\tau|^{2\Delta}}$ is the conformal solution with dimension $\Delta = \frac{1}{p}$ (or $\Delta = 1$ in the spin glass). The first set of terms in the above equation describes corrections to the conformal Green's function which are purely due to the $h_0 = 2$ mode. The first term in the sum over $n$ describes the linear corrections due to the $h_n$ mode. (The values of $h_n$ were given in \eqref{eq:hsolsPM} for the paramagnetic case.) The second term in the sum describes quadratic corrections due to the interaction of an $h_0=2$ and an $h_n$ mode. In principle one can continue in this fashion and formulate a systematic resonance theory, which has been developed in great generality \cite{Kitaev:2017awl,Gu:2019jub,Tikhanovskaya:2020elb}. We will focus on the essential features in order not to introduce too much formalism.

We now discuss how to determine the parameters $\alpha_{i,j}$ (and in fact the operator spectrum $h_n$ once again) by solving the equations of motion at low frequencies (and zero temperature). Generalizations to finite temperature work similarly and are described in the references. 
At zero temperature, we can Fourier transform the ansatz \eqref{eq:qrPertAnsatz} by generalizing \eqref{eq:qrcSolFreq}. This gives:
\begin{equation}
\label{eq:qrcSolFreq2Pert}
\begin{split}
  \hat{q}_r(\omega) = \hat{q}_r^c(\omega)  &\bigg[ 1 + \frac{\alpha_{0,1}}{2\Delta \tan(\pi\Delta)} \, |\omega| + \frac{\alpha_{0,1}^2\alpha_{0,2}}{2\Delta(2\Delta+1)} \, |\omega|^2 
  - \frac{\alpha_{0,1}^3 \alpha_{0,3} \,\Gamma(2\Delta)}{\tan(\pi\Delta) \Gamma(2\Delta+3)} \, |\omega|^3 + \ldots \\
&\quad  - \sum_n \left(\frac{\alpha_{n,1}\cos(\pi\Delta)\Gamma(2\Delta)}{\sin(\pi\Delta+ \frac{\pi}{2} h_n) \Gamma(2\Delta+h_n-1)} \, |\omega|^{h_n-1} 
  + \ldots \right) \bigg] \,.
 \end{split}
\end{equation} 
where $\hat{q}_r^c(\omega) \sim |\omega|^{2\Delta-1}$ was given in \eqref{eq:qrcSolFreq}. To solve the equation of motion, we will have to expand the inverse of this expression for small $\omega$:
\begin{equation}
\begin{split}
\frac{1}{\hat{q}_r(\omega)} &= \frac{1}{\hat{q}_r^c(\omega)} \left\{ 1 -  \frac{\alpha_{0,1}}{2\Delta \tan(\pi\Delta)} \, |\omega| + \left[\left( \frac{\alpha_{0,1}}{2\Delta \tan(\pi\Delta)} \right)^2 - \frac{\alpha_{0,1}^2\alpha_{0,2}}{2\Delta(2\Delta+1)} \right]\, |\omega|^2 + \ldots \right\}
\end{split}
\end{equation}
Similarly, we get for the self-energy:
\begin{equation}
\begin{split}
 \hat{\Lambda}_r(\omega) = \hat{\Lambda}_r^c(\omega)  
  & \bigg\{ 1 +   \frac{\alpha_{0,1}}{2\Delta \tan(\pi(p-1)\Delta)} \, |\omega| + \frac{\alpha_{0,1}^2 (2 \alpha_{0,2} +2-p)}{4\Delta(2(p-1)\Delta+1)} | \omega|^2 
  + \ldots \bigg\}
\end{split}
\end{equation}
with $\hat{\Lambda}_r^c(\omega) \sim |\omega|^{2(p-1)\Delta-1}$ given in \eqref{eq:LambdacSolPM}. The low frequency equation of motion $\hat{q}_r(\omega)^{-1} = - J^2 \hat{\Lambda}_r(\omega)$ can now be solved order by order by matching the coefficients of powers of $|\omega|$. This determines the parameters $\alpha_{i,j}$ for $j\geq 2$. 

To do this explicitly, we set $\Delta = \frac{1}{p}$ according to the results of section \ref{sec:conformal1}. For the second and third order corrections due to the $h_0 =2$ mode we find:
\begin{equation}
   \alpha_{0,2} = \frac{(p+2) \big[ 2(p-1) + p \cos \big(\frac{2\pi}{p}\big) \big]}{8p \sin^2\big(\frac{\pi}{p}\big)}  \,,\quad\;\; \alpha_{0,3} = \frac{(p+2)(p+1)\big[ 6 - 8 p - p \cos \big(\frac{2\pi}{p}\big) \big]}{24 p \sin^2\big(\frac{\pi}{p}\big)} \,,\;\ldots
\end{equation}
For $p=3$ this yields $\alpha_{0,2} = \frac{25}{36}$, $\alpha_{0,3} = -\frac{55}{9}$ and so on.\footnote{ For $p=4$, we get $\alpha_{0,2} =  \frac{9}{4}$, $\alpha_{0,3} = - \frac{65}{4}$, which is consistent with \cite{Tikhanovskaya:2020elb}. It is also interesting to consider the large $p$ limit, where $\alpha_{0,2} \sim \frac{3 p^3}{8\pi^2}$ and $\alpha_{0,3} \sim - \frac{3p^4}{8\pi^2}$.} Note that $\alpha_{0,1}$ is left undetermined by this procedure (the matching at order $|\omega|^{2-2/p}$ is trivial). 

Similarly, we can do the matching for the perturbations due to the $h_n$ modes for $n\geq 1$. Again, $\alpha_{n,1}$ is left undetermined by matching terms of order $|\omega|^{h_n-2/p}$. However, the matching at this order is still non-trivial -- it gives the following constraint on the dimensions $h_n$:
\begin{equation}
\label{eq:spectrumConstr2}
 -2(p-1) \frac{\Gamma \big(\frac{2}{p} - h_n \big) \Gamma\big(\frac{2}{p}+h_n-1\big) \sin\big(\frac{\pi}{p} + \frac{h_n \pi}{2}\big) \cos\big(\frac{\pi}{p} - \frac{h_n \pi}{2}\big)}{\Gamma\big(\frac{2}{p}\big) \Gamma\big( \frac{2}{p}-1\big) \sin \big( \frac{2\pi}{p} \big) } = 1 \,.
\end{equation}
Solving this constraint gives precisely the same operator spectrum as in the previous subsection (see \eqref{eq:hsolsPM} for $p=3$). This is, of course, to be expected, but provides a nice consistency check (note that the constraint \eqref{eq:spectrumConstr2} is not identical to the constraint \eqref{eq:khPM}, but it has the same solutions).
At quadratic order for the dimension $h_n$ perturbation we find:
{\small
\begin{equation}
\alpha_{n,2} =  \frac{(2+(h_n-1)p) \Gamma\big(\frac{2}{p}\big)\Gamma\big(\frac{2}{p}-1\big) \cos^2\!\big(\frac{\pi}{p}\big)+(p-1)(p-2)\Gamma\big(\frac{2}{p}-1-h_n\big)\Gamma\big(\frac{2}{p}+h_n\big)\sin\big(\frac{\pi}{p} - \frac{h_n \pi}{2}\big)\sin\big(\frac{\pi}{p} + \frac{h_n \pi}{2}\big)}{\Gamma\big(\frac{2}{p}\big) \Gamma\big(\frac{2}{p}-1\big) \sin \big(\frac{2\pi}{p}\big) \tan \big( \frac{\pi}{p} + \frac{h_n \pi}{2} \big)+2(p-1) \Gamma\big(\frac{2}{p}-1-h_n\big) \Gamma\big(\frac{2}{p}+h_n\big) \sin\big(\frac{\pi}{p} - \frac{h_n \pi}{2} \big)\sin\big(\frac{\pi}{p} + \frac{h_n \pi}{2} \big)}
\end{equation}
}\normalsize
More explicitly, for $p=3$ we get:
\begin{equation}
  \alpha_{1,2} = 0.170 \, , \qquad \alpha_{2,2} = -0.635 \,,\qquad \alpha_{3,2} = -1.498 \,,\quad \ldots
\end{equation}

Finally, we turn to the coefficients $\alpha_{n,1}$. These are not determined by the above procedure of solving equations of motion order by order. Instead, one needs to determine these parameters numerically by matching numerical solutions of the full equations of motion to the analytical expressions computed here. This has been done in the literature (for example \cite{Tikhanovskaya:2020elb}), so we will not go into details here.

\subsection{OTOC in the conformal paramagnet: maximal chaos}
\label{app:PMototc}

This appendix complements the discussion of quantum Lyapunov exponents in section \ref{sec:OTOCs}. In the conformal limit we can compute the Lyapunov exponent of the corresponding paramagnetic solution analytically. This is completely analogous to the discussions in the SYK model \cite{Maldacena:2016hyu}. We should keep in mind that the conformal paramagnet corresponds to a branch of solutions, which are naively unphysical. Nevertheless these are solutions, and it is interesting to study their properties. It would be instructive to improve our numerical implementation and follow the physical branch of paramagnetic solutions discussed in the main text all the way down to zero temperature.

It will be useful to have the real-time propagators in the conformal limit. We recall the Euclidean results for the conformal solution from section \ref{sec:conformal1}. For the replica symmetric case ($u=0$), the analytically continued ``left-right'' Wightman function and the retarded correlator in the limit $\beta J \gg 1$ are:
\begin{equation}
\begin{split}
  q^>(t-i\beta/2) &\approx A \, \left( \frac{\pi}{\beta \cosh \left( \frac{\pi t}{\beta} \right)} \right)^{2/p}\,,\qquad\quad
   q^R(t) \approx -2A\, \sin\left( \frac{\pi}{p} \right)  \, \left( \frac{\pi}{\beta \sinh \left( \frac{\pi t}{\beta} \right)} \right)^{2/p} \Theta(t)\,,
\end{split}
\end{equation}
where $A$ was given in \eqref{eq:Aparamagnetic}.
The retarded kernel in the conformal limit becomes independent of any couplings:
\begin{equation}
   \tilde{K}'_{\rm ret} \approx \frac{\pi(p-1)(p-2)}{p \beta^2} \, \sin \left( \frac{2\pi}{p} \right)  \left[ \frac{\cosh \left( \frac{\pi t_{34}}{\beta}\right)^{2-p} }{\sinh\left(\frac{\pi t_{13}}{\beta} \right) \sinh\left( \frac{\pi t_{24}}{\beta} \right)} \right]^{2/p} \Theta(t_{13})\, \Theta(t_{24})
\end{equation}
Note that the retarded kernel vanishes for $p=2$, so we can immediately infer that the $p=2$ model is not quantum chaotic. For $p\geq 3$ the conformal eigenfunctions and eigenvalues of the retarded kernel $\tilde{K}'_{\rm ret}$ are thus the same as in the SYK model:
\begin{equation}
  F_h(t_1,t_2) = \frac{e^{-h \frac{2\pi}{\beta} \frac{t_1+t_2}{2}}}{\left[ \cosh \left( \frac{\pi t_{12}}{\beta}\right)\right]^{2/p-h}} \,,\qquad \mathfrak{s}_{r}(h) = \frac{\Gamma\left( 3 - \frac{2}{p} \right) \Gamma\left(\frac{2}{p}-h \right)}{\Gamma\left( 1+ \frac{2}{p} \right) \Gamma \left( 2 - \frac{2}{p} - h \right) } 
\end{equation}
such that $\tilde{K}'_{\rm ret} \cdot F_h = \mathfrak{s}_r(h) F_h$ (understood as an integral equation as in \eqref{eq:OTOCeq}). 
Therefore, the solution of the integral equation with eigenvalue $1$ reduces to the following requirement on the quantum numbers $h$ (related to the Lyapunov exponent via $\lambda_L = -h \frac{2\pi}{\beta}$):
\begin{equation}
  \mathfrak{s}_r(h) = 1 \qquad \Rightarrow \qquad h = -1, \, 0, \ldots 
\end{equation}
with the remaining set of $h$ being irrational functions of $p$. For example, for $p=3$, we get the allowed values $h=-1,\,0,\,1.475,\,2.518,\,3.512, \ldots$. In particular there are no exponentially growing modes except for the $h=-1$ mode, which has maximal Lyapunov exponent in the sense of saturating the chaos bound \cite{Maldacena:2015waa}. This is similar to what we observed numerically away from the conformal limit (c.f., figure \ref{fig:lyapunovEVs}).

\section{Derivation and regularization of thermodynamic quantities}
\label{app:thermo}

In this appendix we give more details on the UV regularization of approximate analytical expressions, and on the derivation of various thermodynamics quantities used in sections \ref{sec:EquilThermo} and \ref{sec:MargThermo}.

\subsection{On-shell action for the free solution}
\label{app:regularization}

We want to derive \eqref{eq:freeregularized} starting from \eqref{eq:freediv} using $\zeta$-function regularization. The on-shell UV action is simply \eqref{eq:freediv} for $J=u=0$ evaluated on the UV solution \eqref{eq:asympSol}:
\begin{equation}
	\frac{S_{\rm eff}^\text{UV}}{nN}=\;\frac{1}{2}\sum_{k=-\infty} ^\infty\left\{1+ \log\left[ \frac{M}{\beta} \left( 2\pi k \right)^2-2i\hat{z}(0) \right] \right\} + i \hat{z}(0)~.
\end{equation}
keeping in mind that $-2i\hat{z}(0)$ is positive and real.
Let us recall the definition for the Riemann zeta function and its derivative for Re$(s)>1$: 
\begin{equation}
	\zeta(s)=\sum_{n=1}^\infty\frac{1}{n^s}~,\qquad\qquad \qquad \zeta'(s)=-\sum_{n=1}^\infty \frac{1}{n^s}\log(n)~.
\end{equation}
Crucially these infinite-sum expressions for $\zeta(s)$ only converge for Re$(s)>1$, but the function is well defined for any complex $s$. 
We start with the first term in $S_{\rm div}(Q_0)$:
\begin{equation*}
	\frac{1}{2}\sum_{k=-\infty} ^\infty1 =\frac{1}{2}\left(1+2 \sum_{k=1}^\infty 1\right)\longrightarrow \frac{1}{2}\left[1+2 \zeta(0)\right]=0~.
\end{equation*}
It remains to evaluate the log term: 
\begin{align}
\frac{1}{2}\sum_{k=-\infty} ^\infty\log\left[ \frac{M}{\beta} \left( 2\pi k \right)^2-2i\hat{z}(0) \right] 
	&=\frac{1}{2}\log  \left(-2i\hat{z}(0) \right)+\log\prod_{k=1}^\infty\left[ 1+ \frac{-2i\hat{z}(0)}{\frac{M}{\beta} \left( 2\pi k \right)^2}\right]+\sum_{k=1}^\infty\log\left[ \frac{M}{\beta} \left( 2\pi k \right)^2\right]\,.\label{eq:threeterms}
\end{align}
To evaluate the middle term in \eqref{eq:threeterms}, we simply use the following identity: 
\begin{equation}
	\frac{\sinh x}{x}=\prod_{k=1}^\infty\left(1+\frac{x^2}{\pi^2k^2}\right)\,.
\end{equation}
Let us now deal with the final term of \eqref{eq:threeterms}. To do this, we will rearrange: 
\begin{align}
\sum_{k=1}^\infty\log\left[ \frac{M}{\beta} \left( 2\pi k \right)^2\right]&=2\left[\sum_{k=1}
^\infty \log\left(2\pi\sqrt{\frac{M}{\beta}}\right)+\sum_{k=1}
^\infty\log(k)\right]\nonumber\\
&\longrightarrow 2\left[\zeta(0)\log\left(2\pi\sqrt{\frac{M}{\beta}}\right)-\zeta'(0)\right]
=-\frac{1}{2}\log\frac{M}{\beta}~.
\end{align}
Putting everything together we find
\begin{align}
\frac{S_{\rm eff}^\text{UV}}{nN}
	&=\log \left[2\sinh\left( \sqrt{\frac{-i\hat{z}(0)\beta}{2M}}\right)\right] + i \hat{z}(0) \,.
\end{align}
This concludes our regularization giving rise to \eqref{eq:freeregularized}.

\subsection{Derivation of internal energy}
\label{sec:AppendixE}

Here, we explain how to derive the exact expression \eqref{eq:Eposition} for the internal energy. We discuss the equilibrium situation here. The marginally stable spin glass state is more subtle (as explained in section \ref{sec:ThermoMarg}), but eventually yields the identical expression \eqref{eq:EpositionMarg}.
Working with the regularized free energy \eqref{eq:freeregularized}, we immediately get the following equation for the energy: 
\begin{align}\label{eq:edef}
	\beta \Et&= J \partial_J(\beta \bar{f}) - M \partial_M (\beta \bar{f}) \nonumber\\
	&= \frac{1}{2} \sum_k\left\{1-\left[ \frac{M}{\beta} \left( 2\pi k \right)^2-2i\hat{z}(0) \right]  \frac{\hat{q}_r(k)}{\beta} \right\} - i \hat{z}(0) \, (1-u)- \frac{(\beta J)^2}{2} \left[ (m-1)u^p + \frac{1}{\beta} \int_0^\beta  d\tau\, q(\tau)^p \right] 
\end{align}
where we wrote explicit Lagrange multipliers again such as to obtain a frequency sum that converges as $|k| \rightarrow \infty$.
This expression is to be evaluated on-shell, meaning we can use the equations of motion to simplify it. By taking \eqref{eq:EOMsigma3}, multiplying by $\hat{q}_r(k)$ and summing over $k$ (while using the spherical constraint), we find that
\begin{align}
\label{eq:replace}
	&\frac{1}{2}\sum_k\left\{1-\left[ \frac{M}{\beta} \left( 2\pi k \right)^2-2i\hat{z}(0) \right]  \frac{\hat{q}_r(k)}{\beta} \right\}= - p \frac{(\beta J)^2}{4\beta} \int_0^\beta d\tau \left( q(\tau)^{p-1} - u^{p-1} \right) (q(\tau) - u)\nonumber\\
	&\qquad\qquad\qquad\qquad\qquad = -p \frac{(\beta J)^2}{4\beta} \int_0^\beta d\tau \left[ \left( q^p - u^p \right) -u \left( q^{p-1} - u^{p-1} \right) \right] + \frac{p}{4}\, u^{p-1} (\beta J)^2  \frac{\hat{q}_r(0)}{\beta} \,. 
\end{align}
Combining this with \eqref{eq:edef} and we obtain: 
\begin{align}
	\beta \Et &= (\beta J)^2 \left\{ \frac{p u}{4\beta}  \int d\tau \left( q^{p-1} - u^{p-1} \right) - \frac{1}{2} m u^p - \frac{p+2}{4\beta} \int d\tau \left( q^p -u^p \right) + \frac{p}{4} \, u^{p-1} \, \frac{\hat{q}_r(0)}{\beta} \right\}- i \hat{z}(0) \, (1-u) \nonumber\\
	&= (\beta J)^2 \left\{ \frac{p}{4\beta}  \int d\tau \left( q^{p-1} - u^{p-1} \right) - \frac{p+2}{4} m u^p - \frac{p+2}{4\beta} \int d\tau \left( q^p -u^p \right)  \right\} \nonumber\\
	&\qquad -   (1-u) \left\{ i \hat{z}(0) + \frac{p(\beta J)^2}{4\beta} \int d\tau \left( q^{p-1} - u^{p-1} \right) + \frac{\beta}{2\hat{q}_r(0)} \right\}  +  \frac{\beta}{2\hat{q}_r(0)}
\end{align}
where we used the first line of \eqref{eq:mvar1} in the last step (we assumed $m \neq 1$ and $u \neq 0$, but one can easily check that the same holds for the paramagnetic phase). Finally, the bracket in the last line of \eqref{eq:replace} vanishes on-shell (c.f. \eqref{eq:zSol}). We thus find the desired expression \eqref{eq:Eposition} for the energy.

\subsection{Low energy expansions in the marginal spin glass phase}
\label{app:ElowEnergy}

We also need the low energy (large $\beta J$) expansion of the various thermodynamic quantities in the marginally stable spin glass. We now discuss some of the relevant calculations.

\subsection*{Moments of the approximate marginal solution}
To evaluate thermodynamic quantities on the approximate low-temperature solution, we will frequently need some identities for the moments of $q^\approx_r(\tau)$. The first moment is simply the zero mode, which we recall here, (see around \eqref{eq:qApproxFinal}):
\begin{equation}
	\int_0^\beta d\tau \,q^\approx_r(\tau)=\frac{16\gamma^2}{M\pi}\int_0^{\frac{\pi}{2}}d\theta\, \cos^2\theta=\frac{4\gamma^2}{M}=\hat{q}_r(0)~,
\end{equation}
which works since $\gamma\equiv\sqrt{\frac{M\hat{q}_r(0)}{4}}$. We can now write an equivalent expression for the second moment: 
\begin{multline}
	\int_0^\beta d\tau \,(q^\approx_r(\tau))^2=\frac{128\gamma^3}{M^2\pi^2}\int_0^{\frac{\pi}{2}}d\theta\int_0^{\frac{\pi}{2}} d\theta'\frac{\cos^2\theta\cos^2\theta'\sin\theta\sin\theta'}{\sin(\theta-\theta')\sin(\theta+\theta')}\\\times\left[\sin\theta \coth\left(\frac{\beta}{2\gamma}\sin\theta'\right)-\sin\theta' \coth\left(\frac{\beta}{2\gamma}\sin\theta\right)\right]
\end{multline}
This is an exact expression, but the evaluation of this integral for arbitrary $\beta$ is hopeless, so we work in the $1/(\beta J)$ expansion. The zero temperature integral can be done: 
\begin{equation}
	\int_0^\beta d\tau \,(q_{r,\beta=\infty}^\approx(\tau))^2=\frac{64 \gamma^3}{15\pi M^2}\,.
\end{equation}
The remaining corrections are computed by: 
\begin{equation}
	\frac{128\gamma^3}{M^2\pi^2}\sum_{n=1}^\infty\int_0^{\frac{\pi}{2}}d\theta\int_0^{\frac{\pi}{2}} d\theta'\frac{\cos^2\theta\cos^2\theta'\sin\theta\sin\theta'}{\sin(\theta-\theta')\sin(\theta+\theta')}\left[2e^{-n\frac{\beta}{\gamma}\sin\theta'}\sin\theta -2e^{-n\frac{\beta}{\gamma}\sin\theta}\sin\theta' \right]
\end{equation}
For large $\beta/\gamma$, each value of $n$ contributes a limiting value of $\frac{128\gamma ^5}{\pi n^2\beta ^2 M^2}$ thus leading to 

\begin{equation}
	\int_0^\beta d\tau \,(q^\approx_r(\tau))^2=\frac{64 \gamma^3}{15\pi M^2}\left(1+5\pi^2\frac{\gamma^2}{\beta^2}+\dots\right)
\end{equation}
Plugging in the low temperature expansion of our parameters \eqref{eq:qr0lowtempexp} and trading $M$ for ${\cal J}$ and $u_*$ using \eqref{eq:ustardef}, we find: 
\begin{equation}
	\frac{1}{\beta}\int_0^\beta d\tau \,(q^\approx_r(\tau))^2=\frac{2}{5}\frac{(1-u_*)u_*^{1-p/2}}{\beta \mathcal{J}}\left(1+\frac{8 u_*^{2-p}[(p+8)u_*-(p-2)]}{9(\beta \mathcal{J})^2(1-u_*)^2\left[(p+2)u_*-(p-2)\right]}+\dots\right)
\end{equation}
Now we see that the leading contribution to this expression is smaller than the first moment \eqref{eq:qr0lowtempexp} by a factor of $(1-u_*)$, which approaches zero as $M{\cal J}$ gets large. We can see this explicitly by working in the double expansion for large $\beta {\cal J}$ and large $M{\cal J}$:
\begin{multline}
\frac{1}{\beta}\int_0^\beta d\tau \,(q^\approx_r(\tau))^2=\frac{8}{15\pi\beta\mathcal{J}\sqrt{M\mathcal{J}}}\left(1+\frac{p-2}{\pi\sqrt{M\mathcal{J}}}+\dots\right)\\+\frac{2}{3(\beta\mathcal{J})^3}\left(\pi\sqrt{M\mathcal{J}}+\frac{28}{15}(p-2)+\dots\right)+\dots
\end{multline}

The cubic integral (which is the highest power required for $p=3$) must be analyzed by numerical fitting, even at strictly zero temperature. We find:
\begin{align}
\int_0^\beta d\tau \,(q^\approx_r(\tau))^3=\frac{1024\gamma^4}{15\pi^2M^3}\left(n_3+n_3'\,\frac{\pi^2\gamma^2}{\beta^2}+n_3''\,\frac{\pi^3\gamma^3}{\beta^3}+n_3'''\,\frac{\pi^4\gamma^4}{\beta^4}\dots\right)~,
\end{align}
with
\begin{equation}
\label{eq:zetacube}
	n_3 \approx 0.106447 ~, \qquad n_3'\approx 0.488334~,\qquad n_3''\approx 0.195977~,\qquad n_3'''\approx0.687666~.
\end{equation}

\subsection*{Internal energy}
We want to compute the low temperature thermodynamic quantities in the spin glass phase. To do this, we will use the approximation that underlies our derivation of the low-temperature solutions:
\begin{equation}\label{eq:SGapprox}
	q(\tau)^n-u^n\approx n u^{n-1}q_r(\tau)+\frac{n(n-1)}{2}\, u^{n-2}q_r(\tau)^2+\dots
\end{equation}
Keeping the lowest term in this expansion, we immediately find from \eqref{eq:EpositionMarg}:
\begin{align}
   \beta \Et =   \frac{1}{2} \frac{\beta}{\hat{q}_r(0)}  + p\frac{(\beta J)^2}{4} \, u^p &\left\{-\frac{p+2}{p}m+\left(\frac{p-1}{u^2}-\frac{p+2}{u}\right)\frac{\hat{q}_r(0)}{\beta}\right.\nonumber\\
   &~~+\frac{p-1}{2}\left(\frac{p-2}{u^3}-\frac{p+2}{u^2}\right)\frac{1}{\beta}\int_0^\beta d\tau\,q_r(\tau)^2\nonumber\\
  &~~\left.+ \frac{(p-1)(p-2)}{6}\left(\frac{p-3}{u^4}-\frac{p+2}{u^3}\right)\frac{1}{\beta}\int_0^\beta d\tau q_r(\tau)^3+\dots \right\}~. 
\end{align}
Note that for $p=3$ (which we usually assume in numerical calculations), the above terms are all there is.
We now want to evaluate this on the low temperature solutions \eqref{eq:mlowtSG}, \eqref{eq:ulowtSG}, \eqref{eq:qr0lowtempexp}, using the perturbative expressions for moments of $q^\approx_r$ derived in the previous paragraphs:
\begin{align}
\label{eq:betaElowTres}
\beta \Et=&-u_*^{\frac{p}{2}-1}\, \beta \mathcal{J}\left[\left(1+\frac{2}{p}\right)u_*-1+\frac{(1-u_*)}{10}\left(p+2-\frac{p-2}{u_*}\right)-n_3\frac{(p-2)(1-u_*)^2}{5}\left(\frac{p-3}{u_*^2}-\frac{p+2}{u_*}\right)\right] \nonumber\\&-\frac{4u_*^{-\frac{p}{2}}}{45\beta\mathcal{J}\left(1-\frac{p+2}{p-2}u_*\right)}\left[8+6 u_*-3p(1-u_*)-4n_3(1-u_*)\left(p+2-\frac{2(p-3)}{u_*}\right)\right.\nonumber\\&\left.\qquad\qquad\qquad\qquad\qquad\qquad+n_3'(p-3-(p+2)u_*)\left(p+2-\frac{p-2}{u_*}\right)\right]+\dots
\end{align}
This gives us a zero temperature energy and the first correction tells us that the specific heat is linear in $T$.
Also note that the following combination simplifies at $T\rightarrow 0$:
\begin{multline}
	m\beta \Et \underset{T\rightarrow0}{=}-(p-2)\left[\left(\frac{p+2}{p}-\frac{1}{u_*}\right)+\frac{(1-u_*)}{10 u_*}\left(p+2-\frac{p-2}{u_*}\right)\right.\\\left. +n_3\frac{(p-2)(1-u_*)^2}{5}\left(\frac{p+2}{u_*^2}-\frac{p-3}{u_*^3}\right)\right]~. 
\end{multline}
Evaluating \eqref{eq:betaElowTres} in the \emph{quantum scaling}, we obtain the main result \eqref{eq:Equantum}.

\subsection*{Entropy}

To compute the entropy note that by definition, it is given by $\St= \beta \Et- \beta\bar{f}$,
thus we can combine \eqref{eq:fSG} and \eqref{eq:EpositionMarg} and obtain: 
\begin{align}
\St=& \,\frac{1}{2}\sum_{k=-\infty}^\infty\log \left[\left(\frac{M}{\beta} \left({2\pi k}\right)^2-2   {i\hat{z}(0)}\right)  \frac{\hat{q}_r(k)}{\beta}  \right] +\frac{\beta}{\hat{q}_r(0)}
 \nonumber\\
&-\left[\frac{(\beta J)^2}{4}\left\lbrace 2p\,m\,  u^{p}+\frac{1+2p}{\beta}\int_0^\beta d\tau\, \left[q(\tau)^{p}-u^{p}\right]-\frac{2p}{\beta}\int_0^\beta d\tau\, \left[q(\tau)^{p-1}-u^{p-1}\right]\right\rbrace\right] \nonumber\\
&-\log\left[2\sinh\sqrt{\frac{-i\hat{z}(0)\beta}{2M}}\right]+\frac{1}{2}\left(\frac{u}{\frac{\hat{q}_r(0)}{\beta}+mu}\right)~.
\label{eq:Sappendix}
\end{align}
To evaluate this expression, we will again use the approximation \eqref{eq:SGapprox}:
\begin{align}
\St\approx & +\frac{\beta}{\hat{q}_r(0)}+\frac{1}{2}\left(\frac{u}{\frac{\hat{q}_r(0)}{\beta}+mu}\right)\nonumber\\
&-p\frac{(\beta J)^2}{2}u^p\left[ m+\frac{1}{u}\left(\frac{2p+1}{2 }-\frac{p-1}{u}\right)\frac{\hat{q}_r(0)}{\beta}+\frac{(p-1)}{2u^2}\left(\frac{2p+1}{2}-\frac{p-2}{u}\right)\frac{1}{\beta}\int_0^\beta d\tau (q^\approx_r(\tau))^2\right.\nonumber\\
&\quad\quad\quad\quad\quad\quad\quad\left.+\frac{(p-1)(p-2)}{6u^3}\left(\frac{2p+1}{2}-\frac{p-3}{u}\right)\frac{1}{\beta}\int_0^\beta d\tau (q^\approx_r(\tau))^3+\dots\right]\nonumber\\
&+\frac{1}{2}\sum_{k=-\infty}^\infty\log \left[\left(\frac{M}{\beta} \left({2\pi k}\right)^2-2   {i\hat{z}(0)}\right)  \frac{\hat{q}_r(k)}{\beta}  \right]-\log\left[2\sinh\sqrt{\frac{-i\hat{z}(0)\beta}{2M}}\right] ~,\label{eq:lowtS}
\end{align}
as well as
 \begin{align}
-2   {i\hat{z}(0)}&=\frac{\beta}{\hat{q}_r(0)}+p\frac{\beta^2 J^2}{2\beta}\int_0^\beta d\tau \left[q(\tau)^{p-1}-u^{p-1}\right]~\nonumber\\
&\approx \frac{\beta}{\hat{q}_r(0)}+ p(p-1)u^{p-2}\frac{(\beta J)^2}{2}\left[\frac{\hat{q}_r(0)}{\beta}+\frac{p-2}{2u\beta}\int_0^\beta d\tau (q^\approx_r(\tau))^2+\frac{(p-2)(p-3)}{12u^2\beta}\int_0^\beta d\tau (q^\approx_r(\tau))^3+\dots\right]\label{eq:lowtempz0}
\end{align}
Since $\frac{\hat{q}_r(0)}{\beta}$ is getting parametrically small in the \emph{quantum limit}, then $-i\hat{z}(0)$ is getting parametrically big, as its leading term is $\frac{\beta}{\hat{q}_r(0)}$.
In order to proceed, we must estimate the sum in the last line of \eqref{eq:lowtS}. We will do this at low temperatures by writing it as a continuous integral, using: 
\begin{equation}
	\hat{q}^{\approx}_r(\omega)=\frac{8\gamma^2}{M}\left(\frac{1}{2}+\gamma^2\omega^2-\sqrt{\gamma^2\omega^2+\gamma^4\omega^4}\right) \qquad \text{with}\quad 
	\gamma\equiv\sqrt{\frac{M\hat{q}_r(0)}{4}}~,\quad \omega \equiv \frac{2\pi k}{\beta}~, 
\end{equation}
as before. This gives
\begin{align}
 \lim_{\beta\rightarrow\infty}\frac{1}{2}\sum_{k=-\infty}^\infty\log \left[\left(\frac{M}{\beta} \left({2\pi k}\right)^2-2   {i\hat{z}(0)}\right)  \frac{\hat{q}_r(k)}{\beta}  \right]\approx&\, 2 \times \beta\int_{0}^{\infty} \frac{d\omega}{2\pi}\frac{1}{2}\log \left[\left(M \omega^2-\frac{2i\hat{z}(0)}{\beta}\right)\hat{q}^{\approx}_r(\omega)\right]\nonumber\\
=&\sqrt{\frac{-i\hat{z}(0)\beta}{2M}} - \frac{\beta}{\pi \gamma}
\label{eq:logSumFinal}
\end{align}
In the quantum scaling, this contribution is ${\cal O}(\Lambda)$, and we are blind to the subleading therms. Combining this with the other contribution from the third line of \eqref{eq:lowtS}, we have : 
\begin{multline}
\frac{1}{2}\sum_{k=-\infty}^\infty\log \left[\left(\frac{M}{\beta} \left({2\pi k}\right)^2-2   {i\hat{z}(0)}\right)  \frac{\hat{q}_r(k)}{\beta}  \right]-\log\left[2\sinh\sqrt{\frac{-i\hat{z}(0)\beta}{2M}}\right]\\\approx\sqrt{\frac{-i\hat{z}(0)\beta}{2M}}- \frac{\beta}{\pi \gamma}	-\log\left[2\sinh\sqrt{\frac{-i\hat{z}(0)\beta}{2M}}\right]\approx - \frac{\beta}{\pi \gamma} -\log\left(1-e^{-2\sqrt{\frac{-i\hat{z}(0)\beta}{2M}}}\right)
~. 
\end{multline}

The above analysis shows that in the \emph{quantum scaling} the leading contributions to the entropy naively are large, of order ${\cal O}(\Lambda)$. However, putting everything together, we find that all leading contributions to the entropy cancel:
\begin{equation}
	\St = 0  + {\cal O}(\Lambda^0)~.  
\end{equation}
In order to compute the ${\cal O}(\Lambda^0)$ contribution, we would need a more systematic way to approximate the sum over logarithms, \eqref{eq:logSumFinal}. While it would be interesting to investigate this further,  the above computation is enough to show that the complexity $\Sigma$ dominates the full entropy-like quantity $m\St +\Sigma$ at low temperatures (c.f., \eqref{eq:mSSigma}).

\section{The quadratic action of fluctuations}
\label{app:kernelDerivation}

In this appendix we discuss the quadratic action for fluctuations of the matrix $Q_{ab}(\tau,\tau')$. In particular, we describe the different types of fluctuations taking into account the replica symmetry breaking ansatz. Our discussion of fluctuations in replica space serves as a review of appendix 3 of \cite{crisanti1992sphericalp}. We also derive the time-dependent, diagonal fluctuations in position space, which encode the leading correction to the four-point function.

\subsection{General second order fluctuations}

We begin by copying the effective action \eqref{eq:effaction} in terms of the replica matrix $Q_{ab}$:
\begin{multline}\label{eq:effactioncopy}
	\frac{S_{\rm eff}}{N}=-\frac{1}{2}\text{Tr}\log \left[Q_{ab}(\tau,\tau')\right]-i\sum_{a=1}^n\int_0^\beta d\tau\,  z^a(\tau)\left(Q_{aa}(\tau,\tau)-1\right)\\
	-\sum_{a,b=1}^n\int_0^\beta \int_0^\beta d\tau\,d\tau'\,\left[\delta_{ab}\,\delta(\tau-\tau')\frac{M}{2}\partial_\tau^2Q_{ab}(\tau,\tau')+\frac{J^2 }{4} Q_{ab}(\tau,\tau')^p\right]~.
\end{multline}
In terms of the Fourier transformed replica matrix
\begin{equation}
	Q_{ab}(\tau,\tau')=\frac{1}{\beta}\sum_{k=-\infty}^\infty e^{\frac{2\pi i\,k}{\beta}(\tau-\tau')}\hat{Q}_{ab}(k)~,
\end{equation}
the effective action is 
\begin{multline}\label{eq:effactionk}
	\frac{S_{\rm eff}}{N}=-\frac{1}{2}\sum_{k=-\infty}^\infty\text{Tr}\log \left[\frac{\hat{Q}_{ab}(k)}{\beta}\right]+i\sum_{a=1}^n \hat{z}^a(0)\left(1-\sum_{k=-\infty}^\infty \frac{\hat{Q}_{aa}(k)}{\beta}\right)\\
	+\sum_{a,b=1}^n\left[\sum_{k=-\infty}^\infty\delta_{ab}\frac{M}{2}\left(\frac{2\pi k}{\beta}\right)^2\hat{Q}_{ab}(k)-\frac{J^2 }{4}\int_0^\beta \int_0^\beta d\tau\,d\tau'\, Q_{ab}(\tau,\tau')^p\right]~.
\end{multline}
Now we want to expand this action around a classical solution $\hat{Q}_{ab}^\star(k)$ such that $\hat{Q}_{ab}(k)=\hat{Q}_{ab}^\star(k)+\hat{r}_{ab}(k)$. To quadratic order in $r_{ab}$ the action is: 
\begin{multline}\label{eq:quadfluct}
	\frac{\delta S_{2}}{N}=+\frac{1}{4}\sum_{k=-\infty}^\infty\text{Tr}\left[\left(\hat{\mathbf{Q}}^{\star-1}(k)\cdot\hat{\mathbf{r}}(k)\right)\cdot \left(\hat{\mathbf{Q}}^{\star-1}(k)\cdot\hat{\mathbf{r}}(k)\right)\right]\\
	-\frac{\mathcal{J}^2 }{4}\sum_{a,b=1}^n\int_0^\beta \int_0^\beta d\tau\,d\tau'\,\left[\sum_{k=-\infty}^\infty\sum_{l=-\infty}^\infty e^{\frac{2\pi i}{\beta}(k+l)(\tau-\tau')}\frac{\hat{r}_{ab}(k)}{\beta}\frac{\hat{r}_{ab}(l)}{\beta} Q^\star_{ab}(\tau,\tau')^{p-2}\right]~,
\end{multline}
where we remind the reader that $\mathcal{J}^2=J^2p(p-1)/2$.
Recall that we parametrized the saddle point solution as follows: 
\begin{equation}
	Q_{ab}^\star(\tau,\tau')=\delta_{a_1b_1}\left[\delta_{a_0b_0}q(\tau,\tau')+u(1-\delta_{a_0b_0})\right]\,,
\end{equation}
where we have split the index $a=a_0\times a_1$ with $a_0=1,\dots,m$ parameterizing the location in the block and $a_1=1,\dots,\frac{n}{m}$ parameterizing which block we are in (see equations \eqref{eq:1rsb} and \eqref{eq:rsbnotation}). In momentum space this implies
\begin{equation}
\label{eq:QstarMat}
	\hat{Q}_{ab}^\star(k)=\delta_{a_1b_1}\left[\delta_{a_0b_0}\hat{q}(k)+\hat{u}(k)(1-\delta_{a_0b_0})\right]\,,\qquad 
	\hat{u}(k)=\beta u\,\delta_{k,0}~.
\end{equation}
It is easy to verify directly that the inverse matrix in the replica indices is: 
\begin{equation}
\label{eq:QstarMatInv}
	(\hat{Q}^{\star})^{-1}_{ab}=\delta_{a_1b_1}\left[\delta_{a_0b_0}\left(\hat{A}(k)+\hat{B}(k)\right)+\hat{B}(k)(1-\delta_{a_0b_0})\right]\,,
\end{equation}
where
\begin{equation}
	\hat{A}(k)=\frac{1}{\hat{q}(k)-\hat{u}(k)}~,\qquad\qquad \hat{B}(k)=\frac{\hat{u}(k)}{\left[\hat{u}(k)-\hat{q}(k)\right]\left[\hat{q}(k)+(m-1)\hat{u}(k)\right]}\,.
\end{equation}
We will also use the following notation to write the 1-RSB matrices: 
\begin{equation}
	\delta_{ab}\equiv\delta_{a_0b_0}\delta_{a_1b_1}~,\quad\;\; \gamma_{ab}=\delta_{a_1b_1}\,,\quad\;\;  \text{with} ~a=a_0\times a_1, \;b=b_0\times b_1
\end{equation}
which means that $\gamma_{ab}$ equals 1 along $m\times m$ block diagonals and 0 otherwise (this matrix is often called $\epsilon_{ab}$ in the literature, but we will not use this notation to avoid confusion with the Levi-Civita symbol). We can thus write \eqref{eq:QstarMat} and \eqref{eq:QstarMatInv} as: 
\begin{equation}
	\hat{Q}_{ab}^\star(k)= \left(\hat{q}(k)-\hat{u}(k)\right)\delta_{ab}+\hat{u}(k)\gamma_{ab}\,,\qquad\quad	
	(\hat{Q}^{\star})^{-1}_{ab}=\hat{A}(k)\delta_{ab}+\hat{B}(k)\gamma_{ab}~.
\end{equation}
Plugging these into the quadratic fluctuation action, we find:
\begin{multline}\label{eq:quadfluctred}
	\frac{\delta S_{2}}{N}=\frac{1}{4}\sum_{k=-\infty}^\infty\text{Tr}\bigg\lbrace\bigg[ \hat{A}(k)^2 \hat{\mathbf{r}}(k)+\hat{A}(k)\hat{B}(k)\left[\hat{\mathbf{r}}(k)\cdot\gamma+\gamma\cdot\hat{\mathbf{r}}(k)\right]+\hat{B}(k)^2(\gamma\cdot\hat{\mathbf{r}}(k)\cdot \gamma)\bigg]\cdot\hat{\mathbf{r}}(k)\bigg\rbrace\\
	-\frac{\mathcal{J}^2 }{4}\beta^{2-p}\sum_{a,b=1}^n\sum_{k,l=-\infty}^\infty\sum_{l_1,\dots,l_{p-2}}\delta_{k+l+ l_1+\dots+l_{p-2},0}\,\hat{r}_{ab}(k)\left[\delta_{ab}\prod_{i=1}^{p-2}\hat{q}(l_i)+(\gamma_{ab}-\delta_{ab})\prod_{i=1}^{p-2}\hat{u}(l_i)\right]\hat{r}_{ab}(l)~.
\end{multline}
We can rewrite this fluctuation action as follows
\begin{equation}
	\frac{\delta S_2}{N}= \sum_{k,l}\sum_{a,b,c,d}\hat{r}_{ab}(k)G_{abcd}(k,l)\hat{r}_{cd}(l)
\end{equation}
with
\begin{multline}
G_{abcd}(k,l)\equiv\frac{1}{4}\left[\hat{A}(k)^2\delta_{ad}\delta_{bc}+\hat{A}(k)\hat{B}(k)\left(\delta_{ad}\gamma_{bc}+\delta_{bc}\gamma_{ad}\right)+\hat{B}(k)^2\gamma_{ad}\gamma_{bc}\right]\delta_{kl}\\-\beta^{2-p}\frac{\mathcal{J}^2}{4}\sum_{l_1,\dots,l_{p-2}}\delta_{k+l+ l_1+\dots+l_{p-2},0}\left[\delta_{ab}\left(\prod_{i=1}^{p-2}\hat{q}(l_i)-\prod_{i=1}^{p-2}\hat{u}(l_i)\right)+\gamma_{ab}\prod_{i=1}^{p-2}\hat{u}(l_i)\right]\delta_{ac}\delta_{bd}~.
\end{multline}
We have mentioned this before, but it bears repeating that $k$ dependence can only be present in the replica-diagonal components. This means that: 
\begin{equation}
	\hat{r}_{ab}=\hat{r}(k)\delta_{ab}+ \beta s_{ab}\delta_{k,0}~.
\end{equation}
We will take 
\begin{equation}\label{eq:fluctansatz}
	\hat{r}(k)=\hat{r}(-k)~,\qquad s_{ab}=s_{ba},\qquad \text{and}\qquad s_{aa}=0 ~~\forall a,
\end{equation}
and where in order to preserve the spherical constraint, 
\begin{equation}
	\frac{1}{\beta}\sum_{k=-\infty}^\infty \hat{r}(k)=0~.
\end{equation}
Using that $\gamma\cdot \gamma = m \gamma$, on this ansatz the quadratic fluctuation action is: 
\begin{align}\label{eq:quadactsplit}
	\frac{\delta S_2}{N}=&~\frac{n}{4}\sum_{k=-\infty}^\infty\hat{r}(k)\left\lbrace\left[\hat{A}(k)+\hat{B}(k)\right]^2+(m-1) \hat{B}(k)^2\right\rbrace \hat{r}(k)\nonumber\\
	&-n\frac{\mathcal{J}^2 }{4}\sum_{k,l=-\infty}^\infty\sum_{l_1,\dots,l_{p-2}}\hat{r}(k)\left[\prod_{i=1}^{p-2}\frac{\hat{q}(l_i)}{\beta}\right]\hat{r}(l)\,\delta_{k+l+ l_1+\dots+l_{p-2},0}\nonumber\\
	&+\frac{\beta}{2}\, \hat{r}(0)\left\lbrace2\hat{A}(0)\hat{B}(0)+m\hat{B}(0)^2\right\rbrace \text{Tr}\left(\gamma\cdot \mathbf{s}\right)\nonumber\\
	&+\frac{\beta^2}{4}\text{Tr}\left\lbrace \mathbf{s}\cdot\left[\hat{A}(0)^2\,\mathbf{s}+\hat{A}(0)\hat{B}(0)\left(\gamma\cdot\mathbf{s}+\mathbf{s}\cdot\gamma\right)+\hat{B}(0)^2\left(\gamma\cdot\mathbf{s}\cdot\gamma\right)\right]\right\rbrace\nonumber\\
	&-\frac{\beta^2\mathcal{J}^2 }{4}\sum_{a,b=1}^n u^{p-2}s_{ab}\gamma_{ab}s_{ab}~.
\end{align}
Notice that the diagonal and off-diagonal fluctuations almost decouple, were it not for the third line in \eqref{eq:quadactsplit}. However, to assess stability of the solutions to the Schwinger-Dyson equations, we can consider the spectrum of dynamical and off-diagonal fluctuations separately. 

\subsection{Static fluctuations}\label{ap:statfluct}
For static fluctuations ($\hat{r}(k)=0$) the fluctuation action is: 
\begin{multline}\label{eq:quadactstatic}
	\frac{\delta S_2}{N}=\frac{\beta^2}{4}\text{Tr}\left\lbrace \mathbf{s}\cdot\left[\hat{A}(0)^2\,\mathbf{s}+\hat{A}(0)\hat{B}(0)\left(\gamma\cdot\mathbf{s}+\mathbf{s}\cdot\gamma\right)+\hat{B}(0)^2\left(\gamma\cdot\mathbf{s}\cdot\gamma\right)\right]\right\rbrace\\-\frac{\beta^2\mathcal{J}^2 }{4}\sum_{a,b=1}^n u^{p-2}s_{ab}\gamma_{ab}s_{ab}~,
\end{multline}
and the eigenvalue equation we would like to solve is: 
\begin{multline}\label{eq:eigstatic}
\hat{A}(0)^2s_{ab}+\hat{A}(0)\hat{B}(0)\left[\left(\mathbf{s}\cdot\gamma\right)_{ab}+\left(\gamma\cdot \mathbf{s}\right)_{ab}\right]+\hat{B}(0)^2\left(\gamma\cdot \mathbf{s}\cdot\gamma\right)_{ab}\\-{\mathcal{J}^2}u^{p-2}\gamma_{ab}s_{ab}=\beta^{-2}\lambda s_{ab}~.
\end{multline}
for $a\neq b$, since $s_{aa}=0~~\forall a$~. The quantity $\lambda$ on the right hand side is the eigenvalue and should not be confused with the auxiliary field introduced in \eqref{eq:deltafunc}, nor the Lyapunov exponent $\lambda_L$.  We now summarize the results from \cite{crisanti1992sphericalp} regarding these fluctuations.  

\paragraph{\uline{Transverse fluctuations}: $\left(\mathbf{s}\cdot\gamma\right)_{ab}=\left(\gamma\cdot \mathbf{s}\right)_{ab}=0~~\forall a,b$}\mbox{}\vspace{.2cm}

The transverse constraint amounts to the fact that, if we break up each row and column into strings of size $m$, each such string must sum to zero independently. Said in another way, each row and each column of every $m\times m$ block of $s_{ab}$ must sum to zero. This removes $2m-1$ (the difference between an $m\times m$ matrix and an $(m-1)\times(m-1)$ matrix) elements from each off-diagonal block, and removes $m$ elements from each diagonal block, since the diagonal blocks are symmetric and have zeros along the diagonal. This leaves 
\begin{equation}
	d_T\equiv\frac{n(n-1)}{2}-\frac{\tfrac{n}{m}\left(\tfrac{n}{m}-1\right)}{2}(2m-1)-\frac{n}{m}\times m
\end{equation}
free elements in $s_{ab}$. 

On this subspace the eigenvalue equation reduces to: 
\begin{equation}
\hat{A}(0)^2s_{ab}-\mathcal{J}^2u^{p-2}\gamma_{ab}s_{ab}=\beta^{-2}\lambda\, s_{ab}~.
\end{equation}
There are two sets of eigenmatrices of this equation. The first set corresponds to $s_{ab}$ being zero along the block diagonals: 
\begin{align}
s_{ab}\gamma_{ab}&=0~~\forall a,b \nonumber\\
\lambda_T^{\rm off-diag}&=\beta^2\hat{A}(0)^2=\left(\frac{\hat{q}(0)}{\beta}- u\right)^{-2}\nonumber\\
\text{deg}&=\frac{\frac{n}{m}\left(\frac{n}{m}-1\right)}{2}(m-1)^2~.\nonumber
\end{align}
The second set corresponds to $s_{ab}$ only having elements along the block diagonals
\begin{align}
s_{ab}\gamma_{ab}&=s_{ab}~~\forall a,b \nonumber\\
\lambda_T^{\rm diag}&=\beta^2\hat{A}(0)^2-{\beta^2\mathcal{J}^2}u^{p-2}=\left(\frac{\hat{q}(0)}{\beta}- u\right)^{-2}-{\beta^2\mathcal{J}^2}u^{p-2}\nonumber\\
\text{deg}&=\frac{n}{2}(m-3)~.\nonumber
\end{align}
We will come to the fact that the degeneracy of these modes is negative for $m<3$ later. The case of marginal stability is diagnosed by the existence of a zero-eigenvalue among these transverse fluctuations: $\lambda_ {T}^{\rm diag}=0$. This gives: 
\begin{equation}
	\frac{1}{\left(\frac{\hat{q}(0)}{\beta}-{u}\right)^{2}}={\beta^2\mathcal{J}^2}u^{p-2}\label{eq:vanishinglambdaT}
\end{equation}
now we can combine this with equation \eqref{eq:EOMu0} which determines the value of $u$ at equilibrium: 
\begin{equation}
	\frac{\beta^2J^2}{2}p\,u^{p-1}-\frac{u}{\left(\frac{\hat{q}(0)}{\beta}-u\right)\left(\frac{\hat{q}(0)}{\beta}+(m-1)u\right)}=0 
\end{equation}
to show that  $\lambda_T^{\rm diag}=0$ suggests:
\begin{equation}
	u=\left[\frac{p-2}{\beta \mathcal{J} m}\right]^{2/p}~,\qquad \frac{\hat{q}(0)}{\beta}=u\left(1+\frac{m }{p-2}\right)=\left[\frac{p-2}{\beta \mathcal{J} m}\right]^{2/p}\left(1+\frac{m }{p-2}\right)\label{eq:mvarMarg}
\end{equation}
These are the values found in \cite{cugliandolo2001imaginary} that give rise to the marginal spin glass state. A solution with these on-shell values implies the existence of marginal fluctuations in replica space, even though the number of such eigenvalues is going to zero in the $n\rightarrow0$ limit. Note, however, that the equilibrium equation of motion for $m$ \eqref{eq:EOMm0} is not satisfied given the parameters \eqref{eq:mvarMarg}, which implies that the marginal spin glass state is not an equilibrium state in this ensemble. 

\paragraph{\uline{``Longitudinal'' fluctuations}: $\left(\gamma\cdot\mathbf{s}\cdot\gamma\right)_{ab}=0~~\forall a,b$  while $\left(\mathbf{s}\cdot\gamma\right)_{ab}\neq0$ and $\left(\gamma\cdot \mathbf{s}\right)_{ab}\neq0$}\mbox{}\vspace{.2cm}

This condition tells us that the sum of all the elements in each $m\times m$ block, taken together, gives zero. This condition removes one element from each $m\times m$ block. Further demanding that these be distinct from the transverse matrices and we are left with
\begin{equation}
	d_L\equiv\left(\frac{n}{m}\right)^2\left(m-1\right)
\end{equation}
free elements in these longitudinal fluctuations. To obtain the eigenvalue equation in this subspace, we will dot \eqref{eq:eigstatic} with $\gamma_{ab}$ from the left and use that $\gamma\cdot\gamma= m\gamma$, this gives: 
\begin{equation}
\hat{A}(0)^2 (\gamma\cdot \mathbf{s})_{ab}+\hat{A}(0)\hat{B}(0)\sum_{c\neq b}\gamma_{ac}\left[\left(\gamma\cdot \mathbf{s}\right)_{cb}+\left(\mathbf{s}\cdot\gamma\right)_{cb}\right]-\mathcal{J}^2\gamma_{ab} (\gamma\cdot \mathbf{s})_{ab}=\beta^{-2}\lambda (\gamma\cdot \mathbf{s})_{ab}~.
\end{equation}
where we used $\sum_{c=1}^n\gamma_{ac}\gamma_{cb}s_{cb}= \gamma_{ab}\left(\gamma\cdot \mathbf{s}\right)_{ab}$, which one can verify by direct computation. Also, unlike \eqref{eq:eigstatic}, we need not restrict to $a\neq b$. In this subspace, there are again two sets of eigenmatrices that satisfy this eigenvalue equation. The first set corresponds to matrices that are block-off-diagonal after dotting with $\gamma$: 
\begin{align}
\gamma_{ab}\left(\gamma\cdot \mathbf{s}\right)_{ab}&=0~~\forall a,b \nonumber\\
\lambda_L^{\rm off-diag}&=\beta^2\hat{A}(0)\left[\hat{A}(0)+m\hat{B}(0)\right]=\left[\left(\frac{\hat{q}(0)}{\beta}- u\right)\left(\frac{\hat{q}(0)}{\beta}+(m-1) u\right)\right]^{-1}\nonumber\\
\text{deg}&=\frac{n}{m}\left(\frac{n}{m}-1\right)(m-1)~.\nonumber
\end{align}
And the second set corresponds to: 
\begin{align}
\gamma_{ab}\left(\gamma\cdot \mathbf{s}\right)_{ab}&=\left(\gamma\cdot \mathbf{s}\right)_{ab}~~\forall a,b \nonumber\\
\lambda_L^{\rm diag}&=\beta^2\hat{A}(0)\left[\hat{A}(0)+(m-2)\hat{B}(0)\right]-{\beta^2\mathcal{J}^2}u^{p-2}=\frac{\frac{\hat{q}(0)}{\beta}+u}{\left(\frac{\hat{q}(0)}{\beta}- u\right)^2\left(\frac{\hat{q}(0)}{\beta}+(m-1) u\right)}-{\beta^2\mathcal{J}^2}u^{p-2}\nonumber\\
\text{deg}&=\frac{n}{m}(m-1)~.\nonumber
\end{align}

\paragraph{\uline{Remaining cluster fluctuations}: $\left(\gamma\cdot\mathbf{s}\cdot\gamma\right)_{ab}\neq0$}\mbox{}\vspace{.2cm}

From a total $n(n-1)/2$ distinct possible fluctuations for $s_{ab}$, we have accounted for all but $\frac{\frac{n}{m}\left(\frac{n}{m}+1\right)}{2}$ of them. This last number corresponds to the number of elements in a symmetric $\frac{n}{m}\times \frac{n}{m}$ matrix. We will take this seriously. 

The first set of eigenvalues are easy to guess, they correspond to picking an off-diagonal $m\times m$ block of $s_{ab}$ and setting all the elements equal and symmetrizing the whole $n\times n$ matrix. These matrices satisfy $\gamma\cdot\mathbf{s}=\mathbf{s}\cdot\gamma=m\,\mathbf{s}$ and $\gamma\cdot\mathbf{s}\cdot\gamma=m^2\,\mathbf{s}$. The eigenvalues are: 
\begin{align}
\lambda_R^{\rm off-diag}&=\beta^2\left[\hat{A}(0)+m\hat{B}(0)\right]^2=\left(\frac{\hat{q}(0)}{\beta}+(m-1) u\right)^{-2}\nonumber\\
\text{deg}&=\frac{\frac{n}{m}\left(\frac{n}{m}-1\right)}{2}~.\nonumber
\end{align}

The final set of eigenvalues are constructed by picking a diagonal block and setting all elements equal except recalling that $s_{aa}=0$ for all $a$. These matrices satisfy $(\gamma\cdot\mathbf{s})_{ab}=(\mathbf{s}\cdot\gamma)_{ab}=(m-1)\,s_{ab}~~\forall a\neq b$ and $(\gamma\cdot\mathbf{s}\cdot\gamma)_{ab}=m(m-1)\,s_{ab}~~\forall a\neq b$. The eigenvalues are: 
\begin{align}
\Lambda_R^{\rm diag}&=\beta^2\left[\hat{A}(0)+(m-1)\hat{B}(0)\right]^2+\beta^2(m-1)\hat{B}(0)^2-{\beta^2\mathcal{J}^2}u^{p-2}\nonumber\\
&=\frac{\left(\frac{\hat{q}(0)}{\beta}\right)^2+(m-1)u^2}{\left(\frac{\hat{q}(0)}{\beta}- u\right)^2\left(\frac{\hat{q}(0)}{\beta}+(m-1) u\right)^2}-{\beta^2\mathcal{J}^2}u^{p-2}\nonumber\\
\text{deg}&=\frac{n}{m}~.\nonumber
\end{align}

\paragraph{\uline{\textbf{Negative Degeneracies:}}}\mbox{}\vspace{.2cm}

For $m=1$ and $m=2$ notice that the degeneracies of $\lambda_{T}^{\rm diag}$ are negative. Have we made a mistake? No, for $m=1$, $\lambda_{T}^{\rm diag}=\lambda_{R}^{\rm diag}$ and the sum of their degeneracies gives zero. For $m=2$, $\lambda_{T}^{\rm diag}=\lambda_{L}^{\rm diag}$ and the sum of \emph{their} degeneracies gives zero. So everything is consistent, as when there are negative degeneracies, these ``eat'' terms with the same eigenvalue.

\subsection{Dynamical fluctuations and four-point functions}\label{ap:dynfluct}

Finally, we shall now derive the quadratic action for time-dependent fluctuations, \eqref{eq:S2eucl}. To do so,  we return to \eqref{eq:quadactsplit} but restrict to dynamical fluctuations where $s_{ab}=0$. 
This means we are dealing with an effective action in momentum space 
\begin{multline}
	\frac{\delta S_2}{nN}=\frac{1}{4}\sum_{k,l=-\infty}^\infty
	\frac{\hat{r}(k)}{\beta}\left\lbrace\left[\frac{m-1}{m}\frac{\beta^2}{\hat{q}_r(k)\hat{q}_r(l)}+\frac{1}{m}\frac{\beta^2}{(\hat{q}_r(k)+m\hat{u}(k))(\hat{q}_r(l)+m\hat{u}(l))}\right]\delta_{k+l,0}\right.\\\left.- (\beta\mathcal{J})^2 \sum_{l_1,\dots,l_{p-2}}\left[\prod_{i=1}^{p-2}\frac{\hat{q}_r(l_i)+\hat{u}(l_i)}{\beta}\right]\,\delta_{k+l+ l_1+\dots+l_{p-2},0}\right\rbrace \frac{\hat{r}(l)}{\beta}\label{eq:quadactiondyn}
\end{multline}
where we have used:
\begin{equation}
	\hat{q}_r(k) \equiv \hat{q}(k) - \hat{u}(k)~, 
\end{equation}
and have used time-reversal symmetry, meaning $\hat{r}(k)=\hat{r}(-k)$ and similarly for $\hat{q_r}(k)$~. The physics of the Euclidean four-point function is encoded in \eqref{eq:quadactiondyn}. 

Some simple manipulations of the above expression yield 
\begin{equation}
	\frac{\delta S_2}{nN}=\frac{1}{4}\sum_{k,l=-\infty}^\infty
	\frac{\hat{r}(k)}{\beta}\left\lbrace \frac{\hat{F}(k,l)}{\beta^2}\delta_{k+l,0}-(\beta\mathcal{J})^2  \sum_{l_1,\dots,l_{p-2}}\left[\prod_{i=1}^{p-2}\frac{\hat{q}_r(l_i)+\hat{u}(l_i)}{\beta}\right]\,\delta_{k+l+ l_1+\dots+l_{p-2},0}\right\rbrace \frac{\hat{r}(l)}{\beta}\label{eq:exptotransform}
\end{equation}
where
\begin{equation}
\frac{\hat{F}(k,l)}{\beta^2}\equiv \left[\frac{\hat{q}_r(k)}{\beta}\frac{\hat{q}_r(l)}{\beta}+\st\,\delta_{k,0}\delta_{0,l}\right]^{-1}\,,\qquad
\st\equiv \frac{mu\left(m u+2\frac{\hat{q}_r(0)}{\beta}\right)\left(\frac{\hat{q}_r(0)}{\beta}\right)^2}{(m-1)\left(mu+\frac{\hat{q}_r(0)}{\beta}\right)^2+\left(\frac{\hat{q}_r(0)}{\beta}\right)^2}~.\label{eq:sdef}
\end{equation}
In the conformal spin glass phase, we simplify $\st$ by using the on-shell relation $\frac{\hat{q}_r(0)}{\beta}= \frac{m u}{p-2}$ (equivalent to the vanishing of the transverse eigenvalue in replica space, \eqref{eq:mvarMarg}). The above definition then takes the form \eqref{eq:sMain}.

\subsubsection*{Relation to Euclidean time expression}
We would like to re-express \eqref{eq:exptotransform} as an integral over Euclidean time in order to derive the position space expressions \eqref{eq:S2eucl} and \eqref{eq:kernelEucl}.
We will transform it into a Euclidean time expression in two steps by discussing the two parts of \eqref{eq:exptotransform} separately, $\frac{\delta S_2}{nN} \equiv A_1+A_2$. 
Let us begin with the first term:
\begin{align}
A_1&\equiv \frac{1}{4}\sum_{k,l=-\infty}^\infty
	\frac{\hat{r}(k)}{\beta}\frac{\hat{F}(k,l)}{\beta^2} \frac{\hat{r}(l)}{\beta}\delta_{k+l,0}~ \nonumber\\
&=\frac{1}{4}\int_0^\beta\frac{d\tau_1}{\beta}\frac{d\tau_2}{\beta}\frac{d\tau_3}{\beta}\frac{d\tau_4}{\beta}r(\tau_1)F(\tau_2,\tau_3)r(\tau_4)\times\beta\delta\left((\tau_1+\tau_2)-(\tau_3+\tau_4)\right)\nonumber\\
&=\frac{1}{4}\int_0^\beta\frac{d\tau_1}{\beta}\frac{d\tau_3}{\beta}\frac{d\tau_4}{\beta}r(\tau_1)F(-\tau_1+\tau_3+\tau_4,\tau_3)r(\tau_4)~.
\end{align}
In the above expressions $F(\tau_1,\tau_2)$ is the Fourier transform of $\hat{F}(k,l)$. Now, having arrived at this last expression, we use time-translation invariance to shift $\tau_3$:
\begin{equation}
	\tau_3\rightarrow \tau_a-\tau_4~.
\end{equation}
This moves the center of integration of the integral over $\tau_3$ by a variable we are integrating over, but since the function is periodic, and we are integrating over an entire period, this will not change the final answer. Thus
\begin{align}
A_1&=\frac{1}{4}\int_0^\beta\frac{d\tau_1}{\beta}\frac{d\tau_a}{\beta}\frac{d\tau_4}{\beta}\, r(\tau_1)F\left(-\tau_1+\tau_a,-\tau_4+\tau_a\right)r(\tau_4)~.
\end{align}
We now perform two additional shifts: 
\begin{equation}
	\tau_1\rightarrow \tau_1-\tau_2~,\qquad \tau_4\rightarrow \tau_4-\tau_3\,,
\end{equation}
such that
\begin{align}
A_1&=\frac{1}{4}\int_0^\beta\frac{d\tau_1}{\beta}\frac{d\tau_a}{\beta}\,\frac{d\tau_4}{\beta}r(\tau_1-\tau_2)F\left(-(\tau_1-\tau_2)+\tau_a,-(\tau_4-\tau_3)+\tau_a\right)r(\tau_4-\tau_3)\nonumber \\
&=\frac{1}{4}\int_0^\beta\frac{d\tau_1}{\beta}\frac{d\tau_2}{\beta}\frac{d\tau_3}{\beta}\, \frac{d\tau_4}{\beta}\frac{d\tau_a}{\beta}r(\tau_1-\tau_2)F\left(-(\tau_1-\tau_2)+\tau_a,-(\tau_4-\tau_3)+\tau_a\right)r(\tau_4-\tau_3)~,
\end{align}
where we used time-translation invariance to introduce auxiliary integrals over $\tau_2$ and $\tau_3$, which trivially integrate to 1.
Furthermore, note that the integral over $\tau_a$ simply shifts the center of integration of an integral that we are performing over an entire period, therefore it should not affect the final answer, and we can drop it.  
Time reversal symmetry of $F$ and $r$ ($r(\tau)=r(-\tau)$, $F(\tau,\tau')=F(-\tau,\tau')=F(\tau,-\tau')$) finally yields:
\begin{equation}
A_1=\frac{1}{4}\int_0^\beta\frac{d\tau_1}{\beta}\frac{d\tau_2}{\beta}\, \frac{d\tau_3}{\beta}\frac{d\tau_4}{\beta}r(\tau_1-\tau_2)F\left(\tau_1-\tau_2,\tau_3-\tau_4\right)r(\tau_3-\tau_4)~,
\end{equation} 
which we will write more compactly as: 
\begin{equation}
\label{eq:A1res}
A_1=\frac{1}{4}\int_0^\beta\frac{d\tau_1}{\beta}\frac{d\tau_2}{\beta}\, \frac{d\tau_3}{\beta}\frac{d\tau_4}{\beta}r(\tau_1,\tau_2)F\left(\tau_1,\tau_2;\tau_3,\tau_4\right)r(\tau_3,\tau_4)~.
\end{equation}

Next, we consider the second term in \eqref{eq:exptotransform}:
\begin{align}
A_2&\equiv -\frac{(\beta\mathcal{J})^2}{4}\sum_{k,l,l_1,\dots,l_{p-2}=-\infty}^\infty
	\frac{\hat{r}(k)}{\beta}\left\lbrace  \left[\prod_{i=1}^{p-2}\frac{\hat{q}(l_i)}{\beta}\right]\,\delta_{k+l+ l_1+\dots+l_{p-2},0}\right\rbrace \frac{\hat{r}(l)}{\beta}\nonumber\\
	&=-\frac{(\beta\mathcal{J})^2}{4}\int_0^\beta\frac{d\tau}{\beta}\,r(\tau)q(\tau)^{p-2}\,r(\tau)~. 
\end{align}
where we simply inverted all Fourier transforms. 
By similar considerations as for $A_1$ above, this expression is identical to the following: 
\begin{align}
A_2&=-\frac{(\beta\mathcal{J})^2}{4}\int_0^\beta\frac{d\tau_1}{\beta}\frac{d\tau_2}{\beta}\frac{d\tau_3}{\beta}\frac{d\tau_4}{\beta}\, r(\tau_1,\tau_2)q(\tau_1,\tau_2)^{\frac{p-2}{2}}q(\tau_3,\tau_4)^{\frac{p-2}{2}}r(\tau_3,\tau_4)\,\beta\delta(\tau_2-\tau_4)\, \beta\delta(\tau_1-\tau_3)~. 
\label{eq:A2res}
\end{align}

Combining \eqref{eq:A1res} and \eqref{eq:A2res}, we find for the Euclidean quadratic action of fluctuations in position space:
\begin{equation}
\frac{\delta S_2}{nN}=\frac{(\beta\mathcal{J})^2}{4}\int_0^\beta\frac{d\tau_1}{\beta}\frac{d\tau_2}{\beta}\frac{d\tau_3}{\beta}\frac{d\tau_4}{\beta}\,\tilde{r}(\tau_1,\tau_2)\left[\tilde{F}\left(\tau_1,\tau_2;\tau_3,\tau_4\right)-\beta\delta(\tau_1-\tau_3)\beta\delta(\tau_2-\tau_4)\right]\tilde{r}(\tau_3,\tau_4)\,,
\end{equation}
where
\begin{equation}
\begin{split}
	\tilde{r}(\tau,\tau') &\equiv q(\tau,\tau')^{\frac{p-2}{2}} \, r(\tau,\tau') \,,\\
	\tilde{F}\left(\tau_1,\tau_2;\tau_3,\tau_4\right) &\equiv \frac{1}{(\beta\mathcal{J})^2}\, q(\tau_1,\tau_2)^{\frac{2-p}{2}}F\left(\tau_1,\tau_2;\tau_3,\tau_4\right)q(\tau_3,\tau_4)^{\frac{2-p}{2}}\,. 
\end{split}
\end{equation}
From the definition \eqref{eq:sdef} we may deduce that 
\begin{equation}
	F\left(\tau_1,\tau_2;\tau_3,\tau_4\right)=\left[{q}_r(\tau_{13}){q}_r(\tau_{24})+\st\right]^{-1}~. 
\end{equation}
By definition, this implies:
\begin{align}
	\int_0^\beta\frac{d\tau_a}{\beta}\frac{d\tau_b}{\beta}\, F\left(\tau_1,\tau_2;\tau_a,\tau_b\right)F^{-1}\left(\tau_a,\tau_b;\tau_3,\tau_4\right) 
	&=\beta\delta(\tau_1-\tau_3)\, \beta\delta(\tau_2-\tau_4)~,
\end{align}
where the right hand side is the identity operator on the space of functions of four times. From the above analysis, we conclude that
\begin{equation}
	\tilde{F}\left(\tau_1,\tau_2;\tau_3,\tau_4\right)= \tilde{K}^{-1}\left(\tau_1,\tau_2;\tau_3,\tau_4\right)
\end{equation}
and
\begin{equation}
\tilde{K}\left(\tau_1,\tau_2;\tau_3,\tau_4\right)\equiv (\beta\mathcal{J})^2\,q(\tau_{12})^{\frac{p-2}{2}}\left({q}_r(\tau_{13}){q}_r(\tau_{24})+\st\right)q(\tau_{34})^{\frac{p-2}{2}}~.
\end{equation}
This concludes the derivation of the expression \eqref{eq:S2eucl} in the main text.

\section{Simplifying the kernel eigenvalue equation in the quantum scaling regime}
\label{app:kernelDer}

In this appendix we provide a few details on the derivation of \eqref{eq:expCondMS2c}.
We begin with the defining equation for extracting the Lyapunov exponent,
\begin{align}
\label{eq:OTOCeqapp}
	f(t_1 - t_2) \, e^{\lambda_L(t_1+t_2)/2} &=\frac{1}{\beta^2}\int dt dt' \, \tilde{K}'_{\rm ret}(t_1,t_2;t,t') \, f(t-t')\, e^{\lambda_L(t+t')/2}\nonumber \\
	&=\frac{(\beta {\cal J})^2  u^{p-2}}{\beta^2}\int dt dt' \,\left[ q_{r\star}^R(t_1-t) \, q_{r\star}^R(t_{2}-t') \right] \nonumber\\
	&\qquad\qquad\qquad\qquad \times
	\left[1+\frac{p-2}{u} \,q^>_{r\star}(t-t'-i\beta/2)+\dots\right] f(t-t')e^{\lambda_L(t+t')/2}~.
\end{align}
where we used the retarded kernel \eqref{eq:kernelLor} and employed the usual expansion \eqref{eq:qExpandRR} for the conformal spin glass regime to give a perturbative expression for the Wightman function.
We can use the various spectral decompositions outlined in the main text to write this as:  
\begin{align}
&	\int_{-\infty}^{\infty} \frac{d\omega}{2\pi}\hat{f}(\omega) \, e^{i\omega(t_1-t_2)+\lambda_L(t_1+t_2)/2}\\
	&\quad =\frac{(\beta {\cal J})^2  u^{p-2}}{\beta^2}\int_{-\infty}^{\infty} \frac{d\omega}{2\pi}\frac{d\omega_1}{2\pi}\frac{d\omega_2}{2\pi}\int_{-\infty}^{t_1} dt\int_{-\infty}^{t_2} dt' \hat{q}_r(-i\omega_1-\epsilon)\hat{q}_r(-i\omega_2-\epsilon)\nonumber\\
	&\quad\quad \times \left[1+\frac{p-2}{u} \,\int_{-\infty}^{\infty} \frac{d\omega'}{2\pi} \, e^{i\omega'(t-t')} \, \frac{ \rho_r(\omega')}{2\sinh\frac{\beta\omega'}{2}}+\dots\right]
	\hat{f}(\omega)e^{-it\left(\omega_1-\omega+i\frac{\lambda_L}{2}\right)-it'\left(\omega_2+\omega+i\frac{\lambda_L}{2}\right)+i\omega_1 t_1+i\omega_2 t_2}
\end{align}
If $\lambda_L>0$ we can perform the integrals over $t$ and $t'$:
\begin{align}
& \int_{-\infty}^{\infty} \frac{d\omega}{2\pi}\hat{f}(\omega) \, e^{i\omega(t_1-t_2)}
	=\int_{-\infty}^{\infty} \frac{d\omega}{2\pi}\int_{-\infty}^{\infty}\frac{d\omega_1}{2\pi}\frac{d\omega_2}{2\pi}\frac{\hat{q}_r(-i\omega_1-\epsilon)\hat{q}_r(-i\omega_2-\epsilon)}{\hat{q}_r(0)^2} \nonumber \\
	&\qquad\quad \times \int_{-\infty}^{\infty} \frac{d\omega'}{2\pi}\left[2\pi\delta(\omega')+\frac{p-2}{u} \, \,  \, \frac{ \rho_r(\omega')}{2\sinh\frac{\beta\omega'}{2}}\right]\frac{e^{i(\omega+\omega')(t_1-t_2)}\hat{f}(\omega)}{\left[\omega'-\left(\omega_1-\omega+i\frac{\lambda_L}{2}\right)\right]\left[\omega'+\left(\omega_2+\omega+i\frac{\lambda_L}{2}\right)\right]}~, 
\end{align}
where we used \eqref{eq:conformalmagicrel} to simplify the overall coefficient, and we observed that the exponentially growing time dependence cancels on both sides.
This expression implies that we can Fourier transform with respect to the time difference $(t_1-t_2)$:
\begin{align}
 \hat{f}(\omega) & =\int_{-\infty}^{\infty}\frac{d\omega_1}{2\pi}\frac{d\omega_2}{2\pi}\frac{\hat{q}_r(-i\omega_1-\epsilon)\hat{q}_r(-i\omega_2-\epsilon)}{\hat{q}_r(0)^2} \nonumber \\
& \qquad \times \int_{-\infty}^{\infty} \frac{d\omega'}{2\pi}\left[2\pi\delta(\omega')+\frac{p-2}{u} \, \,  \, \frac{ \rho_r(\omega')}{2\sinh\frac{\beta\omega'}{2}}\right]\frac{\hat{f}(\omega-\omega')}{\left[\left(\omega-\omega_1-i\frac{\lambda_L}{2}\right)\right]\left[\left(\omega+\omega_2+i\frac{\lambda_L}{2}\right)\right]}~.
\end{align}
This expression is exact. We will now moreover assume that $\hat{q}_r(\omega)$ is an analytic function\footnote{ This is, of course, not true in the quantum scaling limit, where it would seem to develop a branch cut. However, the branch cut is only present in the approximation of the true expression, which is presumed to be smooth and analytic. It is therefore justified to evaluate the integral by the residue theorem in any case.} and evaluate the integral using the residue theorem: 
\begin{align}
	\hat{f}(\omega) &	=\frac{\hat{q}_r\left(-\frac{\lambda_L}{2}-i\omega-\epsilon\right)\hat{q}_r\left(-\frac{\lambda_L}{2}+i\omega-\epsilon\right)}{\hat{q}_r(0)^2} \times \int_{-\infty}^{\infty} \frac{d\omega'}{2\pi}\left[2\pi\delta(\omega')+\frac{p-2}{u} \, \,  \, \frac{ \rho_r(\omega')}{2\sinh\frac{\beta\omega'}{2}}\right]{\hat{f}(\omega-\omega')}~.
\end{align}
This can equivalently be written in the form \eqref{eq:expCondMS2c}, which we set out to derive.

\section{Details on the numerical implementation}
\label{app:numerics}

Here, we describe briefly how one can solve the equations of motion described above numerically. Similar methods were used by various authors before, see for example \cite{Maldacena:2016hyu}.\footnote{ We thank D.\ Stanford for explanations regarding the implementation in the SYK model.} Then, we consider the eigenvalue problem that determines the quantum Lyapunov exponents.

\subsection{Numerical solution of the equations of motion}

We distinguish between the Euclidean (Matsubara) problem, and the Lorentzian solution (which revolves around finding the spectral function):

\paragraph{Euclidean equations of motion:}
To solve the Euclidean equations of motion numerically, we proceed as follows:
\begin{enumerate}
\item Prepare an initial guess $\hat{q}_r^{[0]}(k)$, e.g., the free solution described in \S\ref{sec:freeSol}. (There are slightly more elaborate guesses, which lead to faster convergence.) Improve the guess iteratively by defining a sequence $\{q_r^{[n]}(k) \}_n$, obtained as follows.
\item Update $q_r^{[n]}(k)$ based on \eqref{eq:EOMsigma3}:
\begin{equation}
   \hat{q}_r^{[n]}(k) = (1-x) \hat{q}_r^{[n-1]}(k)  +  \frac{x}{\frac{1}{\hat{q}^{[n-1]}_r(0)} + M \big(\frac{2\pi k}{\beta}\big)^2 - J^2 \, (\hat{\Lambda}^{[n-1]}_r(k)-\hat{\Lambda}^{[n-1]}_r(0)  ) } 
\end{equation}
where $\hat{\Lambda}_r^{[n-1]}$ is built out of $\hat{q}_r^{[n-1]}$ via Fourier transform, according to \eqref{eq:LambdaDefF}.
The weighting factor $x$ is initially set to $0.5$ and divided by 2 anytime the difference (``error'') $\sum_k \big( \hat{q}_r^{[n]}(k)-\hat{q}_r^{[n-1]}(k)\big)^2$ increases (or only decreases by a very small amount) during an iteration. The algorithm stops when the error value is sufficiently small. The zero mode is determined through the spherical constraint: $\hat{q}_r^{[n]}(0) = (1-u)\beta -  \sum_{k\neq 0} \hat{q}^{[n]}_r(k)$.
\item Solve the constraint equations \eqref{eq:mvar1} (or \eqref{eq:mvarMarg} in the marginally stable case) in order to update the values of $u$ and $m$. Use the updated $u$ to also update $\hat{q}_r^{[n]}(0) = (1-u)\beta -  \sum_{k\neq 0} \hat{q}^{[n]}_r(k)$.
Increase the iteration count $n$ and proceed with step 2.
\end{enumerate} 
Using an efficient implementation of fast Fourier transform, we typically work with ${\cal O}(10^{6-7})$ discrete time intervals and the same number of Fourier coefficients.

\paragraph{Lorentzian equations of motion:} 
We can solve the real-time equations of \S\ref{eq:realTime2pt} using a similar algorithm as for the Euclidean case.
The main complication is that we need to discretize continuous expressions (with IR and UV cutoffs). The procedure revolves around iteratively improving the spectral density $\rho_r(\omega)$ from which all correlators can be computed. We proceed as follows:
\begin{enumerate}
\item Prepare an initial guess for $\rho_r^{[0]}(\omega)$ (e.g., in the marginal spin glass regime, the approximate $\rho_r^\approx(\omega)$ works very well). Note that for certain parameter ranges (such as large $\beta J$) the initial guess needs to be chosen very carefully in order to achieve convergence.\footnote{ A good way to approach a solution for problematic values of the couplings, say $M = M_\text{target}$, is as follows: start with some moderate value $M=M^{(0)}$ of the coupling, for which a simple ansatz yields a solution. Then define a sequence $\{M^{(n)}\}_{n=0,\ldots,N}$ whose final value is the target: $M^{(N)} = M_\text{target}$. Run the algorithm for each value of the coupling, each time taking the solution of the previous run as an initial guess. This way, one can approach the target value of the coupling ``adiabatically''. Except in the last iteration of the algorithm, very high precision is not required, as one only wants to find a rough guess for the next iteration.}
\item Compute the Wightman function in the time domain, $\hat{q}_r^{>[n-1]}(t)$, by Fourier transforming the combination $n_B(\omega) \rho_r^{[n-1]}(\omega)$.
\item Compute the retarded self-energy $\hat{\Lambda}_r^{R,[n-1]}(\omega) $ according to \eqref{eq:selfEnergyTime}.
\item Update the retarded two-point function by iterating \eqref{eq:EOMret}: 
\begin{equation}
 \hat{q}_r^{R,[n]}(\omega) = (1-x) \hat{q}_r^{R,[n-1]}(\omega) + \frac{x}{\frac{1}{\hat{q}_r^{R,[n-1]}(0)} + M \omega^2 - J^2 \, \left( \hat{\Lambda}_r^{R,[n-1]}(\omega)  - \hat{\Lambda}_r^{R,[n-1]}(0)  \right)  }
\end{equation}
\item Update the zero mode using the spherical constraint (e.g., \eqref{eq:mvarMargMain} with $m$ and $u$ previously computed using the Euclidean approach). A consistency check on the convergence of the algorithm is to confirm that the {\it computed} value $\hat{q}_r^{R,[n]}(0)$ should be close to the known {\it target} value $\hat{q}_r^{R}(0)$ obtained from solving the constraint equations for $u$.
\item Extract the spectral function $\rho_r^{[n]}(\omega) = 2 \, \text{Im} \, \hat{q}_r^{R,[n]}(\omega)$.
\item Fourier transform to obtain the retarded $q_r^{R,[n]}(t)$. As a measure of the error, compute the deviation $\int dt \; \big|q_r^{R,[n]}(t) - q_r^{R,[n-1]}(t) \big|^2$. If this deviation has increased since the last iteration, decrease the mixing factor $x$. If the deviation is sufficiently small, stop the algorithm.
\item Increase the iteration count $n$ and continue with step 2.
\end{enumerate}

\subsection{Numerical determination of Lyapunov exponents}

\begin{figure}
\begin{center}
\includegraphics[width=0.5\textwidth]{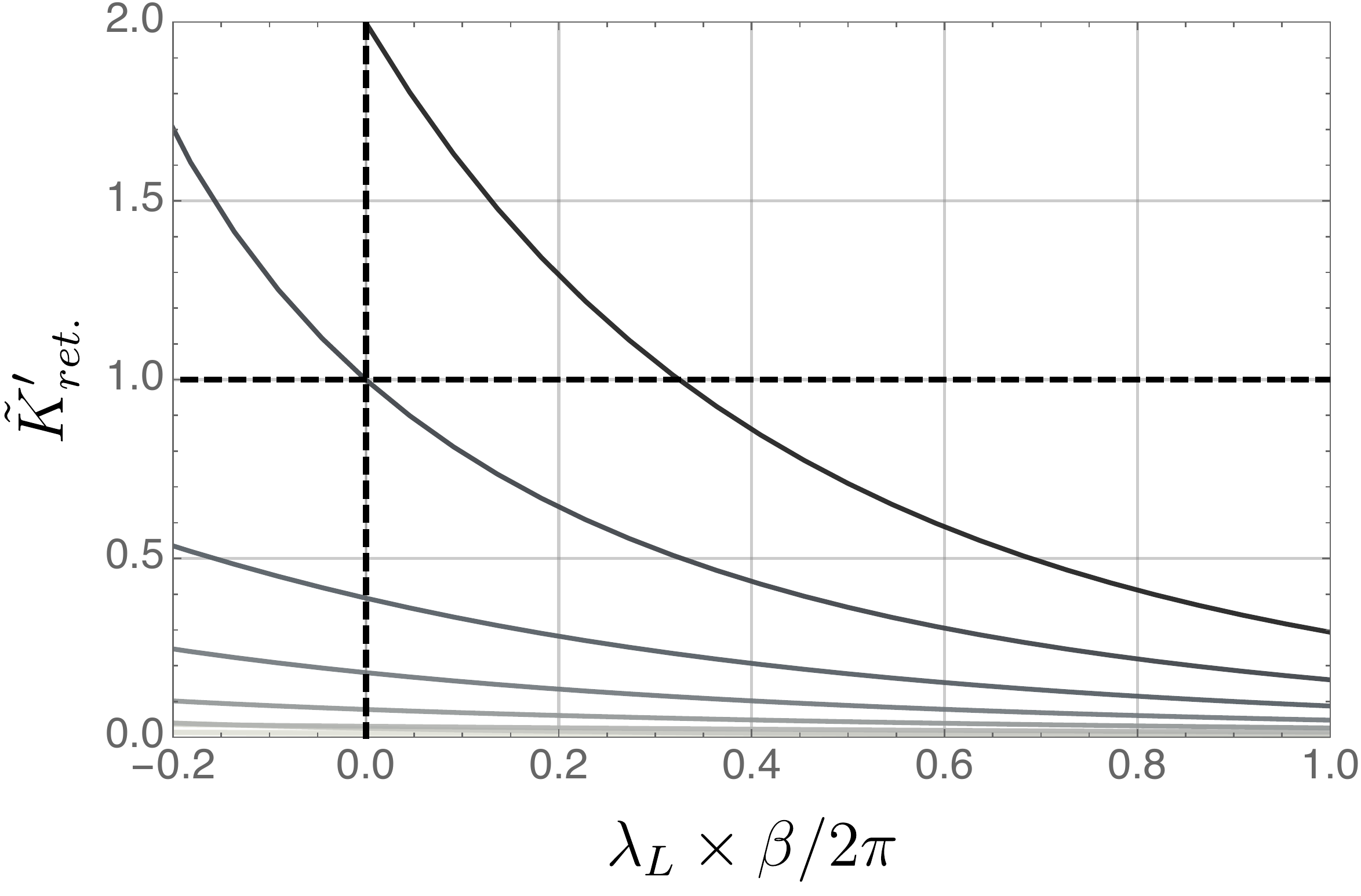}
\end{center}
\vspace{-.5cm}
\caption{{\bf Numerical determination of Lyapunov exponents.} Illustration of the numerical determination of the values of $\lambda_L$ for which the retarded kernel has a unit eigenvalue, i.e., the integral equation \eqref{eq:OTOCeq} (or, equivalently, \eqref{eq:eigValFreq}) has a solution. The exponent $\lambda_L \approx 0.32 \times \frac{2\pi}{\beta}$ is the only {\it positive} exponent for which a unit eigenvalue exists. The plot was produced for $\beta J = 2.51$ and $1/(MJ) = 2.95$.}
\label{fig:lyapunovEVs}
\end{figure}

Numerically, we find the quantum Lyapunov exponents by solving the eigenvalue equation for the retarded kernel, \eqref{eq:OTOCeq}. We discretize time and interpret \eqref{eq:OTOCeq} as a matrix equation. Once the retarded kernel $\tilde{K}'_{\rm ret}$ is found numerically, standard software can determine its eigenvalues. Exponential growth of the OTOC corresponds to the existence of a unit eigenvalue. Figure \ref{fig:lyapunovEVs} shows a sample of the eigenvalue spectrum  as a function of $\lambda_L$. Only a single {\it positive} value of $\lambda_L$ leads to a unit eigenvalue. The next largest allowed exponent is $\lambda_L = 0$ and subsequent exponents are negative (exponentially decaying).

We note that in practice it is more efficient to solve the eigenvalue problem \eqref{eq:OTOCeq} in frequency space. For $p=3$ this takes the following form (c.f., appendix \ref{app:kernelDer}):
\begin{equation}
\label{eq:eigValFreq}
 \frac{1-u \, {\cal J}^2 \, \hat{q}_r^R\left( - \frac{i\lambda_L}{2} + \bar\omega \right)  \hat{q}_r^R\left( - \frac{i\lambda_L}{2} - \bar\omega \right) }{{\cal J}^2 \, \hat{q}_r^R\left( - \frac{i\lambda_L}{2} + \bar\omega \right)  \hat{q}_r^R\left( - \frac{i\lambda_L}{2} - \bar\omega \right) } \times \hat{f}(\bar\omega) =  \int \frac{d\omega}{2\pi} \,  \frac{\rho_r(\omega)}{2\sinh \frac{\beta\omega}{2}} \;\hat{f}(\bar\omega-\omega) \,.
\end{equation}
This equation holds both in the paramagnetic phase (when $u=0$) as well as the spin glass. A similar equation can readily be derived for $p>3$. This eigenvalue problem is efficient to implement numerically for two reasons: first, the only required input is the spectral function $\rho_r(\omega)$. To make this more clear, recall that $\hat{q}_r^R(\omega)$ is given in terms of the spectral function by \eqref{eq:qrRdef}, which is automatically regulated when $\text{Im}(\omega) < 0$ as in the above expressions with $\lambda_L > 0$. And second, the spectral function to a very good approximation has compact support on a finite interval, thus effectively cutting off the integral in \eqref{eq:eigValFreq} and simplifying the discretization.

\bibliographystyle{utphys}
\bibliography{extendedrefs}{}

\end{spacing}
\end{document}